\documentclass[12pt,a4paper,openright,twoside]{report}

\usepackage{graphicx,floatflt}
\usepackage{amsmath,amsfonts,mathbbol}
\usepackage[ngerman, english]{babel}	% mehrsprachiger Textsatz
\usepackage[latin1]{inputenc}		% Eingabekodierung Parameter
\usepackage[T1]{fontenc}		% Schriftenkodierung
\graphicspath{{./images/}}              % location for figures
\usepackage{longtable}
\usepackage{fancyhdr}
\usepackage[bf,footnotesize]{caption}
\usepackage{picinpar}
\usepackage{titlesec}
\usepackage[fit]{truncate}
\usepackage[square]{natbib}

\newcommand{\dcauthorpre}{Herr Dipl.-Phys.} 
\newcommand{\dcauthorsurname}{Sternbeck} 
\newcommand{\dcauthorname}{Andr\'e} 
\newcommand{\dcauthoradd}{geboren am 5.~Juni 1976 in Berlin}

\newcommand{\dctitle}{The infrared behavior of lattice QCD Green's functions} 
\newcommand{\dcsubtitle}{A numerical study of lattice QCD in Landau gauge}  
\newcommand{\dcapprovala}{Prof.~Dr.~M.~M{\"u}ller-Preu\ss{}ker} 
\newcommand{\dcapprovalb}{Prof.~Dr.~R.~Alkofer} 
\newcommand{\dcapprovalc}{Prof.~Dr.~H.~Reinhardt} 

\newcommand{\dcdegree}{doctor rerum naturalium\\(Dr.~rer.~nat.)} 
\newcommand{\dcsubject}{Physik} 
\newcommand{\dcfaculty}{Mathematisch-Naturwissenschaftlichen Fakult\"at I}
\newcommand{\dcuniversity}{Humboldt-Universit\"at zu Berlin}
\newcommand{\dcdean}{Prof.~Thomas Buckhout, PhD}
\newcommand{\dcpresident}{Prof.~Dr.~Christoph Markschies}

\newcommand{\dcdatesubmitted}{3.~Mai 2006}
\newcommand{\dcdateexam}{18.~Juli 2006} 

\newcommand{\dckeydea}{Gluon- und Geist-Propagatoren}
\newcommand{\dckeydeb}{Gitter-QCD}
\newcommand{\dckeydec}{Landau-Eichung}
\newcommand{\dckeyded}{Confinement}

\newcommand{\dckeywordsde}{\vfill \raggedright {\textbf{Schlagw\"orter:}}\\ \dckeydea, \dckeydeb, \dckeydec, \dckeyded \\}

\newcommand{\dckeyena}{Gluon and ghost propagators}
\newcommand{\dckeyenb}{lattice QCD}
\newcommand{\dckeyenc}{Landau gauge}
\newcommand{\dckeyend}{confinement}

\newcommand{\dckeywordsen}{\vfill \raggedright {\textbf{Keywords:}}\\ 
\dckeyena, \dckeyenb, \dckeyenc, \dckeyend \\}

\usepackage{ifpdf}

\ifpdf  %......... hyperref ..................................................
 \usepackage[%
	pdftitle={\dctitle},
	pdfauthor={\dcauthorsurname\ \dcauthorname},
	pdfsubject={Dissertation},
	pdfkeywords={\dckeyena, \dckeyenb, \dckeyenc, \dckeyend},
	pdfpagemode=UseOutlines,
        colorlinks=true,
        linkcolor=black,
        filecolor=black,
        urlcolor=black,
        citecolor=black,
	pdftex=true,               % bitte nicht ändern!
	plainpages=false,          % bitte nicht ändern!
	hypertexnames=false,       % bitte nicht ändern!
	pdfpagelabels=true,        % bitte nicht ändern!
	hyperindex=true]{hyperref} % bitte nicht ändern!
\else
   
\fi

\titleformat{\chapter}
            [display]
	    {\Huge\scshape\flushright}
	    {\filleft\MakeUppercase{\chaptertitlename}\Huge\thechapter}
	    {4ex}
            {\vspace{3ex}\titlerule\vspace{0.5ex}\filleft}[\vspace{0.5ex}\titlerule]

\titlespacing{\chapter}
             {0ex}
             {-3cm}  % beforesep
             {2.0cm}  % aftersep
\titleformat{\section}
            [hang]
	    {\flushleft\Large\bfseries}
	    {\thesection\quad}
	    {0ex}
            {}

\titleformat{\subsection}
            [hang]
	    {\flushleft\large\bfseries}
	    {\thesubsection\quad}
	    {0ex}
            {}

\newcommand{\etal}{\textit{\mbox{et~al.\ }}}          % et al.
\newcommand{\ie}{\textrm{\mbox{i.e.\ }}}              % i.e.
\newcommand{\eg}{\textrm{\mbox{e.g.\ }}}              % e.g.
\newcommand{\vs}{\textrm{\mbox{vs.\ }}}               % vs.
\renewcommand{\Im}{\operatorname{\mathfrak{Im}}}      % Im
\renewcommand{\Re}{\operatorname{\mathfrak{Re}}}      % Re
\newcommand{\Tr}{\operatorname{Tr}}                   % Tr
\newcommand{\identity}{\mathbb{1}}                    % 1
\newcommand{\bc}{\texttt{bc}}                         % bc
\newcommand{\cbc}{\texttt{cbc}}                       % current bc
\newcommand{\fc}{\texttt{fc}}                         % fc
\newcommand{\Ncp}{N_{\textrm{cp}}{}}                  % N_cp
\newcommand{\Fig}[1]{Fig.~\ref{#1}}
\newcommand{\Tab}[1]{Table~\ref{#1}}
\newcommand{\Sec}[1]{Sec.~\ref{#1}}
\newcommand{\App}[1]{App.~\ref{#1}}
\newcommand{\Ch}[1]{Chapt.~\ref{#1}}
\newcommand{\Eq}[1]{Eq.~(\ref{#1})}
\newcommand{\IPR}{\mbox{\texttt{IPR}}}                     % IPR
\newcommand{\GeV}{\textrm{GeV}}
\newcommand{\fm}{\textrm{fm}}
\newcommand{\llgl}{\left\langle}
\newcommand{\rrgl}{\right\rangle}
\newcommand{\QCDSF}{\textsf{QCDSF}}
\newcommand{\MILC}{\textsf{MILC}}
\newcommand{\link}[1]{\texttt{#1}}
\newcommand{\eff}{\textrm{eff}}
\newcommand{\ndf}{\textrm{ndf}}
\newcommand{\Lagr}{\mathcal{L}}
\newcommand{\Leff}{\Lagr_{\textrm{eff}}}
\newcommand{\Linv}{\Lagr_{\textrm{inv}}}
\newcommand{\Lgf}{\Lagr_{\textrm{GF}}}
\newcommand{\Lfp}{\Lagr_{\textrm{FP}}}
\newcommand{\Sgen}{S}
\newcommand{\Sgf}{\Sgen_{\textrm{GF}}}

\newcommand{\Sinv}{\Sgen_{\textrm{inv}}}
\newcommand{\Seff}{\Sgen_{\eff}}

\newcommand{\MS}{\textsf{MS}}
\newcommand{\MSb}{\overline{\MS}}
\newcommand{\MOM}{\textsf{MOM}}

\newcommand{\myover}[2]{\genfrac{}{}{0pt}{1}{#1}{#2}}
\newcommand{\order}[1]{O\left(#1\right)}
\newcommand{\plaq}{\square}
\newcommand{\mc}[3]{\multicolumn{#1}{#2}{#3}}
\newcommand{\FP}{\mbox{FP}\ }
\newcommand{\code}[1]{\mbox{\texttt{#1}}}

\newcommand{\as}{\textrm{as}}
\newcommand{\phys}{\textrm{phys}}
\newcommand{\const}{\textrm{const.}}
\newcommand{\Vindef}{\mathcal{V}}
\newcommand{\Vin}{\Vindef_{\textrm{in}}}
\newcommand{\Vout}{\Vindef_{\textrm{out}}}
\newcommand{\Vphys}{\mathcal{V}_{\phys}}
\newcommand{\Sphys}{S_{\phys}}
\newcommand{\Vo}{\mathcal{V}_{0}}
\newcommand{\MFP}{M}
\newcommand{\ku}{\textsf{u}}

\newcommand{\name}[1]{\textsc{#1}}
\renewcommand{\vec}[1]{\boldsymbol{#1}}
\newcommand{\deltaB}{\delta_{\textrm{B}}}
\DeclareRobustCommand\orbit[2]{\kern0.1em\raise1.0ex\hbox{$\scriptstyle#1$}\kern-0.2em#2}
\newcommand{\abs}[1]{\left\vert #1\right\vert}
\newcommand{\bs}{\operatorname{\mathsf{s}}}
\newcommand{\go}{\textsl{g}_0}
\newcommand{\gr}{\textsl{g}_r}
\newcommand{\mr}{m_r}
\newcommand{\mo}{m_0}
\newcommand{\xio}{\xi_0}
\newcommand{\xir}{\xi_r}
\newcommand{\Cpp}{\textsf{C\protect\raisebox{.18ex}{++}}}

\mathchardef\ordinarycolon\mathcode`\:
 \mathcode`\:=\string"8000
 \begingroup \catcode`\:=\active
   \gdef:{\mathrel{\mathop\ordinarycolon}}
 \endgroup

\newenvironment{chapterintro}[1]{%
  \small \font\yn=cmr17 scaled \magstep4
  \setlength{\parindent}{0pt}
  \setlength{\parskip}{.66\baselineskip}
  \begin{window}[0,l,{\yn #1},{}]}{\end{window}\bigskip\bigskip}

\begin{document}
\selectlanguage{english}

%-Titelblatt------------------------------------------------------

\pagestyle{empty}
\pagenumbering{roman}
% FILE: didi-tit.tex : Version 0.99 vom 12.08.2002: cr
% AUTHOR: 
% Projekt ``Digitale Dissertationen'' 
% Humboldt-Universitaet zu Berlin
% Rechenzentrum und Universitaetsbibliothek
% Unter den Linden 6
% 10099 Berlin
% WWW: http://edoc.hu-berlin.de/e_autoren/latex/
% email: edoc@rz.hu-berlin.de
%
% Das folgende Template ist die Titelseite für digitale Dissertationen 
%
%
%
%-----Generierung der Titelseite-----bitte nicht verändern!------------------

\author{von \\*[2ex] \dcauthorpre{} \dcauthorname{} \dcauthorsurname{} \\ \dcauthoradd}

%----------
\title{\vspace{-6.5cm} 
 \begin{flushright}
   \normalsize HU--EP--06/27\\
 \end{flushright}
\vspace{1.0cm}\dctitle \\ 
\vspace{0.5cm}
\large{\dcsubtitle} \\ 
\vspace{0.5cm} {\Large{DISSERTATION}}\\ 
\vspace{0.5cm} \large{zur Erlangung des akademischen Grades \\ 
\dcdegree\\ im Fach \dcsubject \\ 
\vspace{0.5cm} eingereicht an der \\*[2ex] 
\dcfaculty \\*[1ex]  
der \dcuniversity \\}}
%-----------------
\date{\vspace{0.5cm}
%\raggedright{
\begin{center}
Pr\"asident der Humboldt-Universit\"at zu Berlin:\\
\dcpresident \vspace{-0.3cm}
%}\vspace{0.5cm}\\
\vspace{0.5cm}\\
%
%\raggedright{
Dekan der \dcfaculty:\\
\dcdean \vspace{-0.3cm}
%}
\vspace{1.0cm}\\
\end{center}
\raggedright{
Gutachter:
\begin{enumerate} 
\item{\dcapprovala} \vspace{-0.3cm}
\item{\dcapprovalb} \vspace{-0.3cm}
\item{\dcapprovalc} \vspace{-0.3cm}
\end{enumerate}} \vspace{0.5cm}
%-----------------
\raggedright{
\begin{tabular}{llr}
eingereicht am: &  &\dcdatesubmitted\\ % wenn nicht in der Prüfungsordnung, die Zeile bitte auskommentieren
Tag der m\"undlichen Pr\"ufung: & & \dcdateexam
\end{tabular}}\\ 
}
%-------------------------------------

%-------------------------------------------------------------------------  
%%% Local Variables: 
%%% mode: latex
%%% TeX-master: "../Sternbeck"
%%% End:
	% Bitte KEINE �derungen vornehmen!
\maketitle
\subsection*{Note added to the e-print version}

This Ph.D.\ thesis has been submitted to the Humboldt-University Berlin on May 3rd, 2006. It has been successfully defended on July 18th, 2006. The accepted official version is available online from the 'Dokumenten- und Publikationsserver' (\texttt{http://edoc.hu-berlin.de}) of the Humboldt-University Berlin. Part of chapter~4 has been published in Ref.~\cite{Sternbeck:2005tk} and chapter~6 is based on Ref.~\cite{Sternbeck:2005vs}. All results presented in this thesis represent the research status of May 2006.

\cleardoublepage

%-Zusammenfassung / Abstract--------------------------------------

\pagestyle{plain}
% abstract.tex
%------------------------------------------------------------------------

\addcontentsline{toc}{chapter}{\abstractname}
\begin{abstract}
 \setcounter{page}{3}
   Within the framework of lattice QCD we investigate different
   aspects of QCD in Landau gauge using Monte Carlo simulations. In
   particular, we focus on the low momentum behavior of gluon and
   ghost propagators. The gauge group is that of QCD, namely
   $SU(3)$. For our study of the lattice gluodynamic, simulations were
   performed on several lattice sizes ranging from $12^4$ to $48^4$ at
   the three values of the inverse coupling constant $\beta=5.8$, 6.0 and 6.2. 

   Different systematic effects on the gluon and ghost propagators are
   studied. We demonstrate that the ghost dressing function
   systematically depends on the choice of Gribov copies at low
   momentum, while the influence on the gluon dressing function is not
   resolvable. Also the eigenvalue distribution of the Faddeev-Popov
   operator is sensitive to Gribov copies. 

   We show that the influence of dynamical Wilson fermions on the 
   ghost propagator is negligible at the momenta available to us. For
   this we have used gauge configurations which were generated with
   two dynamical flavors of clover-improved Wilson fermions. On the
   contrary, fermions affect the gluon propagator at large and
   intermediate momenta, in particular where the gluon
   propagator exposes its characteristic enhancement compared to the
   free propagator.  

   We also analyze data for both propagators obtained on asymmetric
   lattices. By comparing these results with data obtained on
   symmetric lattices, we find that both the gluon and the ghost
   propagator suffer from systematic effects at the lowest on-axis
   momenta available on asymmetric lattices.

   We compare our data with the infrared exponents predicted in
   studies of truncated systems of Dyson-Schwinger equations for
   the gluon and ghost propagators. We cannot confirm neither the
   values for both exponents nor the relation which is proposed to
   hold between them. In any case, we demonstrate that the infrared 
   behavior of gluon and ghost propagators, as found in this thesis,
   is consistent with different criteria for confinement. In fact, we
   verify that our data of the ghost propagator and also of the
   Kugo-Ojima confinement parameter satisfy the Kugo-Ojima confinement
   criterion. The Gribov-Zwanziger horizon condition is satisfied by
   the ghost propagator. Also the gluon propagator seems to vanish in the
   zero-momentum limit. However, we cannot judge without doubt on
   the existence of an infrared vanishing gluon propagator.
   Furthermore, explicit violation of reflection positivity by the
   transverse gluon propagator is shown for the quenched and
   unquenched case of $SU(3)$ gauge theory. 

   The running coupling constant given as a renormalization-group-invariant
   combination of the gluon and ghost dressing functions does not
   expose a finite infrared fixed point. Rather the data are in favor of an
   infrared vanishing coupling constant. This behavior does not change
   if the Gribov ambiguity or unquenching effects are taken into account.
   We also report on a first nonperturbative computation of the
   $SU(3)$ ghost-gluon-vertex renormalization constant. We find that
   it deviates only weakly from being constant in the momentum
   subtraction scheme considered here. 

   We present results of an investigation of the spectral
   properties of the Faddeev-Popov operator at $\beta=5.8$ and 6.2
   using the lattice sizes $12^4$, $16^4$ and $24^4$. For this we have
   calculated the low-lying eigenvalues and eigenmodes of the
   Faddeev-Popov operator. The larger the
   volume the more eigenvalues are found accumulated close to
   zero. Using the eigenmodes for a spectral representation of the
   ghost propagator it turns out that for our smallest lattice
   only 200 eigenvalues and eigenmodes are sufficient to saturate the
   ghost propagator at lowest momentum. We associate exceptionally
   large values occurring occasionally in the Monte Carlo 
   history of the ghost propagator at larger $\beta$ to extraordinary
   contributions of the low-lying eigenmodes.  

{\dckeywordsen}
\end{abstract}

\selectlanguage{ngerman}
\addcontentsline{toc}{chapter}{Zusammenfassung (german)}
\begin{abstract}
 \setcounter{page}{5}
   Diese Arbeit untersucht im Rahmen der Gittereichtheorie
   verschiedene Aspekte der QCD in der Landau-Eichung,
   insbesondere solche, die mit den Gluon- und Geist-Propagatoren
   zusammenhängen. Die Eichgruppe ist die der QCD, $SU(3)$, und wir
   untersuchen die Propagatoren bei kleinen Impulsen. Für unsere
   Untersuchungen der reinen Gluodynamik haben wir zahlreiche 
   Monte-Carlo Simulationen auf diversen Gittergrössen
   durchgeführt. Die Gittergrössen variieren im Bereich von $12^4$
   bis $48^4$. Als inverse Kopplungskonstanten haben wir die Werte
   $\beta=5.8$, 6.0 und 6.2 gewählt.

   Wir analysieren den Einfluss unterschiedlicher systematischer
   Effekte auf das Niedrigimpulsverhalten der Gluon- und
   Geist-Propagatoren. Wir zeigen, dass der Formfaktor des
   Geist-Propagators bei kleinen Impulsen systematisch von der Wahl der
   Eichkopien (Gribov-Kopien) abhängt. Hingegen können wir einen
   solchen Einfluss auf den Gluon-Propagator nicht feststellen. Ebenfalls
   wird die Verteilung der kleinsten Eigenwerte des Faddeev-Popov-Operators
   durch die Wahl der Gribov-Kopien beeinflusst.

   Wir zeigen außerdem, dass der Einfluss dynamischer Wilson-Fermionen
   auf den Geist-Propagator für die untersuchten Impulse vernachlässigbar
   ist. Dazu haben wir Eichkonfigurationen betrachtet, die mit einer
   $N_f=2$ clover-verbesserten Wirkung erzeugt worden sind. Für den
   Gluon-Propagator können wir jedoch einen deutlichen Einfluss für große
   und mittlere Impulse feststellen, insbesondere in dem Impulsbereich,
   wo der Gluon-Propagator im Vergleich zum freien Fall seine
   charakteristische Erhöhung aufweist.

   Zusätzlich wurden beide Propagatoren auf
   asymmetrischen Gittern gemessen. Der Vergleich dieser Daten mit
   denen, die auf symmetrischen Gittern gewonnen wurden, zeigt, dass
   die Asymmetrie deutliche systematische Effekte im Bereich kleiner
   Impulse verursacht. Besonders deutlich wird das für die
   Daten, die bei Impulsen in Richtung der elongierten
   Gitterlänge gemessen worden sind.

   \sloppy Weiterhin vergleichen wir unsere Daten mit den  
   Infrarot-Exponenten, die in Studien von abgeschnittenen (truncated) Systemen
   von Dyson-Schwinger-Gleichungen für den Gluon- und
   Geist-Propagator vorhergesagt wurden. Im Rahmen unserer Messungen
   können wir weder die Werte der Exponenten noch die
   vorhergesagte Beziehung zwischen beiden bestätigen. In jedem 
   Falle können wir aber zeigen, dass das in dieser Arbeit gefundene
   Niedrigimpulsverhalten im Einklang mit verschiedenen Kriterien für
   \emph{Confinement} (Einschluss von Farbladungen) ist. Wir zeigen, dass 
   unsere Daten sowohl für den Geist-Propagator als auch für den
   Kugo-Ojima-Confinement-Parameter das Kugo-Ojima-Confinement-Kriterium
   erfüllen. Außerdem ist die Gribov-Zwanziger-Horizontbedingung für
   den Geist-Propagator erfüllt. Der Gluon-Propagator scheint
   im Grenzfall verschwindender Impulse zu Null zu streben. Dennoch 
   können wir nicht endgültig darüber urteilen, ob dies der Fall ist. 
   Wir zeigen zusätzlich, dass der transversale Gluon-Propagator
   explizit die Reflektions-Positivität verletzt. Das gilt sowohl mit
   als auch ohne den Einfluss dynamischer Fermionen.

   Wir berechnen die laufende (effektive) Kopplung, die sich als eine
   renormierungsgruppeninvariante Kombination der Gluon- und
   Geist-Formfaktoren ergibt. Unsere Ergebnisse zeigen deutlich,
   dass im Bereich kleiner Impulse die laufende Kopplung kleiner
   wird und so vermutlich kein endlicher Infrarot-Fixpunkt im
   Grenzfall Impuls Null angestrebt wird. Dieses Verhalten ist
   unabhängig vom Einfluss der Gribov-Kopien oder von der Hinzunahme
   dynamischer Fermionen. Wir präsentieren 
   außerdem eine erste nichtstörungstheoretische Berechnung der
   Renormierungskonstante 
   des $SU(3)$ Ghost-Gluon-Vertex. Wir zeigen, dass in dem
   untersuchten Renormierungsschema keine wesentliche
   Abweichung von einem konstanten Verhalten gefunden wird. 

   Wir berichten außerdem über Untersuchungen zu spektralen
   Eigenschaften des Faddeev-Popov-Operators bei $\beta=5.8$ and
   6.2. Dazu haben wir eine Reihe der kleinsten Eigenwerte und
   Eigenvektoren dieses Operators auf den Gittergrößen $12^4$,
   $16^4$ und $24^4$ berechnet. Wir sehen, dass sich umso mehr
   Eigenwerte nahe Null konzentrieren, je größer das physikalische
   Volumen ist. Anhand einer spektralen Entwicklung des Geist-Propagators
   können wir zeigen, dass für unser kleinstes Gitter ca.~200 
   Eigenwerte und Eigenvektoren genügen, um den Wert des Geist-Propagators beim
   kleinsten Impuls zu reproduzieren. 
   
   Wir zeigen ferner, dass die selten auftretenden, exzeptionell
   großen Messwerte, die für den Geist-Propagator 
   im Verlauf der Monte-Carlo Simulation bei größeren $\beta$ Werten
   gefunden werden, durch außerordentlich starke Beiträge der niedrigsten
   Eigenmoden zu den entsprechenden Fourierkomponenten hervorgerufen
   werden.
{\dckeywordsde}
\selectlanguage{english}
\end{abstract}

%-------------------------------------------------------------------------  
%%% Local Variables: 
%%% mode: latex
%%% TeX-master: "Sternbeck"
%%% End:

\cleardoublepage
\selectlanguage{english}

%-Inhaltsverzeichnis----------------------------------------------

\pagestyle{fancy}
\renewcommand{\chaptermark}[1]{\markboth{\chaptername\ \thechapter \quad #1}{}}
\renewcommand{\sectionmark}[1]{\markright{\thesection\ #1}}
\lhead[\scshape\thepage]{\truncate{0.9\textwidth}{\scshape\rightmark}}
\rhead[\truncate{0.9\textwidth}{\scshape\leftmark}]{\scshape\thepage} 
\chead{} \lfoot{} \cfoot{} \rfoot{}

\setcounter{tocdepth}{2}
\tableofcontents
\cleardoublepage

%-Hauptteil-------------------------------------------------------

\pagenumbering{arabic}
\chapter*{Introduction}
\addcontentsline{toc}{chapter}{Introduction}
\markboth{Introduction}{}
\markright{Introduction}{}

{\font\yn=cmr17 scaled \magstep4
 \setlength{\parindent}{0pt}
 \setlength{\parskip}{.66\baselineskip}
\begin{window}[0,l,{\yn A},{}]
  t present we are reasonably confident that the physics of strong
  interaction, \ie the rich field of \emph{hadron} physics, is
  completely described by a quantized nonabelian gauge  
  field theory which is based on the gauge group of $SU(3)$ color
  symmetry. This theory is called \emph{Quantum Chromodynamics} (QCD).
  Its fundamental constituents are \emph{quarks} and \emph{gluons}. Quarks
  are spin 1/2 fermion fields carrying fractional electric charge and the
  gluons are nonabelian spin 1 gauge fields which interact with the
  quarks as well as among themselves. Due to its nonabelian nature the
  renormalization group tells us that QCD is asymptotically free at
  large Euclidean momentum. In this regime perturbative QCD is
  relevant and theoretical predictions have been successfully
  confronted with experiments. The experimental successes of QCD and
  the partial progress towards a full understanding of the theory form
  the basis for our present belief that QCD is the right theory
  describing all strong interaction physics.
\end{window}}
\medskip

Beyond perturbation theory, however, QCD is still not completely
understood, even though --- as far as we know --- it is not
in conflict with any existing phenomenology of the strong
interaction. Note that in contrast to QED the elementary fields in
QCD, the quarks and gluons, do not describe existing particles and
thus a particle interpretation in QCD has to be completely divorced from 
its elementary degrees of freedom. According to QCD all strongly interacting
particles, the \emph{hadrons}, are colorless bound states of quarks.
This phenomenon is called \emph{confinement}, but the mechanism which 
confines quarks and gluons has to be established yet from first
principles. Moreover, due to the complexity of QCD, a full description
of hadronic states and processes directly in terms of QCD presents an exciting
challenge since many years. 

Many hadronic features have been investigated in the
framework of phenomenological models (see \eg 
\cite{Vogl:1991qt,Klevansky:1992qe,Ebert:1994mf}) which mimic the
essential properties of QCD, namely asymptotic freedom at short distance
and confinement at large distances. This approach represents a rather
practical point of view and is sufficient if one is just interested in
the effective theory of hadrons at low energies. But if QCD is
\emph{the} theory of strong interactions a coherent description
directly based on the dynamics of confined quarks and gluons should be
possible. 

For such a description a complete picture for all propagators and
vertex functions of QCD should be available. These Green's functions 
may then serve as input into bound state calculations based on the
\emph{Bethe-Salpeter} equations for mesons or the \emph{Faddeev} equations for
baryons. But also from a purely theoretical point 
of view a consistent picture of all QCD Green's functions is
interesting. In particular, their infrared momentum behavior provides
insight into the mechanism of quark and gluon confinement
\cite{Alkofer:2000wg}. To give just one example: The realization of
the \emph{Kugo-Ojima confinement scenario} 
\cite{Kugo:1979gm,Kugo:1995km} in QCD in covariant gauges 
is encoded in the infrared behavior of the ghost 2-point
function. Therefore, the investigation of QCD Green's
function at low momentum is important for a coherent description of
hadronic states and processes and also for an understanding of
confinement. 

%{The infrared behavior of QCD Green Functions is interesting from a
%purely theoretical point of view as well as with respect to
%phenomenological applications in hadron physics. On the one hand, it
%provides insight into the confinement mechanisms of quarks and
%gluons. On the other hand, it serves as input into those equations
%which determine hadrons as bound states of colored constituents, the
%Bethe-Salpeter equations for mesons and the Faddeev equation for
%baryons.}

%Lattice could also provide input for phenomenological models

%. Anyway a full description of
%hadrons and their processes in terms of the dynamics of confined
%quarks should be possible if hadronic systems are describable within QCD.

The infrared momentum region corresponds to strong coupling rather
than weak coupling and hence perturbation theory is of no avail in
studying QCD at low 
momentum. Genuinely nonperturbative approaches have to be used to
explore QCD in this area. The Euclidean space, discretized, lattice
gauge theory provides one possibility to study nonperturbative aspects
of QCD by using Monte Carlo (MC) simulations. Another approach is
given by solving truncated systems of the Dyson-Schwinger equations
(DSEs) of QCD. The DSEs are infinite towers of coupled nonlinear
integral equations relating different Green's functions of QCD to each
other. They are directly derived from a generating functional whose
existence beyond perturbation theory still has to be assumed.  In any
case, studying DSEs involves the introduction of a gauge condition
which is not necessary in the standard lattice approach to QCD. 

DSE studies have been performed in recent years with growing intensity
(see \cite{Roberts:1994dr,Roberts:2000aa,Alkofer:2000wg,Maris:2003vk}
for an overview). In particular, for the case of Landau gauge it has
been shown \cite{vonSmekal:1997is,vonSmekal:1998yu} that contributions
of ghost fields are crucial for a consistent description of the
infrared behavior of Landau gauge gluodynamics. In former studies
\cite{Mandelstam:1979xd,Atkinson:1981er,Atkinson:1981ah,Brown:1988bn},
ghost fields have always been neglected. 
 
Different truncations have been employed since then to study
the infrared behavior of gluon, ghost and quark
propagators and the corresponding vertex functions. Truncations are
essential to manage the infinite towers of DSEs. The solutions
presented first in \cite{vonSmekal:1997is,vonSmekal:1998yu} and later in
\cite{Atkinson:1998zc,Atkinson:1997tu,Bloch:2001wz,Bloch:2002eq} and
\cite{Fischer:2002eq, Fischer:2002hn,Fischer:2003zc} all favor the picture of
an infrared diverging ghost propagator being intimately connected with
an infrared vanishing gluon propagator. In fact, both propagators are
proposed to follow power laws at low momentum with intertwined
infrared exponents \cite{Lerche:2002ep,Zwanziger:2001kw}. Such an
infrared behavior is in 
agreement with the \emph{Gribov-Zwanziger horizon condition}
\cite{Gribov:1977wm,Zwanziger:1993dh,Zwanziger:2001kw,Zwanziger:2003cf}
as well as with the \emph{Kugo-Ojima confinement criterion}
\cite{Kugo:1979gm,Kugo:1995km}. Note that their satisfaction is crucial
for the realization of confinement in QCD in Landau gauge. Unquenching
effects on the infrared  
behavior are found to be small \cite{Fischer:2003rp}.  Moreover,
dynamical chiral symmetry breaking and gluon confinement have been
confirmed from solutions of truncated DSEs
\cite{Alkofer:2000wg,Fischer:2003rp,Alkofer:2003jj}.  

Most of these DSE studies are done in Landau gauge. In this gauge,
the ghost-gluon vertex was shown to not suffer from ultraviolet
divergences at any order in perturbation theory
\cite{Taylor:1971ff,Marciano:1977su}. Assuming this to hold beyond
perturbation theory, it allows for a definition
of a nonperturbative running coupling constant that
is solely given in terms of the gluon and ghost propagators and has a
finite infrared fixed point, provided the mentioned infrared power laws hold
\cite{vonSmekal:1997is,vonSmekal:1998yu}. In a recent DSE study
\cite{Alkofer:2004it} of vertex functions, the infrared fixed point
has been confirmed, too. It has also been shown  
that this coupling constant enters directly the kernels of the DSEs for the
gluon, ghost and quark propagators \cite{Bloch:2001wz,Bloch:2002eq}. 

Even though Monte Carlo simulations of lattice QCD provide an alternative
possibility to study QCD at a nonperturbative level, at present, they
cannot compete with the DSE approach concerning the accessible region
of low momenta. However, lattice QCD is a \emph{first principle}
approach to QCD that does not require us to ``simplify'' the
theory. Unlike truncations of DSEs, the approximations involved in
lattice QCD are systematically removable. This possibility of
controlling the systematic errors makes this approach invaluable
\cite{Bowman:2005zi}.  Therefore, lattice simulations may provide an
independent check whether the results obtained in the DSE approach are
realized in lattice QCD, at least in the region of momenta available
at present. Furthermore, lattice QCD enables us to study different 
models for confinement (see \eg \cite{Greensite:2003bk}) by mutilating the
theory such that confinement is explicitly lost. For example, removing
vortices changes the infrared behavior of the lattice ghost propagator
in Landau gauge such that it does not satisfy anymore 
aforementioned criteria for confinement \cite{Gattnar:2004bf}.

In recent years, different groups have investigated different aspects
of lattice Landau gauge QCD. Some have studied the gauge
group $SU(2)$, others $SU(3)$. In particular, the \name{Adelaide Group}
has provided an impressive account on numerical data for
the  $SU(3)$ gluon
\cite{Leinweber:1998im,Leinweber:1998uu,Bonnet:2000kw,Bonnet:2001uh} 
and quark propagators 
\cite{Skullerud:2000un,Skullerud:2001aw,Bonnet:2002ih,Bowman:2002bm,
Zhang:2003fa,Zhang:2004gv,Bowman:2005vx,Parappilly:2005ei} 
and for the quark-gluon vertex 
\cite{Skullerud:2002ge,Skullerud:2003qu}.  
Their data are based on quenched and unquenched $SU(3)$ gauge
configurations where the latter were generated with the AsqTad quark
action by the $\MILC$ collaboration. 

\newpage

For the $SU(3)$ ghost propagator there were not so many data
available until a few years ago, even though this propagator is
expected to be related to the gluon propagator as mentioned above. The
first lattice study of the $SU(2)$ and $SU(3)$ ghost propagators in
Landau gauge was given in \cite{Suman:1995zg} and there were several
studies in Landau gauge which have confirmed the anticipated behavior
for the case of $SU(2)$
\cite{Bloch:2003sk,Langfeld:2001cz,Gattnar:2004bf,Bloch:2002we}. 
Similar investigations for the $SU(3)$ case at even lower  
momenta were not available at that time. 

During the last three years, we have tried to bridge this gap by
investigating the $SU(3)$ ghost and gluon propagators (and related
objects) on quenched and unquenched $SU(3)$ gauge configurations. Our set of
unquenched configurations were generated with clover-improved Wilson
fermions by the $\QCDSF$ collaboration. We have found that for
momenta lower than used in the $SU(2)$ studies (see above) qualitative
differences to the anticipated infrared behavior of ghost and gluon
propagators and of the running coupling constant appear. 

At the same time other groups have performed similar
investigations focussing on different interesting aspects. See, for example,
\cite{Furui:2003jr,Furui:2004cx,Furui:2005bu,Furui:2005mp,Furui:2006rx}
for investigations of the gluon, ghost and quark propagators and of the
running coupling constant using quenched and unquenched configurations 
(provided by the $\MILC$ collaboration). Studies of the $SU(3)$ gluon
propagator at very low momentum can be found in
\cite{Oliveira:2004gy,Silva:2005hb,Oliveira:2005hg,Silva:2005hd}.
There the infrared exponent has been determined using
lattices much elongated in time direction. A study of the $SU(3)$
ghost propagator at large momentum can be found in \cite{Boucaud:2005}.

Furthermore, recent DSE studies
\cite{Fischer:2002eq,Fischer:2002hn,Fischer:2005ui,Fischer:2005nf}
show that the infrared
behavior of the gluon and ghost dressing functions and of the running
coupling constant is changed on a torus. In particular, the running
coupling decreases at low momenta. These findings agree with lattice data as
shown in this thesis, but they contradict results obtained from DSE
studies in the continuum. It is still unknown what is the reason for this
disagreement. Note that a solution to this problem has been proposed in
\cite{Boucaud:2005ce,Boucaud:2006if}.

\medskip

It is the intention of this thesis to give a summary of our
results obtained within the last three years. Some were already
published, others are just finished and being written up.

We have structured this thesis as follows: In the first chapter we 
introduce the path integral formulation of QCD and discuss the
special problems related to the necessity of gauge fixing the
action. A brief introduction to the BRST formalism is given and the 
renormalization program is recalled. In \Ch{ch:infrared} we discuss
some aspects of nonperturbative QCD and introduce criteria for
confinement which are available for QCD in Landau gauge. The
lattice formulation of QCD in this gauge and a
definition of all observables analyzed in this thesis is given in
\Ch{chap:latticeQCD}. We present our results for the ghost and gluon
propagators in \Ch{ch:prop_results}. Different systematic effects
are analyzed and their influence on the infrared behavior of the
gluon and ghost propagators is discussed. After this, we show that our
data for both propagators satisfy necessary criteria for
confinement. In \Ch{ch:spec_FP_operator} spectral 
properties of the FP operator are analyzed. Finally, we draw
our conclusions and give an outlook. The appendix contains some notes
on algorithms and 
performance. In particular, we compare two popular gauge-fixing
algorithms and show that the final ranking of gauge functional values
is already visible at an intermediate iteration state. We also demonstrate
how the inversion of the FP operator can be accelerated considerably.

%===============================================================================
%%% Local Variables: 
%%% mode: latex
%%% TeX-master: "../Sternbeck"
%%% End:

%-- Kapitel ---------------------------------------------------------

\titleformat{\chapter}
            [display]
	    {\Huge\scshape\flushright}
	    {\vspace{4ex}\titlerule\vspace{0.5ex}\filleft\LARGE
             \scshape{\chaptertitlename}% 
             \quad\thechapter\filleft}
	    {0.5ex}
            {\titlerule\vspace{3ex}\filleft}[\vspace{-1ex}]

%----------------------------------------------------------------------------
\chapter{The various colors of QCD}
\label{ch:intro}

\begin{chapterintro}{T}
 his chapter briefly reviews the Euclidean formulation of QCD in the
 continuum, mainly in order to fix notations used subsequently.
 Starting with the classical Lagrangian density and its  
 quantization, the problems encountered by fixing to
 covariant gauges are discussed. A short summary of the BRST formalism and
 renormalization is given.
\end{chapterintro}

%----------------------------------------------------------------------------
\section{Quantization of QCD}

The success of quark-models in describing hadrons as bound states of
quarks, but also of quark-parton models in deep-inelastic lepton-hadron
scattering, to name but a few, suggests that the strong interaction
should be described by a theory where the color symmetry of each
quark flavor is a gauge symmetry and which is also asymptotically free at high
energy-momentum transfers or short distances. Since asymptotic freedom is
inherent in non-abelian gauge theories, and experiments like, for
instances, the pion-decay $\pi^0\rightarrow 2\gamma$ suggest the
gauge group to be $SU(3)$, we are reasonably confident at present
that the strong interaction is completely described by a quantized
non-Abelian gauge field theory based on the $SU(3)$ gauge group.  

%--------------------------------------------------------------------------
\subsection{The classical QCD Lagrangian}
\label{sec:class_lagr}

A common way to setup a quantum field theory is to define first a
Lagrangian density 
\begin{equation}
  \label{eq:genLagr}
  \Lagr\equiv\Lagr[\Phi_1(x),\ldots,\Phi_n(x);\partial_{\mu}\Phi_1(x),\ldots,\partial_{\mu}\Phi_n(x)]
\end{equation}
that is a functional of several fields $\Phi_1(x),\ldots,\Phi_n(x)$
and their derivatives necessary to host (in a consistent
way) all the features and 
symmetries observed in experiments. This Lagrangian density or its
space-time integral, the action 
\begin{equation}
  \label{eq:genS}
  \Sgen[\Phi]\equiv\int
d^4x\,\Lagr\;,
\end{equation}
is then used subsequently for a quantization of the theory choosing one of the
well-known quantization methods, namely the
\emph{Canonical operator formalism}, the \emph{Stochastic formalism} or the
\emph{Functional-integral formalism} \cite{Muta:1998vi}. 

The most general form of the QCD Lagrangian density 
that not only accommodates all those mentioned properties of QCD, but also
is renormalizable in any order of perturbation theory can be 
written (in Euclidean space) as \footnote{Here and in the following, a
  sum over repeated indices is understood if not otherwise
  stated. We will see later that there is always the freedom to have
  multiplicative renormalization constants or to add BRST-exact terms,
  like \eg gauge-fixing and ghost terms in covariant gauges to this
  density.} 
\begin{equation}
  \label{eq:Linv}
  \Linv = \frac{1}{4} F^a_{\mu\nu}F^{\mu\nu,a} -
  \bar{\psi} (\gamma_{\mu}D_{\mu}-\mo) \psi\,. 
\end{equation}
Conceived in general terms, this Lagrangian density describes the
interaction of the quark and antiquark fields, $\psi$ and
$\bar{\psi}$, with the 
self-interacting gluon or gauge fields $A_{\mu}=A^a_{\mu}(x)T^a$. The latter are
hidden in both the definition of the field-strength tensor 
\begin{equation}
  \label{eq:fieldstrength}
  F^a_{\mu\nu} = \partial_{\mu} A^a_{\nu} - \partial_{\nu}
  A^a_{\mu} - \go f^{abc} A^b_{\mu} A^c_{\nu}
\end{equation}
(here in the adjoint representation, \ie $a=1,\ldots,N_c^2-1$) and the
covariant derivative 
\begin{equation}
  \label{eq:covD_fund}
  D^{kl}_{\mu} = \partial_{\mu}\delta^{kl} + i \go A^a_{\mu}(T^a)^{kl}
\end{equation}
given in the fundamental representation (\ie $k,l=1,\ldots,N_c$) of the Lie group
$SU(N_c=3)$ with the eight hermitian generators $T^a$. Beside being
hermitian, these generators satisfy $\Tr\big(T^aT^b\big)=\delta^{ab}/2$ and
$[T^a,T^b]=if^{abc}T^c$ where $f^{abc}$ are the structure constants of
the Lie algebra $\mathfrak{su}(3)$. The bare coupling constant is
labeled $\go$. 
For the sake of completeness, we also remind on the covariant derivative
in the adjoint representation:
 \begin{equation}
  \label{eq:covD_adj}
  D^{ab}_{\mu} = \partial_{\mu}\delta^{ab} + \go f^{abc} A^c_{\mu}.
\end{equation}
The quark fields
\begin{displaymath}
  \psi  \equiv \psi^{\alpha,l}_f(x)
\end{displaymath}
and antiquark fields $\bar{\psi}\equiv\psi^{\dagger}\gamma_0$, of flavor
\mbox{$f=1,\ldots,N_f$} are anti-commuting spinor fields that transform
under the fundamental representation of the $SU(N_c=3)$ color group, \ie
the color index runs over $l=1,\ldots,N_c$. The Dirac matrices
$\gamma_{\mu}$ act upon the spinor indices $\alpha = 1,\ldots,4$ of
the quark fields. The bare mass $\mo$ is a free parameter (for each
flavor) of the theory as is $\go$.

By definition, the Lagrangian density in \Eq{eq:Linv} is
invariant under local $SU(3)$ gauge transformations
\begin{subequations}
 \label{eq:gauge_fin}
 \begin{eqnarray}
 \label{eq:gauge_fin_A}
  A_{\mu} &\rightarrow& \orbit{\omega}{A_{\mu}} = g_{\omega} A_{\mu}
  g^{\dagger}_{\omega} 
  + \frac{i}{\go}  \,g_{\omega} \partial_{\mu} g^{\dagger}_{\omega},\\
 \label{eq:gauge_fin_psi}
  \psi &\rightarrow&\; \orbit{\omega}{\psi} = g_{\omega}\psi,\\
 \label{eq:gauge_fin_barpsi}
  \bar{\psi} &\rightarrow&\; \orbit{\omega}{\bar{\psi}} =
  \bar{\psi}g^{\dagger}_{\omega} 
\end{eqnarray}
\end{subequations}
of gluon, quark and antiquark fields.
Here $g_{\omega}$ is an element of the group $SU(3)$. It can be
parameterized by a set of real-valued functions $\omega^a(x)$, \ie
\begin{equation}
  \label{eq:def_gaugetrafo}
  g_{\omega}\equiv g_{\omega}(x) = e^{-i\go\cdot\omega^a(x) T^a}\quad\in SU(3).
\end{equation}

In subsequent discussions we will frequently refer to the
\emph{infinitesimal} form of those local transformations. What is
usually meant by that notion is the following. If the field
$\Phi_k=\{\bar{\psi},\psi,A\}$ transforms under a local gauge
transformation $\Phi_k\rightarrow \orbit{\omega}{\Phi}_k$ as given in
\Eq{eq:gauge_fin} then the corresponding infinitesimal transformation is
defined by \cite{Collins:1984np}:  
\begin{displaymath}
  \delta\Phi_k(x) \equiv \left. \omega^{b}\frac{\partial}{\partial\omega_b}
   \orbit{\omega}{\Phi}_k\right|_{\omega=0} 
  =: \omega^b\delta_{b}\Phi_k(x)\;.
\end{displaymath}
Using \Eq{eq:def_gaugetrafo} the infinitesimal local gauge
transformations of the gluon and fermion fields take the form: 
\begin{subequations}
  \label{eq:gauge_inf}
  \begin{eqnarray}
 \label{eq:gauge_inf_A}
  \delta_{\omega}A^a_{\mu} &=& \partial_{\mu}\omega^a+\go f^{abc}
  \omega^bA^c_{\mu}\equiv D^{ab}_{\mu}\omega^b\\
 \label{eq:gauge_inf_psi}
  \delta_{\omega}\psi &=& -i\go \omega^aT^a\psi  \\
 \label{eq:gauge_inf_barpsi}
  \delta_{\omega}\bar{\psi} &=& + i\go \omega^a\bar{\psi} T^a 
\end{eqnarray}
\end{subequations}

The invariance of the Lagrangian density $\Linv$ under local
gauge transformations, causes some extra difficulties for the
quantization using either the functional-integral or the canonical
formalism. For example, the definition of a functional-integral over
gauge fields in the continuum requires a gauge condition to be
introduced. As a consequence additional terms are added to $\Linv$. In the 
resulting Lagrangian density, $\Leff$, the gauge invariance is
explicitly lost, but its particular form --- it is \emph{BRST
  invariant} (see below) --- guarantees that expectation values of
gauge-invariant observables are actually independent of the gauge
condition used. 

We note in passing that on the lattice such a
gauge condition is superfluous, as long as gauge-invariant observables
are studied. Therefore, the gauge-invariant action
\begin{equation}
  \label{eq:Sinv}
  \Sinv = \int d^4x\; \Linv 
\end{equation}
is sufficient for a lattice discretization. See
\Ch{chap:latticeQCD} for a particular lattice discretization as used
in this study. 

%-----------------------------------------------------------------------
\subsection{Functional-integral quantization of QCD}

So far the theory is a classical field theory. To quantize it one chooses
one of the well-known quantization methods, namely the \emph{Canonical
  operator formalism}, the \emph{Stochastic formalism} or the
\emph{Functional-integral formalism}. Indeed, all three methods
should lead to the same physical predictions. However, the choice
depends on the feasibility of the method for a particular topic. 

A quantum field theory is completely characterized by the infinite
hierarchy of $n$-point functions or Green's functions. These are
correlation functions of the fields $\Phi_i(x)$ and the three mentioned
formalisms differ in how Green's functions are calculated.
For example, in the canonical approach the fields
are regarded as operators for which canonical commutation relations hold. The
Green's functions are calculated as vacuum 
expectation values of time ordered products of those operators. 
The stochastic formalism introduced by
\name{Parisi} and \name{Wu} \cite{Parisi:1980ys} starts from the
classical equation of motion. The fields are regarded as stochastic
variables. See \cite{Damgaard:1987rr} for a comprehensive account on
that subject. 

The Functional-integral approach was introduced by Feynman
\cite{Feynman:1948ur}. There the fields are taken to be \mbox{c-numbers} and
the Lagrangian density takes its classical form. The Green's functions
are given by functional integrations of products of fields 
over all of their (weighted) possible functional forms.
The present study focuses on the lattice regularization of QCD in
Euclidean space. Since this approach relies on the functional
integral formalism we demonstrate briefly the general
concept.\footnote{Note that 
  due to the work of \name{Kugo} and \name{Ojima} \cite{Kugo:1979gm} a
  consistent quantization of non-abelian gauge fields is also
  available in the covariant canonical operator formalism
  \cite{Muta:1998vi}. Some of their results, namely the Kugo-Ojima
  confinement scenario will also be investigated in this study. For
  the covariant canonical operator formalism see also the book by
  \name{Nakanishi} and \name{Ojima} \cite{Nakanishi:1990qm}.} 

%------------------------------------------------------------------------
\subsubsection{Functional-integral formalism: Illustration of the
  general concept}

The functional-integral formalism introduces generating functionals
$Z$, $W$, and $\Gamma$ which generate, respectively, the \emph{full},
\emph{connected} and \emph{one-particle irreducible} (1PI) Green's functions. 
To get acquainted with the general concept let us assume that for the generic
 Lagrangian density (\Eq{eq:genLagr}) of $n$ different fields $\Phi_i$
the generating functional
\begin{equation} 
 \label{eq:genZ}
  Z[J] =\int [\mathcal{D}\Phi] \exp\left\{-\int d^4x\, \big(\Lagr[\Phi(x)] +
  J^a_i(x)\Phi^a_i(x)\big)\right\} 
\end{equation}
for the full Green's functions can be defined, \ie there exist a
well-defined measure $[\mathcal{D}\Phi]$. Then a full Green's function
$\langle\Phi^{a_1}_1(x_1)\cdots\Phi^{a_n}_n(x_n)\rangle$ is
given by functional derivatives with respect to the
sources $J^a_i(x)$, \ie
\begin{displaymath}
  \langle\Phi^{a_1}_1(x_1)\cdots\Phi^{a_n}_n(x_n)\rangle =\left.
  \frac{\delta^n Z[J]}{J^{a_1}_1(x_1)\cdots
    J^{a_n}_n(x_n)}\right|_{J^{a_1}_1, \ldots, J^{a_n}_n = 0}\;.
\end{displaymath}
Together with \Eq{eq:genZ} and the generic action $\Sgen[\Phi]$
(\Eq{eq:genS}) this yields
\begin{displaymath}
  \langle\Phi^{a_1}_1(x_1)\cdots\Phi^{a_n}_n(x_n)\rangle
  = \frac{1}{Z[0]}\int
  [\mathcal{D}\Phi]\;\Phi^{a_1}_1(x_1)\cdots\Phi^{a_n}_n(x_n) 
  \,e^{-\Sgen[\Phi]}\;.
\end{displaymath}

%---------------------------------------------------------------------------
\subsubsection{Gauge orbits and gauge conditions}

For a quantization of QCD within the functional formalism it is 
necessary to define the generating functional $Z[J]$ that generates all the
Green's functions of the theory. In particular, the definition of a
path-integral over gluon fields needs special care, because 
it is ill-defined if done naively.

In fact, choosing a particular gauge field $\orbit{0}{A}_{\mu}(x)$ there
are infinitely many others $\orbit{\omega}{A}_{\mu}$ which are related
to this by local gauge transformations as defined in
\Eq{eq:gauge_fin_A}. The set of all those is usually referred to as the
\emph{gauge orbit} of $\orbit{0}{A}_{\mu}$, because each element
$\orbit{\omega}{A}_{\mu}$ in the orbit is obtained by acting upon
$\orbit{0}{A}_{\mu}$ with a local gauge transformation $g_{\omega}(x)$.
The Lagrangian $\Linv$ is invariant 
under such a transformation by definition and so all (infinite) elements
of one particular orbit give rise to the same value of $\Linv$. This spoils
a naive integration over all gluon fields, because an integral of kind
\begin{displaymath}
  \int [DA] \;e^{-\Sinv}= \int [D\orbit{0}{A}]\; e^{-\Sinv} \int [D\omega]
\end{displaymath}
is divergent. Here $\Sinv$ denotes the action in \Eq{eq:Sinv}, but for 
simplicity we have dropped fermionic fields. The
integration over the gluon fields must be defined such that it
restricts to gauge-inequivalent configurations, \ie they must belong to
different gauge orbits.

This can be achieved by choosing a gauge condition  
\begin{equation}
  \label{eq:gauge_condition}
  \mathcal{F}[\orbit{\omega}{A};x]\Big|_{\omega=\bar{\omega}}=0
\end{equation}
at each point $x$ in space-time. If this condition is satisfied for
only one representative on each gauge orbit, \ie the solution
$\bar{\omega}$ is unique, then it is called an \emph{ideal
  gauge condition} \cite{Williams:2003du}. The set of those
representatives is called the 
\emph{fundamental modular region} $\Lambda$. It is a hypersurface
defined by \Eq{eq:gauge_condition} in the space of all gauge fields.
If we can define an integration over this region, the integral
\begin{displaymath}
  \int_{\Lambda} [DA] \;e^{-\Sinv}
\end{displaymath}
does not suffer from local gauge invariance, as does a naive
integration. 

If the gauge condition (\Eq{eq:gauge_condition}) is
ambiguous, it is termed \emph{non-ideal} and an integration beyond
perturbation theory may become ill-defined. The different solutions to a
non-ideal gauge condition belong to the same orbit and are
called \emph{Gribov copies} in honor of its discoverer
\cite{Gribov:1977wm}. In the following,  we assume the gauge condition
to be ideal. Note that even popular non-ideal
gauge conditions, like the Coulomb or Landau gauge, are sufficient
within the framework of perturbation theory. This is because in
perturbation theory only small fluctuations of $A^a_{\mu}(x)$ around
zero are necessary and with respect to infinitesimal gauge
transformations  
\begin{eqnarray*}
  g_{\omega}(x;\tau) = \identity + i\tau\omega^a(x)T^a +
  O(\tau^2)\qquad(\tau\ll 1).
\end{eqnarray*}
even non-ideal gauge conditions are unique. The problems of Gribov
copies and nonperturbative quantization will be discussed in
\Sec{sec:nonperturbative}.\footnote{Gauge-fixing is also necessary in 
  canonical quantization, but not for stochastic quantization. Therefore the
latter has the advantage to do not suffer from Gribov copies.
However, it is more complicated than the other two methods. For the
standard lattice approach to QCD gauge-fixing is also not
necessary. See also \Ch{ch:infrared} and \ref{chap:latticeQCD}.}

%----------------------------------------------------------------------------
\subsection{The Faddeev-Popov method}
\label{sec:FPmethod}

From ordinary calculus of discrete $n$-dimensional vectors it is known that
\begin{equation}
\label{eq:identity_nvec} 
  1 = \int \left[\prod^n_i df_i\right] \delta^n(\vec{f}) =  \int\left[\prod^n_i
  d\omega_i\right] \delta^n\left(\vec{f}(\vec{\omega})\right) \left| 
    \det\frac{\partial f_i}{\partial\omega_j} \right|
\end{equation}
where the determinant in the last expression is the Jacobian
determinant that arise due to the substitution rule for integrals with
multiple variables.
If $f$ is invertible near $\omega$ then its Jacobian determinant
at $\omega$ is non-zero (inverse function theorem).

If we assume in the following that \Eq{eq:gauge_condition}
represents an ideal gauge condition, the identity
\Eq{eq:identity_nvec} may be generalized to an identity 
for functional integrals
\begin{equation}
  \label{eq:FP_one}
    \identity = \int [D\omega]\;
    \delta\left(\mathcal{F}[\orbit{\omega}{A}]\right)
    \Delta_{\textrm{FP}}[A]
\end{equation}
which was first proposed by \name{Faddeev} and \name{Popov}
\cite{Faddeev:1967fc}. In this relation, the Jacobian determinant%
\footnote{In general the absolute
  value of the FP determinant has to be considered. However, the
  assumption of an ideal gauge condition guarantees the determinant to
  be nonzero. So it cannot change sign which cancels anyway due to
  normalization. See also the discussion in \Sec{sec:Gribov_problem_non}.} 
\begin{equation}
 \label{eq:detFP}
 \Delta_{\textrm{FP}}[A]:=\det\MFP[A] 
\end{equation}
is known as the \emph{Faddeev-Popov} (FP) determinant
\cite{Faddeev:1967fc} of a matrix
\begin{equation}
  \label{eq:FP_matrix_cont}
   \MFP^{ab}_{xy}[A] := \left.\frac{\delta
       \mathcal{F}[\orbit{\omega}{A}^a,x]}{\delta\omega^b(y)}\right|_{\omega=0}
\end{equation}
that represents the change of $\mathcal{F}$ under local
gauge transformation at $\omega=0$. Inserting the identity in
\Eq{eq:FP_one} now in the naive integration over gluon fields we end up with
\begin{equation}
 \label{eq:int_DA_detFP_deltaFa}
  \int [DA]\; \Delta_{\textrm{FP}}[A]\; \delta(\mathcal{F}[A])\;e^{-\Sinv[A]}.
\end{equation}
This represents an integration over the fundamental modular region,
but if and only if the gauge condition \Eq{eq:gauge_condition} is unique.

%--------------------------------------------------------------------------
\subsection{An effective Lagrangian density in covariant gauge}

A popular gauge condition for practical calculations is given by the
family of \emph{covariant gauges} specified by the condition
\begin{equation}
  \label{eq:covariant_gauge_cond}
  \mathcal{F}^a[A] := \partial_{\mu}A^a_{\mu}(x) - B^a(x) = 0.
\end{equation}
Here $B^a(x)$ is an arbitrary function\footnote{In Minkowski space
  $B^a$ would transform as a Lorentz scalar.}. Since the
work of \name{Gribov} \cite{Gribov:1977wm} it is 
well-known that local gauge conditions of this type are ambiguous with
respect to finite gauge  transformations (see also
\cite{Singer:1978dk, Williams:2003du}), and so, in our notation, belongs
to the class of non-ideal gauge conditions. With respect to
infinitesimal gauge transformations, however, they are unique and may 
be treated as ideal ones. 

The family of covariant gauges turns out to be useful in perturbative
expansions in many applications. It also allows us to
represent the delta function in the functional integral
\Eq{eq:int_DA_detFP_deltaFa} as a (functional) 
integral over the fields $B$. Since these fields are arbitrary we can use a
Gaussian weight of width $\xio$ to integrate over, \ie 
\begin{displaymath}
  \int DB^a\; \exp\left\{-\frac{1}{2\xio}\int d^4x\; (B^a)^2(x)\right\}
  \delta\left(\partial_{\mu}A^a_{\mu}(x) - B^a(x)\right)
  =: e^{-\Sgf[A;\xio] }\;.
\end{displaymath}
Here $\Sgf[A;\xio]$ is defined as the (Euclidean) space-time integral
of
\begin{equation}
  \label{eq:Lgf}
  \Lgf^{\xi} = \frac{1}{2\xio} \left[ \partial_{\mu}A_{\mu}(x)\right ]^2
\end{equation}
$\Lgf^{\xi}$ is known as the \emph{gauge-fixing} term which is added 
to the invariant Lagrangian density $\Linv$. It serves as a substitute for
the delta-function that specifies the hypersurface in the functional
integral \Eq{eq:int_DA_detFP_deltaFa}. The parameter $\xio$ is
the \emph{gauge parameter} that specifies the particular gauge
condition in the family of covariant gauges. 
The special case of $\xio=0$ is known as the \emph{Landau} or
\emph{Lorentz gauge} and $\xio=1$ as the \emph{Feynman gauge}.

%---------------------------------------------------------------------------
\subsubsection{Never call a ghost stupid --- A few good ghosts can help}

\setlength{\unitlength}{1mm}
\begin{picture}(0,0)
  \put(121,8){\includegraphics[width=1cm]{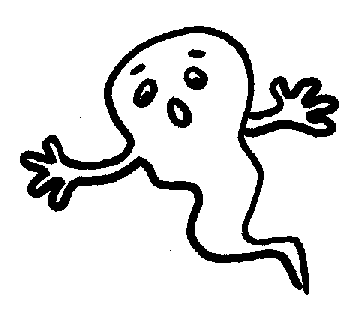}}
\end{picture}
Although the FP determinant (\Eq{eq:detFP}) is not a local function of
the gauge fields, the functional integration in
\Eq{eq:int_DA_detFP_deltaFa} can be extended such that the FP
determinant is expressed by an additional (local) term, $\Lfp$, added
to $\Linv$, too. This is done by the 
familiar device of integrating\footnote{For any finite $N$ it holds that the
  determinant of a matrix $M$ can be expressed as a functional integral over
  anti-commuting Grassmann numbers, \ie 
  \begin{displaymath}
  \det M = \left[\prod^N_i \int d\bar{c}_idc_i\right] e^{-\bar{c}_j M_{ji}c_i}.
  \end{displaymath}
} 
over ghost and anti-ghost fields $c$ and $\bar{c}$ which are
independent Grassmann valued fields. With the definition of the FP
matrix $M$ (\Eq{eq:FP_matrix_cont}) we obtain in covariant gauge
\begin{eqnarray}
\nonumber
  \MFP^{ab}_{xy}[A] &=&
  \left.\frac{\delta(\partial_{\mu}\orbit{\omega}{A}^a_{\mu}(x) -
      B^a(x))}{\delta\omega^b(y)}\right|_{\omega=0} 
  = \left.\partial^x_{\mu} \frac{\delta
    \orbit{\omega}{A}^a_{\mu}(x)}{\delta\omega^b(y)}\right|_{\omega=0}\\
  &=& - \partial^x_{\mu}D^{ab}_{x,\mu}[A]\delta^4(x-y)\;.
\end{eqnarray}
Here $D^{ab}_{x,\mu}$ denotes the covariant derivative in the adjoint
representation. Using this, the term  $\Lfp$, known as the \emph{ghost
  term}, takes the form
\begin{equation}
 \label{eq:Lfp}
  \Lfp = - (\partial_{\mu}\bar{c}^a)(\partial^{\mu}\delta^{ab} +
  \go f^{abc}A^c_{\mu})c^b\;.
\end{equation}

%------------------------------------------------------------------------
\subsubsection{Generating functional for QCD in covariant gauge}

In summary, we arrive at an effective Lagrangian density
\begin{equation}
  \label{eq:eff_Lagrangian}
  \Leff = \Linv + \Lgf + \Lfp^{\xi}
\end{equation}
where the individual terms $\Linv$, $\Lgf$ and $\Lfp^{\xi}$ are defined in
\Eq{eq:Linv}, (\ref{eq:Lgf}) and (\ref{eq:Lfp}), respectively.
$\Leff$  can be used to define the (Euclidean)
generating functional (see \eg \cite{Alkofer:2000wg})
\begin{align}
 \label{eq:Z_QCD}
\nonumber
  Z[j^a,\bar{\eta},\eta,&\sigma,\bar{\sigma}] 
  =\int [\mathcal{D}A][\mathcal{D}\psi][\mathcal{D}\bar{\psi}]
 [\mathcal{D}c][\mathcal{D}\bar{c}]\\
&  \cdot \exp\left\{-\int d^4x\; \left(\Leff^r -
        A^a_{\mu}j^a_{\mu} - \bar{\eta}\psi -
        \bar{\psi}\eta - \bar{\sigma}c -
        \bar{c}\sigma\right)\right\}.
\end{align}
Here $\bar{\eta}$, $\eta$, $\sigma$ and $\bar{\sigma}$ refer to
the Grassmannian sources, respectively, for the quark,
anti-quark, ghost and anti-ghost fields as introduced above.
In covariant perturbation theory this generating functional is used
for the calculation of Euclidean Green's functions as power series
expansions of the interaction terms in $\Leff$. Actually, for this 
the \emph{renormalized} effective Lagrangian density $\Leff^r$ given in
\Eq{eq:renLeff} must be used instead. Otherwise 
perturbative expansions beyond tree level would be rendered
meaningless by divergent mathematical expressions. We have
indicate this already in \Eq{eq:Z_QCD} by giving the suffix $r$ to
$\Leff$. The explicit form of $\Leff^r$ and the renormalization
program is discussed in
\Sec{sec:regularization_and_renormalization}. 

Note also that the existence of the generating functional beyond
perturbation theory rather has the status 
of being postulated than confirmed. So far only the continuum
limit of a lattice formulation of quantum field theory provides a safe
definition of the measure in the Euclidean generating functional, and
thus the Euclidean Green's functions as its moments
\cite{Alkofer:2000wg}.

Under the assumption of its existence, vacuum expectation values
of observables are obtained from the generating
functional as functional derivatives. For a general observable denoted
as $\mathcal{O}$ this yields 
\begin{equation}
 \label{eq:O_continuum}
  \langle\mathcal{O}\rangle \propto \int
   [\mathcal{D}A][\mathcal{D}\bar{\psi}][\mathcal{D}\psi][\mathcal{D}\bar{c}]
   [\mathcal{D}c]\;\mathcal{O}\; e^{- \Seff[A,\bar{\psi},\psi,\bar{c},c]}
 \end{equation}
where $\Seff$ denotes the effective action, \ie the space-time
integral of the effective Lagrangian density. Note
that this density also depends on the gauge parameter $\xio$.

%---------------------------------------------------------------------------
\subsubsection{Gauge independence and gauge invariance}

For gauge-invariant observables it can be shown (see \eg
\cite{Weinberg:1996jk}) that functional integrals of type as given in
\Eq{eq:O_continuum} are independent (within broad limits) of the
gauge-fixing functional $\mathcal{F}$, \ie of the
gauge condition. The different types only result in irrelevant
constant factors which are normalized away in the ratio of functional
integrals. On the contrary, vacuum expectation values for
gauge-variant observables dependent on the gauge condition.
We found it worth to quote in this context a note from
Collins's book \cite{Collins:1984np}:
\begin{quote}\small
  It is important to distinguish the concepts of gauge invariance and
  gauge independence. Gauge invariance is a property of a classical
  quantity and is invariance under gauge transformations. Gauge
  independence is a property of a quantum quantity when quantization
  is done by fixing the gauge. It is independence of the method of
  gauge fixing. Gauge invariance implies gauge independence, but only
  if the gauge fixing is done properly \cite[p.31f]{Collins:1984np}. 
\end{quote}

%----------------------------------------------------------------------
\subsection{The BRST formalism}
\label{sec:brst_intro}

In the last section we ended up with an effective Lagrangian density
\Eq{eq:eff_Lagrangian} that is no longer local gauge invariant. Local
gauge invariance spoils a naive integration over gauge fields and thus
a gauge has to be fixed before the functional-integral formalism is
applicable for quantization. 
However, it is a fundamental physical requirement that gauge-fixing is
done in such a way that matrix elements between physical states are
independent of the actual choice of gauge condition. The class of
effective Lagrangians $\Leff^{\xi}$ generated by the FP formalism
above, can be shown to fulfill this requirement. They all
yield the same unitary $S$-matrix\footnote{See \eg
  \cite{Weinberg:1996jk} for a more details}.

If phrased in a modern language of quantum field theory, namely 
the \emph{BRST formalism}, the effective Lagrangian must be 
\emph{BRST invariant} in order to have a renormalizable theory 
yielding a unitary $S$-matrix. This formalism takes BRST
invariance as a \emph{first} principle and can be
even used as a substitute for the FP method, in particular there where
the FP method fails.\footnote{For example, the BRST symmetry is a
  basis for developing the canonical operator formalism. See the book
  by \name{Nakanishi} and \name{Ojima} \cite{Nakanishi:1990qm} for a
  comprehensive account on that.}

The BRST formalism goes back to the discovery of \name{Becchi}, \name{Rouet}
and \name{Stora} \cite{Becchi:1975nq,Becchi:1976nq} who first noted 
(independent also \name{Tyutin} \cite{Tyutin:1975je,Iofa:1976je}), that
even if $\Leff$ is no longer locally gauge-invariant, it is invariant under a
special type of global symmetry transformation. This symmetry is
a supersymmetry that involves ghost fields $c^a(x)$ in an
essential way. Remember, in the FP approach ghost fields are merely a technical
device to express the FP determinant in terms of a path integral.
In the BRST formalism, however, they serve as
parameters $\omega^a(x)=\delta\lambda c^a(x)$ of infinitesimal
local gauge transformations of gauge and fermion fields. Here
$\delta\lambda$ is an (infinitesimal) $x$-independent Grassmann number. 
In fact, a BRST transformation is isomorphic to an infinitesimal gauge
transformation (\Eq{eq:gauge_inf}). They are written in the form
\begin{displaymath}
  \deltaB\Phi= \delta\lambda\bs\Phi\qquad\textrm{where}\quad
  \Phi=\{A^a_{\mu},\psi\}.
\end{displaymath}
Here $\bs$ denotes the \emph{BRST operator} that acts upon the
gauge and fermion fields according to 
\begin{subequations}
\label{eq:BRST}
\begin{align}
\label{eq:BRST_A}
\bs A^a_{\mu}(x) &= D^{ab}_{\mu}c^b(x),\\
\label{eq:BRST_psi}
  \bs\psi(x) &= -i\go T^a c^a(x)\psi(x)\;.
\end{align}
Upon the ghost field $c^a(x)$ the BRST operator $\bs$ is defined to act as
\begin{equation}
  \label{eq:BRST_c}
  \bs c^a(x) = - \frac{\go}{2} f^{abc} c^b(x) c^c(x) \;.
\end{equation}
This and the Jacobi identity for the structure constants $f^{abc}$
suffice to show that the BRST operator $\bs$ is nilpotent
\cite{Kugo:1981hm}, \ie
\begin{displaymath}
  \bs(\bs\Phi) = 0. 
\end{displaymath}

In addition to the ghost fields, the BRST formalism introduces
antighost $\bar{c}^a$ and Nakanishi-Lautrup auxiliary fields $B^a$
\cite{Nakishi:1966di} that transform as
\begin{eqnarray}
  \label{eq:BRST_bar_c}
  \bs \bar{c}^a(x) &=& iB^a(x)\;,\\
  \label{eq:BRST_B}
  \bs B^a(x) &=& 0\;.
\end{eqnarray}
\end{subequations}
The introduction of the auxiliary fields $B^a$ linearizes the BRST
transformations and renders the operator $\bs$ to be nilpotent also
off-shell.

\subsubsection{The BRST charge}

Since the BRST symmetry is a global symmetry\footnote{Note that the
  invariance of $\Linv$ is a trivial consequence of its gauge
  invariance. After renormalization (see
  \Sec{sec:regularization_and_renormalization}) the BRST symmetry is
  still a global symmetry of $\Leff^r$ (\Eq{eq:renLeff}) supposed the BRST
  transformations are substituted by the renormalized ones given in 
  \Eq{eq:rBRST_delta} and (\ref{eq:rBRST}). } 
of $\Leff$ there exists a corresponding \emph{Noether
current} $J_B$ that is conserved, \ie $\partial_{\mu}J_{B\mu}=0$. Its
explicit form (see \eg\cite{Nakanishi:1990qm}) is not of interest for
the following discussions. But the existence of a corresponding
unbroken charge $Q_B$ is important (see below and the discussion concerning
the Kugo-Ojima confinement criterion in \Sec{seq:KOscenario}). 

In general, the charge $Q$ corresponding to a current $J_{\mu}$ is defined as the
spatial integral of $J_0$. It is a generator of
the global symmetry, even if the integral is not
convergent.\footnote{If the integral is not convergent, the charge
  $Q$ is an ill-defined operator, and hence neither eigenstates nor
  expectation values of $Q$ can be considered. However, for any
  local quantity $\Phi(y)$ the (anti-)commutator
\begin{displaymath}
  [iQ,\Phi(y)]_{\mp}\equiv\int d^3x [J_0(x),\Phi(y)]_{\mp}
\end{displaymath}
can be considered, because the integrand vanishes for sufficiently
large $\vec{x}$. It follows that $Q$ is an generator of infinitesimal
symmetry transformations. See \cite[p13f.]{Nakanishi:1990qm} for
details.} If this is the case, however, then the 
charge is ill-defined and is a generator of a spontaneously broken
global symmetry. 

It has been argued by \name{Kugo} and \name{Ojima} \cite{Kugo:1979gm}
that the BRST charge $Q_B$ is an unbroken charge and so we
can consider its eigenstates. In particular, states $\Psi_i$ belonging
to the physical state space $\Vphys$ are assumed to be BRST singlet
states of $Q_B$, \ie they are annihilated by $Q_B$
\begin{displaymath}
  Q_B|\Psi_i\rangle = 0\;.
\end{displaymath}
This assumption plays an important role in the Kugo-Ojima confinement
scenario to be introduced in \Sec{seq:KOscenario}. It is also 
related to the requirement of gauge-independence for physical matrix
elements as we discuss now.

\subsubsection{Gauge-fixing and the BRST formalism}

Within the BRST formalism gauge-fixing is neatly performed by
considering the BRST invariance as a first principle, \ie a
Lagrangian density has to be BRST invariant to have a renormalizable theory
that yields a unitary $S$-matrix. Since the BRST operator $\bs$ is
nilpotent, BRST--exact (or BRST--coboundary) terms --- those are of the
form $\bs(*)$ ---  can freely be added to the gauge invariant
Lagrangian density $\Linv$, \ie 
\begin{displaymath}
  \Leff = \Linv[\bar{\psi},\psi,A] +
  \bs\mathcal{T}[\bar{\psi},\psi,A,c,\bar{c},B] \;.
\end{displaymath}
This will not change physics in any order of perturbation theory, but
it can be used to represent the sum $\Lgf^{\xi}+\Lfp$ of the gauge-fixing
and compensating ghost terms \cite{Kugo:1981hm}. In fact, one can show
that $\Lgf^{\xi}+\Lfp$ is of the form
\begin{displaymath}
  \Lgf^{\xi}+\Lfp = \bs\mathcal{T}[\mathcal{F}^a[A],\bar{c}^a]
\end{displaymath}
where $\mathcal{F}^a$ was defined for covariant gauges in
\Eq{eq:covariant_gauge_cond}.

Any change $\Delta\mathcal{T}$ in the functional $\mathcal{T}$, for
example in $\mathcal{F}^a$, must
not change any  matrix element $\langle\Psi_1|\Psi_2\rangle$ of
physical states \cite{Weinberg:1996jk}, \ie  
\begin{equation}
 \label{eq:change_in_matrix_element}
 0 = \langle\Psi_1|\bs\Delta\mathcal{T}|\Psi_2\rangle\;.
\end{equation}
With $Q_B$ being a generator of the BRST symmetry 
\begin{displaymath}
  i\bs(*) = [Q_B,*]_{\mp}
\end{displaymath}
we obtain that \Eq{eq:change_in_matrix_element} can only hold
for arbitrary changes in $\mathcal{T}$ if physical states are in the kernel
of $Q_B$, \ie 
\begin{displaymath}
  0 = \langle\Psi_1|[Q_B,\Delta\mathcal{T}]|\Psi_2\rangle 
 \quad\Longleftrightarrow\quad
      \langle\psi_1|Q_B = Q_B|\psi_2 \rangle = 0 
\end{displaymath}

Note that the BRST symmetry of the full quantum Lagrangian is a basis
for developing the canonical operator formalism. It also is very
useful for deriving the Slavnov-Taylor-identities (STI). These are
used for the proof of renormalizability of QCD.

%----------------------------------------------------------------------------

%------------------------------------------------------------------------
\section{Regularization and renormalization}
\label{sec:regularization_and_renormalization}

The quantum field theory of the strong interaction 
introduced so far is still incomplete, because perturbative
expansions of Green's functions beyond tree level would be rendered
meaningless by divergent mathematical expressions. In particular,
loop integrals produce ultraviolet divergences when the cutoff for the  
internal momentum integral is send to infinity. This is known since
the early days of QED (see \eg \cite{Oppenheimer:1930fh}).

Fortunately, QCD is a \emph{renormalizable} theory to any finite order in
perturbation theory.\footnote{The first proof for non-abelian 
  gauge theories to be renormalizable was given
  by 't~\!\!\name{Hooft} and \name{Veltman}
  \cite{'tHooft:1971fh,'tHooft:1971rn} using Slavnov-Taylor 
  identities (STI) \cite{Slavnov:1972fg,Taylor:1971ff}. Modern proofs
  take advantage of the BRST formalism.} 
That is, all divergences may be absorbed into a change of the
normalization of the Green's functions and a suitable redefinition of all
parameters appearing in the Lagrangian. The renormalized theory then
yields only finite expressions at any order of perturbation theory.

\subsection{Regularization}

For this to work, QCD needs to be regularized prior to
renormalization, using, for example, the \emph{Pauli--Villar}, the
\emph{dimensional} or the \emph{lattice regularization}. Actually, the 
latter is the only known nonperturbative regularization of QCD. In
this regularization, the lattice spacing $a$ serves as a 
regularization parameter that renders all momentum loop integrations
finite via a gauge invariant ultraviolet cutoff
$\Lambda=a^{-1}$. Consequently,  arbitrary $n$-point functions
calculated within the lattice approach due not suffer from
ultraviolet divergences as long as $a>0$. We shall briefly 
introduce the lattice regularization of QCD in \Sec{sec:latticeQCD}. 
A list of references for detailed information can also be found there.

\subsection{Renormalization}

After a suitable regularization has been carried out, for example by
setting a cutoff $\Lambda$, the renormalization program
introduces so called $Z$-factors which absorb the finite (because of
the cutoff), but potentially divergent part in each of
the fundamental two and three-point 
functions, \ie those with tree-level counterpart in the Lagrangian
density \cite{Alkofer:2000wg}. These are the inverse gluon, ghost and quark
propagators as well as the three-gluon, four-gluon, ghost-gluon and
quark-gluon vertices. The corresponding $Z$-factors are denoted by 
$Z_3$, $\widetilde{Z}_3$, $Z_2$, $Z_1$, $Z_4$, $\widetilde{Z}_1$ and $Z_{1F}$
respectively, and are formally introduced by writing the renormalized
Lagrangian density as (see \eg \cite{Alkofer:2000wg})
\begin{eqnarray}
 \Leff^r \,  &=&  \,  Z_3 \, \frac{1}{2} \, A^a_\mu
 \left( -
 \partial^2 \delta_{\mu\nu} \, - \, \left( \frac{1}{Z_3\xi_r} - 1 \right)
 \partial_\mu  \partial_\nu  \right) A^a_\nu  \nonumber\\
 && + \, \widetilde Z_3 \, \bar c^a \partial^2 c^a \, + \, \widetilde Z_1 \,
 \gr f^{abc} \,  \bar  c^a \partial_\mu
 \left(A_\mu^c c^b\right) \, - \, Z_1 \, \gr f^{abc} \, \left(\partial_\mu A_\nu^a\right) \, A_\mu^b
 A_\nu^c  \nonumber \\
&& + \,Z_4 \,  \frac{1}{4} \gr^2 f^{abe} f^{cde} \, A^a_\mu A^b_\nu A^c_\mu
 A^d_\nu  \, + \, Z_2 \, \bar \psi \big( -\gamma_{\mu}\partial_{\mu} +
 Z_m m_r \big) \psi \nonumber \\
&& \, - \,
 Z_{1F} \,  i \gr  \, \bar \psi  \gamma_\mu  T^a \psi \, A^a_{\mu}   \;. 
 \label{eq:renLeff}
\end{eqnarray}
An additional factor, $Z_m$, is necessary to adjust the mass of the
quark propagator to the pole mass and $\gr$ refers to the renormalized
coupling constant. The latter is related to the bare parameter $\go$
by considering, for example, the ghost-gluon vertex. Using this,
\begin{equation}
 \label{eq:g_glghgh_v}
  \gr := \frac{Z^{1/2}_3 \widetilde{Z}_3}{\widetilde{Z}_1}\go\;.
\end{equation}
Of course, any other vertex function could be used instead to define
$\gr$. If all renormalization constants were independent then
each vertex would define its own renormalized coupling constant. For
example, using the three-gluon or the quark-gluon vertex this is
\begin{displaymath}
  g_{AAA} := \frac{Z_3^{3/2}}{Z_1}\go \qquad\textrm{or}\qquad 
  g_{\psi\bar{\psi}A} := \frac{Z_3^{1/2} Z_2}{Z_{1F}}\go\;.
\end{displaymath}
But in order to guarantee that they all define the same coupling
constant, \ie $\gr=g_{AAA}=g_{\psi\bar{\psi}A}$, the $Z$-factors are
constrained by the \emph{Slavnov-Taylor identities} (STI)
\cite{Slavnov:1972fg,Taylor:1971ff} giving:
\begin{equation}
  \label{eq:STI}
  \frac{Z_1}{Z_3} = \frac{\widetilde{Z}_1}{\widetilde{Z}_3} = 
  \frac{Z_{1F}}{Z_2} =  \frac{Z_4}{Z_1} =: Z_g Z^{1/2}_3 \;.
\end{equation}
Then the renormalized coupling constant $\gr$ is universal and can be
related to the bare coupling constant $\go$ by $Z_g\gr=\go$ where
$Z_g$ is defined in \Eq{eq:STI}. Obviously, the renormalization
constants are not independent of each other and the STIs allow us to
express the constants for the vertices by the field renormalization
constants $Z_3$, $\widetilde{Z}_3$, $Z_2$ and an
independent one~$Z_g$. 

Comparing \Eq{eq:renLeff} and (\ref{eq:eff_Lagrangian}), we see that
the renormalized Lagrangian (\Eq{eq:renLeff}) is related to its bare
expression (\Eq{eq:eff_Lagrangian}) by rescaling the fields
\begin{displaymath}
  A^a_{\mu} \rightarrow Z_{3}^{1/2}A^a_{\mu},\qquad 
  \psi     \rightarrow Z_2^{1/2}  \psi,\qquad
  c^{a}    \rightarrow \widetilde{Z}^{1/2}_3 c^{a}
\end{displaymath}
and by redefining the parameters appearing in the Lagrangian:
\begin{displaymath}
  \go =  Z_g\gr, \qquad
  \mo  =  Z_m m_r, \qquad
  \xio = Z_3 \xi_r\;.
\end{displaymath}

In an analogous manner, this translates to the BRST
transformations introduced in \Sec{sec:brst_intro}. To be specific, the
renormalized Lagrangian density $\Leff^r$ is BRST invariant and
all considerations made previously remain 
valid if the (infinitesimal) parameter of the BRST transformation is replaced by
\begin{equation}
 \label{eq:rBRST_delta}
  \delta\lambda_r := Z_3^{-1/2}\widetilde{Z}_3^{-1/2}\delta\lambda
\end{equation}
and the renormalized fields transform as (see \eg \cite{Alkofer:2000wg})
\begin{subequations}
 \label{eq:rBRST}
 \begin{eqnarray}
  \label{eq:rBRST_A}
  \bs_r A^a_{\mu}(x) &=& \widetilde{Z}_3D^{ab}_{\mu}c^b(x)\;,\\
  \label{eq:rBRST_psi}
  \bs_r \psi(x) &=& -\widetilde{Z}_1i\gr T^a c^a\psi(x)\\
  \label{eq:rBRST_c}
  \bs_r c^a(x) &=& -\widetilde{Z}_1\frac{\gr}{2}f^{abc}c^bc^c(x)\;,\\
  \label{eq:rBRST_bar_c}
  \bs_r \bar{c}^a(x) &=& \frac{1}{\xir}\partial_{\mu}A_{\mu}^a(x)\;.
\end{eqnarray}
\end{subequations}

%-----------------------------------------------------------------------------
\subsection{The $\MOM$ scheme}

After this rather formal rescaling and renaming of fields and
parameters, any unrenormalized, but regularized $n$-point or Green's
function $G^n_{\textrm{reg}}$ is related to their renormalized one (in
momentum space) through 
\begin{equation}
 \label{eq:ren_Green}
  G_r(p_1,\ldots,p_n; \gr,\mr,\xir) = Z_{G}\cdot
  G_{\textrm{reg}}(p_1,\ldots,p_n; \Lambda,\go,\xio,\mo)  
\end{equation}
where $Z_{G}$ refers to the corresponding product of $Z$-factors
that appear in the Green's function. To give two simple examples: for the
gluon two-point function in momentum space
$D^{ab}_{\mu\nu}(p):=\langle A^a_{\mu}(p)A^b_{\nu}(-p)\rangle$ 
this is $Z_{G}=Z_3$, whereas for the ghost-gluon vertex
$Z_{G}$ is given by $Z_{G}=\widetilde{Z}_1=Z_g\widetilde{Z}_3Z_3^{1/2}$.

The $Z$-factors have to be determined such that the renormalized
expression on the left hand side of \Eq{eq:ren_Green} is finite. 
The way this is done is defined by the \emph{renormalization scheme}.
There are different renormalization schemes. In each scheme the
divergent part is absorbed into the $Z$-factors, but they differ in
how much of the finite part is absorbed, too \cite{Muta:1998vi}. 
Common renormalization schemes are the subtraction
schemes $\MS$, $\MSb$ or $\MOM$. The latter type is considered in this
thesis, even though there are infinite many different $\MOM$ schemes.

A $\MOM$ scheme defines the $Z$-factors such that the fundamental
two-point and three-point functions equal their corresponding
tree-level expressions at some momentum $\mu^2$, the renormalization
point. 

The two-point functions that are relevant in this thesis are
the gluon propagator $\langle A^a_{\mu}(x)A^b_{\nu}(y)
\rangle$ and the ghost propagator $\langle c^a(x)\bar{c}^b(y)\rangle$
in Landau gauge. Actually, we are interested in the Fourier transform of these
two expressions. In momentum space the gluon propagator in Landau gauge
has the following tensor structure: 
\begin{equation}
  \label{eq:gluonprop_dress}
  D^{ab}_{\mu\nu}(p,\mu) = 
  \delta^{ab}\left(\delta^{\mu\nu}-\frac{p_{\mu}p_{\nu}}{p^2}\right) 
  \frac{Z(p^2,\mu^2)}{p^2}
\end{equation}
where $Z$ denotes the form factor or the \emph{dressing function} of
the gluon propagator. It expresses the deviation of $D^{ab}_{\mu\nu}(p)$ from
its tree-level form ($Z\equiv1$). For the ghost propagator
the corresponding tensor structure is given by
\begin{equation}
 \label{eq:ghostprop_dress}
  G^{ab}(p,\mu)   = \delta^{ab}\;\frac{J(p^2,\mu^2)}{p^2} .
\end{equation}
Here $J$ denotes the dressing function of the ghost propagator.

In a $\MOM$ scheme the renormalization constants, for instance of the gluon and
ghost fields $Z_3$ and $\widetilde{Z}_3$, are defined by requiring
the renormalized expressions to equal their tree-level form at some
(large) momentum $\mu^2$. That is, $Z_3$ is defined as
\begin{equation}
  \label{eq:Z3}
  D^{ab}_{\mu\nu}(p;\Lambda,\go,\mo,\xio)\Big|_{p^2=\mu^2} =: Z_3\; \delta^{ab}
  \left(\delta^{\mu\nu}-\frac{p_{\mu}p_{\nu}}{\mu^2}\right)\frac{1}{\mu^2}   
\end{equation}
where $D^{ab}_{\mu\nu}$ denotes the unrenormalized gluon propagator. 
$\widetilde{Z}_3$ is given by
\begin{equation}
  \label{eq:tildeZ3}
     G^{ab}(p;\Lambda,\go,\mo,\xio)\Big|_{p^2=\mu^2}  =:
     \widetilde{Z}_3\;\delta^{ab}\frac{1}{\mu^2} 
\end{equation}

Therefore, a renormalization constant can be determined by
calculating the corresponding unrenormalized (regularized) Green's
function. Its value depends on the renormalization point
$\mu^2$ and also on the bare parameters of the regularized theory.
For example: In our lattice simulations we have calculated
the bare (quenched) gluon propagator using the bare parameters:
$\go(\Lambda^2)=1$, $1/\mo=0$ and $\xio=0$. Requiring
\Eq{eq:Z3} to hold at some momentum $\mu^2$, we 
have fixed $Z_3$. The renormalized gluon propagator (at $\mu^2$) 
is then obtained via multiplicative renormalization according to
\Eq{eq:ren_Green}. 

%-----------------------------------------------------------------------
\section{The renormalization group}

Obviously, a renormalized Green's function depends on the 
subtraction point $\mu$ whose choice is not unique. Also the
renormalized parameters $\gr$, $\mr$ and $\xir$ depend
(via the corresponding $Z$-factors) on $\mu$. Keeping $\go$,
$\mo$, $\xio$ and $\Lambda$ fixed, we could, of course, had chosen another 
point, say $\mu'$. This would yield a new renormalized  
Green's function with the new values $\gr(\mu')$, $\mr(\mu')$ and
$\xi_r(\mu')$. Even though both renormalized Green's functions are
different, they are related by a finite multiplicative
renormalization, \ie
 \begin{equation}
 \label{eq:G_zG}
   G_r\big(p_i;\gr(\mu'),\mr(\mu'),\xir(\mu'),\mu'\big) = z(\mu',\mu)\,
   G_r\big(p_i;\gr(\mu),\mr(\mu),\xir(\mu),\mu\big) 
\end{equation}
where $z$ is a finite number depending on $\mu$ and $\mu'$. The finite 
renormalization of Green's functions forms an Abelian group
called the \emph{renormalization group} (RG). Physically measurable
quantities are invariant under renormalization group
transformations, \ie they are independent of the subtraction point
$\mu$. Green's functions, are generally not renormalization-group invariant. 
An analytic expression of this property is given by
the renormalization group equation \cite{Muta:1998vi}.   

\subsection{The renormalization group equation}

The \emph{renormalization group equation} is best derived by
noting that the unrenormalized Green's function does not depend
on the renormalization point $\mu$ if all bare parameters ($\go$,
$\mo$, $\xio$ and $\Lambda$) are fixed, \ie
\begin{equation}
 \label{eq:RGE_bare}
  0 = \mu\frac{d}{d\mu}G(p_i; \go,\xio,\mo,\Lambda)\;.
\end{equation}
On the contrary, the renormalized Green's function
depends on $\mu$ not only explicitly, but also implicitly due to the 
renormalized parameters. By using the chain rule for differentiation,
\Eq{eq:RGE_bare} yields for the renormalized Green's function 
\begin{equation}
 \label{eq:RGE_ren}
  \left(\mu\frac{\partial}{\partial\mu}
    + \beta\frac{\partial}{\partial \gr} 
    + \beta_{\xi}\frac{\partial}{\partial \xir}
    - \gamma 
    + \mr\gamma_{m}\frac{\partial}{\partial \mr}
   \right)G_r = 0. 
\end{equation}
Here a sum over the different fermion flavors is implied and
the (dimensionless) RG functions are defined as 
(see \eg \cite{Muta:1998vi})
\begin{subequations}
  \label{eq:callan_symanzik}
\begin{eqnarray}
  \label{eq:beta_funct}
  \beta\left(\gr,\frac{\mr}{\mu},\xi_r\right) &:= &\left. \mu
    \frac{\partial \gr}{\partial 
    \mu}\right|_{\go,\mo,\xio,\Lambda\, \textrm{fixed}} \\
  \label{eq:gamma_m_funct}
  \gamma_{m}\left(\gr,\frac{\mr}{\mu},\xi_r\right) &:=&
  \left.\frac{\mu}{\mr} \frac{\partial \mr}{\partial \mu} 
  \right|_{\go,\mo,\xio,\Lambda\, \textrm{fixed}}\\
  \label{eq:gamma_funct}
  \gamma\left(\gr,\frac{\mr}{\mu},\xi_r\right) &:=&\left.\mu
    \frac{\partial\ln Z_{G} }{\partial \mu}
  \right|_{\go,\mo,\xio,\Lambda\, \textrm{fixed}}\\
  \label{eq:beta_xi_funct}
  \beta_{\xi}\left(\gr,\frac{\mr}{\mu},\xi_r\right) &:=&\left. \mu
    \frac{\partial \xir}{\partial \mu} 
   \right|_{\go,\mo,\xio,\Lambda\, \textrm{fixed}}.
\end{eqnarray}
\end{subequations}
The RG equation expresses how the renormalized Green's function, in
particular their parameters, change under a variation of the
renormalization point $\mu$. In the following we shall assume that 
$\mu\gg\mr$ always holds. Thus approximately, the RG functions do not
depend on $\mr$.\footnote{Such an approximation corresponds to using a 
mass-independent renormalization scheme, like for example the 
$\MSb$ scheme.} The $\beta$-function has also been proven to be 
gauge independent \cite{Marciano:1977su}, \ie
\begin{displaymath}
  \beta(\gr,\xir) = \beta(\gr)\;. 
\end{displaymath}
To get rid of the gauge dependence of the other RG
functions, we shall restrict ourselves in the following to the Landau
gauge, because the Landau gauge is a fixed point under the
renormalization group. To see this note that in general covariant
gauge we obtain for the RG function $\beta_{\xi}$
(\Eq{eq:beta_xi_funct}) depending on the gauge parameter ($Z_3\xir = \xio$) 
\begin{displaymath}
  \beta_{\xi}(\gr,\xir) = \left.\mu \frac{\partial \xir}{\partial \mu}
   \right|_{\go,\xio,\Lambda\,\textrm{fixed}} = \left. -\frac{\xi_0}{Z_3^2}\,\mu
   \frac{\partial Z_3}{\partial
     \mu}\right|_{\go,\xio,\Lambda\,\textrm{fixed}} = - \xir\,\mu 
 \frac{\partial \ln Z_3}{\partial \mu}\;. 
\end{displaymath}
Therefore, given the initial condition $\xio=0$ the function
$\beta_{\xi}$ vanishes completely in Landau gauge.  

We are left with the three RG equations $\beta(\gr)$, $\gamma_m(\gr)$ and
$\gamma(\gr)$ (see \Eq{eq:beta_funct}, (\ref{eq:gamma_m_funct}) and
(\ref{eq:gamma_funct})) which in our approximation only depend on the
renormalized coupling constant $\gr(\mu)$. They express how the
renormalized parameters $\gr(\mu)$, $\mr(\mu)$ 
and the renormalized Green's function change under a variation
of $\mu$. In fact, given the initial values $\gr(\mu)$ and $\mr(\mu)$
at a renormalization point $\mu$, the values $\gr(\mu')$ and
$\mr(\mu')$ at $\mu'$, are determined by the solution of the
differential equations (\ref{eq:beta_funct}) and
(\ref{eq:gamma_m_funct}), respectively. That is
\begin{align}
 \label{solution_gmu}
  \frac{m(\mu')}{m(\mu)} &= \exp\left\{\int^{\gr(\mu')}_{\gr(\mu)} dh 
  \frac{\gamma_m(h)}{\beta(h)}  \right\}\;,\\
 \label{solution_mmu}
  \frac{\mu'}{\mu} &=\exp\left\{\int^{\gr(\mu')}_{\gr(\mu)}
    \frac{dh}{\beta(h)}\right\}\;.
\end{align}
Similarly, the change of the renormalized Green's function under the
RG transformation $\mu\rightarrow\mu'$ is obtained. To see this, consider
\Eq{eq:G_zG}. The RG equation (\ref{eq:RGE_ren}) tells us that
\cite{Collins:1984np} 
\begin{displaymath}
  \mu'\frac{d}{d\mu'}\ln z = \mu'\frac{d}{d\mu'} 
  \ln\left\{\frac{G_r(\mu')}{G_r(\mu)}\right\} = -\gamma(\gr).
\end{displaymath}
The function $\gamma$ is known as the \emph{anomalous dimension} for
reasons that become clear in \Sec{sec:anomalous}. A full solution to
this RG equation is given by \Eq{eq:G_zG} where \cite{Collins:1984np}
\begin{equation}
  \label{eq:z}
  z(\mu',\mu) =
  \exp\left\{\int^{\gr(\mu')}_{\gr(\mu)} dh\frac{\gamma(h)}{\beta(h)}\right\} \;.
\end{equation}
The values of $\gr(\mu')$ and $\mr(\mu')$ are given in \Eq{solution_gmu} and
(\ref{solution_mmu}), respectively.

Of course, the explicit form of the RG functions are generally
unknown, but approximations can be made by taking a finite number of
terms in the perturbation series for the RG functions $\beta$,
$\gamma_m$ and $\gamma$. Since this is an expansion in $\gr$ it is
only valid at sufficiently small $\gr$. We shall see
in \Sec{sec:theor_running_coupling} that for QCD this is realized in the
asymptotic region of large Euclidean momenta.  

Therefore, the most important application of the RG equation in QCD 
is to study the asymptotic behavior of Green's functions at large
Euclidean momentum \cite{Collins:1984np}.

%----------------------------------------------------------------------
\subsection[Perturbative expansion of the  $\beta$--function]%
{Perturbative expansion of the $\boldsymbol{\beta}$--function}
\label{sec:theor_beta_function}

For the $\beta$-function (\ref{eq:beta_funct}) a power expansion in the
coupling constant $\gr$ can be calculated by choosing one of
the four vertex functions which define~$\gr$. Taking, for example, the
three-gluon vertex the renormalized coupling constant $\gr$ is related to
the bare coupling as
\begin{displaymath}
  \gr = Z^{3/2}_{3} Z^{-1}_1 \go\,.
\end{displaymath}
The renormalization constants, $Z_3$ and $Z_1$, are defined at a
subtraction point \mbox{$p^2=\mu^2$} in terms of the bare (transverse)
gluon propagator and the bare three-point vertex, respectively.
Extracting both $Z$-factors to two-loop order, the
solutions can be plugged into the RG equation \Eq{eq:beta_funct} for
$\gr(\mu)$. After some algebra this gives an expansion for the
$\beta$--function (see 
\eg \cite{Weinberg:1996jk})
\begin{equation}
 \label{eq:beta_fct}
  \beta[\gr(\mu)] = - \beta_0
  \frac{\gr^3(\mu)}{16\pi^2} - \beta_1\frac{\gr^5(\mu)}{128\pi^4} 
+ \order{\gr^7(\mu)}
\end{equation}
that holds at small $\gr(\mu)$. The first two coefficients are given by
\begin{subequations}
 \label{eq:beta_coef}
\begin{eqnarray}
  \label{eq:beta0}
  \beta_0 &=& 11 - \frac{2}{3}N_f\,, \\
 \label{eq:beta1}
  \beta_1 &=& 51 - \frac{19}{3}N_f \,.
\end{eqnarray}
\end{subequations}
Here $N_f$ is the number of quark flavors with masses below the
energies of interest.\footnote{Note that in each energy range between
  any two successive quark masses we have a different value of $N_f$,
  and also a different $\mathsf{\Lambda}$ (see
  \Sec{sec:theor_running_coupling}), chosen to make $g(\mu)$ continuous
  at each quark mass.\cite[p.157]{Weinberg:1996jk}} 

Since the determination of the $Z$-factors generally depends on the 
renormalization scheme used, the explicit form of $\beta(\gr)$ depends
on the gauge and on how the running coupling is precisely
defined. However, it can be shown that the first two coefficients,
$\beta_0$ and $\beta_1$, are renormalization-scheme independent,
whereas those of higher loop-expansions are scheme-depended. 

\subsection{The running coupling constant}
\label{sec:theor_running_coupling}

Using the expansion of the $\beta$ function we can solve the RG
equation \Eq{eq:beta_funct} for the coupling constant $\gr(\mu)$ to
the given order. The general solution to \Eq{eq:beta_funct} takes the
form given in \Eq{solution_gmu}. As mentioned above, it describes
the variation of $\gr$ under the change $\mu\rightarrow\mu'$ 
keeping the bare parameters $\go$, $\mo$ and $\Lambda$ fixed. This is
usually termed as the \emph{running} of the coupling constant $\gr$
changing the energy or the momentum scale~$\mu$.  

It is common practice to parameterize the running coupling
constant\footnote{In the literature sometimes the notation 
  $\bar{g}[\gr(\mu),\ln(\mu/\mu')]$ is used for the running coupling
  constant \cite{Gross:1973ju,Gross:1973id,Politzer:1973fx}. It is the
  same as $\gr(\mu')$, because $\bar{g}$ is defined to be the value of
  the coupling constant renormalized at $\mu'$ (what we call
  $\gr(\mu')$) if it is known to have the value $\gr(\mu)$ at $\mu$
  \cite{Celmaster:1979km}.} $\gr(\mu)$ 
by introducing a RG-invariant mass parameter $\mathsf{\Lambda}$ being the
integration constant of a solution to the differential equation in
\Eq{eq:beta_funct}. It is defined by
\begin{align}
 \nonumber
  \mathsf{\Lambda} := \mu \exp\left\{\frac{-1}{2b_0\gr^2(\mu)}\right\}&\cdot
   \left[b_0 \gr^2(\mu)\right]^{-b_1/(2b_0^2)}\\*[0.1cm]
  \label{eq:Lambda}
  & \cdot\exp\left\{-\int^{\gr(\mu)}_0
      \!\!dh\left[\frac{1}{\beta(h)} + \frac{1}{b_0h^3} -
        \frac{b_1}{b_0^2h} \right]\right\}.
\end{align}
Here $b_0$ and $b_1$ are nothing but the first two coefficients of the
$\beta$-function given in \Eq{eq:beta_fct}, \ie
\begin{equation}
 \label{eq:b0_b1}
  b_0 \equiv \frac{\beta_0}{16\pi^2} \quad\textrm{and}\quad b_1 \equiv
  \frac{\beta_1}{128\pi^4}\;.
\end{equation}
Specifying $\gr(\mu)$ at one value of $\mu$ is exactly
equivalent to fixing $\mathsf{\Lambda}$ \cite{Collins:1984np}. Thus 
if the $\beta$-function were known we could calculate
$\mathsf{\Lambda}$ from the knowledge of $\gr(\mu)$ at one value
$\mu$ and vice versa. Note that renormalization has introduced a
new parameter $\mathsf{\Lambda}$ of dimension mass into the theory
that is not present in the bare Lagrangian density. This is called 
\emph{dimensional transmutation}.  The parameter should be determined
by comparing experimental data with QCD predictions. 

The definition of $\mathsf{\Lambda}$ in \Eq{eq:Lambda} is
scheme dependent, since the coupling constant $\gr(\mu)$ may have even
different meanings in different renormalization schemes. For example,
on the lattice a scale is given by the lattice spacing~$a$ and the
coupling is given by the bare coupling constant $\go^2(a)$ depending
on~$a$. The scheme dependent parameter $\mathsf{\Lambda}$ on the
lattice is called $\mathsf{\Lambda}_{\textrm{LAT}}$. In the $\MSb$ scheme the
$\mathsf{\Lambda}$-parameter is usually called
$\mathsf{\Lambda}_{\MSb}$, whereas in $\MOM$ scheme it is called
$\mathsf{\Lambda}_{\MOM}$. The knowledge of the ratio allows to relate
results obtained in different renormalization schemes. For
example the ratio of $\mathsf{\Lambda}_{\MOM}$ and
$\mathsf{\Lambda}_{\MSb}$ is given by (see \eg
\cite{Celmaster:1979km,Montvay94}) 
\begin{displaymath}
  \frac{\mathsf{\Lambda}_{\MOM}}{\mathsf{\Lambda}_{\MSb}} = 2.895655.
\end{displaymath}

Up to two-loop order the coefficients of the $\beta$-function are
scheme independent. Using these coefficients an expression for the
running coupling constant can be given that is valid in any renormalization
scheme as long as $\mu\gg\mathsf{\Lambda}$ is fulfilled in that
scheme. Defining 
\begin{equation}
  \label{eq:alpha_s_allgem}
  \alpha_s(\mu) := \frac{\gr^2(\mu)}{4\pi}\;,
\end{equation}
the solution for the running coupling constant $\alpha_s(\mu)$ in
two-loop order is (see \eg \cite{Collins:1984np,Eidelman:2004wy})  
\begin{equation}
\label{eq:alpha_s_twoloop}
  \alpha_s(\mu) =\frac{4\pi}{\beta_0\ln(\mu^2/\mathsf{\Lambda}^2)}\Bigg[
     1 - \frac{2\beta_1}{\beta^2_0}
     \frac{\ln[\ln(\mu^2/\mathsf{\Lambda}^2)]}{\ln(\mu^2/\mathsf{\Lambda}^2)}
       \Bigg]  
     + \order{\frac{\ln^2[\ln(\mu/\mathsf{\Lambda})]}{\ln^3
     (\mu/\mathsf{\Lambda})}}.
\end{equation}
This two-loop result for $\alpha_s(\mu)$ will be used in
\Sec{sect:running_coupling}. 

We note in passing that a similar expression can be derived for the
\emph{running mass}, but since this will not be a subject of this
thesis we refer to standard textbooks for it.

\subsection{The anomalous dimension}
\label{sec:anomalous}

It is interesting to have a look once more at \Eq{eq:G_zG} and
(\ref{eq:z}). Assuming a renormalized Green's function $G_r$
has been computed at the momentum $\lambda p$ using a large value for
the scale factor $\lambda$. Setting $\mu'=\lambda\mu$, from
\Eq{eq:G_zG} one knows that
\begin{displaymath}
  G_r(\lambda p_i;\gr,\mr,\mu) = z(\mu,\lambda\mu) G_r(\lambda
  p_i;\gr(\lambda \mu),\mr(\lambda \mu),\lambda \mu) \;. 
\end{displaymath}
Under the assumption $\mr(\lambda\mu)$ does not get large and using
dimensional analysis\footnote{If one scales the momenta
  $p_i\rightarrow\lambda p_i$ then using dimensional arguments,
  a Green's function behaves as: 
  \begin{displaymath}
    G_r(\lambda p_i,\gr,\mu) = \mu^D f(\lambda^2 p_i\cdot p_j/\mu^2) 
  \end{displaymath}
where $D$ is the dimension of $G$ and $f$ is dimensionless. This is
because $G$ is Lorentz invariant, and hence can only be a function of
the various dot products $p_i\cdot p_j$ \cite{Kaku:1993ym}. Hence
\begin{displaymath}
  G_r(\lambda p_i,\gr,\lambda\mu) = \lambda^D \mu^D f(p_i\cdot
  p_j/\mu^2) = \lambda^D G_r(p_i,\gr,\mu). 
\end{displaymath}
} one obtains \cite{Collins:1984np}
\begin{align}
 \nonumber
  G_r(\lambda p_i;\gr,\mr,\mu) &\approx z\cdot G_r(\lambda p_i;\gr(\lambda
  \mu),0,\lambda \mu) \\
 \label{eq:Gscale_lambda}
 & = \lambda^{D} \cdot z(\mu,\lambda\mu)\cdot G_r(p_i;\gr(\lambda
  \mu),0, \mu)
\end{align}
where $D$ is the dimension of $G$.
This makes it evident that the relevant coupling constant is the 
effective coupling at the scale of the momenta involved. A change
in all external momenta using a common scale factor $\lambda$ is 
equivalent to changing the coupling constant. We see further from
\Eq{eq:Gscale_lambda} that the overall scale factor
is not just $\lambda^{D}$, as one might expect from naive dimensional
analysis (\ie $\beta(\gr)\equiv0$), but that it includes an extra factor
$z$ defined in \Eq{eq:z}. This is the reason why $\gamma$ is called
the \emph{anomalous dimension}. Note that this dimension arises from
the fact that a 
scale changes the renormalization point, and $G$ is not necessarily
invariant under this operation.  

To lowest order in perturbation theory the anomalous dimension
$\gamma(\gr)$ is given by the expansion
$\gamma(\gr) = c_0\gr^2 + \order{\gr^4}$  \cite{Collins:1984np}
where $c_0$ is the zeroth-order coefficient of the anomalous
dimension. Using $c_0$ and the corresponding coefficient $b_0$
(\Eq{eq:b0_b1}) of the $\beta$-function, we obtain from the definition of
$z(\lambda)=z(\lambda\mu,\mu)$ (\Eq{eq:z}) to lowest order 
in perturbation theory \cite{Collins:1984np}
\begin{align}
  z(\lambda\mu,\mu) &\simeq \exp\left\{
    \int^{\gr(\lambda\mu)}_{\gr(\mu)}\frac{c_0}{b_0} \frac{dh}{h}
  \right\} = \left[\frac{\gr(\lambda\mu)}{\gr(\mu)}\right]^{c_0/b_0}
  \left[1+\order{\gr^2}\right]\\
  & \propto
  \left[\ln\lambda\right]^{-\delta}\qquad
  \lambda\rightarrow\infty 
\end{align}
where $\delta:=c_0/(2b_0)$. For the gluon and ghost propagators in the
quenched case ($N_f=0$) these exponents are $\delta_D=13/22$ and
$\delta_G=9/44$, respectively. Therefore, the corresponding dressing
functions, $Z$ and $J$, behave in the far ultraviolet momentum region
like
\begin{displaymath}
  Z(p^2) \sim
  \left(\ln\frac{p^2}{\mathsf{\Lambda}^2}\right)^{-\delta_D}
  \quad\textrm{and}\quad 
  J(p^2) \sim \left(\ln\frac{p^2}{\mathsf{\Lambda}^2}\right)^{-\delta_G}\,.
\end{displaymath}

%===============================================================================
%%% Local Variables: 
%%% mode: latex
%%% TeX-master: "Sternbeck"
%%% End:

\chapter{Infrared QCD and criteria for confinement}
\label{ch:infrared}

\begin{chapterintro}{W}
 e discuss briefly some problems and recent developments concerning
 nonperturbative quantization with a particular focus on the infrared
 region of QCD. We give a summary of results obtained by using the
 Dyson-Schwinger (DS) approach. Thereby we restrict
 ourselves to those results which are of relevance for the discussion of
 our lattice results presented in subsequent sections. In the second part
 of this chapter we introduce different criteria for confinement
 that can be formulated for QCD in covariant gauges. 
 Note that in \Ch{ch:confcriteria} we shall try to confirm these
 criteria for lattice QCD in Landau gauge.
\end{chapterintro}

%-----------------------------------------------------------------------------
\section{Nonperturbative approaches to QCD}
\label{sec:nonperturbative}

It should be clear from the discussion of the renormalization group that 
it is possible to define an effective (running) coupling constant $\gr$
which is a function of the momentum. This functional dependence is
intimately related to the momentum dependence of the vertex functions
of QCD. For sufficiently large Euclidean momenta $q$, the coupling
constant becomes small 
and thus perturbation theory seems an appropriate calculation tool in
this asymptotic regime. On the other hand, $\gr$ grows large with
decreasing Euclidean momentum and diverges at a certain
momentum, the so called \emph{Landau pole}. Its position signals the
breakdown of perturbation theory. Even though significant 
progress has been achieved in recent years in the 
perturbative calculation of higher order corrections to
renormalization group functions, like, for example, for the $\beta$ 
function or the anomalous dimension\footnote{For details on the
  progress made and also for recent results see, for example, the work by
  \name{Chetyrkin} 
  \cite{Chetyrkin:1999pq,Chetyrkin:1997dh,Chetyrkin:2004mf} and 
  references therein.}, perturbation theory only applies to the region of large
Euclidean momenta. Note that there is no unique definition of
\emph{the} running coupling constant, since its definition depends on
the renormalization scheme employed \cite{Celmaster:1979km}. 

%-----------------------------------------------------------------------------
\subsection{Brief remarks on nonperturbative methods}

In any case, the infrared
region corresponds to strong coupling rather than weak coupling, and
hence is of interest for studying confinement of QCD.
Since perturbation theory is of no avail in studying QCD at low
momentum, this region can be explored only in genuinely
nonperturbative approaches. One such approach is the vast field of
Monte Carlo simulations of lattice QCD. It is a \emph{first principle} 
approach that contains the full nonperturbative structure of QCD and
has the striking feature of being manifestly gauge invariant.
(For some details and a list of references see \Ch{chap:latticeQCD}). 

However, lattice simulations are limited by the enormous
computational effort they require and thus its application to a wider 
range of unresolved problems connected with QCD can only be extended when
computer technology continues to improve. Moreover, lattice calculations are
always afflicted with uncertainties in extrapolating to the infinite 
volume and continuum limit which is necessary to connect with the physical
world. Therefore, it is worthwhile to pursue other approaches that
preserve features of QCD that the lattice formulation lacks. For example,
lattice simulations cannot made definite statements about the far infrared
region of QCD due to finite lattice volumes. Although, we did our best
in this thesis to do so.  

In this context another complementary nonperturbative approach based
on the infinite tower of Dyson-Schwinger equations (DSE) has gained
much attention in recent years. In \Sec{sec:dse} we give a brief overview
about the recent developments that are relevant for this thesis, but
before we would like to stress a particularly important point concerning
the problem of Gribov copies in the nonperturbative regime.

%----------------------------------------------------------------------------
\subsection{The problem of Gribov copies}
\label{sec:Gribov_problem_non}

The Dyson-Schwinger equations of QCD are derived from a generating
functional. In the continuum, this 
requires gauge-fixing for the definition of an integration
measure. For several reasons the gauge condition is mostly chosen within
the family of covariant gauges which are known to be not unique
beyond perturbation theory due to the existence of Gribov
copies. Remember that in the asymptotic regime, 
where perturbative QCD is relevant, the problem of Gribov copies could
be safely ignored. This is because all Gribov copies
$\{\orbit{g}{A}_{\mu}\}$ of a particular (gauge-fixed) gluon field
$A_{\mu}$ carry an additional $1/\gr$ dependence --- they are related
by a particular local gauge transformation (\Eq{eq:gauge_fin}) --- and
hence can be neglected within the framework of perturbation
theory. Therefore, we can use the FP method 
or, even more elegantly, the BRST formalism for the class of covariant
gauges to obtain a quantized gauge theory which is manifestly
Lorentz covariant and for gauge-invariant observables gauge-independent.
Moreover, there are elegant BRST proofs of multiplicative
renormalizability and unitarity to any order of perturbation theory
\cite{Baulieu:1983tg}.

Beyond perturbation theory we have to face the problem of Gribov
copies. Actually, it represents the main impediment to nonperturbative
gauge-fixing. As pointed out in  \cite{Williams:2003du}, there is no
known Gribov-copy-free gauge fixing which is a local function of gluon
fields $A_{\mu}$. Therefore, one has to adopt either 
a nonlocal Gribov-copy free gauge or attempt to maintain local BRST
invariance at the expense of admitting Gribov
copies \cite{Williams:2003du}. Concerning the latter 
point, there is, however, the well-known \emph{Neuberger problem} of
pairs of Gribov copies with opposite sign giving expectation values
$0/0$ \cite{Neuberger:1986xz, Testa:1998az}. It is still  
unknown whether local BRST invariance for QCD can be maintained in the
nonperturbative regime \cite{Williams:2003du}.

Of course, it is desirable to overcome the Neuberger problem and
elevate the BRST formalism to the nonperturbative level. A first
successful step in that direction was done recently 
for the massive Curci-Ferrari model \cite{Kalloniatis:2005if}. Also in
a recent work \cite{Ghiotti:2005ih}, a generalization of the FP method
was given that is valid beyond perturbation theory and, most notably,
circumvents the Neuberger problem. In fact, in \cite{Ghiotti:2005ih}
a path integral representation of the absolute value of the FP determinant in  
terms of auxiliary bosonic and Grassmann fields has been
presented.  Remember that
usually the absolute value of the FP determinant is dropped, but this
cannot be done beyond perturbation theory using a gauge condition that
suffers from the Gribov ambiguity. The resulting gauge-fixing
Lagrangian density is local and enjoys a larger \emph{extended} BRST
and anti-BRST symmetry, though it cannot be represented as a BRST
exact object \cite{Ghiotti:2005ih}.

%------------------------------------------------------------------------
\subsection{Nonperturbative quantization in Landau gauge}

Moreover, progress has been made concerning the
nonperturbative quantization in Landau gauge. It 
fact, it has been argued by \name{Zwanziger} that an exact
nonperturbative quantization of continuum gauge theory is provided by
the FP method in the Landau gauge if the integration is restricted to the
Gribov region $\Omega$ \cite{Zwanziger:2003cf}. That is, the vacuum
expectation value of a gluonic observable $\mathcal{O}$ is given by  
\begin{equation}
 \label{eq:Gribov_expval}
    \langle\mathcal{O}\rangle_{\Omega} = \frac{1}{Z} \int_{\Omega}
  \delta(\partial_{\mu}A_{\mu})\det \MFP[A] \mathcal{O}(A)
  e^{-S_{\textrm{YM}}[A]}\;. 
\end{equation}
Here we have indicated the restriction to the Gribov region by giving a
suffix $\Omega$ to the expectation value.

To remind the reader, the Gribov region $\Omega$ is defined as the region
in the space of \emph{transverse} gluon fields where the FP operator,
\mbox{$\MFP[A] \equiv -\partial_{\mu}D_{\mu}[A]$}, is
positive, \ie all its (nontrivial) eigenvalues are positive. 
\begin{displaymath}
  \Omega := \left\{A_{\mu}: \;\partial_{\mu}A_{\mu}=0;\;
 -\partial_{\mu}D_{\mu}[A] > 0 \right\}\;.
\end{displaymath}
It can be shown that the Gribov region is convex, it is bounded in
every direction and it contains the origin $A_{\mu}=0$ (see \eg
\cite{Zwanziger:2003cf}). The boundary $\partial\Omega$ of the Gribov
region is called 
the \emph{Gribov horizon}. It consists of transverse gluon fields for
which the lowest nontrivial eigenvalue of the FP operator vanishes.
Thus the FP determinant, being the product of all eigenvalues
$\det\MFP=\prod_n\lambda_n$, is positive 
inside the Gribov horizon and vanishes on it.

Due to the fact that the Gribov region is
not free of Gribov copies \cite[and references therein]{Zwanziger:2003cf},
\Eq{eq:Gribov_expval} was generally abandoned to be used for an
exact quantization in favor of an integration over the \emph{fundamental
modular region}~$\Lambda$ \cite{Zwanziger:1993dh}
\begin{equation}
  \label{eq:Lambda_expval}
  \langle\mathcal{O}\rangle_{\Lambda} = \frac{1}{Z} \int_{\Lambda}
  \delta(\partial_{\mu}A_{\mu})\det \MFP[A] \mathcal{O}(A)
  e^{-S_{\textrm{YM}}[A]} \;. 
\end{equation}
(See Ref.~\cite{Zwanziger:2003cf} for a discussion). The fundamental
modular region is the set of unique representatives of 
every gauge orbit, \ie its interior is free of Gribov copies. The
boundary of the fundamental modular region, $\partial\Lambda$,
contains different points that are Gribov copies of each other, but 
which have to be identified topologically \cite{vanBaal:1991zw}. Like
$\Omega$, the fundamental modular region is convex, bounded in every
direction and included in the Gribov region
\cite{Zwanziger:1982na,Zwanziger:1991ac}. The boundary of the Gribov
region, \ie the Gribov horizon $\partial\Omega$, touches
$\partial\Lambda$ at so called \emph{singular boundary points}. 

It is difficult to give an explicit description how to restrict the
functional integration to the fundamental modular region. One usually
introduces the \emph{gauge functional}
\begin{align}
 \label{eq:func_cont}
  F_A[g] &= \int d^4x \Tr \left[
  \orbit{g}{A}_{\mu}\orbit{g}{A}_{\mu}^{\dagger}\right]\\
 \nonumber %\label{eq:func_cont_expansion}
   &= F_A[\identity] - i \int d^4x \Tr\omega\partial_{\mu}A_{\mu} +
   \int d^4x \Tr \omega\partial_{\mu}D_{\mu}\omega + \order{\omega^3}
\end{align}
to characterize the different regions. Here $\orbit{g}{A}_{\mu}$
denotes a gluon field $A_{\mu}$ locally gauge-transformed by $g$ (see
\Eq{eq:gauge_fin_A}). The gauge transformation is parameterized by
$\omega$ as given in \Eq{eq:def_gaugetrafo}. The Gribov region $\Omega$
corresponds to the set of all relative minima of $F_A[g]$ with respect
to local gauge transformations $g$. Those minima which are absolute
characterize the fundamental modular region
\cite{Zwanziger:2003cf}. 

Returning to \Eq{eq:Gribov_expval} and (\ref{eq:Lambda_expval}), it is
of advantage if the integration needs not to be restricted to the
fundamental modular region. That this could be the case, has been argued
recently by \name{Zwanziger} in \cite{Zwanziger:2003cf}. In fact, in
this reference it is stated that functional integrals are
dominated by the common boundary of $\Lambda$ and $\Omega$. Thus, in
the continuum, expectation values of correlation functions over the
fundamental modular region $\Lambda$ are equal to those over the
Gribov region $\Omega$, \ie 
\begin{equation}
  \label{eq:Lambda_exp_Gribov_exp}
  \langle\mathcal{O}\rangle_{\Lambda} = \langle\mathcal{O}\rangle_{\Omega}\;,
\end{equation}
even though the Gribov region is not free of Gribov copies. The 
Gribov copies inside $\Omega$ are argued to not affect expectation
values in the continuum.

This is fortunate for many reasons. For example, in Monte Carlo
simulations of lattice QCD gauge-fixing is usually implemented by an
iterative maximization (or minimization depending on conventions) of a
lattice analogue to the gauge functional\footnote{See
  \Eq{eq:functional} for a definition of 
  the lattice gauge functional as used in this study.}
in \Eq{eq:func_cont}. Therefore, lattice
gauge-fixing automatically restricts to the Gribov region. Note, however,
that it has also been pointed out in \cite{Zwanziger:2003cf} that
on a finite lattice the distinction between the fundamental modular region
and the Gribov region cannot be ignored, but hopefully in the
thermodynamic limit the relation \Eq{eq:Lambda_exp_Gribov_exp} holds.
We have found indications in our data \cite{Sternbeck:2005tk} that
support this assumption to be true. See \Sec{sec:gribov_problem} for
more details.

For our discussion in the next section it is worthwhile to mention that if
the functional integral is cut off at the (first) Gribov horizon, both
the (Euclidean) gluon and propagators have to be positive
\cite{Zwanziger:2002ia}. This is fulfilled by the solutions obtained for
truncated systems of Dyson-Schwinger equations for the gluon and ghost
propagators summarized in the next section. Therefore,
they automatically restrict to the Gribov region, even though 
no direct restriction to $\partial\Omega$ has been done
\cite{Zwanziger:2003cf}.

%------------------------------------------------------------------------------
\section{The Dyson Schwinger equations of QCD}
\label{sec:dse}

Before we start to introduce the lattice approach to QCD
(\Ch{chap:latticeQCD}) and report on the results we have obtained
in studying some infrared properties of Landau gauge gluodynamics
(\Ch{ch:prop_results} -- \ref{ch:spec_FP_operator}), we summarize
results achieved in 
recent years within the DSE approach to QCD. Those results presented
here will then be compared to our lattice data in subsequent chapters. 

The Dyson-Schwinger equations (DSEs) of QCD are infinite towers of
coupled nonlinear integral equations relating different Green's
functions of QCD to each other (see below for two
examples). Generally speaking, the DSEs follow from 
the observation that an integral of a total derivate vanishes. The DSEs can
be directly derived using the generating functional given in
\Eq{eq:Z_QCD} where the existence of a well-defined integration
measure is assumed. 

\begin{figure}
  \centering
  \includegraphics[width=9cm]{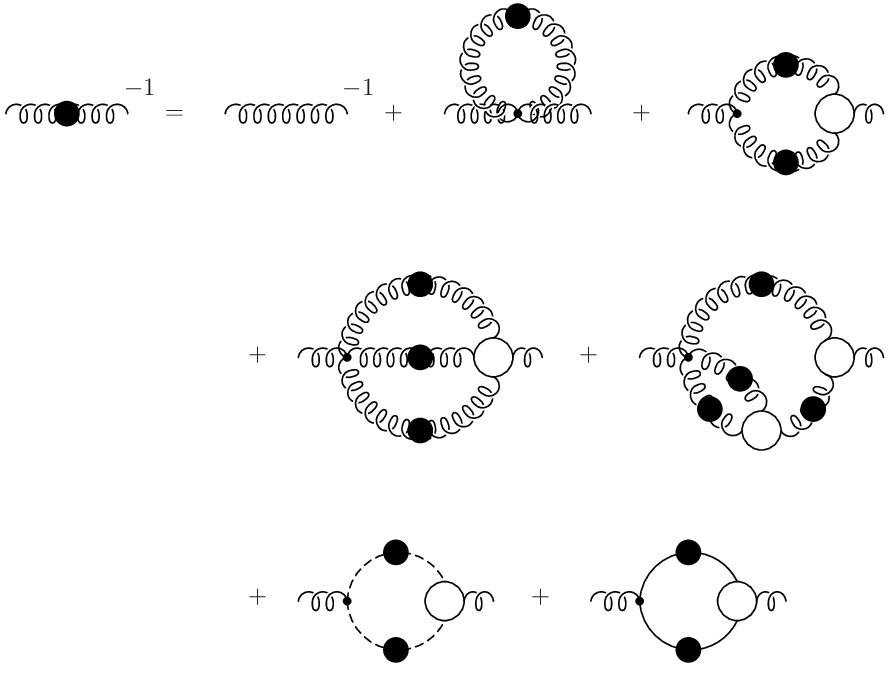}\\*[1cm]
  \caption{Diagrammatic representation of the gluon DSE. 
    Adapted from Ref.~\cite{Alkofer:2000wg}.} 
  \label{fig:DSE_gluon}
\end{figure}

The relevant DSEs for the subsequent discussion are those for the ghost
and gluon propagators. Therefore, these are briefly recalled, but for
an explicit derivation of all of the DSEs of QCD we refer to
Ref.~\cite{Alkofer:2000wg} (see also \cite{Roberts:1994dr}). Following
\cite{Alkofer:2000wg}, the DSE of the (inverse)
ghost propagator takes the form 
\begin{displaymath}
  (G^{-1})^{ab}(k) = -\delta^{ab}\widetilde{Z}_3 k^2 + \gr^2f^{acd}
  \widetilde{Z}_1\int \frac{d^4q}{(2\pi)^4} ik_{\mu} G^{ce}(q)
  \Gamma^{efb}_{\nu}(q,k) D^{df}_{\mu\nu}(k-q)
\end{displaymath}
in momentum space. $G$ denotes the full ghost propagator, $D$ the full gluon
propagator and $\Gamma^{efb}_{\nu}$ the full ghost-gluon vertex. For
other notations used herein we refer to \Ch{ch:intro}. In
\Fig{fig:DSE_ghost} we also give a pictorial representation of the
ghost DSE. This DSE contains the inverse of the tree-level propagator
(dashed line), the tree-level 
ghost-gluon vertex (small filled circle) and the full ghost-gluon vertex (open
circle). The latter is coupled to a fully dressed ghost and gluon propagator
(dashed and wiggled lines carrying a filled circle, respectively).
\begin{floatingfigure}[r]{8cm}
  \includegraphics[width=7cm]{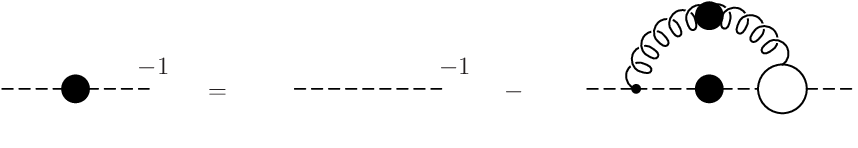}
  \caption{The ghost DSE \cite{Alkofer:2000wg}.}
  \label{fig:DSE_ghost}
\end{floatingfigure}
The gluon DSE is obtained in a similar manner as the ghost DSE, but in
comparison to that, it is much more complex. Therefore, we only
give a pictorial representation of the gluon DSE in
\Fig{fig:DSE_gluon}. Please refer again to Ref.~\cite{Alkofer:2000wg}
for an explicit expression.
As in \Fig{fig:DSE_ghost}, wiggled lines refer to the gluon propagator
in \Fig{fig:DSE_gluon}, whereas a line and a dashed line refer to the quark
and ghost propagator, respectively. Lines carrying a full circle
correspond to fully dressed propagators. Open circles denote full
vertex functions. 

For the DSE of the quark propagator a similar diagram
as for the ghost DSE can be given. See
Ref.~\cite{Alkofer:2000wg} for both an explicit and a pictorial
representation.

%------------------------------------------------------------------------------
\subsection{Infrared behavior of ghost and gluon propagators in 
 Landau gauge} 

If the full solution of QCD's Dyson-Schwinger equations were available
it would provide us with a solution to QCD. However, this has not been
obtained so far. The main impediment is
that the infinite tower of coupled nonlinear
integral DSEs has to be truncated in order to be manageable. Thereby,
the challenge is to use suitable truncations that respect as
much as possible the symmetries of the theory. Doing so, solutions of the
truncated systems might provide deep insights into many
phenomena in hadron physics, because they may serve as input into bound
state calculations based on the \emph{Bethe-Salpeter} equations for
mesons or the \emph{Faddeev} equations for baryons. 

Studying DSEs has become a subject of growing interest in recent years. See
Refs.~\cite{Roberts:1994dr,Roberts:2000aa,Alkofer:2000wg,Maris:2003vk} for a 
comprehensive overview. In particular, in Landau gauge or general
covariant gauges considerable progress has been 
made in studying the low-momentum region for the coupled system of
quark, gluon and ghost propagators. In contrast to earlier attempts
\cite{Mandelstam:1979xd,Atkinson:1981er,Atkinson:1981ah,Brown:1988bn} 
where contributions of ghost fields were neglected, the more recent 
attempts, initiated by \name{von Smekal} \etal
\cite{vonSmekal:1997is,vonSmekal:1998yu}, 
have shown that the inclusion of ghost fields is important for the
generation of a consistent infrared behavior of QCD
\cite{Bloch:2001wz}. In fact, it has been shown in
\cite{vonSmekal:1997is,vonSmekal:1998yu} that in the infrared momentum
region a diverging ghost propagator is intimately related to a
vanishing gluon propagator, both following a power law at low momentum.

We have given explicit expressions for the
gluon and ghost propagators in \Eq{eq:gluonprop_dress}
and (\ref{eq:ghostprop_dress}), respectively. The corresponding
renormalized dressing functions have been denoted by $Z$ and
$J$. According to \cite{vonSmekal:1997is,vonSmekal:1998yu}
these dressing functions are predicted to follow the power laws 
\begin{subequations}
 \label{eq:infrared-gh_gl}
\begin{eqnarray}
  \label{eq:infrared-gh}
  Z(q^2,\mu^2) &\propto& (q^2/\mu^2)^{\kappa_D} \\
 \label{eq:infrared-gl}
  J(q^2,\mu^2) &\propto& 
  (q^2/\mu^2)^{-\kappa_G} \;
\end{eqnarray}
\end{subequations}
in the limit $q^2 \to 0$ with infrared exponents, $\kappa_D$
and $\kappa_G$, satisfying 
\begin{equation}
  \label{eq:kappaD_kappa_G_DSE}
  \kappa_D = 2\kappa_G\;.
\end{equation}
This has been confirmed later by \name{Atkinson} and \name{Bloch}
\cite{Atkinson:1997tu,Atkinson:1998zc} at the level in which the
vertices in the DSEs are taken to be bare. Also investigations of the
DSEs in flat Euclidean space-time performed by \name{Fischer} \etal
\cite{Fischer:2002eq,Fischer:2003zc,Fischer:2002hn} without 
\emph{angular approximations}, as used in the former studies, 
support such an infrared behavior. 
The value of $\kappa_G$ depends on the truncation used, but in Landau
gauge it has been argued that $\kappa_G \approx 0.595$
should be expected \cite{Lerche:2002ep,Zwanziger:2001kw}. 
Thus the ghost propagator is supposed to diverge stronger than
$1/q^2$ and the gluon propagator to be vanishing in the infrared regime. 

These findings are, as we shall see in \Sec{sec:kugo_ojima_def} and
\ref{sec:horizon_condition}, in agreement with the \emph{Kugo-Ojima
confinement criterion} \cite{Kugo:1979gm} as well as with the
\emph{Zwanziger-Gribov horizon condition}
\cite{Zwanziger:2003cf,Zwanziger:1993dh,Gribov:1977wm}. According to
the latter condition, a diverging ghost and a vanishing gluon
propagator result from restricting the gluon fields to the Gribov
region $\Omega$ \cite{Zwanziger:2003cf}.

Note, however, that quite recently \name{Fischer} \etal
have investigated DSEs on a torus. They have found
quantitative differences with the infinite volume results at
small momenta \cite{Fischer:2005ui,Fischer:2005nf}. Moreover,
\name{Bloch} has developed truncation schemes for the DSEs 
of the ghost, gluon and quark propagators
\cite{Bloch:2001wz,Bloch:2002eq} which preserve 
multiplicative renormalizability, in contrast to those
schemes used in the references cited above. Furthermore, \name{Bloch}
has demonstrated 
that a definite conclusion about the existence of infrared 
power-behaved gluon and ghost propagators cannot be reached yet.

We shall show in \Sec{sec:infrared_ghost_gluon} that our lattice data
and also those by others do not confirm \Eq{eq:kappaD_kappa_G_DSE}, at
least for the lattice momenta available at present. The 
values of the infrared exponents have been found to be different.
Therefore, we think it is still an open question whether the infrared
behavior as favored by current DSE studies is realized for lattice QCD in
Landau gauge. However, if it is realized then it has interesting
consequences for the running coupling constant discussed next.

%----------------------------------------------------------------------------
\subsection{A nonperturbative running coupling constant}
\label{sec:theo_running_coup}

From the renormalization of the ghost-gluon vertex, the renormalized
coupling constant at the renormalization point $\mu$ is defined by
\begin{equation}
  \label{eq:alpha_s_mu_ghg}
  \alpha_s(\mu^2) = \alpha(\Lambda^2)\, 
  \frac{Z_3(\Lambda^2,\mu^2)
  \widetilde{Z}^2_3(\Lambda^2,\mu^2)}{\widetilde{Z}^2_1(\Lambda^2,\mu^2) }  \;.
\end{equation}
Here $\alpha(\Lambda^2):=\go^2(\Lambda^2)/4\pi$ denotes the bare
coupling constant that depends on the ultraviolet cutoff $\Lambda$ and
the multiplicative constants, $Z_3$, $\widetilde{Z}_3$ and
$\widetilde{Z}_1$, refer to the renormalization constants of the 
gluon and ghost propagators and of the ghost-gluon vertex,
respectively. Long time ago, it has been shown by \name{Taylor}
\cite{Taylor:1971ff} that to any order in perturbation theory  
the ghost-gluon vertex in Landau gauge is finite (see
\Sec{sec:finite_ghg} for some discussions). Therefore, one may chose 
\begin{equation}
 \label{eq:Z_1=1}
  \widetilde{Z}_1(\Lambda^2,\mu^2) = 1.
\end{equation}

Based on this and by using the relation between the bare and
renormalized ghost and gluon dressing functions
\begin{subequations}
 \label{eq:ren_dressing_to_bare}
 \begin{eqnarray}
   \label{eq:ren_gh_dressing_to_bare}
   J(q^2,\mu^2) &=& \widetilde{Z}^{-1}_3(\Lambda^2,\mu^2)\; 
   J_B(\Lambda^2,q^2)\\
   \label{eq:ren_gl_dressing_to_bare}
   Z(q^2,\mu^2) &=& Z^{-1}_3(\Lambda^2,\mu^2)\; Z_B(\Lambda^2,q^2)
 \end{eqnarray}
\end{subequations}
where $J(\mu^2,\mu^2)=Z(\mu^2,\mu^2) = 1$, one can show that the
product \cite{vonSmekal:1997is,vonSmekal:1998yu}
\begin{equation}
  \label{eq:running_coup}
  \alpha_s(q^2) := \alpha_s(\mu^2) Z(q^2,\mu^2) J^2(q^2,\mu^2),
\end{equation}
is renormalization group invariant and defines a running coupling
constant within the context of DSEs
\cite{Alkofer:2000wg,Alkofer:2004it}. In fact, with
\Eq{eq:alpha_s_mu_ghg}, (\ref{eq:Z_1=1}) and  
(\ref{eq:ren_dressing_to_bare}) it holds that
\begin{displaymath}
  \alpha_s(\mu^2) Z(q^2,\mu^2) J^2(q^2,\mu^2) =
  \alpha(\Lambda^2) Z_B(\Lambda^2,q^2)J^2_B(\Lambda^2,q^2).
\end{displaymath}
Obviously, the right hand side does not depend on the renormalization
point $\mu^2$ and thus the combination of (renormalized) ghost and gluon
dressing functions on the left hand side is
renormalization-group-invariant. Evaluating the left hand side once at
an arbitrarily chosen $\mu^2$ and once at $\mu^2=q^2$ one obtains the
product given in \Eq{eq:running_coup} \cite{Alkofer:2004it}.

The definition of a running coupling constant by
\Eq{eq:running_coup} has 
been first derived in \cite{vonSmekal:1997is,vonSmekal:1998yu}. Later it
has been shown by \name{Bloch} \cite{Bloch:2001wz,Bloch:2002eq} that
after a reformulation of the DSEs for the ghost, gluon and quark  
propagators this coupling constant enters directly the kernel of the
DSEs.

Assuming multiplicative renormalizability to hold beyond
perturbation theory, and assuming also the power
laws (\Eq{eq:infrared-gh_gl}) for the dressing function 
being realized at infrared momenta according to the relation
(\ref{eq:kappaD_kappa_G_DSE}), then $\alpha_s(q^2)$ has a
\emph{finite} infrared fixed point
\cite{vonSmekal:1997is,vonSmekal:1998yu}. The precise value of
$\alpha_s(0)$ depends on $\kappa_G$, but under certain assumption it has
been shown to be \cite{Lerche:2002ep}
\begin{displaymath}
 \alpha_s(0) \approx 8.915/N_c \qquad\textrm{for}\ SU(N_c).
\end{displaymath}

However, we would like to stress already here that our lattice data
for $\alpha_s(q^2>0)$ do \emph{not} indicate the existence of a finite
value at zero momenta (see \Sec{sect:running_coupling}). We have also
found no indications for deviations from 
$\widetilde{Z}_1$ being  constant (\Sec{sec:z1_results}). Our lattice
  data presented in this thesis are in agreement with other lattice
  studies and, most notably, also agree qualitatively with recent
  investigation of DSEs on a torus \cite{Fischer:2005ui,Fischer:2005nf}. 
This, once more, puts the proposed form for infrared power laws into question.

%------------------------------------------------------------------------------
\subsection{The finiteness of the ghost-gluon vertex}
\label{sec:finite_ghg}

\begin{figure}[t]
  \centering
  \includegraphics[width=100\unitlength]{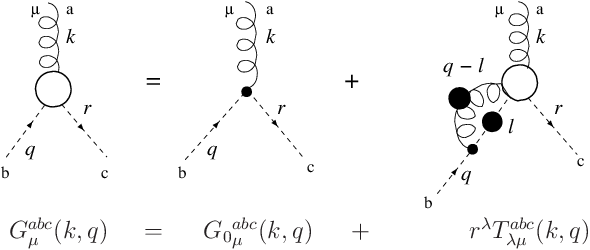}
  \caption{The Dyson-Schwinger equation of the ghost-gluon
    vertex.}
  \label{fig:DSE-ghg}
\end{figure}

Arguments were given first by \name{Taylor} \cite{Taylor:1971ff} that
in Landau gauge the renormalization constant of the 
ghost-gluon vertex $\widetilde{Z}_1=1$. He showed that in this  
gauge the ghost-gluon vertex is finite in the ultraviolet momentum
region and stays bare for a vanishing incoming ghost
momentum. To motivate this we recall arguments given in
Refs.~\cite{Marciano:1977su,Alkofer:2004it} using the
DSE for the full ghost-gluon vertex. A pictorial
representation of this DSE is shown in \Fig{fig:DSE-ghg}. 

Considering this figure, the argument goes at follows
\cite{Marciano:1977su,Alkofer:2004it}: The bare  
ghost-gluon vertex in the interaction diagram (rightmost diagram) is
proportional to the internal loop momentum $l_\mu$. Since in Landau
gauge the gluon propagator $D_{\mu \nu}(l-q)$ is transverse, it holds that 
$l_\mu D_{\mu \nu}(l-q) = q_\mu D_{\mu \nu}(l-q)$ and thus the 
interaction diagram vanishes in the limit $q_\mu \rightarrow
0$ \cite{Alkofer:2004it}. This implies (see \Fig{fig:DSE-ghg})
\begin{displaymath}
  T^{abc}_{\lambda\mu}(k,q) = g_{\lambda\mu}(A^{abc}q^2 +
  B^{abc}k^2 + C^{abc}r^2) + D^{abc}k_{\lambda}k_{\mu} + \textrm{etc.}
\end{displaymath}
as $k^2$, $r^2$ and $q^2\rightarrow0$ so that
$r^{\lambda}T^{abc}_{\lambda\mu}(k,q)$ vanishes
\cite{Marciano:1977su}. Therefore, the full vertex equals the
bare vertex at the subtraction point \mbox{$p^2=k^2=r^2=0$} and there is
no renormalization. This, however, is not true if one renormalizes at
another point $\mu^2>0$, even though it remains true to lowest order
in perturbation theory \cite{Marciano:1977su}. Note also that this
argument would be invalidated if the two-ghost--two-gluon
scattering kernel $T^{abc}_{\lambda\mu}(k,q)$ had an infrared
divergence \cite{Alkofer:2004it}. 

We shall show in \Sec{sec:z1_results} that our lattice results for the 
renormalization constant $\widetilde{Z}_1$ do not support the
existence of such a divergence in the particular $\MOM$
scheme $p^2=r^2$ and $k^2=0$. This was also seen by others
\cite{Cucchieri:2004sq} in 
a similar lattice study of quenched $SU(2)$ gauge theory. Furthermore,
semiperturbative calculations of $\widetilde{Z}_1$ using either
$p^2=r^2=k^2=\mu^2$ or $p^2=r^2; k^2=0$ show that
such a divergence is absent. The analytical study 
presented in \cite{Alkofer:2004it} also shows that the dressing of the
ghost-gluon vertex remains finite if all external momenta vanish.

We also refer to the recent study \cite{Boucaud:2005ce} where arguments 
in favor of a singular infrared behavior of the ghost-gluon
vertex function is given. 

%-------------------------------------------------------------------------
\section{Criteria for confinement in linear covariant gauges}

Despite the success of QCD in describing strong interaction processes
at high energies, there is still the unsolved and theoretically
demanding problem what kind of mechanism confines QCD
degrees of freedom, the quark and gluon states, and keeps them from
being observed in the physical particle spectrum. In contrast to QED,
where a one-to-one correspondence between basic fields and stable
particles can be assumed, in QCD it cannot. Only hadronic amplitudes
are physical and only color singlets can contribute to the
physical state space of QCD. 

For QCD in covariant gauges there is the additional subtlety that it
requires a state space with indefinite metric. In order to
allow for a quantum mechanical interpretation there has to
be some mechanism which generates a subspace of (colorless) physical
states that has a positive semi-definite metric.

In the following three criteria for confinement are briefly
introduced, namely the \emph{Kugo--Ojima confinement scenario}, the
\emph{Gribov--Zwanziger horizon 
condition} and the \emph{violation of reflection positivity} of propagators
corresponding to confined particles.  These criteria are
proposed to be sufficient to indicate confinement; and we will see
in succeeding chapters that our lattice simulations show evidence for 
those criteria to be satisfied in lattice QCD in Landau gauge.

%-------------------------------------------------------------------------
\subsection{The Kugo--Ojima confinement scenario}
\label{seq:KOscenario}

In this section we temporarily switch from the functional formalism to
the covariant operator formalism which is more suitable in the present
context. This formalism takes fields as operators, rather than as
c-numbers, and they satisfy (anti-) commutation relations. Also the notion of
BRST symmetry is important for a formulation of a covariant operator
formalism. For a comprehensive account of this subject we refer to the book
\cite{Nakanishi:1990qm} by \name{Nakanishi} and
\name{Ojima} where most of the material summarized below is
exhaustively discussed.

%--------------------------------------------------------------------------
\subsubsection{Requirements for a physical S-matrix to exist}

Covariant quantum gauge theories require state spaces
$\Vindef$ with indefinite metric. If one can show, however, that (a)
the Hamiltonian operator of a theory is hermitian so that a 
$S$-matrix $S$ exists in 
$\Vindef$ satisfying $\langle S\Psi_1|S\Psi_2\rangle =
\langle\Psi_1|\Psi_2\rangle$ and (b) there is a subspace
$\Vphys\subseteq\Vindef$ which not only is invariant under time
evolution, but also (c) has a positive semi-definite inner product
$\langle\cdot|\cdot\rangle$, then a physical $S$-matrix $\Sphys$ can
be defined in the completed quotient space\footnote{The overline
  denotes the completion of this space, \ie all the limiting
  states of Cauchy sequences are incorporated in this, too. For a proof
  and further details see \cite[A.2]{Nakanishi:1990qm}.} 
\begin{displaymath}
  \mathcal{H}_{\phys} \equiv \overline{\Vphys/\Vo}\;.
\end{displaymath}
$\mathcal{H}_{\phys}$ denotes a Hilbert space with positive definite inner
product, and $\Sphys$ is unitary with respect to this Hilbert space
structure. The subspace \mbox{$\Vo\subset\Vphys$} contains the zero-norm
states of the positive semi-definite subspace $\Vphys$
and is orthogonal to it, \ie $\Vo\perp\Vphys$ \cite[A.2]{Nakanishi:1990qm}.

It can be shown that the first two requirements, (a) and (b), are
satisfied automatically for QCD in covariant gauges if the Lagrangian
density $\Leff^r$ (\Eq{eq:renLeff}) is hermitian\footnote{For this the ghost
  fields must satisfy $c^{a\dagger}=c^a$ and 
  $\bar{c}^{a\dagger}=\bar{c}^a$ \cite{Nakanishi:1990qm}.} 
and the physical subspace $\Vphys$ is defined as the kernel of the
BRST--charge $Q_B$, \ie
\begin{equation}
 \label{eq:Vphys}
  \Vphys = \ker Q_B \equiv \{\Psi\in\Vindef : Q_B\Psi = 0\}
\end{equation}
where $Q_B$ is assumed to be an unbroken generator of BRST symmetry.

The proof of condition (c), the positive semi-definiteness of the
inner product in $\Vphys$, however, is a nontrivial problem and
requires a detailed analysis of the inner product structure of both
the total state space $\Vindef$ and the physical subspace $\Vphys$
\cite{Nakanishi:1990qm}.

%--------------------------------------------------------------------------
\subsubsection{Representations of the BRST algebra}

Such an analysis has been done by \name{Kugo} and \name{Ojima}
\cite{Kugo:1979gm} long ago from the viewpoint of the
BRST--algebra given by
\begin{eqnarray*}
  \left\{Q_B,Q_B\right\} &=& 0, \qquad
  \left[ iQ_c, Q_B \right] = Q_B, \qquad
   \left[Q_c,Q_c\right] = 0.
\end{eqnarray*}
Here $Q_c$ denotes the FP--ghost charge which is assumed to be
unbroken\footnote{See also our discussion concerning the BRST--charge in
  \Sec{sec:brst_intro}.} as it is assumed for~$Q_B$. 

The
BRST--algebra has two types of irreducible representations, namely 
singlet and doublet representations. While a doublet consists of  
a so-called \emph{parent state} $|\pi\rangle$ and a \emph{daughter
state} $|\delta\rangle\equiv Q_B|\pi\rangle\neq0$ (\ie the BRST
transform of $|\pi\rangle$), a BRST--singlet is a state that
is annihilated by $Q_B$ without having a corresponding parent state in
$\Vindef$. 

\name{Kugo} and \name{Ojima} have shown that to each
BRST--doublet there exists always another being FP-conjugate to it,
\ie with opposite eigenvalues of~$iQ_c$. This 
FP-conjugate pair of BRST--doublets is called a \emph{BRST--quartet}.
Any state in $\Vindef$ can be classified to be either a
BRST--singlet or to belong to a BRST--quartet and it can be shown that
this exhausts all possible representations of the BRST--algebra in spaces with
indefinite inner product. We will see below that this classification
is important, in that under certain conditions, colorless asymptotic
states are BRST--singlets which then can be identified with physical particle
states. On the other hand colored asymptotic states are members of a
BRST--quartet and, therefore, do not appear in $S$-matrix elements.

To see this note first that daughter states $|\delta\rangle$
and BRST--singlet states (with no parents) belong to the physical space
$\Vphys$ as defined in \Eq{eq:Vphys}. However, daughter states are
orthogonal to all states $\Psi\in\Vphys$
\begin{displaymath}
  \langle\Psi|\delta\rangle = \langle\Psi|Q_B|\pi\rangle = 0
\end{displaymath}
and so cannot contribute to any element of the physical $S$-matrix 
$S_{\phys}$. All the physical content is contained in the physical Hilbert space
\begin{equation}
 \label{eq:Hphys}
  \mathcal{H}_{\phys} \equiv \overline{\Vphys/\Vo} \simeq \Vindef_s.
\end{equation}
which is isomorphic to the space $\Vindef_s$ of BRST--singlet states. Here
\begin{displaymath}
  \Vo = \operatorname{im} Q_B = \{|\delta\rangle\in\Vindef: |\delta\rangle =
  Q_B|\pi\rangle,|\pi\rangle\in\Vindef\} 
\end{displaymath}
is the set of all zero-norm (daughter) states.

%-------------------------------------------------------------------------
\subsubsection{Confinement of colored asymptotic fields}

Under the assumption of $Q_B$ and $Q_c$ to be unbroken,
the classification of representations of the BRST--algebra can be
translated into the properties of creation and annihilation operators
of \emph{asymptotic fields}. Intuitively, these are understood to
create asymptotic particle states which are observable in scattering
experiments long before and long after collisions. 

Asymptotic states constitute two Fock spaces, $\Vin$ and $\Vout$. 
By the postulate of \emph{asymptotic completeness} they are equal to
the whole Hilbert space~$\Vindef$, \ie 
\begin{displaymath}
  \Vin = \Vindef = \Vout.
\end{displaymath}
This is an important postulate and in particular it guarantees that
any operator is expressible in terms of asymptotic fields
\cite{Nakanishi:1990qm}. For any two states
$|f\rangle,|g\rangle\in\Vindef$ an asymptotic field 
$\phi^{\as}$ is defined as the \emph{weak limit}
\begin{displaymath}
  \lim_{x^0\rightarrow\mp\infty}\;\langle f| \phi^{(r)}(x) - \phi^{\as}(x)|
  g \rangle = 0  
\end{displaymath}
of the corresponding renormalized operator $\phi^{(r)}=Z^{-1/2}\phi$.

Due to confinement there must be some mechanism that causes colored
asymptotic fields, if any, not to contribute to any physical
$S$-matrix element. Such a mechanism, the \emph{quartet mechanism}, has
been proposed by \name{Kugo} and \name{Ojima} \cite{Kugo:1979gm} 
for covariant gauge theories. 

They have analyzed the total state space $\Vindef$ 
as the Fock space of asymptotic fields and showed that any
asymptotic field is either a BRST--singlet or a quartet member,
because no other irreducible representation exists as they have shown
(see above). %[p193].
If combined with the fact that under certain conditions (specified below) the
charge~$Q^a$ of global gauge transformation is BRST--exact\footnote{An
   BRST-exact operator $A$ is a BRST variation $\delta_B$ of another
   operator $B$, i.e. $A$ is of the form: $A=\delta_B
   B\equiv\{iQ_B,B\}$.}
and can be written as (see next subsection)
\begin{equation}
  \label{eq:charge_Qa}
  Q^a = \left\{Q_B, \mathcal{C}\right\}\qquad\textrm{where}
  \quad\mathcal{C} := \int d^3x (D_{0}\bar{c})^a(x).
\end{equation}
then for any physical states $|\Psi_1\rangle,|\Psi_2\rangle\in\Vphys$
it holds that 
\begin{displaymath}
 \langle\Psi_1|Q^a|\Psi_2\rangle = 0.
\end{displaymath}
The charge~$Q^a$ vanishes in the physical Hilbert space
$\mathcal{H}_{\phys}$ defined in \Eq{eq:Hphys}. Consequently, if there
were colored asymptotic fields, they would belong to the BRST--quartet
representations. BRST--singlets, on the other hand, are necessarily
colorless and thus can be identified with physical particles. This is
known as confinement by the quartet mechanism. 

One example is the so called \emph{elementary quartet} that can be easily
deduced from the BRST transformation of $A_{\mu}$ and
$\bar{c}^a$ (\Eq{eq:BRST} or (\ref{eq:rBRST})). Obviously, it consists
of the parent states 
$|A^a_{\mu}\rangle$ and $|\bar{c}^a\rangle$ and of the daughter states 
$|D^{ab}_{\mu}c^b\rangle$ and $|B^a\rangle$. As shown in \cite{Kugo:1995km} the
corresponding (massless) asymptotic states --- they describe
longitudinally polarized gluons, ghost and antighosts
--- also form a BRST--quartet representation and are therefore not
observable in the physical spectrum.

It is also expected that the quartet mechanism applies to
transverse gluon and quark states, as far as they exist
asymptotically. A violation of (reflection) positivity (see
\Sec{sec:violatio_of_pos}) for such states  
entails these are not observable either
\cite{Nakanishi:1990qm,Alkofer:2000wg}. In \Sec{sec:pos_vio} it is  
shown that the lattice gluon propagator in Landau gauge
violates reflection positivity explicitly which thus supports this expectation.

%-------------------------------------------------------------------------
\subsubsection{The Kugo-Ojima confinement criteria}
\label{sec:kugo_ojima_def}

Color confinement by the quartet mechanism can only take place if
there is an unbroken and BRST--exact color charge $Q^a$.

In general, $Q^a$ is a generator of the global gauge symmetry that,
in addition to the BRST symmetry, is left in the gauge-fixed
effective Lagrangian density~$\Leff$. The corresponding symmetry
transformations are given in \Eq{eq:gauge_fin} if there the parameters
$\omega^a$ are taken to be space-time independent parameters. Being a
global symmetry there exist \emph{Noether currents} $J^a_{\mu}$ which
are conserved , \ie 
\begin{equation}
  \label{eq:current_conservation}
  \partial_{\mu}J^a_{\mu}=0\,  .
\end{equation}

As pointed out first by \name{Ojima} \cite{Ojima:1978hy}, these 
currents enter the equation of motion for the gauge 
fields in the form
\begin{equation}
 \label{eq:gauge_current}
  gJ^a_{\mu} + \partial_{\nu}F^{a}_{\nu\mu} = \{Q_B ,D_{\mu}\bar{c}^a\}.
\end{equation}
This equation is usually referred to as the \emph{quantum Maxwell
  equation} in the non-abelian case, because for any physical states
$|\psi_1\rangle,|\psi_2\rangle\in\mathcal{V}_{\phys}$ the classical
Maxwell-type equation $\langle\psi_1|(\partial_{\nu}F_{\nu\mu}^a +
gJ^a_{\mu})|\psi_2\rangle = 0$ holds.

Since the conservation law in \Eq{eq:current_conservation} allows  
adding arbitrary terms of the form $\partial_{\mu}f^a_{[\mu\nu]}$ to
$J^{a}_{\mu}$ with $f^a_{[\mu\nu]}$ being a local antisymmetric
tensor, one could be tempted to define the global charge operators
$Q^a$ as the spatial integral of the current
\begin{displaymath}
  J^{'a}_{\mu} = J^{a}_{\mu} + \frac{1}{g}\partial_{\nu}F^a_{\nu\mu}
\end{displaymath}
such that the BRST--exact expression in \Eq{eq:charge_Qa} is
retrieved. But this naive definition of $Q^a$ is ill-defined due to
massless one-particle contributions to $J^a_{\mu}(x)$,
$\partial_{\nu}F^{a}_{\mu\nu}$ and $\{Q_B,D_{\mu}\bar{c}^a\}$ which
cause the integral to not converge \cite{Kugo:1995km}.

If however these contributions are consistently incorporated in the
definition of $Q^a$, a well-defined expression is obtained
(see \cite{Kugo:1995km} for details). This is important, since with
the Goldstone theorem\footnote{In Ref.~\cite{Kugo:1979gm} various
  versions of the \emph{Goldstone theorem} are given which altogether
  state that the following conditions concerning a conserved current
  $J_{\mu}$ and its global charge $Q$ are equivalent:
  (1) $Q = \int d^3x J_0$ is a well-defined charge;
  (2) $Q$ does not suffer from spontaneous symmetry breaking;
  (3) $J_{\mu}$ contains no discrete massless spectrum:
    $\langle0|J_{\mu}\Psi(p^2=0)\rangle=0$.}
this automatically implies that condition (B) formulated in the original work
\cite{Kugo:1979gm} of \name{Kugo} and \name{Ojima}, namely  
\begin{quote}
  (B)\qquad \emph{$Q^{a}$ is not spontaneously broken},
\end{quote}
is satisfied in the indefinite metric space~$\Vindef$. If furthermore
a certain parameter $\ku^{ab}$, the \emph{Kugo-Ojima confinement
  parameter}, turns out to satisfy condition
\begin{quote}
  (A)\qquad $\ku^{ab}=-\delta^{ab}$,
\end{quote}
then the color charge $Q^a$ takes the BRST--exact form as given in
\Eq{eq:charge_Qa} and color confinement by the quartet mechanism 
\begin{displaymath}
  \langle \Psi_1|Q^a|\Psi_2\rangle = 0 \qquad \Psi_1,\Psi_2\in\Vphys
\end{displaymath}
takes place. But this holds only if condition (A) is
fulfilled. Otherwise we cannot identify BRST--singlet states in the
physical Hilbert space $\mathcal{H}_{\phys}$ with color singlets.

To investigate whether condition (A) is realized, the Kugo-Ojima
confinement parameter $\ku^{ab}$ can be obtained as the zero--momentum limit 
\begin{equation}
 \label{eq:lim_u}
  \ku^{ab}:=  \lim_{p^2\rightarrow 0} u^{ab}(p^2)
\end{equation}
of a function $u^{ab}(p^2)$ which itself may be defined through the
correlation function (see, for instance, \cite{Kugo:1995km} or
 \cite{Alkofer:2000wg})
\begin{equation}
  \label{eq:def_u}
  \int d^4x e^{ip(x-y)} \left\langle D^{ae}_{\mu} c^e(x) \go f^{bcd}
  A^d_{\nu}(y) \bar{c}^{c}(y)\right\rangle =: \left( \delta^{\mu\nu} -
    \frac{p_{\mu}p_{\nu}}{p^2}\right) u^{ab}(p^2).
\end{equation}

To our knowledge a direct determination of $\ku^{ab}$ in terms of a
lattice calculation of $u^{ab}(p^2)$ has never been done. There are a
few explorative lattice studies
\cite{Nakajima:1999dq,Nakajima:2000mp,Furui:2003jr,Furui:2004cx}, but
these are based on data of the ghost renormalization function
$\widetilde{Z}_3$ (see below for the relation between
$\widetilde{Z}_3$ and $\ku$.). The major problem is that in a lattice simulation
$u^{ab}(p^2)$ can be calculated only at finite momenta $p$ and the
data then have to be extrapolated to $p=0$ for which a suitable ansatz
has to be chosen.  

%------------------------------------------------------------------------------
\subsubsection{The ghost propagator in the infrared is related to $\boldsymbol{u(p)}$}

In Landau gauge the calculation of the correlation function
in \Eq{eq:def_u} can be even circumvented, because it has been shown
 \cite{Kugo:1995km} that in this gauge the ghost dressing
 function $J$ is related to $u^{ab}(p^2):=\delta^{ab}u(p^2)$ according to 
\begin{equation}
  \label{eq:ghost_ku_q}
  J(p^2) = \frac{1}{1+u(p^2)+ p^2v(p^2)}
\end{equation}
and $v(p^2)$ is an arbitrary function (see \cite{Kugo:1995km} for a
definition). In the zero-momentum limit this yields
\begin{equation}
  \label{eq:ghost_ku_0}
  J(0) = \frac{1}{1+u(0)}\;.
\end{equation}

Therefore, if condition (A) is realized in QCD then the ghost dressing
function in Landau gauge should diverge for vanishing momenta. Turning the
argument around, if the ghost propagator is found to be more singular than a
simple pole this may serve as a sufficient criterion for the Kugo-Ojima
confinement scenario to be realized.

Based on this arguments the infrared behavior of the ghost propagator
as extracted from the corresponding truncated systems of DSE, has been
stated to realize the Kugo-Ojima confinement scenario
for QCD in Landau gauge. In support of this, in
\Sec{sec:infrared_ghost_gluon} we will 
show that also the ghost propagator as calculated in lattice
simulations diverges stronger than a simple pole, albeit with different
exponent compared to the one found in DSE studies. 

More importantly, however, in \Sec{sec:ko_results} we present data for the
Kugo-Ojima confinement parameter $u^{ab}(p)$ at finite 
momentum $p$ and compare these data to those of the ghost
propagator. The data indicate $u^{ab}(p)$ to become proportional to
$-\delta^{ab}$ in the limit of vanishing momenta. So the Kugo--Ojima confinement
scenario seems to be realized for QCD in Landau gauge.

%------------------------------------------------------------------------------
\subsection{The Gribov--Zwanziger horizon condition}
\label{sec:horizon_condition}

In the previous section we have seen that the infrared behavior of the
ghost propagator is intimately connected to one condition of the Kugo--Ojima
confinement criteria. In fact, if condition (A) is fullfilled then the
ghost dressing function must be infrared divergent.

In coincidence with this, there is another condition, namely the
\emph{Gribov--Zwanziger horizon condition}
\cite{Gribov:1977wm,Zwanziger:1992qr}, that also requires the ghost
dressing function to diverge at vanishing momentum. In fact, the
horizon condition states that 
the ghost propagator $G$ in Landau gauge diverges stronger than $1/p^2$
in the zero-momentum limit\footnote{But only, if the 
  restriction to the Gribov region is done properly.}, \ie 
\cite{Zwanziger:2001kw}
\begin{equation}
  \label{eq:horizon_condition}
  \lim_{p^2\rightarrow0} \left[p^2G(p)\right]^{-1} = 0.
\end{equation}
Basically, this limit is a consequence of the expectation that the 
infrared modes of the gauge fields are very close to the Gribov
horizon\footnote{By definition, the Gribov horizon
  occurs where the lowest nontrivial eigenvalue of the FP operator
  vanishes. For typical configurations on large Euclidean volumes this
  operator is expected to have a high density of eigenvalues near zero
  \cite{Zwanziger:2003cf}.} 
$\partial\Omega$ and hence give rise to an accumulation of small
non-zero eigenvalues of the FP operator
\cite{Zwanziger:1993dh,Zwanziger:1992qr,Zwanziger:1991gz}. Since the
ghost propagator, essentially, is the inverse of this operator, it
must diverge in the infrared.

Remarkably, the horizon condition as given in
\Eq{eq:horizon_condition} allows us to 
renormalize the ghost propagator at $p=0$ in the form of a
nonperturbative formula for the corresponding renormalization constant
$\widetilde{Z}_3$ (see \cite{Zwanziger:2003cf} for details). This
formula disagrees with the usual perturbative expression for
$\widetilde{Z}_3$, but it satisfies the perturbative renormalization-group flow
equation. If used together with the DSE for the
ghost propagator it even yields an infrared anomalous dimension $\kappa_G$  
for it, such that it behaves like \cite{Zwanziger:2003cf}
\begin{displaymath}
  G(p) \sim
  \frac{1}{p^2}\left(\frac{\mu^2}{p^2}\right)^{\kappa_G} 
\end{displaymath}
in the infrared. This is in agreement with the infrared behavior of the
ghost and gluon propagators extracted from their Dyson-Schwinger
equations \cite{vonSmekal:1997is,Alkofer:2000wg,Lerche:2002ep}.

Furthermore, it has been argued by \name{Zwanziger}
\cite{Zwanziger:1991gz,Zwanziger:1991ac} that the gluon propagator in
Landau gauge vanishes in the infrared, \ie
\begin{displaymath}
  \lim_{p^2\rightarrow0} D(p) = 0,
\end{displaymath}
because the infrared components $A(k)$ of the gluon field are suppressed
by the proximity of the Gribov horizon in infrared
directions \cite{Zwanziger:2001kw}.

In our study we will not only check for the infrared limits of both 
the gluon and ghost propagators as given above, but we will also show
that the accumulation of near-to-zero eigenvalues of the FP operator 
increases with enlarging the physical volume (see
\Sec{sec:gribov_zwanziger_results}, \Sec{sec:ko_results} and
\Sec{eq:lowlying_eigenvalues}). Therefore, gauge  
configurations at the Gribov horizon seem to dominate the infrared
properties of lattice Landau gauge theory in the thermodynamic limit.

%---------------------------------------------------------------------------
\subsection{Violation of reflection positivity as a criterion for
  confinement}
\label{sec:violatio_of_pos}

The mechanism for confinement by Kugo and Ojima introduced in
\Sec{sec:kugo_ojima_def}
relies on the existence of an unbroken BRST symmetry
beyond perturbation theory. This, however, has not been proven yet and
thus the Kugo-Ojima scenario should not necessarily apply to
QCD. Nevertheless, the numerical results that we will discuss in
\Sec{sec:ko_results}, but also recent studies of truncated systems of
Dyson-Schwinger equations for the ghost and gluon propagators (see
next section) favor the Kugo-Ojima confinement scenario to be realized
for QCD in Landau gauge.

In any case, there is another particular criterion for confinement
that has been proposed in recent years (see
\eg \cite{Alkofer:2000wg,Alkofer:2003jj}) and is focused on in this
study too, namely the \emph{violation of reflection positivity}. In
fact, reflection positivity is an essential part of the famous
\emph{Osterwalder-Schrader axioms}
\cite{Osterwalder:1973dx,Osterwalder:1974tc} for Euclidean
quantum field theory\footnote{Reflection positivity of lattice gauge
  theory assures that gauge-invariant excitations have a physical
  spectrum \cite{Zwanziger:1991ac}.}. Arbitrary partial sums of Euclidean
$n$-point functions have to fulfill those axioms to ensure
their analytic continuation to the physically interesting functions in
Minkowski space. They thus guarantee the
reconstruction of a G\aa{}rding-Wightman relativistic quantum field
theory. This is important in order to arrive at a physical interpretation.

For the purpose of our study there is no need to go into detail
about all the Osterwalder-Schrader axioms\footnote{The justification
  of this axioms is far from being trivial. For a comprehensive account
on this the reader is referred to the books by Haag \cite{Haag:1992hx}
or Glimm and Jaffe \cite{Glimm:1987ng}}.  We will rather focus on
the notion of reflection positivity which states that an 
Euclidean Green's function (Schwinger function) $\Delta$ has to satisfy
\begin{align}
  \label{eq:positivity}
  \nonumber
  \sum_{n,m}\int
  &\left[\prod_{i=1}^{n}d^4x_i\right]\left[\prod_{j=1}^{n}d^4y_j\right]\; 
   f^*(\Theta x_1,\ldots,\Theta x_n) \\ 
  &\cdot\Delta(\Theta x_1,\ldots,\Theta
  x_n,y_1,\ldots,y_n) f(y_1,\ldots,y_n)\; \ge\; 0.
\end{align}
Here $f$ refers to a complex valued test
function\footnote{Those functions belong to the Laurent Schwartz space
  of infinitely often differentiable functions, decreasing together with
  their derivatives faster than any power as $x$ moves to infinity in
  any direction \cite{Haag:1992hx}.} with support
for positive (Euclidean) times, \ie $f(x_1,\ldots,x_n)=0$ for any
$x^4_i<0$; and $\Theta$ is the reflection operator that acts on $x_i$
according to: $\Theta x_i = (\textbf{x}_i,-x^4_i)$. 

In particular for a generic (Euclidean) 2-point function
\mbox{$\Delta(x,y)=\Delta(x-y)$}, reflection positivity is a necessary and
sufficient condition (see \cite{Alkofer:2000wg}) for the 
existence of a K\"allen-Lehmann representation
\cite{Kaellen:1955cy,Lehmann:1954xi}. This is a spectral
representation of 2-point functions\footnote{This
  representation is very helpful for 
  describing the analytic structure of propagators. Combined with the
  positivity requirements it yields bounds on their asymptotic
  behavior and the magnitude of renormalization constants. See
  e.g. \cite{Weinberg:1995mt,Peskin:1995ev} for more details.}
with positive, but generally unknown, spectral density
$\rho(m^2)$ which in momentum space takes the form
\cite{Alkofer:2000wg,Aiso:1997au,Cucchieri:2004mf} 
\begin{equation}
  \label{eq:Dp}
  \Delta(p) = \int_0^{\infty} dm^2\;\frac{\rho(m^2)}{p^2+m^2}
  \qquad\textrm{with}\quad\rho(m^2)\ge 0.
\end{equation}
The absence of a K\"allen-Lehmann representation for a particular
2-point function is a sufficient condition for confinement of the
corresponding particle, because
then it cannot be interpreted in terms of stable particle states.

Considering the temporal correlator $C(t,\textbf{p}^2)$, the absence
of such a representation can be even formulated more straightforward.
In fact, $C(t,\textbf{p}^2)$ is the Fourier transform of
\Eq{eq:Dp} and takes the form\footnote{Note, $\Delta(p)$ is an even
  function of $p_4$ which simplifies the Fourier transform.}
\cite{Alkofer:2000wg} 
\begin{displaymath}
  C(t,\textbf{p}^2) =\frac{1}{\pi}\int^{\infty}_0\!\!dp_4\cos(p_4t)
     \int_{0}^{\infty}\!\!dm^2
     \frac{\rho(m^2)}{p_4^2 + \omega^2}
   = \int^{\infty}_0\!\! dm^2 \rho(m^2)\frac{\pi}{2\omega}\,e^{-\omega t}
\end{displaymath}
where $\omega^2=m^2+\textbf{p}^2$. After substitution this gives
\begin{equation}
 \label{eq:C_t}
   C(t,\textbf{p}^2) = \int_{\sqrt{\textbf{p}^2}}^{\infty}
    d\omega\;\rho(\omega^2-\textbf{p}^2)\, e^{-\omega t}.
\end{equation}
One clearly sees that if the spectral density $\rho$ is a positive function 
then the temporal correlator $C(t,\textbf{p}^2)$ is positive as
well, but it does not hold in general the other way.
If on the contrary $C(t,\textbf{p}^2)$ is found to be
negative for a certain range in $t$, i.e.
\begin{equation}
  \label{eq:viol_cond}
  C(t,\textbf{p}^2) < 0 
\end{equation}
there cannot be a positive spectral density and thus
reflection positivity is violated. This is an indication for
confinement \cite{Alkofer:2000wg,Cucchieri:2004mf}. 

For the gluon two-point function reflection-positivity
violation, of course, is expected to happen. However, it has never be shown 
explicitly in lattice simulations for the case of $SU(3)$. 
In three-dimensional pure $SU(2)$ gauge theory numerical evidence for
reflection-positivity violation of the lattice Landau gluon
propagators has been given \cite{Cucchieri:2004mf}. This study shows
that also the $SU(3)$ gluon propagator in Landau gauge violates
reflection positivity for the quenched and unquenched case. 

\medskip
We note in passing, that if the gluon propagator $D(p)$ in momentum space
would be infrared vanishing, as expected from the proximity of
the Gribov horizon (see the previous section),
then it would violate reflection positivity \emph{maximally}. This can
be seen from 
\begin{displaymath}
  0 = D(p=0) = \int d^4x\, D(x).
\end{displaymath}
This can only happen if the gluon propagator in
coordinate space, $D(x)$, contains positive as well as negative
contributions of equal integrated strength \cite{Alkofer:2003jj}.

The ghost propagator violates reflection positivity trivially which
can be deduced already from its bare expression. Hence ghosts are
explicitly unphysical indeed. The unphysical spin--statistic
relation of ghost fields already suggests this.

%===============================================================================
%%% Local Variables: 
%%% mode: latex
%%% TeX-master: "Sternbeck"
%%% End:

%-----------------------------------------------------------------------------
\chapter{QCD Green's functions in lattice Landau gauge}
\label{chap:latticeQCD}

\begin{chapterintro}{T}
  his chapter very briefly introduces the lattice regularization of QCD. We
  concentrate on the Wilson formulation of QCD with and without clover
  improved Wilson fermions as employed for this study. After specifying
  some general aspects we focus on lattice QCD in Landau gauge and
  define all relevant gauge-variant observables which are analyzed in
  the following chapters.  
\end{chapterintro}

%-----------------------------------------------------------------------------
\section{Basics of lattice QCD}
\label{sec:latticeQCD}

We know that the very definition of a quantum field theory, as
QCD, requires a regularization which often breaks some of the
underlying symmetries of the classical theory and introduces a
new scale into the theory \cite{Hasenfratz:1998bb}. At present the
only known regularization of QCD beyond perturbation theory
is the lattice
regularization which discretizes Euclidean space-time into a lattice
of points. Thereby, the lattice spacing $a$ serves as a regulator
of the theory that renders all ultraviolet divergences, usually
encountered in QCD, finite. Unlike field theory with a naive
ultraviolet cutoff, the lattice formulation maintains exact gauge
invariance \cite{Muta:1998vi,Bowman:2005zi}. Furthermore, the
lattice regularization offers the possibility to investigate
nonperturbative aspects of QCD in terms of numerical Monte Carlo (MC)
simulations. Therefore, it is a valuable tool for cross-checking results
obtained using other nonperturbative methods, for example the DS
approach to QCD.

It is beyond the scope of this thesis to adequately describe all
aspects of lattice QCD. Therefore, we shall recall only some
basic points important for the following. For a comprehensive
introduction on the subject we refer to standard textbooks
\cite{Creutz:1984mg,Montvay94,Rothe:1997kp,Smit2002ug} as well as to
the lectures 
\cite{Kogut:1982ds,Lepage:1998dt,Gupta:1997nd,Davies:2002cx,
  Luscher:2002pz,Davies:2005as}, to name but a few. Most of the
material summarized below can be found there. To find out about the
most recent developments in the field the proceedings
\cite{Lat2003,Lat2004,Lat2005} of the yearly \name{Lattice Conference}s
provide a good starting point. 

\newpage
\subsection{The lattice and its fields}

\begin{floatingfigure}[r]{3cm}
\hspace{-1.3em}\includegraphics[width=2.8cm]{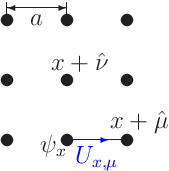}
\end{floatingfigure}
  Very briefly, lattice QCD is a discretization of QCD in Euclidean
space. It replaces the four-dimensional space-time continuum through a
hypercubic lattice and restricts fermion and antifermion fields, $\psi_x$ and
$\bar{\psi}_x$ respectively, to dwell on the lattice sites~$x$. Rather than
specifying the gauge fields by gluon fields 
$A_{\mu}(x)$, on the lattice gauge fields
are associated with links joining adjacent lattice sites,
$x$ and $x+\hat{\mu}$. Here and in the following
$\hat{\mu}$ is a unit vector in the $x_{\mu}$ direction of
space-time. The lattice spacing $a$ is the distance between adjacent
lattice sites. 
The gauge fields --- also known as \emph{link
  variables} or even just as \emph{links} --- are usually denoted 
by $U_{x,\mu}$ and take values in a compact Lie group, here
$SU(3)$. They are the lattice version of the parallel transport matrix
between adjacent sites and therefore are related to the continuum gluon fields
$A_{\mu}(x)$ by the line integral (see \eg \cite{Leinweber:1998uu})
\begin{equation}
 \label{eq:U_exp_int_A}
  U_{x,\mu} \equiv \textsf{P} \exp\left\{ i\go\int^{1}_0 
      A_{\mu}(x+at\hat{\mu})\, dt \right\}\; \simeq\; e^{ia\go
      A_{\mu}(x+\hat{\mu}/2)} + \order{a^3}. 
\end{equation}
Here $\textsf{P}$ denotes path ordering of the gluon fields along the
integration path such that gluon fields
$A_{\mu}(x+at\hat{\mu})$ with larger $t$ stand to the left of those
with smaller $t$. The gauge fields $U$ are taken in place of the gluon
fields for the purpose of maintaining explicit gauge invariance on the
finite lattice \cite{Bowman:2005zi}. A gauge invariant formulation of
lattice QCD directly in terms of the gluon fields is not possible
\cite{Lepage:1998dt}. 

Quantization of lattice QCD is done in the functional integral
formalism, \ie expectation values of different observables 
are given in terms of path integrals over the gauge and quark field
variables \cite{Montvay94}
\begin{equation}
 \label{eq:latt_expectation_value}
  \langle \mathcal{O}\rangle = 
     \frac{1}{Z}\int [\mathcal{D}U,\mathcal{D}\bar{\psi},\mathcal{D}\psi]\,
     \mathcal{O}[\psi,\bar{\psi},U]\; e^{-S_{QCD}[U,\bar{\psi},\psi]}\;.  
\end{equation}
Here the partition function $Z$ is chosen such that $\langle
\identity\rangle$ = 1 and $\mathcal{O}[\psi,\bar{\psi},U]$ denotes an
arbitrary function of the field variables, for instance, $n$ pairs of
quark fields and a product of link variables
\begin{displaymath}
  \mathcal{O}[\psi,\bar{\psi},U] = \psi^{a_1}_{x_1}\cdots
  \bar{\psi}^{b_1}_{y_1}\cdots U_{z_1\mu_1}\cdots  \;.
\end{displaymath}
For simplicity, the indices $a_i$ and $b_i$ denote the set of all
internal symmetry indices, like color and spinor degrees of
freedom. The color indices of link variables are hidden. Lattice sites
are denoted by $x_i$, $y_i$ or $z_i$ and $\mu_i$ refers to one
particular direction on the lattice. 

In contrast to any formulation in the continuum, the lattice action
$S_{QCD}$ maintains exact gauge invariance. Since the gauge fields take
values only in the group $SU(3)$, \ie they are restricted to a
compact manifold, the functional integration measure is well-defined
and gauge-fixing is not necessary if gauge-invariant observables
$\mathcal{O}[\psi,\bar{\psi},U]$ are considered. However, in this
thesis gauge-variant quantities are investigated, in particular in
Landau gauge. Therefore, gauge-fixing becomes necessary here as well. In
\Sec{sec:lattice_landau_gauge} this is discussed in more detail.

\medskip

%---------------------------------------------------------------------
\subsection{The Wilson action with clover-improved fermions}

Apart from the requirement of local gauge invariance the definition of
$S_{QCD}$ is not unique. There are infinite ways to define a lattice
action of QCD, but any definition has to be such that it takes the
classical continuum form for vanishing lattice spacing~$a$. This freedom allows
for a clever reduction of systematic errors caused by finite lattice
spacings. Any reasonable formulation will give the same continuum
theory up to finite renormalizations of the gauge coupling and the
quark masses \cite{Luscher:2002pz}. 

Common actions used in the literature consist of two parts, a gauge
$S_G$ and a fermionic part $S_F$, \ie they are of the general form 
\begin{equation}
  \label{eq:qcd_action}
  S_{QCD} = S_G[U] + S_F[U,\bar{\psi},\psi]\;.
\end{equation}
Both parts depend on the gauge fields $U\equiv\{U_{x,\mu}\}$ and the
fermionic part contains in addition bilinear expressions in the fermion fields
$\bar{\psi}$ and $\psi$ constructed such that the whole action 
is manifestly gauge invariant. 

For the gauge part we have employed the standard Wilson
gauge action \cite{Wilson:1974sk} throughout this study. It is given
by the sum 
\begin{equation}
  S_G[U] :=  \beta \sum_{x} \sum_{1 \le\mu<\nu\le 4}\left( 
    1 - \frac{1}{N_c}\Re\Tr \plaq_{x,\mu\nu}\right)
\end{equation}
over traces of plaquettes denoted here by
\begin{equation}
  \label{eq:plaq}
  \plaq_{x,\mu\nu} :=
  U_{x,\mu}U_{x+\hat{\mu},\nu}U^{\dagger}_{x+\hat{\nu},\mu}U^{\dagger}_{x,\nu}\; .
\end{equation}
It represents a square of four links on the lattice.
The parameter $\beta$ is defined such that $S_G$ takes its classical
continuum form in the limit $a\rightarrow0$, \ie it is defined as
\begin{equation}
  \label{eq:beta}
  \beta := \frac{2N_c}{\go^2}
\end{equation}
where $N_c=3$ for $SU(3)$ and $\go$ is the bare coupling constant.

For the fermionic part $S_F$ there are various definitions in use
which either belong to the family of \emph{Wilson} or of
\emph{Staggered fermion} actions or start therefrom. For the
purpose of this study fermions of the Wilson type have been employed.
In fact, our choice are \emph{clover--improved} Wilson fermions
\cite{Sheikholeslami:1985ij,Luscher:1996sc}.  The
corresponding part of the action can be written as
\begin{equation}
 \label{eq:clover}
  S_F[U,\bar{\psi},\psi] =  a^4\sum_{f,x,y} \bar{\psi}^f_x\,
  Q_{xy}\, \psi^f_y 
\end{equation}
where for each flavor $f=1,\ldots,N_f$ the fermion matrix $Q$ is
defined to act upon the (Grassmann valued) fermion fields,
$\bar{\psi}_x$ and $\psi_y$, according to
\begin{align}
     \sum_{xy} \bar{\psi}_x Q_{xy} \psi_y := 
       \sum_x  \Bigg\{
            \frac{1}{a}\bar{\psi}_x \psi_x
             &- {\frac{\kappa}{a}} \sum_{\mu} \bar{\psi}_x \,
             U^{\dagger}_{x-\hat{\mu},\mu}
                   \left[1 + \gamma_{\mu}\right] \psi_{x-\hat{\mu}}
                                                  \nonumber \\
        &  - \frac{\kappa}{a} \sum_{\mu} \bar{\psi}_x \,
            U_{x,\mu} \left[1 - \gamma_{\mu}\right] \psi_{x+\hat{\mu}}
                                                  \nonumber \\
       &   - \frac{\kappa a }{2} \,c_{sw} \,\go\,  \sum_{\mu\nu}
            \bar{\psi}_x\,\sigma_{\mu\nu} F_{x,\mu\nu}^{\textrm{clover}}
                      \psi_x  \Bigg\}
\label{eq:fermion_matrix}
\end{align}
(see \eg \cite{Gockeler:2004wp}). The fermion fields are normalized
 such that they correspond to the 
continuum fields by rescaling $\psi\rightarrow1/\sqrt{2\kappa}\,\psi$. 
Due to the last term in \Eq{eq:fermion_matrix}, $S_F$ is sufficient to remove all
$\order{a}$ errors for on-shell quantities\footnote{For gauge
  dependent quantities it is an open question whether further, gauge
  non-invariant (but BRST invariant) terms must be added
  \cite{Skullerud:2000un}. However, one usually assumes any 
  such term to be small.}. 
The value of the parameter $c_{sw}$ depends on $\go$ and has to be tuned
appropriately. The clover field-strength tensor is given by
\begin{equation}
   F^{\textrm{clover}}_{x,\mu\nu} := \frac{1}{8i\go a^2} \sum_{\pm\mu,\pm\nu}
       \Big( \plaq_{x,\mu\nu} - \plaq^{\dagger}_{x,\mu\nu} \Big)\,,
\end{equation}
where the definition of the plaquette (see \Eq{eq:plaq}) has been extended
such that the $\mu$, $\nu$ directions can be negative 
\cite{Gockeler:2004wp}. The hopping parameter $\kappa$ is related to 
the (subtracted) bare quark mass via
\begin{equation}
   m = \frac{1}{2a} \left( \frac{1}{\kappa} -
     \frac{1}{\kappa_c} \right)
\label{bare_qm_def}
\end{equation}
where $\kappa_c$ is defined as the value of $\kappa$ at which the pion
mass vanishes. In this study we consider only the cases of either
infinite heavy quarks, \ie no quarks ($N_f=0$) or $N_f=2$ mass
degenerate quarks. The former case is known as the \emph{quenched}
approximation of QCD, whereas the latter is an approximation to the
real, \emph{unquenched} world of two light and a number of heavier
quarks that do respond to the gauge field.

%-----------------------------------------------------------------------------
\subsection{Vacuum expectation values from MC simulations}

After having defined the lattice action we now come back to the
definition of vacuum expectation values in
\Eq{eq:latt_expectation_value} and outline the way Monte Carlo (MC)
simulations are employed to calculate them. First note that the
integral over the fermionic variables can be done instantly due to
their Grassmann nature. We obtain \cite{Montvay94}
\begin{align}
 \nonumber
  \langle \mathcal{O}\rangle &\equiv \left\langle \psi_{x_1}
  \bar{\psi}_{y_1}\cdots \psi_{x_n}\bar{\psi}_{y_n}
  \mathcal{G}[U]\right\rangle \\*[1ex]
  \label{eq:expvalue}
  &= \frac{1}{Z} \int [\mathcal{D}U]\, e^{-\Seff[U]}
   \; \mathcal{G}[U] \sum_{z_1,\ldots,z_n}  
  \epsilon^{z_1,\ldots,z_n}_{x_1,\ldots,x_n}\;
  Q^{-1}_{z_1,y_1}[U]\cdots Q^{-1}_{z_n,y_n}[U] 
\end{align}
where the effective action $\Seff[U]$ (see \Eq{eq:eff_wilson_action})
depends only on gauge fields. $\mathcal{G}[U]$ denotes an arbitrary
function depending on link variables only. The tensor
\mbox{$\epsilon^{z_1,\ldots,z_n}_{x_1,\ldots,x_n}:=1$} 
(\mbox{$\epsilon^{z_1,\ldots,z_n}_{x_1,\ldots,x_n}:=-1$}) if 
\mbox{$z_1,\ldots,z_n$} is an even (odd) permutation of the lattice sites
\mbox{$x_1,\ldots,x_n$}; or zero else. This tensor multiplied with
elements of the inverse fermion matrix $Q^{-1}$ appears due to the 
integration over (Grassmann valued) fermion fields present
in the observable. It is absent if vacuum
expectation values of pure gluonic observables $\mathcal{G}[U]$ are considered. 

Since the fermionic part $S_F$ is usually of the form as given
in \Eq{eq:clover}, the effective action can be written as 
\begin{equation}
 \label{eq:eff_wilson_action}
  \Seff[U] = S_G[U] - \log\det Q[U]
\end{equation}
where in our case the fermion matrix $Q$ is defined in \Eq{eq:fermion_matrix}.
The determinant in \Eq{eq:eff_wilson_action} is known as the \emph{fermion
determinant}. In the quenched approach to QCD this determinant is set
to equal one. This simplifies numerical calculations of expectation values
enormously, however, at the expense neglecting quark loops.

To evaluate the remaining integral in \Eq{eq:expvalue} we
can perform numerical MC simulations which allow us to
estimate vacuum expectation values as statistical averages. Note
that after discretization the generating functional $Z$ of our lattice theory
corresponds to a partition function of a statistical system.

In a typical MC simulation of lattice QCD, sets 
of gauge field configurations $U^{(1)}, U^{(2)}\ldots$ are successively
generated by an appropriately chosen Markov process. Ideally, this process
samples a large number of, say $N$, independent configurations being a 
realization of the \emph{Boltzmann} weight
\begin{displaymath}
  \frac{1}{Z}\exp\left\{-\Seff[U]\right\}\;.
\end{displaymath}
This is known as \emph{important sampling}. On each configuration
$U^{(i)}$ the observable of interest  
$\mathcal{O}$ is measured such that the sample average 
\begin{displaymath}
  \label{eq:MC_average}
  \langle\mathcal{O} \rangle_U := \frac{1}{N}\sum^{N}_{i=1}
  \mathcal{O}[U^{(i)}]
\end{displaymath}
is an estimator of the ensemble average
$\left\langle\langle\mathcal{O}\rangle_U\right\rangle$. The latter 
is equal to the expectation value $\langle\mathcal{O} \rangle$. 
The difference between the ensemble average and its 
estimator $\langle\mathcal{O} \rangle_U$ is important, because the latter
is just an average over a finite sample of
$N$ gauge fields and therefore is naturally afflicted with a
statistical error of $\sigma_{\mathcal{O}}/\sqrt{N}$. Only in the
limit $N\rightarrow\infty$ they would match. Here
$\sigma_{\mathcal{O}}$ refers to the (usually unknown) standard
deviation of the observable $\mathcal{O}$. It can be estimated for
example by the \emph{jackknife} (\eg \cite{Wu:1986yt}) or the
\emph{bootstrap} method \cite{efron}. 

In addition to the statistical error an estimate is also afflicted by
systematic errors due to the finite volume~$V$ and the finite lattice
spacing~$a$, respectively. To estimate those, MC simulations are
usually performed on different lattice sizes and at different setups
of the parameter $\beta$ (and eventually $\kappa$). The latter two are
related to the lattice spacing and the quark mass. With respect to all
these effect, ideally the continuum value is obtained by taking the
multiple limit
\begin{displaymath}
  \langle\mathcal{O} \rangle = \lim_{V\rightarrow\infty}
  \lim_{\myover{a\rightarrow0}{V=\const}}
  \lim_{N\rightarrow\infty}
  \;\langle\mathcal{O} \rangle_U \;.
\end{displaymath}
This limit, in this order, corresponds to the prescription of an
axiomatic field theory.

%-----------------------------------------------------------------------------
\section{The Landau gauge on the lattice}
\label{sec:lattice_landau_gauge}

The major focus of this thesis is to investigate the infrared behavior
of the gluon and ghost propagators and other observables in Landau
gauge. For their calculation gauge-fixing is necessary as well. In the
following we discuss how the Landau gauge condition is imposed in 
our lattice simulations and then we define all observables
relevant for this study.

Gauge-fixing in the continuum usually includes a parameter, the gauge
parameter $\xi$, that causes the measure of the functional integral to
peak around a particular gauge field on the gauge orbit. Lattice
gauge-fixing achieves the same by a two-step process.
First an ensemble of lattice gauge field configurations is generated
using standard MC methods. Since the lattice
action $\Seff$ as defined in \Eq{eq:eff_wilson_action},
is invariant under gauge transformation
\begin{equation}
  \label{eq:gaugetrafo}
  U_{x,\mu} \rightarrow {}^{g}U_{x,\mu}=g_x\,
  U_{x,\mu}\,g^{\dagger}_{x+\hat{\mu}},
\end{equation}
the ensemble generated does not satisfy any gauge 
condition. Then in a second step, for each such generated
configuration $U\equiv\{U_{x,\mu}\}$ a gauge transformation
$g=\{g_{x}\}$ is chosen such that ${}^{g}U_{x,\mu}$ satisfies the
(lattice version of the) gauge condition.  In this way, a particular
configuration on the gauge orbit of $U_{x,\mu}$ is chosen.

\subsection{The gauge functional}

For the particular case of Landau gauge, one usually searches for a
gauge transformation $g=\{g_{x}\}$, keeping $U$ fixed, that maximizes a certain
functional, the gauge functional $F_U[g]$. This typically reads 
\begin{equation}
  F_{U}[g] = \frac{1}{4V}\sum_{x}\sum_{\mu=1}^{4}\Re\Tr \;{}^{g}U_{x,\mu}
  \label{eq:functional}
\end{equation}
where $V$ denotes the lattice volume. Obviously, in the trivial case
of $U=\identity$ the largest value $F_{U}[g]=3$ is obtained for
$g=\identity$. For any other $U$, choosing a maximum of $F_{U}[g]$
makes all $^{g}U_{x,\mu}$ on average as close to unity as
possible. A continuum analog of the gauge functional has been given in
\Eq{eq:func_cont}. 

The functional $F_{U}[g]$ has many different local maxima which can
be reached by inequivalent gauge transformations $g$, the number of
which increases with the lattice size. As the inverse coupling constant
$\beta$ is decreased, increasingly more of those maxima become accessible
by an iterative gauge fixing process starting from a given (random)
gauge transformation $g$. In this study we have employed
two popular algorithms for gauge-fixing: \emph{over-relaxation}
\cite{Mandula:1990vs} and \emph{Fourier--accelerated gauge-fixing} 
\cite{Davies:1987vs}.\footnote{For a comparison of
  both algorithms see \App{sec:fag_vs_rlx}}
The different gauge copies corresponding to 
the maxima reached are called \emph{Gribov copies}, due to their resemblance to
the Gribov ambiguity in the continuum \cite{Gribov:1977wm}. All Gribov
copies $\{{}^{g}U\}$ belong to the same gauge orbit spanned by the Monte
Carlo configuration $U$. They all satisfy the differential Landau gauge
condition (lattice transversality condition)
$(\nabla_{\mu}\orbit{g}{A}_{\mu})(x) = 0$ where
\begin{equation}
  (\nabla_{\mu}A_{\mu})(x)\equiv(\nabla\cdot A)(x) := \sum_{\mu=1}^4
  \Big[A_{\mu}( x+ \hat\mu/2)-A_{\mu}( x- \hat\mu/2)\Big] \;.
  \label{eq:transcondition}
\end{equation}
Here $A_\mu(x+\hat{\mu}/2)$ is
the non-Abelian (hermitian) lattice gauge potential which may be
defined at the midpoint of a link
%\begin{equation}
% A_\mu(x+\hat{\mu}/2)
%  :=\frac{1}{2ia\go}\left(U_{x,\mu} - U^{\dagger}_{x,\mu}\right)
%   -\frac{\identity}{6ia\go}\Tr\left(U_{x,\mu} - U^{\dagger}_{x,\mu}\right)\, .
%\label{eq:A-definition}
%\end{equation}
\begin{equation}
 A_\mu(x+\hat{\mu}/2)
  :=\frac{1}{2i}\left(U_{x,\mu} -  U^{\dagger}_{x,\mu}\right)
   -\frac{\identity}{6i}\Tr\left(U_{x,\mu} -  U^{\dagger}_{x,\mu}\right)\, .
\label{eq:A-definition}
\end{equation}
In this way it is accurate to $\order{a^2}$. Note that this association
between lattice and continuum gauge fields is not unique, but the
one chosen here represents the maximally local choice for such an assignment
\cite{Mandula:1987rh}. The bare gauge coupling $\go$ is related to the
inverse lattice coupling via $\beta=6/\go^2$ in the case of $SU(3)$
(see \Eq{eq:beta}).

In the following, we will drop the label~$g$
for convenience, \ie we assume $U$ to satisfy the Landau gauge
condition such that $g\equiv\identity$ maximizes the functional in
\Eq{eq:functional} relative to the neighborhood of the identity. To
simplify notation we will also use a more compact notation
\begin{displaymath}
    A_{x,\mu} := A_\mu(x+\hat{\mu}/2)
\end{displaymath}
for the lattice gluon fields, but it is always understood that they dwell
at the midpoint of a link.\footnote{This one should keep in mind if the
Fourier transform of the gluon field has to be
calculated.} Additionally, we give the adjoint 
expression of a lattice gluon field
\begin{equation}
  \label{eq:A_a-definition}
  A^a_{x,\mu} := A^a_{\mu}(x+\hat{\mu}/2) = 2\cdot \Im\Tr\{T^a U_{x\mu}\}
\end{equation}

%--------------------------------------------------------------------------
\subsection{The Faddeev-Popov operator}

Before we can go further and introduce the observables relevant for
this study, it is necessary to give first a lattice expression
for the Faddeev-Popov (FP) operator in Landau gauge. This operator can
be easily derived by considering a one-parameter subgroup of the local
$SU(3)$ gauge group defined by (see \eg \cite{Zwanziger:1990by}) 
\begin{displaymath}
g_{\omega}(\tau,x) = \exp\left\{i\tau\omega^c_x T^c\right\}  
\qquad \tau,\omega_x^c\in\mathbb{R}\;.
\end{displaymath}
The generators $T^c$ of the $SU(3)$ group have been defined in
\Sec{sec:class_lagr}. In fact, if we assume $U$ to represent a local
maximum of the gauge 
functional then for any $\tau$ it holds that $F_U[\identity]\ge
F_{U}[g_{\omega}(\tau)]$ for all $\omega$. Consequently, at $\tau=0$ the first
derivative of the one-parameter function
$f_{\omega}(\tau):=F_{U}[g_{\omega}(\tau)]$ with respect to $\tau$
should vanish. One can easily show that
\begin{displaymath}
  0 =\left.\frac{\partial}{\partial \tau} f_{\omega}(\tau)\right|_{\tau=0}
    = \frac{1}{2}\sum_{x,c} \omega^c_x \sum_{\mu}\big[
    A^c_{x-\hat{\mu},\mu} - A^c_{x,\mu}\big]\; , 
\end{displaymath}
and so any maximum of the gauge functional automatically satisfies the lattice  
Landau gauge condition. Then the second derivative of $f_{\omega}(\tau)$ at
$\tau=0$ defines a symmetric quadratic form 
\begin{displaymath}
 \left.\frac{\partial^2}{\partial \tau^2}
    f_{\omega}(\tau)\right|_{\tau=0} = \sum_{x,y,c,d}\omega^{c}_{x}\,
  M^{cd}_{xy}\,\omega^{d}_{y}
\end{displaymath}
whose kernel 
\begin{equation}
  M^{ab}_{xy} = A^{ab}_{x}\,\delta_{x,y} - \sum_{\mu} 
  \left(  B^{ab}_{x,\mu}\,\delta_{x+\hat{\mu},y}
  + C^{ab}_{x,\mu}\,\delta_{x-\hat{\mu},y}\right)
  \label{eq:FPoperator}
\end{equation}
with
\begin{subequations}
\label{eq:FP_ABC_def}
\begin{eqnarray}
  A^{ab}_{x} &=& \mbox{$\sum_{\mu}$}\Re\Tr\left[
    \{T^a,T^b\}(U_{x,\mu}+U_{x-\hat{\mu},\mu}) \right],\\
  B^{ab}_{x,\mu} &=& 2\cdot\Re\Tr\left[ T^b\,T^a\, U_{x,\mu}\right],\\
  C^{ab}_{x,\mu} &=& 2\cdot\Re\Tr\left[ T^a\,T^b\, U_{x-\hat{\mu},\mu}\right]\;.
\end{eqnarray}
\end{subequations}
is the Hessian of $F_{U}[g]$.  $M$ defines a real symmetric matrix
that in the case of $U$ satisfying $\nabla\cdot A = 0$ equals the FP
operator
\begin{displaymath}
  M[U] = -\nabla\cdot D[U] =
  -D[U]\cdot\nabla \quad\Longleftrightarrow\quad  \nabla\cdot A = 0
\end{displaymath}
where $D[U]$ refers to the covariant derivative
\cite{Zwanziger:1993dh}. A lattice definition for $D[U]$ can be found, for
instance, in the same reference. After some algebra one can show that
in the adjoint representation this can be written in the form
\begin{equation}
 \label{eq:covD}
  \Big(D_{\mu}[U]\Big)^{ab}_{xy} =
  2\Re\Tr\left[T^bT^aU_{x,\mu}\right]\delta_{x+\hat{\mu},y} - 
  2\Re\Tr\left[T^aT^bU_{x,\mu}\right]\delta_{x,y} \;.
\end{equation}

%----------------------------------------------------------------------------
\subsection[Defining $\Gamma$, $\Omega$ and $\Lambda$ on the
lattice]{Defining $\boldsymbol{\Gamma}$,  $\boldsymbol{\Omega}$ and  $\boldsymbol{\Lambda}$ on the lattice}

On the lattice the Gribov ambiguity of the Landau gauge condition
finds its expression in the ambiguity to find a local maxima of the
gauge functional in \Eq{eq:functional}. Therefore, terms
like transversal plane, Gribov region and fundamental modular region
also translate to the lattice formulation. 

The transversal plane is constituted by all (gauge transformed)
configurations $U$ that satisfy the lattice Landau gauge
condition $\nabla\cdot A(U)=0$ using the definition
(\ref{eq:transcondition}), \ie
\begin{displaymath}
  \Gamma := \left\{ U : \nabla\cdot A(U)=0 \right\}\;.
\end{displaymath}
The subset of $\Gamma$ whose elements in addition give rise to a
semipositive definite FP operator $M$ is called the Gribov region
\begin{displaymath}
  \Omega := \left\{ U : U\in\Gamma, M[U] \ge 0\right\}.
\end{displaymath}
Of course any element in $\Omega$ is a local maximum of the gauge
functional, but only those which are global maxima constitute the
fundamental modular region 
\begin{displaymath}
  \Lambda := \left\{U: F_U(\identity) \ge F_U[g]\quad\textrm{for all}~g\right\}.
\end{displaymath}
For a finite lattice it has been proven that the interior of $\Lambda$
consists of non-degenerate absolute maxima (or minima depending on the
definition). Gribov copies may only occur on the boundary
$\partial\Lambda$ \cite{Zwanziger:1993dh}. 

%--------------------------------------------------------------------------
\section{Lattice definition of our observables} 

\subsection{The (inverse)  FP operator in momentum space}
\label{sec:inverse_of_FP}

In subsequent sections we shall derive expressions for the numerical
calculation of the ghost propagator and the ghost-gluon-vertex
renormalization constant in momentum space. For theses purposes the following
Fourier transform 
\begin{equation}
 \label{eq:M_inv_k}
   \left(\mathcal{M}^{-1}\right)^{ab}(k) =
   \frac{1}{V} \sum_{x,y} e^{-ik\cdot x}
  \left(M^{-1}\right)^{ab}_{xy}\, e^{ik\cdot y}
\end{equation}
of the inverse FP operator $M^{-1}$ is of interest. Here and in the
following the scalar product
\begin{equation}
  \label{eq:k_x}
  k\cdot x \equiv \sum^4_{\mu=1} 2\pi \frac{k_{\mu}x_{\mu}}{L_{\mu}}
\end{equation}
of lattice momentum $k$ and lattice site $x$ is understood. $L_{\mu}$
denotes the lattice extension in direction $\mu$.

Due to its eight trivial zero eigenvalues the inverse $M^{-1}$ needs
to be defined with 
care. If we are just interested in non-zero momenta $k$
we are automatically in a subspace orthogonal to the space spanned by
the (space-time constant) zero modes. Hence, for non-zero momenta we can apply,
for example, the conjugate gradient method to solve the sparse linear system
\begin{equation}
[M\psi_b]^{cz}\equiv \sum_{a,x} M_{cz,ax} \psi^{ax}_{b} = \xi^{cz}_b(k)
  \label{eq:Mpsi}
\end{equation}
using a fixed source $\xi_b$ with $8V$ complex components
$\xi^{cz}_b(k):=\delta^{cb}e^{ik\cdot z}$. Here $c$ and
$z$ label the vector components of $\xi_b$, while index $b$ specifies
the different sources on the right hand side of \Eq{eq:Mpsi}, \ie
which of the color components are non-zero. The solution $\psi_{b}$ to
this linear system can then be used to write the Fourier transform in
\Eq{eq:M_inv_k} as the following scalar product in space-time\footnote{For our
  convenience, $\psi_{b}$ carries the index $b$ as well, in order to
  trace back  afterwards the color index of non-zero components of
  $\xi_b$.}
\begin{equation}
  \left(\mathcal{M}^{-1}\right)^{ab}(k) = \frac{1}{V} \sum_{x}
  e^{-ik\cdot x}\cdot\psi^{ax}_b(k)\;.
  \label{eq:FT_M_scalar_prod}
\end{equation}
With \Eq{eq:Mpsi} it is clear that $\psi^{ax}_b$ represents
the $8V$ vector components with respect to the matrix multiplication
of $M^{-1}$ with $\xi_b$, \ie 
\begin{displaymath}
 \psi^{ax}_{b}(k) = \sum_{y}M^{-1}_{ax,by}\, e^{ik\cdot y}\qquad   (k>0). 
\end{displaymath}

For the numerical calculation of $\psi^{ax}_{b}$ we rather solve the
two independent linear systems
\begin{eqnarray}
  \label{eq:Mcos}
   \left[M\textsf{c}_b(k)\right]^{cz} &=& \delta^{cb} \cos(k\cdot z)\\
\label{eq:Msin}
   \left[M\textsf{s}_b(k)\right]^{cz} &=& \delta^{cb} \sin(k\cdot z),
\end{eqnarray}
than that given in \Eq{eq:Mpsi}, because
$\psi^{ax}_{b}=\textsf{c}^{ax}_{b} + i\,\textsf{s}^{ax}_{b}$. With
this notation we can write the Fourier transform in \Eq{eq:FT_M_scalar_prod} as
\begin{align}
\label{eq:FT_M_cos_sin}
\nonumber
  \left(\mathcal{M}^{-1}\right)^{ab}(k) = \frac{1}{V} \sum_{x,y}\quad
\cos(k\cdot x)\textsf{c}^{ax}_{b}(k) &+ \sin(k\cdot x)\textsf{s}^{ax}_{b}(k)\\
+ i\left[\cos(k\cdot
  x)\textsf{s}^{ax}_{b}(k)\right.&\left.-\sin(k\cdot
  x)\textsf{c}^{ax}_{b}(k)\right].  
\end{align}

We shall see subsequently that the calculation of both,
$\textsf{c}_{b}$ and $\textsf{s}_{b}$, is even not always
necessary depending on the observable considered. For example, 
for the calculation the ghost-gluon-vertex renormalization constant we
only have to solve \Eq{eq:Msin}. This, as we shall see later, relies
on the fact that the \mbox{FP} operator is symmetric and thus
\begin{eqnarray}
 \nonumber
 \sum_{x}  \cos(k\cdot x)\cdot \textsf{s}^{ax}_{b}(k) 
 &=& \sum_{x,y} \cos(k\cdot x) M^{-1}_{ax,by} \sin(k\cdot y) \\ \nonumber
 &=& \sum_{y,x} \cos(k\cdot y) M^{-1}_{ay,bx} \sin(k\cdot x) \\ \nonumber
 & \stackrel{(M=M^T)}{=}& \sum_{y,x} 
  \sin(k\cdot x)M^{-1}_{bx,ay} \cos(k\cdot y)\\
\label{eq:M_cos_sin}
 \sum_{x}  \cos(k\cdot x)\cdot \textsf{s}^{ax}_{b}(k) &=& \sum_{x}
 \sin(k\cdot x)\cdot \textsf{c}^{bx}_{a}(k) \, .
\end{eqnarray}

%-----------------------------------------------------------------------------
\subsection{The gluon and ghost propagator}
\label{sec:gluon_ghost_latt_def}

%--------------------------------------------------------------------------
\subsubsection{The gluon propagator}

Studying nonperturbatively gauge-dependent quantities on the lattice 
the gluon propagator is perhaps the simplest object to start with.
Given the definition for the lattice gluon fields $A^a_{x,\mu}\equiv
A_\mu(x+\hat{\mu}/2)$ in \Eq{eq:A-definition}, the gluon propagator is
estimated in lattice simulations by the MC average of the corresponding
two-point function
\begin{displaymath}
  D^{ab}_{\mu\nu}(x,y) = D^{ab}_{\mu\nu}(x-y) = \left\langle
    A^a_{x,\mu}A^b_{y,\nu}\right\rangle_U \;.
\end{displaymath}
In this study we are in particular interested in the Fourier transform
of this two-point function which on the lattice is given by
\begin{equation}
 \label{eq:latt_gluon_munu}
  D^{ab}_{\mu\nu}(q(k)) = \frac{1}{V} \left\langle \sum_{x,y}
  A^a_{x,\mu}A^b_{y,\nu}e^{ik\cdot (x+\hat{\mu}/2)}
  e^{-ik\cdot (y+\hat{\nu}/2)} \right\rangle_U\;.
\end{equation}
Note that the term $A^a_{x,\mu}A^b_{y,\nu}$
in general is not translational invariant. However, we have imposed
this invariance for the vacuum expectation value by summing over all
differences $(x-y)$ available on the lattice. If we assume that also on
the lattice the gluon propagator in Landau gauge has the
continuum tensor structure 
\begin{equation}
 \label{eq:latt_gluon_tensor_structure}
  D^{ab}_{\mu\nu}(q) = \delta^{ab}\left(
    \delta^{\mu\nu}-\frac{q_{\mu}q_{\nu}}{q^2} \right)D(q^2)
\end{equation}
then all the physical information is contained in the scalar function
\begin{equation}
 \label{eq:latt_gluon}
 D(q^2) = \frac{1}{N_c^2-1}\sum_{a\mu} D^{aa}_{\mu\mu}(q)
\end{equation}
where $N_c=3$ denotes the number of colors of $SU(3)$. 

%-----------------------------------------------------------------------------
\subsubsection{Connecting a lattice momentum to its continuum counterpart}

Before we proceed with the ghost propagator we can define already
here how the lattice momentum $k$ is related to its continuum
counterpart $q$. In fact, it is well-know that due to lattice
artifacts the tree-level gluon propagator $D^0$ does not simply reproduce 
its continuum expression, but rather has the form
\begin{equation}
 \label{eq:gluon_tree_level_latt}
  D^{0ab}_{\mu\nu}(k) = \delta^{ab}
  \left(\delta^{\mu\nu}-\frac{q_{\mu}(k)q_{\nu}(k)}{q^2(k)}
  \right)\frac{1}{q^2(k)}.
\end{equation}
where $q_{\mu}(k)$ is defined as
\begin{equation}
 \label{eq:def_p}
  q_{\mu}(k_{\mu}) := \frac{2}{a} \sin\left(\frac{\pi k_{\mu}}{L_{\mu}}\right).
\end{equation}
Therefore, one usually employes \Eq{eq:def_p} for translating a
lattice momentum $k_{\mu}$ to the corresponding continuum momentum $q_{\mu}$.
Even more importantly, in Ref.~\cite{Leinweber:1998uu} it has been
verified that the lattice gluon propagator in Landau gauge 
shows \emph{scaling} at $\beta=6.0$ and $6.2$ in the entire
range of studied momenta $q^2$ if these are defined according to
\Eq{eq:def_p}. Also systematic effects with respect to the tensor
structure of the gluon propagator (see \Eq{eq:latt_gluon_tensor_structure})
are reduced. So the definition in \Eq{eq:def_p} is a reasonable definition of
$q_{\mu}$ though it becomes worse at larger $k_{\mu}$. Since we are
interested mainly in the infrared properties of the gluon propagator
we will use this definition of $q_{\mu}$ in our study.

%--------------------------------------------------------------------------
\subsubsection{The ghost propagator}

Beside the gluon propagator we are also interested in the Landau gauge
ghost propagator. Given the lattice definition of the FP operator in
\Eq{eq:FPoperator} this propagator (in momentum space) can be estimated 
in lattice simulations by
\begin{eqnarray*}
  G^{ab}(q^2(k)) &=& \frac{1}{V} \left\langle
    \sum_{x,y}\left(M^{-1}\right)^{ab}_{xy}e^{ik\cdot (x-y)}\right\rangle_U\;.
\end{eqnarray*}
Although the inverse of the FP operator itself is not translational
invariant, the vacuum expectation value (here the ensemble average)
has to be. Therefore, a sum over all possible differences $(x-y)$ at
the same momentum $k$ is taken here again.

Since the continuum ghost propagator is of the form
$G^{ab}(q)=\delta^{ab}G(q^2)$ we are interested in the scalar function 
\begin{equation}
 \label{eq:ghost}
  G(q^2(k)) = \frac{1}{N_c^2-1}\sum_{a}G^{aa}(q^2(k)) = \frac{1}{N^2_c-1}\llgl
    \Tr\mathcal{M}^{-1}(k)\rrgl_U
\end{equation}
where in the last step the expression $(\mathcal{M}^{-1})^{aa}(k)$ as defined
\Eq{eq:M_inv_k} has been used. With respect to the discussion in
\Sec{sec:inverse_of_FP}, the trace can be calculated by using
\Eq{eq:FT_M_cos_sin} and (\ref{eq:M_cos_sin}) which finally yields
\begin{displaymath}
  \Tr\mathcal{M}^{-1}(k) = \sum_{a,x,y} \cos(k\cdot x)\cdot\textsf{c}^{ax}_{a}+
  \sin(k\cdot x)\cdot\textsf{s}^{ax}_{a}
\end{displaymath}
where $\textsf{c}^{ax}_{a}$ and $\textsf{s}^{ax}_{a}$ are solutions to
the two independent linear systems given in \Eq{eq:Mcos} and
(\ref{eq:Msin}). 

In our lattice simulations we solved the two linear systems by applying
the pre-conditioned CG algorithm (PCG) where as pre-conditioning matrix
we used the inverse Laplacian operator $\Delta^{-1}$ with diagonal
color substructure. This significantly has reduced the amount of computing
time as it is discussed in more detail in \App{sect:pcg}.

\subsubsection{Eigenmode expansion of the ghost propagator}

For a better understanding of the infrared behavior it is interesting to
analyze the ghost propagator also by exploiting the  
spectral representation of the inverse FP operator for a given gauge
field $U$ in terms of its real (ascendent) eigenvalues
$\lambda_i$ and its (normalized) eigenvectors $\vec{\phi}_i(x)$
in coordinate space
\begin{equation}
  [M^{-1}(U)]^{ab}_{xy} =
   \sum_{i=1}^N \phi_i^a(x) \frac{1}{\lambda_i}
          \phi_i^b(y)\; .
 \label{eq:Def-ghost-x-by-spectrum}
\end{equation}
Here $\phi_i^a(x)$ are the components of $\vec{\phi}_i(x)$.
Taking the Fourier transformed vectors $\vec{\Phi}_i(k)$
at lattice momentum $k_{\mu}$ and averaging over a Monte Carlo (MC)
generated ensemble of gauge field configurations one can compute the
ghost propagator from truncated mode expansions
\begin{equation}
  G_n(q^2(k)) = \langle G(k|n) \rangle_{U}
  \label{eq:Def-ghost-q-by-spectrum}
\end{equation}
where
\begin{equation}
  G(k|n) = \frac{1}{8}~\sum_{i=1}^n
  \frac{1}{\lambda_i}\,\vec{\Phi}_i(k)\cdot\vec{\Phi}_i(-k)
  \label{eq:contribution}
\end{equation}
denotes the contribution of the eigenvalues and eigenmodes on a given
gauge field configuration. Here the vector and scalar product notation
refers to the color indices. The Fourier momenta $k_{\mu}$
are related to the physical momenta $q_{\mu}(k_{\mu})$ by
\Eq{eq:def_p}.

In fact, if the whole eigenvalue spectrum and all eigenvectors were
known the ghost propagator would be determined completely. However,
this is numerical too demanding. Nevertheless, restricting the sum in
\Eq{eq:contribution} to the $n$ lowest eigenvalues and eigenvectors
($n \ll N=8V-8$), we can figure out to what extent this sum
saturates the full value $G(q^2)$ determined using
\Eq{eq:ghost}. We will see in \Ch{ch:spec_FP_operator} that at lowest
momentum the low-lying part of the FP spectrum gives the major
contribution to $G$.  

\newpage
%-----------------------------------------------------------------------------
\subsection{The Kugo-Ojima confinement parameter}

The next observable we are interested in is the Kugo-Ojima confinement
parameter. According to \Sec{sec:kugo_ojima_def} this parameter
$\ku^{ab}=u^{ab}(0)$ is defined as the zero-momentum limit of
some function $u^{ab}(p^2)$ which itself has been introduced in \Eq{eq:def_u}
using a particular correlation function. On the lattice this requisite
correlation function is of the form
\begin{eqnarray*}
  \mathcal{U}^{ab}_{\mu\nu}(k) &:=& \sum_{x,y}\sum_{c,d,e}
  e^{-ik\cdot (x-y)} \left\langle D^{ae}_{\mu} c^e_x  f^{bcd} A^d_{y\nu}
    \bar{c}^{c}_y\right\rangle_U\\
  &=&  \left\langle \sum_{x,y}\sum_{c,d,e} e^{-ik\cdot x} D^{ae}_{\mu}
    \,\left(M^{-1}\right)^{ec}_{xy}\, f^{bcd} A^d_{y\nu} e^{ik\cdot y}
  \right\rangle_U\;. 
\end{eqnarray*}
Here the (adjoint) lattice covariant derivative $D_{\mu}$ acts upon a
color-space vector as given in \Eq{eq:covD} and the lattice gluon
fields $A^d_{y,\nu}$ are defined according to
\Eq{eq:A_a-definition}. With respect to the 
Lorentz-structure as given in \Eq{eq:def_u} the function
$u^{ab}(q^2)$ is obtained by the sum 
\begin{equation}
 \label{eq:ku_trU}
  u^{ab}\left(q^2(k)\right) = \frac{\go}{N_d - 1}\sum_{\mu}
    \mathcal{U}^{ab}_{\mu\mu}(k)
\end{equation}
over Lorentz indices where $\go$ denotes the bare coupling
constant and $N_d=4$ is the number of dimensions. For the
actual lattice calculation of $u^{ab}(q^2)$ 
we note that the correlation function $\mathcal{U}^{ab}_{\mu\nu}(k)$
can be written as a scalar product of a plane wave and a vector
$\psi_{b,\nu}$ multiplied by the covariant derivative
\begin{displaymath}
  \mathcal{U}^{ab}_{\mu\nu}(k) =  \left\langle \sum_{x} e^{-ik\cdot x}
    \left[D_{\mu}\psi_{b,\nu}(k)\right]^{ax}\right\rangle_U\;. 
\end{displaymath}
Here $\psi_{b,\nu}$ is the solution of the
linear system $M\psi_{b,\nu}(k) = \phi_{b,\nu}(k)$ with
$M$ being the FP~matrix and a source $\phi_{b,\nu}(k)$ defined
by the components 
\begin{displaymath}
  \phi^{cy}_{b,\nu}(k) =\sum_{d} f^{bcd} A^d_{y\nu}\;e^{ik\cdot y}\ .
\end{displaymath}
As for the ghost propagator the pre-conditioned conjugate-gradient
algorithm (see appendix \ref{sect:pcg}) has been employed to extract
$\psi_{b,\mu}$ for each non-zero momentum $k$, separately. Then by
using \Eq{eq:ku_trU} the function $u^{ab}(q^2(k))$ is determined.

%-----------------------------------------------------------------------------
\subsection{Renormalization of propagators}
\label{sec:renorm_latt_obs} 

Quantities like gluon or ghost propagators as they come out from typical
lattice simulations have to be renormalized yet. Assuming
multiplicative renormalization the renormalized continuum 
propagators $D_R$ and $G_R$ are related to the bare, dimensionless
lattice propagators via 
\begin{align*}
  a^2D(a^2q^2) &= Z_3(\mu^2,a^2)D_{R}(q^2;\mu^2)\quad\textrm{and}\\
  a^2G(a^2q^2) &= \widetilde{Z}_3(\mu^2,a^2)G_{R}(q^2;\mu^2).
\end{align*}
The renormalization constant $Z_3$ and $\widetilde{Z}_3$ are determined by
imposing a renormalization condition at some chosen renormalization
scale $\mu^2$. In this study we apply the
$\MOM$ scheme for renormalization according to which the renormalized
propagators equal there tree-level form at some momentum $\mu^2$. Here
for the scalar functions of gluon and ghost propagators it holds that 
\begin{align}
 \label{eq:ren_cond_D}
  D_R(q,\mu)\Big|_{q^2=\mu^2} &= \frac{1}{\mu^2}\\
  \label{eq:ren_cond_G}
  G_R(q,\mu)\Big|_{q^2=\mu^2} &= \frac{1}{\mu^2}.
\end{align}

In the subsequent chapter we consider in the majority of cases not the
propagators themselves but their dressing functions $Z$ and $J$ (see
\Eq{eq:gluonprop_dress} and (\ref{eq:ghostprop_dress})). 
They describe the deviation of the propagators from
their tree-level forms. With \Eq{eq:ren_cond_D} and (\ref{eq:ren_cond_G})
it is clear that at the renormalization point $\mu^2$ the dressing
functions equal one.

%-----------------------------------------------------------------------------
\subsection{The ghost-gluon-vertex renormalization constant}
\label{sec:ghost-gluon-vertex_latt}

Apart from the renormalization constants of the gluon and ghost
propagators we are also interested in $\widetilde{Z}_1$, the
renormalization constant of the ghost-gluon vertex. In Landau gauge,
the most general tensor structure of this vertex with gluon momentum
$s$ and ghost momenta $q$ and $t$ is given by (see \eg
\cite{Schleifenbaum:2004id}) 
\begin{displaymath}
  \Gamma^{abc}_{\nu}(s;q,t) = i\go\left[q_{\nu} \left(f^{abc} +
  A^{abc}\left(s^2;q^2,t^2\right)\right) +
s_{\nu}B^{abc}\left(s^2;q^2,t^2\right)\right] .
\end{displaymath}
Here $A^{abc}$ and $B^{abc}$ are scalar functions which describe the
deviation from the tree-level from. They are assumed to
have the same color structure as in perturbation theory, \ie 
$A^{abc} =: f^{abc}A(s^2;q^2,t^2)$ and $B^{abc}=: f^{abc}B(s^2;q^2,t^2)$.

In a $\MOM$ scheme the renormalized vertex
$\Gamma_R=\widetilde{Z}_1\Gamma$ equals its tree-level 
expression $\mathsf{\Gamma}$ at a renormalization point $\mu^2$. Therefore, if
we consider the particular renormalization scheme $\widetilde{\MOM}$
which is defined by subtracting the vertex function at the asymmetric
point $t^2=q^2=\mu^2$ and $s^2=0$, then the renormalization constant
$\widetilde{Z}_1$ is defined by
\begin{displaymath}
  \Gamma^{abc}_{\nu}(0;q,t)\Big|_{q^2=t^2=\mu^2} =
  \widetilde{Z}^{-1}_1\cdot\mathsf{\Gamma}^{abc}_{\nu}(0;q,t)\Big|_{q^2=t^2=\mu^2}
\end{displaymath}
where the tree-level expression is given by
\begin{displaymath}
  \mathsf{\Gamma}^{abc}_{\nu}(0;q,t) = i\go\,f^{abc}q_{\nu}\;.
\end{displaymath}

Since in this renormalization scheme the tensor structure of the
ghost-gluon vertex boils down to  
\begin{displaymath}
  \Gamma^{abc}_{\nu}(0;q) =
  i\go\,f^{abc}q_{\nu} \Gamma(0;q^2) 
\end{displaymath}
where $\Gamma(0;q^2)=1+A(0;q^2,q^2)$ 
we arrive at the (inverse) vertex renormalization constant given by
\cite{Cucchieri:2004sq} 
\begin{displaymath}
  \widetilde{Z}^{-1}_1\left(\mu^2=q^2\right) = \frac{1}{\go
    N_c(N_c^2-1)}\frac{1}{q^2}\sum_{\nu=1}^4 
  q_{\nu}\sum_{abc}f^{abc}\Im \Gamma^{abc}_{\nu}(0;q). 
\end{displaymath}

On the lattice, the definition for $\widetilde{Z}^{-1}_1$ can be derived in a
similar manner as in the continuum 
(see \eg \cite{Cucchieri:2004sq}). Only the tree-level expression of
the vertex is different. Using the tree-level expression known from lattice
perturbation theory the definition of the ghost-gluon-vertex
renormalization constant is given by \cite{Cucchieri:2004sq}
\begin{equation}
  \widetilde{Z}^{-1}_1\left(q^2(k)\right) = 
  \frac{c}{a^2q^2(k)}\sum_{\nu}\tan\left(\frac{\pi
        k_{\nu}}{L_{\nu}}\right)\sum_{a,b,c} f^{abc}\, 
  \Im\Gamma^{abc}_{\nu}(0,q(k))
 \label{eq:Z1_latt} 
\end{equation}
where the constant $c:=2/(\go N_c(N_c^2-1))$ and $q_{\nu}(k_{\nu})$ is defined in
\Eq{eq:def_p}. Here $L_{\nu}$ refers to the number of lattice points
in direction $\nu$ and $k_{\nu}$ takes values in the interval
$(L_{\nu}/2,L_{\nu}/2]$ as usual.

The vertex $\Gamma^{abc}_{\nu}$ can be obtained by amputating the
external ghost and gluon legs from the three-point function
of gluon, ghost and anti-ghost fields. In the $\MOM$ scheme considered
here this yields \cite{Cucchieri:2004sq}
\begin{equation}
 \label{eq:Gamma^abc_mu}
  \Gamma^{abc}_{\nu}(0,q) =
  \frac{G^{abc}_{\nu}(0,q)}{D(0)G^2(q^2)}
\end{equation}
where $D$ and $G$ refer to gluon and ghost propagators, respectively, 
and $G^{abc}_{\nu}$ is given on the lattice by the MC average
\begin{equation}
  G^{abc}_{\nu}(0,q(k)) =
  \left\langle \mathcal{A}^a_{\nu}(0)
    \left(\mathcal{M}^{-1}\right)^{bc}(k)\right\rangle_{U} \;.
  \label{eq:G^{abc}_{mu}}
\end{equation}
In this expression $\mathcal{A}^a_{\nu}(0)$ denotes the Fourier transform of the
gluon fields $A^a_{x,\nu}$ at zero momentum, \ie
\begin{displaymath}
 \mathcal{A}^a_{\nu}(0)= \frac{1}{V}\sum_{x}A^{a}_{x,\nu} 
\end{displaymath}
with $A^{a}_{x,\nu}$ defined in \Eq{eq:A-definition}. 
$\mathcal{M}^{-1}$ is the inverse of the FP operator in momentum
space considered at non-zero momenta (see \Eq{eq:M_inv_k}). 
If combined with the definition in \Eq{eq:Z1_latt} the
renormalization constant $\widetilde{Z}^{-1}_1(q^2)$ can be estimated by
the MC average
\begin{equation}
 \label{eq:Z^{-1}}
  \widetilde{Z}^{-1}_1(q^2) = \frac{c}{a^2q^2D(0)G^2(q^2)}\sum_{\nu}
  \tan\left(\frac{\pi k_{\nu}}{L_{\nu}}\right) \cdot \mathsf{G}_{\nu}(q(k))
\end{equation}
where $c:=2/(\go N_c(N_c^2-1))$ and $\mathsf{G}_{\nu}$ represents the
average 
\begin{equation}
 \label{eq:mathsf_G_mu}
  \mathsf{G}_{\nu}(q(k))=\sum_{a,b,c} f^{abc}\, \Im G^{abc}_{\nu}(0,q)
 =  \left\langle \sum_{a}\mathcal{A}^a_{\nu}(0)\cdot
     \phi^{a}(k)\right\rangle_{U}.
\end{equation}
$\mathcal{A}^a_{\nu}(0)$ is the same as defined above and
$\phi^{a}(k)$ is given by
\begin{eqnarray}
\nonumber
  \phi^{a}(k) &=& \sum_{b,c} f^{abc}\, \Im \sum_{x,y} e^{-ik\cdot x}
  M^{-1}_{bx,cy}\, e^{ik\cdot y}\\
\nonumber
 &=& \sum_{b,c} f^{abc}\, \sum_{x,y}\left(\cos(k\cdot x)\cdot \textsf{s}^{bx}_{c}
     - \sin(k\cdot x)\cdot\textsf{c}^{bx}_{c}\right)\\
\label{eq:phi^a}
 &=&  \sum_{x,b,c} 2f^{abc} \cos(k\cdot x)\cdot \textsf{s}^{bx}_{c}.
\end{eqnarray}
In the derivation we have used the antisymmetry of $f^{abc}$,
\Eq{eq:FT_M_cos_sin} and (\ref{eq:M_cos_sin}). We see that if we 
calculate the ghost propagator anyway we can use the set of solutions
$\textsf{s}_c$ ($c=1,\ldots,8$), obtained at an intermediate step, to extract the
ghost-gluon-vertex renormalization constant
$\widetilde{Z}^{-1}(q^2)$ at the same momenta as the ghost propagator.

%===============================================================================
%%% Local Variables: 
%%% mode: latex
%%% TeX-master: "Sternbeck"
%%% End:

%--------------------------------------------------------------------
\chapter{Results for lattice QCD Green's functions}
\label{ch:prop_results}

\begin{chapterintro}{I}
  n this chapter results for lattice QCD in Landau gauge, both in the
  quenched and unquenched cases, are presented. We start with a
  discussion of systematic effects due to finite volumes, 
  discretization and the problem of Gribov copies. For this
  we restrict our attention to the quenched approximation. The
  infrared behavior of the gluon and ghost dressing functions is
  analyzed then, along with a discussion of unquenching
  effects. Subsequently, we report on results for the running coupling
  constant and the renormalization constant of the ghost-gluon vertex.
\end{chapterintro}

\section{General prerequisites}

\subsection{Specification of our lattice samples}

In this study we have analyzed pure $SU(3)$ gauge configurations, all
thermalized with the standard Wilson gauge action at three values of
the inverse coupling constant $\beta=5.8$, $6.0$ and $6.2$. The
different lattice sizes studied are given in \Tab{tab:stat_que}. For
thermalization an update cycle of one heatbath and four
micro-canonical over-relaxation steps was used.

In addition, we have analyzed dynamical $SU(3)$ gauge configurations
provided to us by the $\QCDSF$ collaboration\footnote{We thank
  the $\QCDSF$ collaboration, in particular \name{Gerrit Schierholz}
  and \name{Dirk Pleiter}, for giving us access to their configurations via the
  International Lattice Data Grid (ILDG).}. Those 
configurations were generated using the same gauge action,
supplemented with the interaction with two flavors of clover-improved
Wilson fermions. A definition of that action was given in
\Eq{eq:eff_wilson_action}. The different pairs of couplings $(\beta,\kappa)$
are specified in \Tab{tab:stat_dyn} together with the lattice sizes
that have been used.

\subsection{Gauge--fixing}

Each gauge configuration $U$ was transformed into Landau gauge by
searching for a local gauge transformation $g\equiv\{g_x\}$ that
maximizes the functional $F_U[g]$ defined in \Eq{eq:functional}. For
this purpose either one of the two algorithm commonly used, namely the 
\emph{over-relaxation} (RLX) \cite{Mandula:1990vs} method or
\emph{Fourier--accelerated  
  gauge-fixing} (FAG) \cite{Davies:1987vs} were employed.
Both algorithms are iterative in nature and each iteration-cycle
increases the functional $F_U[g]$ until a (local) maximum is
reached\footnote{In \App{sec:fag_vs_rlx} we compare both
  algorithms and analyze how the iteration numbers scale with
  the lattice size.}. As stopping 
criterion not the functional itself, but 
the violation of transversality (see \Eq{eq:transcondition}) was
used. In fact, the iteration process stopped as soon as
\begin{equation}
  \max_x \Re\Tr\left[(\nabla_{\mu} {}^g\!\! A_{x,\mu})(\nabla_{\mu}
  {}^g\!\! A_{x,\mu})^{\dagger}\right] < 
  \varepsilon := 10^{-14}
  \label{eq:stop_crit}
\end{equation}
was fulfilled at each lattice site, \ie the lattice average was even
lower. For some configuration we used 
$\varepsilon=10^{-13}$ which is also appropriate.

%---------------------------------------------------------------------------
\subsection{The \fc{}-\bc{} strategy}
\label{sec:fcbc_strategy}

A subset of our quenched gauge configurations (see \Tab{tab:stat_que}) were
gauge-fixed more than once in order to investigate the influence of the Gribov
ambiguity of gauge-dependent observables. In \cite{Sternbeck:2005tk}
we have investigated such a dependence, following the strategy of
choosing for each gauge 
configuration $U$ the \emph{first} (\fc{}) and the \emph{best} (\bc{})
gauge copy among $\Ncp$ copies. Each copy has been fixed to 
Landau gauge always starting from a new random gauge copy of $U$. As
\emph{best} we have considered that copy with largest functional value
among the $\Ncp$ 
gauge copies. Of course, the first gauge copy is as good as any other
arbitrarily selected gauge copy. 

For each set of \fc{} and \bc{} copies we then have measured the ghost
and gluon propagators as well as the eigenvalue spectrum of the FP
operator (see \Ch{ch:spec_FP_operator}) to determine the systematic
effect caused by the Gribov ambiguity. In the following we call
this particular way of studying the dependence on Gribov copies as the
\fc{}-\bc{} \emph{strategy}. Below we shall also motivate why we think
that our values for $\Ncp$ given in \Tab{tab:stat_que} are
sufficient for this purpose. Of course, the more gauge 
copies one gets to inspect, the bigger the likeliness that the copy
labeled as \bc{} actually represents the absolute maximum of the
functional in \Eq{eq:functional}. But as it is discussed in more detail
below, the expectation value of gauge variant quantities, evaluated on
\bc{} representatives, is converging more or less rapidly with
increasing number~$\Ncp$.

\begin{table*}
  \centering
  \begin{tabular}{c@{\qquad}c@{\qquad}c@{\qquad}c@{\qquad}cc@{\qquad}c}
\hline\hline\rule{0pt}{2.5ex}
    no. & $\beta$ & lattice & $a^{-1}$ [$\GeV$] & $a$ [$\fm$] & \#conf
    & $\Ncp$ \\ 
\hline\rule{0pt}{2.5ex}
    S-1 & 5.8  & $16^4$  & 1.446 & 0.1364 & 40 & 30   \\
    S-2 & \vdots  & $24^4$  &\vdots  & \vdots & 25 & 40     \\
    F-1 & \vdots   & $24^4$   & \vdots & \vdots & 40 & 30   \\
   S-3  & 5.8   & $32^4$  & 1.446  & 0.1364 & 34 & 1    \\*[0.3cm]
   S-4 & 6.0  & $16^4$  & 2.118  & 0.0932 & 40 & 30   \\
    S-5 & \vdots   & $24^4$ & \vdots  & \vdots & 30 & 40   \\
    S-6   & \vdots  & $32^4$  & \vdots &\vdots & 40 & 1 \\
    S-7   &\vdots & $48^4$  &\vdots  & \vdots & 20 & 1 \\
    A-1  &\vdots  & $24^3\!\times\!\phantom{1}48$ & \vdots & \vdots & 30 & 1 \\
    A-2  & \vdots & $32^3\!\times\!\phantom{1}64$  & \vdots &\vdots & 40 & 1\\
    A-3   & \vdots & $16^3\!\times128$ &\vdots  & \vdots & 30 & 1 \\
    A-4   & 6.0  & $24^3\!\times128$ & 2.118  & 0.0932  & 30 & 1 \\*[0.3cm]
    F-2 & 6.2  & $12^4$ & 2.914 & 0.0677 & 150 & 20\\
    F-3 & \vdots  & $16^4$ & \vdots  & \vdots  & 100 & 30\\
    S-8 & \vdots  & $16^4$ &\vdots  & \vdots  & 40 & 30 \\
    F-4 & \vdots & $24^4$ & \vdots & \vdots  & 35 & 30\\
   S-9 & \vdots   & $24^4$ & 2.914 & 0.0677  & 30 & 40 \\
\hline\hline
  \end{tabular}
\caption[The $\beta$ values and lattice sizes used in our
  simulations of the quenched case.]{The $\beta$ values and lattice
    sizes used in simulations of the quenched case. Also
    numbers of configurations 
  used are given. $\Ncp$ specifies the number of different random gauge copies
  considered for each configuration. The 5th (and 4th) row lists the (inverse)
  lattice spacings corresponding to $\beta$. Labels in the first row are
  used in the text.}
  \label{tab:stat_que}
\end{table*}
\begin{table*}
\centering
\begin{tabular}{cccccccc}
\hline\hline\rule{0pt}{2.5ex}
  no. & $\beta$ & $\kappa$ & $\kappa_c$ & $ma$ & $a$ [$\fm$] &
  $a^{-1}$ [$\GeV$] & \#conf \\ 
\hline\rule{0pt}{2.5ex}
  D-1   & 5.29 & 0.13500 & 0.13641(9) & 0.03828 & 0.0957 & 2.063 &
  90 \\
  D-2  & 5.29 & 0.13550 & 0.13641(9) & 0.02462 & 0.0898 & 2.196 & 60 \\
  D-3 & 5.29 & 0.13590 & 0.13641(9) & 0.01376 & 0.0850 & 2.320 & 55 \\
  D-4  & 5.25 & 0.13575 & 0.13625(7) & 0.01352 & 0.0904 & 2.183 & 60 \\
\hline\hline
\end{tabular}
\caption[The $\beta$ and $\kappa$ values of all dynamical
  gauge configurations used in this study.]{The $\beta$ and
    $\kappa$ values of all dynamical 
  gauge configurations used in this study. The $\kappa_c$ values are 
  taken from Ref.~\cite{Gockeler:2005rv} where also the values for
  the Sommer scale in lattice units $r_0/a$ are specified. The latter
  were used to assign physical units to 
  $a$. We also give values for \mbox{$ma=1/2(1/\kappa-1/\kappa_c)$}. 
  The lattice size is $16^3\times32$ for the first (D-1) and $24^3\times48$
  for the other three sets (D-2, D-3, D-4). The numbers of
  configurations used are given in the last row. For all sets
  $\Ncp=1$. Labels in the first row are used in the text.}
\label{tab:stat_dyn}
\end{table*}

%-----------------------------------------------------------------------------

\subsection{Selection of momenta}
\label{sec:cylindercut}

Our implementation of the lattice gluon propagator first constructs
the lattice gauge fields $A^a_{x,\mu}$ according to
\Eq{eq:A_a-definition} and then Fourier-transforms these using a
Fast-Fourier transformation (FFT) algorithm\footnote{For all FFTs
  we have employed the \code{FFTW}-library \cite{FFTW98},
  see also online: \link{http://www.fftw.org}.}.
Because a FFT provides us with all lattice momenta $k_{\mu}$ at once, the
data for the gluon propagator $D(q^2(k))$ have been determined for all
momenta available.

It is obvious from \Eq{eq:def_p} that different lattice momenta
$k_{\mu}$ give rise to the same value $q^2(k)$. Naively, one would
average over all data of $D(q^2(k))$ at different $k_{\mu}$ but same
$q^2(k)$. This however leads to systematic errors due to finite volume
and discretization effects. It has been shown
\cite{Leinweber:1998uu} that both of these 
systematic errors can be reduced by applying two cuts on the
data, \ie only a subset of momenta is used. But within this subset, all data
with same $q^2(k)$ are then averaged over all different realizations
of $k_{\mu}$ and configurations. 

One of these cuts, known as the \emph{cylinder cut}
\cite{Leinweber:1998uu}, reduces errors due to finite lattice
spacings. Conceived in general terms, it selects data of
$D(k)=D(q^2(k))$ with $k$ lying in a cylinder with radius of one momentum unit
along one of the (lattice) diagonals
$\hat{n}=1/2(\pm1,\pm1,\pm1,\pm1)$. To be specific, 
\begin{equation}
 \label{eq:cylindercut}
  \left[\sum_{\mu=0}^4 \left(\frac{k_{\mu}}{L_{\mu}}\right)^2\right] -
  \left[ \sum_{\mu=0}^4\frac{k_{\mu}n_{\mu}}{L_{\mu}}\right]^2 \le \frac{1}{L^2_s}
\end{equation}
where $L_s$ is the extension in spatial direction. In the special case
of $L_T=L_s$, \Eq{eq:cylindercut} reduces to $\sum_{\mu}k^2_{\mu} -
(\sum_{\mu}k_{\mu}\hat{n}_{\mu})^2\le 1$.  In agreement with
\cite{Leinweber:1998uu} this recipe has drastically reduced lattice
artifacts for the gluon propagator, in particular for larger momenta.

The other of the two cuts is known as the \emph{cone cut}
\cite{Leinweber:1998uu}. It addresses finite volume errors by
removing all data $D(k)$ with one or more vanishing momentum
components $k_{\mu}$. We have applied this cut to our data only
for smaller lattice volumes.  In the next section we shall show in more 
detail how finite volume effects have influenced our data at lower
momenta. In particular, if asymmetric lattice geometries are used the
cone cut is necessary.

For the ghost propagator and some other quantities considered in this
thesis, the numerical calculation involves an individual inversion of the FP
operator for each vector~$k$. Due to limited computing time we did not had
the chance to obtain data for all momenta allowed by the cuts. However,
our selection of different~$k$ were guided by either cuts.

%-------------------------------------------------------------------------
\subsection{Mapping to physical units}
\label{sec:physical_units}

The $\beta$ values chosen for the quenched case allow us to apply
the results of Ref.~\cite{Necco:2001xg}, giving a
parameterization of the functional dependence of the
lattice spacing $a$, or better of $\ln(a/r_0)$, on $\beta$. Using such
parameterization and the Sommer scale $r_0=0.5~\fm$
\cite{Sommer:1993ce} we obtain for 
the three values $\beta=5.8$, 6.0 and 6.2  
~$a^{-1}=1.446~\GeV$, $2.118~\GeV$ and $2.914~\GeV$ or 
$a=0.1364~\fm$, $0.09315~\fm$ and $0.0677~\fm$, respectively. In our opinion,
this mapping between $a$ and $\beta$ is more appropriate as
another one formerly used by us and others (see
\cite{Sternbeck:2004xr,Sternbeck:2004qk,Silva:2004bv,Leinweber:1998uu}). 

For our sets of unquenched gauge configurations the
values of $r_0/a$ were provided to us by the $\QCDSF$ collaboration (see
Table~II in Ref.~\cite{Gockeler:2005rv}). Using again $r_0=0.5~\fm$ we
can assign to each pair of $\beta$ and $\kappa$ a lattice spacing
in physical units. These are given in \Tab{tab:stat_dyn} together with other
specifications.

Finally, with the help of \Eq{eq:def_p} we can then map lattice
momenta~$k$, or better $a^2q^2(k)$, to physical momenta $q^2$. If not
otherwise stated, $q^2$ is always given in GeV$^2$.

%-----------------------------------------------------------------------------
\section{Systematic effects on gluon and ghost
  propagators at low momentum}

In this section different systematic effects on the gluon and ghost
propagators are discussed. We start with effects due to finite
volume and finite lattice spacings, followed by a warning to refrain
from using quite asymmetric lattice geometries. Finally,
attention is paid to the dependence on Gribov copies.

%----------------------------------------------------------------------------
\subsection{Finite volume and discretization effects}
\label{sec:systematic}

Precisely because we have applied the above-mentioned cuts to our data, it is 
quite natural to analyze here the different systematic effects on the
gluon and ghost propagators of changing either the lattice spacing $a$
or the physical volume~$V$. However, due to the preselected set of momenta for
the ghost propagator and the three chosen $\beta$ values, our study
is partial and limited to a region of intermediate
momenta. For the gluon propagator this has 
been done in more detail by other authors (see \eg~\cite{Bonnet:2001uh}).

%----------------------------------------------------------------------------
\subsubsection{Finite volume effects}

\begin{figure}[tb]
  \centering
  \includegraphics[width=\textwidth]{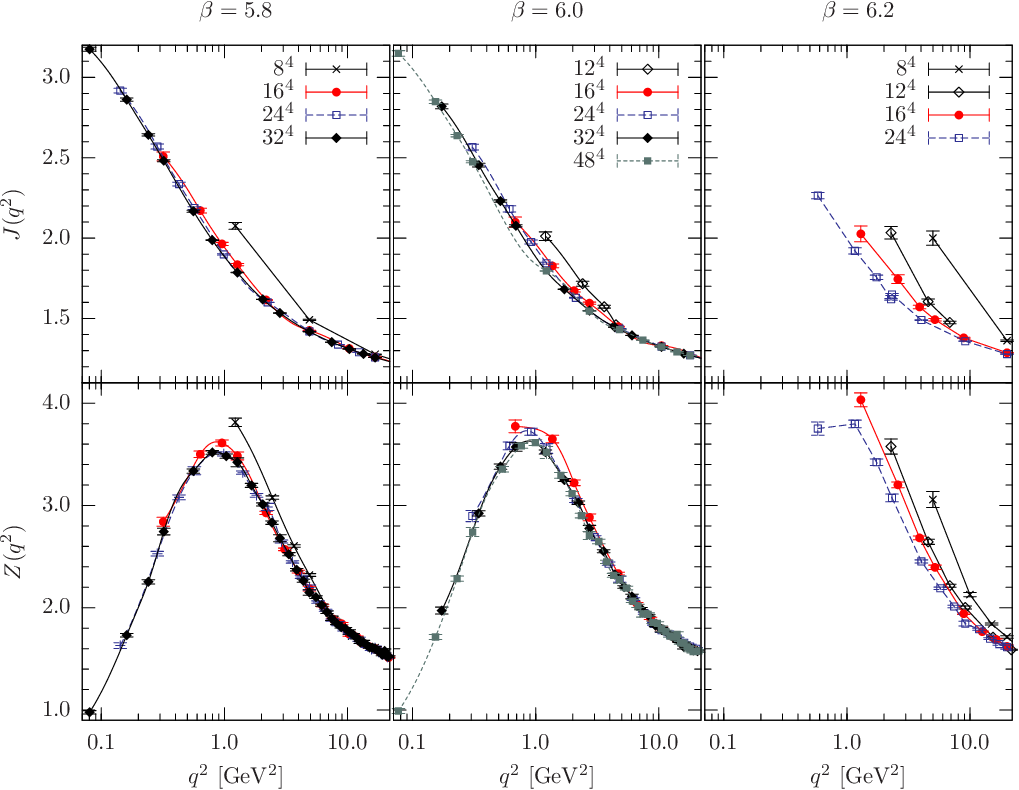}
  \caption{The ghost (upper panels) and gluon (lower panels)
  dressing functions for different lattice sizes as functions of
  momentum $q^2$. From left to right the panels show data at
  $\beta=5.8$, 6.0 and 6.2. Only data on \fc{} gauge copies are shown
  here. Lines are drawn to guide the eye.} 
  \label{fig:gh_gl_V}
\end{figure}

Keeping first the lattice spacing fixed we have found that both the ghost
and gluon dressing functions calculated at the same physical momentum $q^2$
decrease as the lattice size is increased. This is illustrated
for various momenta in \Fig{fig:gh_gl_V}. There both
dressings functions versus the physical momentum are shown for
different symmetric lattice sizes at $\beta=5.8$, 6.0 and 6.2. In
contrast to our  
study \cite{Sternbeck:2005tk} here we show data obtained on \fc{}
gauge copies, because the larger lattice sizes were manageable
only without repeating gauge-fixing several times.

In this figure we have not dropped data with vanishing momentum
components $k_{\mu}$ (\ie those excluded by the cone cut) to emphasize
the influence of a finite volume on those (low) momenta. We also show
data from simulations on a $8^4$ and $12^4$ lattice. One clearly sees
that the lower the momenta the larger the effect due to the finite
volume. In comparison with $\beta=5.8$ and 6.0 this is even more
drastic at $\beta=6.2$. At this $\beta$ the lattice spacing is about
$a=0.06~\fm$. Thus the largest volume considered at $\beta=6.2$ is
about $(1.4~\fm)^4$, which is even smaller than the physical volume of
a $16^4$ lattice at $\beta=5.8$.

In summary we can state that for both dressing functions finite volume
effects are clearly visible at volumes smaller than $(2.2~\fm)^4$,
which corresponds to a $16^4$ lattice at $\beta=5.8$. The effect grows
with decreasing momentum or decreasing lattice size (see the right panels in
\Fig{fig:gh_gl_V}). At larger volumes, however, the data for $q>1~\GeV$
coincide within errors for the different lattice sizes (left and middle
panels). For $q<1~\GeV$ we find only small finite volume effects
for both dressing functions at the lowest momentum if data
obtained on a $24^4$, $32^4$ and a $48^4$ lattice at $\beta=5.8$
and 6.0 are considered.

\begin{figure}[bt]
  \centering
  \includegraphics[width=0.8\textwidth]{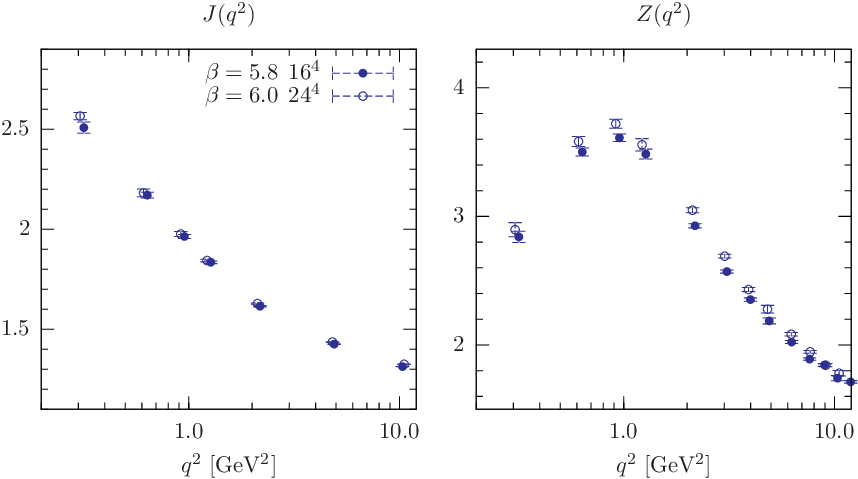}
  \caption{The ghost (left) and gluon (right) 
    dressing functions are shown as functions of momentum~$q^2$ at
    (approximately) fixed physical volume \mbox{$V\approx(2.2~\fm)^4$}.
    The data at $\beta=5.8$ (6.0) correspond to a
    lattice spacing of about $a=0.136~\fm$ ($0.093~\fm$). Only
    data obtained on \fc{} gauge copies are shown here.}
  \label{fig:gh_gL_a}
\end{figure}

%----------------------------------------------------------------------------
\subsubsection{Discretization errors}

Based on our chosen $\beta$ values and lattice sizes we can pick up
equal physical volumes (with different coarseness) only
approximately. Hence also the physical momenta are only approximately
the same if the ghost and gluon dressing functions are compared at
different $\beta$, \ie at different lattice spacings. Therefore, it is
difficult to analyze the systematic effect of changing~$a$ if for both
dressing functions this leads to small hidden variations in~$q^2$. Consequently,
in \Fig{fig:gh_gL_a} we show data for the ghost and gluon dressing
functions obtained at approximately the same physical volume
\mbox{$V\approx(2.2~\fm)^4$} for two different~$a$ as functions
of~$q^2$. This allows us to disentangle by inspection a change of data due to
varying~$a$, given the physical dependence of the propagators on~$q^2$. 
Looking at \Fig{fig:gh_gL_a} one concludes that the gluon dressing
function at the same physical momentum and volume increases with
decreasing the lattice spacing. A similar effect (beyond error bars)
is not observable for the ghost dressing function.

%----------------------------------------------------------------------------
\subsection{Asymmetric lattices cause strong systematic errors}

Besides discretization and finite volume errors, there are also
systematic effects involved using asymmetric lattice geometries. 
It could be tempting to use an asymmetric lattice
such that in one direction, for example in time directions, the
lattice extension is much longer than in all other
directions. Going this way, one would tend to believe that much lower
momenta can be studied compared to using symmetric lattice geometries at same
computational costs. This approach has been pursued for example in
\cite{Silva:2005hb,Oliveira:2005hg,Silva:2005hd,Oliveira:2004gy} aiming at the 
infrared behavior of the gluon propagator, in particular, at the
determination of an infrared exponent for this propagator.

\begin{figure}[tb]
  \centering
  \mbox{\includegraphics[width=0.48\textwidth]{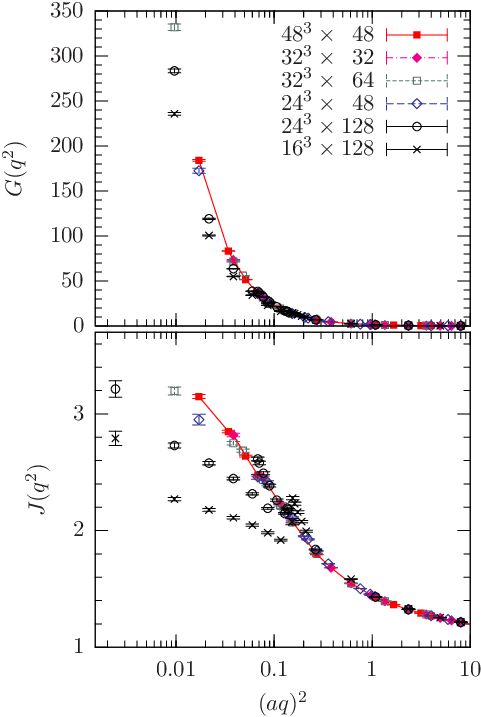}\quad
        \includegraphics[width=0.48\textwidth]{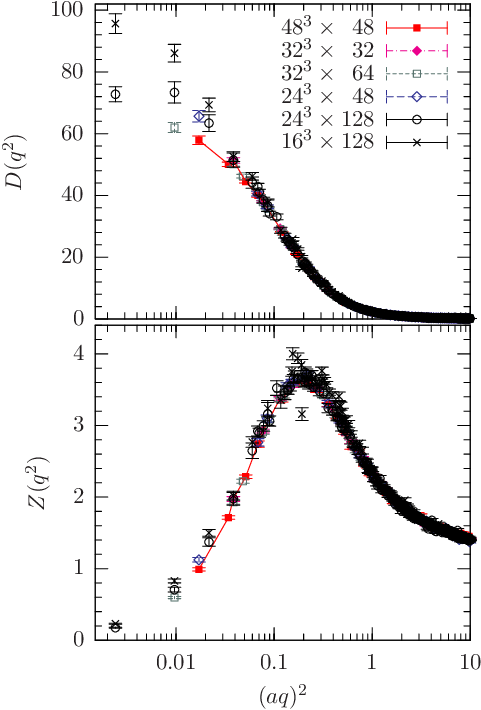}}
  \caption{The ghost $G$ and gluon propagators $D$
    (upper panels) and the corresponding dressing functions (lower panels) are
    shown at $\beta=6.0$ as functions of the momentum in lattice units.
    The data were obtained using different lattice geometries.
    Lines connecting data points for symmetric lattice geometries are
    drawn to guide the eye.}
  \label{fig:gh_gl_qq_asym}
\end{figure}

However, as a comparison of our data obtained either for symmetric or
asymmetric lattice geometries suggests: \emph{``There is no 
  free lunch!''}. 
The more asymmetrically a lattice has been chosen, the larger
are the systematic errors encountered by that. To make this point
more clear, we have calculated both the ghost and gluon propagators
not only on symmetric lattices as discussed above, but also on
lattices of sizes $32^3\times64$, $24^3\times48$, $16^3\times128$ and 
$24^3\times128$ all at the same $\beta=6.0$ (see runs A-1 to A-4 in
\Tab{tab:stat_que}). The corresponding data are plotted in
\Fig{fig:gh_gl_qq_asym} together with those obtained on a $32^4$ and a
$48^4$ lattice at the same value of $\beta$ (S-6, S-7 in
\Tab{tab:stat_que}). The lower panels of this figure show the
corresponding dressing functions. 

In \Fig{fig:gh_gl_qq_asym} we clearly see that data obtained on a
$16^3\times128$ lattice suffer under the largest systematic errors in
the low momentum region compared to the other data. In fact, 
the data on a $32^3\times64$ lattice, shown using open squares in this
figure, exhibit the smallest asymmetry effects.
There only the data point at the lowest on-axis momentum for the
ghost propagator is lower in tendency than  
an extrapolation including the lowest momentum 
data point referring to a $48^4$ lattice would
give. For the gluon propagator this deviation is even smaller. Also
for data on a $24^3\times48$ lattice, shown using open diamonds, the asymmetry
effect is small. There only the data point at the lowest on-axis momentum
$k=(0,0,0,1)$ is lower (larger) than the corresponding result for the
ghost (gluon) propagator on a $48^4$ lattice at the same momentum
$a^2q^2(k)$. At larger momenta this effect is negligible within errors
for these particular sets of data ($24^3\times48$ and $32^3\times64$ lattice). 

Considering instead data generated on the lattices sizes $16^3\times128$ and
$24^3\times128$, the results at the six or seven lowest momenta are 
significantly different to what is expected from the data obtained on
symmetric lattices. A data inspection yields that
the distorted momenta are only the on-axis momenta
\mbox{$k=(0,0,0,k_t)$} allowed by a cylinder cut, but forbidden by a cone
cut (see \Sec{sec:cylindercut} for a definition of these cuts).

Similar systematic effects due to asymmetric lattice geometries have
been reported recently for the gluon propagator in three-dimensional
pure $SU(2)$ gauge theory \cite{Cucchieri:2006za}, but in fact already
in the well-known reference \cite{Leinweber:1998uu} for pure $SU(3)$
gauge theory. Therefore, extractions of an infrared exponent
for the gluon propagator using only on-axis momenta on asymmetric
lattices (see \eg the references mentioned above) should be taken with
reservations.

We note in passing that we also see in \Fig{fig:gh_gl_qq_asym} points 
from asymmetric lattice geometries at not so small momenta which are
slightly larger (lower) compared to those for the ghost (gluon)
propagator on symmetric lattices. With respect to our discussion of
finite volume effects above, we can, however, interpret this
observation more generally as a finite volume effect caused by the
lower spatial volumes of these asymmetric lattices.  

%--------------------------------------------------------------------------
\subsection{Facing the problem of Gribov copies}
\label{sec:gribov_problem}

Apart from the systematic effects inherent in any lattice study, it is
known that the presence of Gribov copies may also causes a systematic
error on the data of gauge-variant quantities. Below it is shown that
this is indeed the case for the ghost propagator, while the influence
on the gluon propagator seems to be negligible (\ie hidden in the
statistical noise). In \cite{Sternbeck:2005tk} we have investigated
this following the \fc{}-\bc{} strategy introduced in
\Sec{sec:fcbc_strategy}.  For each set of \fc{} and \bc{} copies we
have measured the ghost and gluon propagator to determine the
systematic error caused through the Gribov ambiguity.

%-------------------------------------------------------------------------
\subsubsection{Estimating the number of gauge copies}

\begin{figure*}
\centering
\mbox{\includegraphics[height=0.4\textheight]{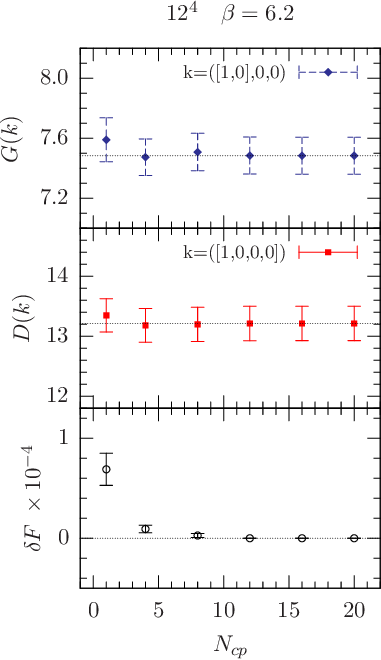}\qquad\quad
  \includegraphics[height=0.4\textheight]{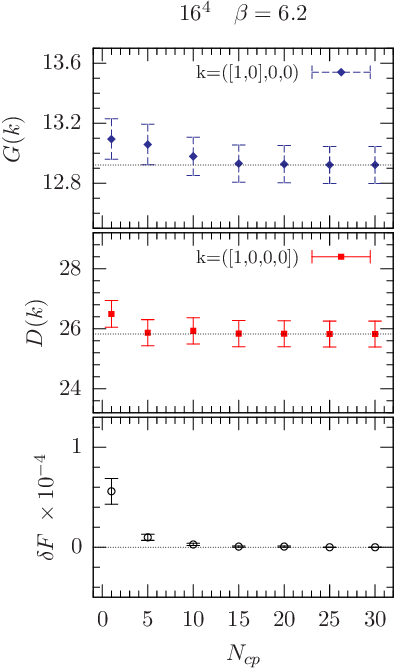}}
\caption{The upper panels show the ghost propagator $G(k)$ as average
 over two realizations $k=(1,0,0,0)$ and $k=(0,1,0,0)$ of the
 smallest lattice momentum, measured always on the \emph{best} gauge
 copy among $\Ncp$ copies. In the middle panel the same
 dependence is shown for the  gluon $D(k)$ propagator,
 however, as average over all four permutations of $k=(1,0,0,0)$.
 The lower panels show the relative difference $\delta F =
 1-F^{\cbc}/F^{\bc}$ of the corresponding current best functional
 values $F^{\cbc}$ to the value $F^{\bc}$ of the overall best
 copy. The data are obtained at $\beta=6.2$ using the lattice sizes $12^4$
 (left) and $16^4$ (right).}
\label{fig:gh_vs_cp_12_16}
\bigskip
\centering
\mbox{\includegraphics[height=0.4\textheight]{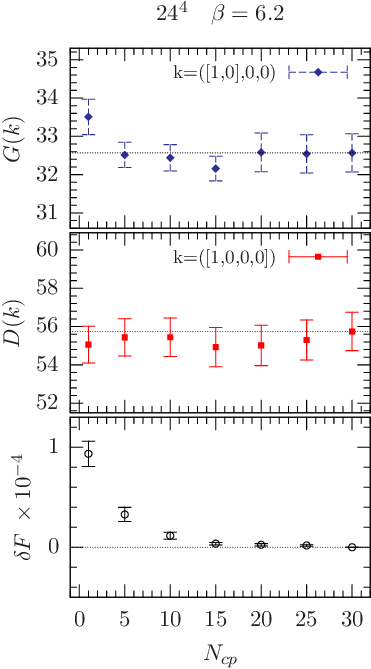}\qquad\quad
  \includegraphics[height=0.4\textheight]{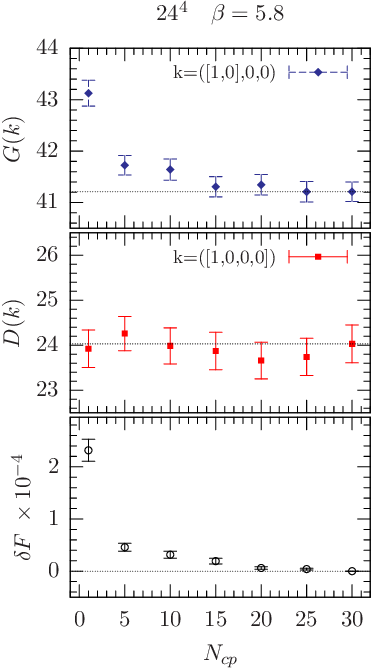}}
\caption{The same as in \Fig{fig:gh_vs_cp_12_16}, however, the data 
  refer to $\beta=5.8$ and 6.2 on a $24^4$ lattice.}
\label{fig:gh_vs_cp_24}
\end{figure*}

In pursuing the \fc{}-\bc{} strategy, at some stage one has to
figure out how large $\Ncp$ has to be approximately. This does not mean that
we can be confident that the set of \emph{best} copies then in any
respect represents the real ones, \ie those which give rise to the 
absolute maximum of the gauge functional $F_U[g]$ (\Eq{eq:functional})
for each $U$. It turns out, however, that for each gauge-variant
observable there exists a certain $\Ncp$ that warrants the convergence
of that particular observable.

Numerically, it turns out that the dependence of the ghost propagator on
the choice of the best copy is most severe for the smallest
momentum. In addition, this sensitivity depends on the lattice size and
$\beta$. Therefore we studied first the dependence of the ghost
and gluon propagators at lowest momentum (ignoring the cone cut) on
the (same) best copies as function of the number of gauge copies
$\Ncp$ under inspection. This was done at \mbox{$\beta=6.2$} where we
used $12^4$, $16^4$ and $24^4$ lattices. The number of thermalized
configurations used for these three lattice sizes are given in
[\Tab{tab:stat_que}: F-2 to F-4]. To check the dependence on $\beta$ also a
simulation at $\beta=5.8$ on a $24^4$ lattice was performed
[\Tab{tab:stat_que}: F-1]. 

The results of this investigation are shown in
\Fig{fig:gh_vs_cp_12_16} and \ref{fig:gh_vs_cp_24}. While there the ghost
propagator is shown as an average over the two realizations
\mbox{$k=(1,0,0,0)$} and \mbox{$k=(0,1,0,0)$} of the smallest lattice
momentum $a^2q^2(k)$, the gluon propagator
has been averaged over all four non-equivalent realizations. Note
that $D(k)=D(-k)$. 

In \Fig{fig:gh_vs_cp_12_16} and \ref{fig:gh_vs_cp_24} it is clearly
visible that the expectation value of the gluon propagator does not
change within errors as $\Ncp$ increases, independently from the lattice
size and $\beta$.  Contrarily, the ghost propagator at $\beta=5.8$ on
a $24^4$ lattice saturates (on average) if calculated on the best
among $\Ncp=15$ gauge copies. At $\beta=6.2$ the number of necessary
gauge fixings attempts reduces to $5\le\Ncp\le 10$ on a $16^4$ and
$24^4$ lattice. On the $12^4$ lattice a small impact of Gribov copies
is visible, namely $1<\Ncp\le5$ is sufficient for convergence.  The
lower panels of \Fig{fig:gh_vs_cp_12_16} and \ref{fig:gh_vs_cp_24}
show the average of the relative 
difference $\delta F = 1-F^{\cbc}/F^{\bc}$ of the corresponding
(current best) functional value $F^{\cbc}$ to the value $F^{\bc}$ of
the overall best copy after $\Ncp=20$, respectively $\Ncp=30$,
attempts. This may serve as an indicator how large $\Ncp$ has to be, on
average, for the chosen algorithm to find a maximum of $F$ close
to the global one. In \Tab{tab:stat_que} the respective numbers $\Ncp$
for the different simulations are given.

%----------------------------------------------------------------------------
\subsubsection{Gribov copies may cause systematic errors}

Having specified the necessary amount of gauge copies we focus now
on the influence the Gribov ambiguity might have on the gluon and
ghost propagators. To investigate this a combined study of the gluon
and ghost propagators on the same sets of \fc{} and \bc{}
representatives of our thermalized gauge field configurations has
been performed (see \cite{Sternbeck:2005tk}). This
has allowed us to assess the importance of the Gribov copy problem for
the ghost propagators in the low-momentum region.
\begin{figure}
  \centering
  \mbox{\includegraphics[width=0.47\textwidth]{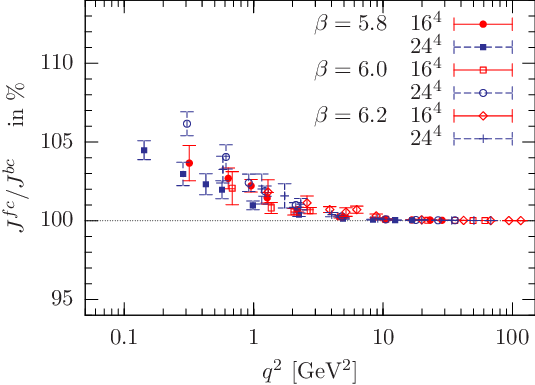}\quad
  \includegraphics[width=0.47\textwidth]{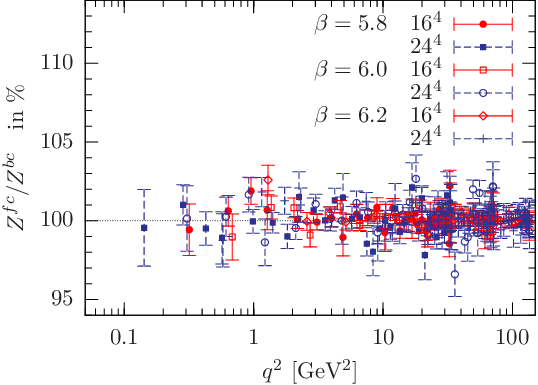}}
  \caption{The ratios $J^{\fc} / J^{\bc}$ and $Z^{\fc} / Z^{\bc}$ of
    ghost (l.h.s.) and gluon (r.h.s.) dressing functions,
    respectively, \vs momentum $q^2$. $J^{\fc}$ and $Z^{\fc}$ refer to
    values determined on first (\fc{}) gauge copies, whereas $J^{\bc}$ and
    $Z^{\bc}$ to those on best (\bc{}) copies. The 
    ratios were calculated using \emph{jackknife}.}
  \label{fig:fcbc_ratios}
\end{figure}

To be specific we have generated different sets of quenched gauge
configuration using the inverse coupling constants $\beta=5.8$,
6.0 and 6.2 and the lattice sizes $16^4$ and $24^4$. The corresponding
entries in \Tab{tab:stat_que} are S-1, S-2, S-4, S-5, S-8 and S-9. 
Following our \fc-\bc{} strategy, each gauge configuration $U$ has
been fixed to Landau gauge a number of $\Ncp$ times using
over-relaxation and starting always from an random gauge copy of
$U$. Then for each $U$ the \emph{first} (\fc{}) and the \emph{best}
(\bc{}) gauge copy (among $\Ncp$ copies) form the two ensembles for
measurements of the ghost and gluon propagator.
\begin{figure}
  \centering
  \includegraphics[width=0.9\textwidth]{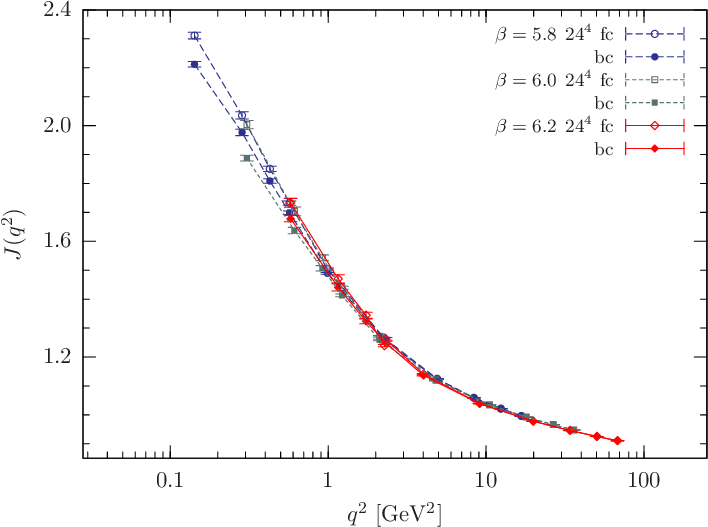}
  \caption{The ghost dressing function determined on first
    (\fc{}) and best (\bc{}) gauge copies \vs momentum $q^2$. 
    Only data on a $24^4$ lattice are shown here at three different values
    of $\beta$. Lines connecting data points belonging to the same
    $\beta$ are drawn to guide the eye.}
  \label{fig:gh_dress_qq_bc_fc}
\bigskip
  \centering
  \includegraphics[width=0.9\textwidth]{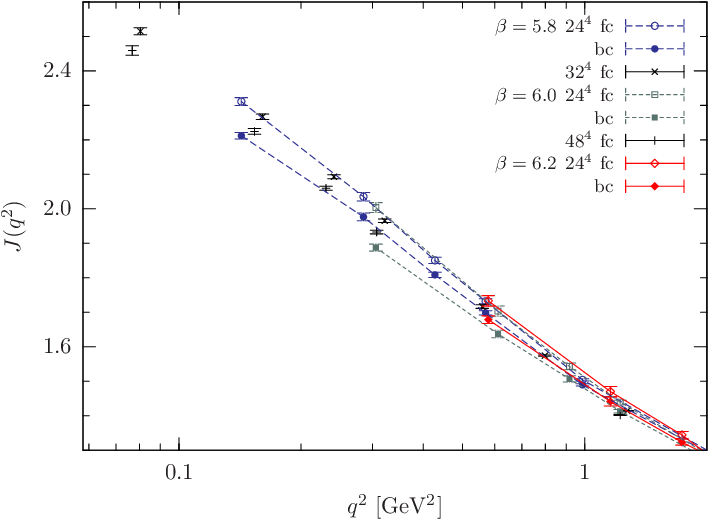}
  \caption{The same as in \Fig{fig:gh_dress_qq_bc_fc}, but 
    focussing on a lower range of momenta. In addition, $\fc$ data on a $32^4$
    lattice at $\beta=5.8$ and on a $48^4$ lattice at $\beta=6.0$ are
    shown.} 
  \label{fig:gh_dress_qq_bc_fc2}
\end{figure}

In \Fig{fig:fcbc_ratios} we illustrate the effect of the 
different Gribov copies by plotting the ratios $J^{\fc}/J^{\bc}$ and
$Z^{\fc}/Z^{\bc}$ of the ghost and gluon dressing functions,
respectively. $J^{\fc}$ and $Z^{\fc}$ refer to values determined on
first (\fc{}) gauge copies, whereas $J^{\bc}$ and $Z^{\bc}$ to those
on best (\bc{}) copies. Obviously, in \Fig{fig:fcbc_ratios} there is
no influence visible for the gluon propagator  
within the statistical noise. On the contrary, for the ghost propagator the
Gribov problem can cause $\order{5\%}$ deviations in the low-momentum region
($q<1~\mbox{GeV}$). For better gauge copies the ghost dressing function
becomes less singular in the infrared. This can also be seen in
\Fig{fig:gh_dress_qq_bc_fc} and \ref{fig:gh_dress_qq_bc_fc2} where
both \fc{} and \bc{} data for the ghost propagator are shown. 

A closer inspection of the data in \Fig{fig:fcbc_ratios} and
\ref{fig:gh_dress_qq_bc_fc2} indicates that the influence of Gribov 
copies on the ghost propagator becomes weaker for \emph{increasing}
the lattice size. In fact, comparing 
in \Fig{fig:fcbc_ratios} ratios for the ghost dressing function at
$q<1~\GeV$, the rise at $\beta=6.0$ is obviously larger than that 
at $\beta=5.8$. In both cases the data are from simulations on a
$24^4$ lattice. Thus, it seems that by increasing the physical volume
(lower~$\beta$) the effect of the Gribov ambiguity gets smaller if the
same physical momentum is considered. This we can also deduce from
\Fig{fig:gh_dress_qq_bc_fc} and \ref{fig:gh_dress_qq_bc_fc2}. There
obviously the difference between $J^{\fc}$ and $J^{\bc}$ at lower
momentum reduces when $\beta$ 
decreases. We think this observation is not biased by a too 
small number $\Ncp$ of inspected gauge copies since, judging from
\Fig{fig:gh_vs_cp_12_16} and \ref{fig:gh_vs_cp_24}, $\Ncp=40$ as
specified in \Tab{tab:stat_que} seems to be on the safe side. However,
in a future project one should definitely check the ratios on a $32^4$
lattice at $\beta=6.2$ to eliminate the last doubts.

We conclude: the ghost propagator systematically
depends on the choice of Gribov copies, while the impact on the gluon
propagator is not resolvable within our statistics. However, there
are indications that the dependence on Gribov copies decreases with
increasing physical volume. This is also in agreement with the data
listed in the two lattice studies
\cite{Bakeev:2003rr,Cucchieri:1997dx} of the $SU(2)$ ghost 
propagator $G$, while it is not explicitly stated there. In fact, in
Ref.~\cite{Bakeev:2003rr} the ratio $G^{\fc}/G^{\bc}$ at $\beta=2.2$
on a $8^4$ lattice is larger than that on a $16^4$ lattice at the
same physical momentum. In \cite{Bogolubsky:2005wf} similar
indications have been found for $SU(2)$ taking non-periodic
$Z(2)$ transformations into account.

Note also that this observation is in agreement
with a recent claim by \name{Zwanziger} according to which in the
infinite volume limit averaging over gauge configurations in the Gribov region
should lead to the same result as averaging over configurations in
the fundamental modular region \cite{Zwanziger:2003cf}.

%---------------------------------------------------------------------------
\section{The infrared behavior of gluon and ghost propagator in the 
  quenched and unquenched case}
\label{sec:infrared_ghost_gluon}

After discussing different systematic effects on the gluon and
ghost propagators or on the corresponding dressing functions we
concentrate now on the infrared behavior. Thereby, however, we shall 
study a further systematic effect, namely the change of data at lower
momenta due to unquenching the theory. As we shall see below, this
change is negligible for the ghost propagator, but conspicuously
large for the gluon propagator at not so small momenta.

%-----------------------------------------------------------------------------
\subsection{The gluon propagator}
\label{sec:gluon_results}

From the material presented above it is clear that the gluon propagator in
Landau gauge does not suffer from the Gribov ambiguity, at least in
the momentum range considered. Therefore, if we restrict ourselves in the 
following to consider only $\fc{}$ gauge copies, the systematic
effect involved will stay most probably within statistical errors. This
has the advantage of allowing us to use data obtained on larger lattice sizes
even though they have been collected only from the ensemble of $\fc{}$
gauge copies.

Just as a reminder, in our simulations the gluon propagator
was determined for all momenta at once. However, only a
subset of data was used for the final analysis. For details see
\Sec{sec:cylindercut} where our selection of momenta is
specified. The way how lattice momenta are connected to physical ones
is discussed in \Sec{sec:physical_units}. Beside this, it is also
clear that all data obtained in lattice simulations have to be subject
to a renormalization, discussed in
\Sec{sec:renorm_latt_obs}. In the following 
the data presented for the gluon and ghost dressing functions, $Z$ and $J$, are
renormalized such that at the renormalization point $\mu=4~\GeV$ 
they are fixed to $Z(\mu^2,\mu^2)=J(\mu^2,\mu^2)=1$. 

\subsubsection{The quenched gluon propagator}

\begin{figure*}
  \centering
    \includegraphics[width=0.8\textwidth]{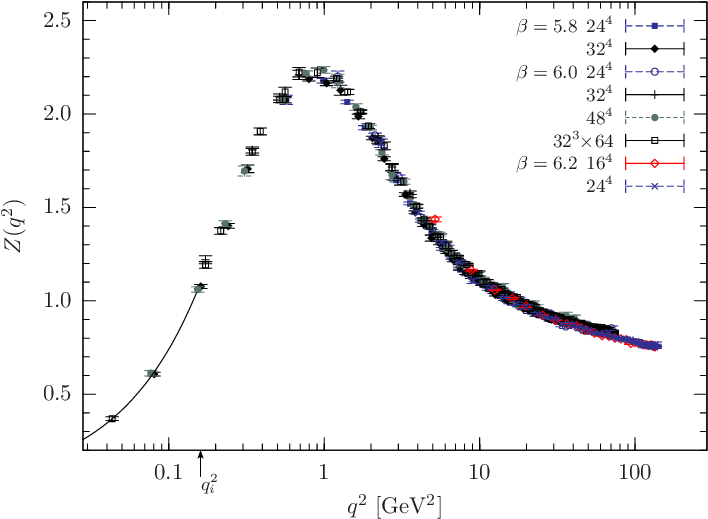}
  \caption{The dressing function of the gluon propagator renormalized
    at $\mu=4$~GeV is shown as a function of momentum $q^2$ using
    various lattice sizes. All data have been obtained from
    \emph{first} gauge copies in quenched simulations (see
    \Tab{tab:stat_que}). The line refers 
  to an infrared fit on the data as explained in the
  text. The momentum $q^2_i$ marks the largest
  momentum used for this fit.} 
     \label{fig:gl_dress_qq_fc}
\vspace{0.7cm}
    \includegraphics[width=0.8\textwidth]{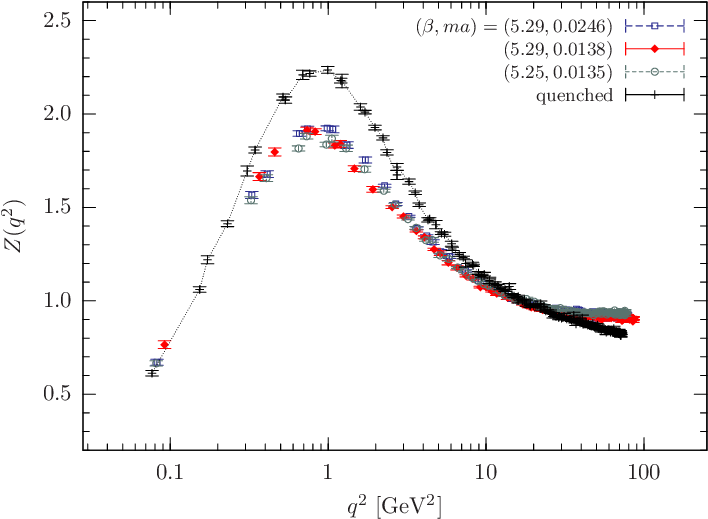}
  \caption{The same as in \Fig{fig:gl_dress_qq_fc}, including now data
    obtained on \emph{first} gauge copies of unquenched
    configurations (see D-2 to D-4 in \Tab{tab:stat_dyn}). For comparison
    some quenched data at $\beta=6.0$ [\Tab{tab:stat_que}: S-6, S-7]
    have been included in this figure as well. A dotted line connecting
    quenched data is drawn to guide the eye.}
     \label{fig:gl_dress_qq_dyn}
\end{figure*}

In \Fig{fig:gl_dress_qq_fc} the dressing function of the gluon
propagator is collected from data at different $\beta$ values as a
function of the 
(physical) momentum $q^2$. Based on the experiences gained in previous
sections only data obtained on larger lattice sizes are
presented there. Beside of a cylinder cut applied to all data, a cone cut 
has been imposed to data related only to the lattice sizes $16^4$ and $24^4$.
Since finite volume errors on the larger lattices are small, it is not
necessary to impose this cut there. 

Obviously, the data surviving all cuts lie along a smooth curve which
not only is nonperturbatively enhanced around $q=1~\GeV$ as
expected, but also decreases in a sufficiently large range of low
momenta. Therefore, it appears natural trying to fit the ansatz
\begin{equation}
  \label{eq:infr_gluon_fit}
  f_D\left(q^2\right) = A_D\cdot\left(q^2\right)^{\kappa_D}
\end{equation}
to the data at lower momenta. As mentioned above, from studies of
truncated DSE for the gluon propagator an infrared exponent
$\kappa_D=2\kappa$ with $\kappa \approx 0.595$ is expected to describe
the infrared behavior \cite{Lerche:2002ep,Zwanziger:2001kw}.  

We have tried to fit this ansatz to the data by imposing a upper-momentum cut
$q^2<q^2_i$ that has been varied. The parameters $A_D$ and $\kappa_D$
extracted and the values of $q^2_i$ used are listed in
\Tab{tab:kappa_fit}. As can be seen in this table shifting $q^2_i$ to
larger momenta a fit with 
ansatz (\ref{eq:infr_gluon_fit}) becomes worse. The parameters of our
best fit to the data have been given in bold letters and in
\Fig{fig:gl_dress_qq_fc} we have plotted the corresponding fitting function
$f_D(q^2)$. Although this fit supports the conjecture of an infrared
vanishing gluon propagator our value of $\kappa_D=0.83(2)$ is lower than
expected from the DSE studies. However, there is a tendency for
$\kappa_D$ to rise as the interval of momenta, contributing to the
fit, is shrinking towards the infrared. Therefore, we cannot exclude the
expected infrared behavior to become realized at much lower momenta.  

%--------------------------------------------------------------------------
\subsubsection{The unquenched gluon propagator}

Beside of our measurements in the quenched approximation we have also 
determined the gluon propagator on gauge-fixed copies (gauge-fixed
only once) of dynamical gauge configurations specified in \Tab{tab:stat_dyn}.
The results of the corresponding dressing functions are shown in
\Fig{fig:gl_dress_qq_dyn}. With 
respect to finite volume effects reported above for the quenched case
we have plotted in this figure only data obtained on a $24^3\times48$
lattice, nevertheless for three different pairs of $\beta$ and
$\kappa$ (see D-2, D-3 and D-4 in \Tab{tab:stat_dyn}). Note, the data
related to the $16^3\times32$ lattice (D-1) turn out to exhibit similar finite
volume effects as reported for quenched configurations.

In order to perform a comparison with our quenched data with the
greatest of ease some of the quenched data have been included in this figure as
well, namely the results at $\beta=6.0$ associated with a $32^4$ and a
$48^4$ lattice. The lattice spacing associated with this $\beta$ is
comparable to those of the dynamical configurations. As above the dressing
functions were renormalized at $\mu=4~\GeV$.

As can be seen from \Fig{fig:gl_dress_qq_dyn}, the unquenching effect
becomes clearly visible for the gluon dressing function, in particular
around $q^2\simeq 1\GeV^2$. There the non-perturbative enhancement,  
characteristic to this function, is drastically reduced compared
to the quenched data ($ma=\infty$). It also becomes softer as the quark
mass is decreasing, even though this is a small effect. This has been observed
also in recent lattice computations of the gluon propagator using
configurations generated with dynamical AsqTad-improved staggered
quarks \cite{Bowman:2004jm} as well as from unquenching studies for
the ghost and gluon propagators within the 
DSE approach \cite{Fischer:2005wx,Fischer:2005en,Fischer:2003rp}.
We refer also to recent lattice studies with dynamical Kogut-Susskind
and Wilson fermions \cite{Furui:2005bu,Furui:2005he}.
The difference between quenched and unquenched data in the ultraviolet
asymptotic tail is consistent with what is expected from perturbation
theory (see \Sec{sec:anomalous}).

Unfortunately, the amount of data collected for momenta in the range
$q^2<1\GeV^2$ does not allow for a fit to the ansatz given in
\Eq{eq:infr_gluon_fit}. Intuitively, one may guess, looking at the
low--momentum tendency of the unquenched data points in
\Fig{fig:gl_dress_qq_dyn}, that in the infrared the unquenched gluon
dressing function might match that of the quenched case.  

%-----------------------------------------------------------------------------
\subsection{The ghost propagator}
\label{sec:ghost_results}

In the discussions of different systematic effect we have found that
the Gribov ambiguity has an influence on the ghost
propagator. Nevertheless, in \Fig{fig:gh_dress_qq_bc_fc2} we have also
seen that even though the effect is visible in the data, the
systematic error one would encounter by neglecting the Gribov copy
problem is not that large. Since at present our data collected on
$\fc{}$ gauge copies cover much lower momenta, we analyze in the
following only those. Incidentally, this also allows for a fair
comparison of our quenched and unquenched data.

%-----------------------------------------------------------------------------
\subsubsection{The quenched ghost propagator}

The ghost dressing function has been determined on the same 
gauge configurations as that of the gluon propagator discussed
above. However, the list of momenta studied is much shorter than that
of the gluon propagator. Anyway, we have tried to cover as much as
possible the whole range of momenta with special focus on the infrared region.
The selection of momenta was guided by the cylinder cut and 
for smaller lattices sizes sharpened by the cone cut.

\begin{figure*}
  \centering
  \setlength{\unitlength}{1mm}
  \begin{picture}(0,0)
    \put(10,10){\includegraphics[width=0.7cm]{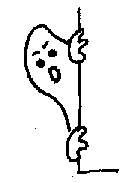}}
  \end{picture}
    \includegraphics[width=0.8\textwidth]{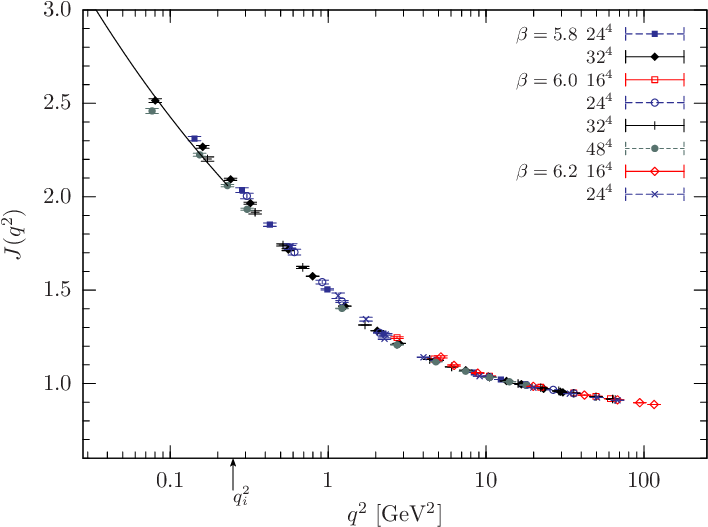}
  \caption{The dressing function of the ghost propagator renormalized
    at $\mu=4~\GeV$ is shown as a function of momentum $q^2$ using
    data from various lattice sizes. All data have been obtained from
    \emph{first} gauge copies in quenched simulations (see
    \Tab{tab:stat_que}). The line refers 
    to an infrared fit to the data as explained in the
    text. The momentum $q^2_i$ marks the largest
    momentum used for this fit.}
     \label{fig:gh_dress_qq_fc}
\vspace{0.7cm}
    \includegraphics[width=0.8\textwidth]{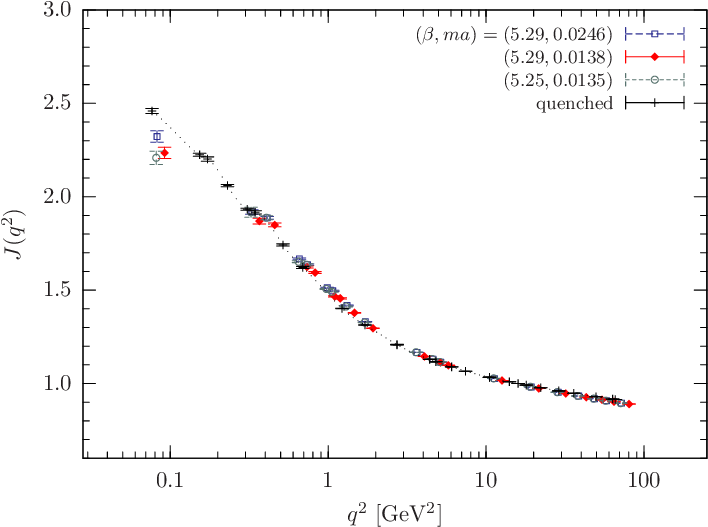}
  \caption{The same as in \Fig{fig:gh_dress_qq_fc}, however, most of
    the data are obtained on \emph{first} gauge
    copies of unquenched configurations 
    using a $24^3\times48$ lattice (see D-2 to D-4 in \Tab{tab:stat_dyn}).
    For comparison, some quenched data at $\beta=6.0$
    [\Tab{tab:stat_que}: S-6, S-7] have been included into this figure
    as well. A dotted line connecting quenched data is drawn to 
  guide the eye. The unquenched data at lowest momentum refer to the
  lattice momentum $k=(0,0,0,1)$ on a $24^3\times48$ lattice and are
  most probably affected by lattice-asymmetry effects.} 
     \label{fig:gh_dress_qq_dyn}
\end{figure*}

Our results for the quenched ghost dressing function are shown in
\Fig{fig:gh_dress_qq_fc}. As for the gluon dressing function we 
used $\mu=4~\GeV$ as renormalization point such that
$J(\mu^2,\mu^2)=1$. The data all lie along a smooth curve. Thus we
may conclude that both cylinder and cone cut work well also for the ghost
propagator. Note that in contrast to the gluon dressing function here a
cone cut has not been imposed on a $24^4$ lattice.

As expected the ghost dressing function seems to diverge
with decreasing momenta and we have tried to fit the infrared
behavior again by an ansatz motivated through the mentioned studies of the
ghost propagator within the DSE approach. To be specific, we fitted the ansatz
\begin{equation}
  \label{eq:infr_ghost_fit}
  f_G\left(q^2\right) = A_G\cdot\left(q^2\right)^{-\kappa_G}
\end{equation}
to the data at lower momenta. From the mentioned studies one usually
expects to get $\kappa_G \approx 0.595$
\cite{Lerche:2002ep,Zwanziger:2001kw}. Even more importantly, one
expects $\kappa_G=2\kappa_D$ to hold. See below for a discussion
about that.

The ansatz (\ref{eq:infr_ghost_fit}) has been fitted to our data by
imposing an upper momentum cut $q^2<q^2_i$. However, in
contrast to the gluon dressing function, here the parameter $A_G$ and
$\kappa_G$ turn out to be more robust against shifting
$q^2_i$. This can be seen in \Tab{tab:kappa_fit} where the parameters
obtained are given. In \Fig{fig:gh_dress_qq_fc} we show the
corresponding fitting function. Even though data at the
lowest three momenta available on a $24^4$ lattice are shown in
\Fig{fig:gh_dress_qq_fc} they do not contribute to the fit. We also
have used data only at $\beta=6.0$, because including data
at $\beta=5.8$ in the fit makes it even worse.

Our fits to the data suggest that \mbox{$\kappa_G=0.20(1)$} which is far away
from values larger than 0.5. Therefore, we cannot confirm the
expected infrared exponent \mbox{$\kappa \approx 0.595$}  
on the basis of our data. Also the not so small values found for
$\chi^2/\ndf$ indicate that the power ansatz (\ref{eq:infr_ghost_fit}) seems
to be not reasonable for the range of momenta considered. Note also
that the linear rise of the data in \Fig{fig:gh_dress_qq_fc} (momentum
log-axis) for decreasing momenta suggest the ghost dressing
function might diverge logarithmically at least in the considered range
of momenta. However, we do not know arguments that would support
this.

Together with our results for the gluon dressing function we can
neither confirm the expected values of the infrared exponents
nor that there the relation 
\begin{displaymath}
  \kappa_D - 2\kappa_G = \Delta
\end{displaymath}
holds with $\Delta=0$, at least from the data available to us. Our fits yield
\mbox{$\Delta\approx0.43$}. Furthermore, one should remark here that even
though we have fitted the Ans\"atze (\ref{eq:infr_ghost_fit}) and
(\ref{eq:infr_gluon_fit}) to the corresponding dressing functions, the
fits are quite unstable. For the ghost dressing function
the fitting function does even not really describe the tendency of data
towards the infrared (see \Fig{fig:gh_dress_qq_fc}).

\begin{table}
  \centering
  \begin{tabular}{r@{\qquad}rrr@{\qquad}@{\quad}rrr}
   \hline\hline\rule{0pt}{2.5ex}
    & \mc{3}{c}{ghost} & \mc{3}{c}{gluon}\\
    \mc{1}{c}{$q_i^2$} & \mc{1}{c}{$A_G$} & \mc{1}{c}{$\kappa_G$} 
  & \mc{1}{c@{\qquad}}{$\chi^2/\ndf$} & \mc{1}{c}{$A_D$} &
    \mc{1}{c}{$\kappa_D$} & \mc{1}{c}{$\chi^2/\ndf$} \\
    \hline\rule{0pt}{2.5ex}
%    0.10    &  --   &   --   &   --   & 5.0(10)  & 0.82(8) & 2.4\\
    \textbf{0.16}  & \mc{1}{c}{--}  & \mc{1}{c}{--}  &
    \mc{1}{c@{\qquad}}{--} & \textbf{5.0(3)}   & \textbf{0.83(2)} & 
    \textbf{1.2}\\ 
    0.20     & \mc{1}{c}{--}  & \mc{1}{c}{--}  &
    \mc{1}{c@{\qquad}}{--} & 5.1(3) & 0.84(3) & 3.2 \\ 
    \textbf{0.25}    & \textbf{1.55(5)} & \textbf{0.20(2)} & \textbf{3.8}  &  4.4(2)  & 0.77(3) 
    & 8.6 \\
    0.35    & 1.53(3) & 0.20(1) & 4.0     & 4.0(1) & 0.72(2)
    & 9.8\\
    0.55    & 1.53(1) & 0.20(1)  & 3.1     & 3.3(1) & 0.62(3)
    & 38.0 \\
    0.70   & 1.51(1)  & 0.21(1)  & 4.3     & \mc{1}{c}{--} & \mc{1}{c}{--} 
    & -- \\
   \hline\hline
  \end{tabular}
  \caption{Parameter extracted corresponding to the fitting Ans\"atze 
    (\ref{eq:infr_ghost_fit}) and (\ref{eq:infr_gluon_fit}) for the ghost 
    and gluon dressing function, respectively. The first row specifies the
    upper-momentum cutoff $q^2_i$, \ie $q^2<q^2_i$. Bold letters
    indicate parameters used for the fitting function shown in
    \Fig{fig:gh_dress_qq_fc} and \ref{fig:gl_dress_qq_fc}, respectively.} 
  \label{tab:kappa_fit}
\end{table}

\subsubsection{The unquenched ghost propagator}

In \Fig{fig:gh_dress_qq_dyn} we present
our full QCD result for the ghost dressing functions. There we have
not discard data at the lowest on-axis momentum in order to illustrate the
systematic effect encountered through the asymmetric
lattice geometry. For comparison, selected data of the quenched case
(\ie for infinite quark mass) are also shown, namely those on
$32^4$ and $48^4$ lattice at $\beta=6.0$. As in the quenched case the dressing
function has been renormalized such that $J(\mu^2,\mu^2)=1$ at $\mu = 4~\GeV$.

In contrast to the gluon dressing function, in \Fig{fig:gh_dress_qq_dyn}
the unquenching effect is very small. In the ultraviolet the
unquenched data are slightly lower and at lower momenta they are
slightly larger than the quenched data, but altogether the effect
stays within error bars and thus can be neglected in the region of
momenta considered here. Negligible unquenching effect are in
agreement with what is expected from unquenching studies for the ghost
and gluon 
propagators within the DSE approach
\cite{Fischer:2005wx,Fischer:2003rp}. We refer also to
recent lattice studies with dynamical Kogut-Susskind and Wilson
fermions \cite{Furui:2005bu,Furui:2005he}. Thus, even though we
cannot confirm the predicted infrared exponent (from the DSE studies), we
affirm that the influence of fermions on the infrared behavior of the ghost
propagator is negligible. 

\newpage
%-----------------------------------------------------------------------------
\section{The running coupling and the ghost-gluon vertex}

After extensively reporting on results obtained for the gluon and
ghost dressing functions, now we focus on the running coupling
constant $\alpha_s(q^2)$. As explained in \Sec{sec:theo_running_coup},
$\alpha_s(q^2)$ may be defined by a renormalization-group-invariant
combination of the gluon and ghost dressing functions (\Eq{eq:running_coup})
\begin{equation}
 \label{eq:running_coup2}
  \alpha_s(q^2) = \alpha_s(\mu^2)Z(q^2,\mu^2)J^2(q^2,\mu^2) .
\end{equation}
Remember that this definition relies on the assumption that the 
ghost-gluon vertex in Landau gauge stays bare also beyond perturbation
theory.  Below we give numerical evidence 
confirming $\widetilde{Z}_1\approx1$ in a $\MOM$ scheme where the gluon
momentum equals zero. This we show for both the quenched and
unquenched case of $SU(3)$. Similar results indicating this,
directly \cite{Cucchieri:2004sq} and indirectly
\cite{Bloch:2003sk}, were presented in studies of the quenched $SU(2)$
gauge theory, but also in semiperturbative calculations within the
DSE approach \cite{Schleifenbaum:2004id}.

\subsection{Results for the running coupling constant}
\label{sect:running_coupling}

Based on our data for the renormalized gluon and ghost dressing
functions, we have estimated the product
in \Eq{eq:running_coup2} using the \emph{bootstrap}
method with drawing 500 random samples.  Since the
ghost-gluon-vertex renormalization constant $\widetilde{Z}_1$ has been
set to one, 
there is an overall normalization factor which has been fixed by
fitting the data for $q^2>q^2_c$ to the well-known perturbative results of the
running coupling $\alpha_{\texttt{2-loop}}$ at 2-loop order (see also
\cite{Bloch:2003sk}). Defining \mbox{$x\equiv q^2/\Lambda^2_{\texttt{2-loop}}$},
the 2-loop running coupling is given by
\begin{equation}
  \label{eq:2loop}
  \alpha_{\texttt{2-loop}}(x) = \frac{4\pi}{\beta_0\ln x}
  \left\{1 - \frac{2\beta_1}{\beta^2_0}\frac{\ln(\ln x)}{\ln x} \right\}.
\end{equation}
The $\beta$-function coefficients $\beta_0$ and $\beta_1$ have been
defined in \Eq{eq:beta0} and (\ref{eq:beta1}). They are
independent of the renormalization prescription. The value of
$\Lambda_{\texttt{2-loop}}$ 
has been fixed by the same fit. The lower bound $q^2_c$ has been
chosen such that an optimal value for $\chi^2/{\ndf}$ has been achieved. 

\begin{figure*}
\centering
  \includegraphics[width=0.8\textwidth]{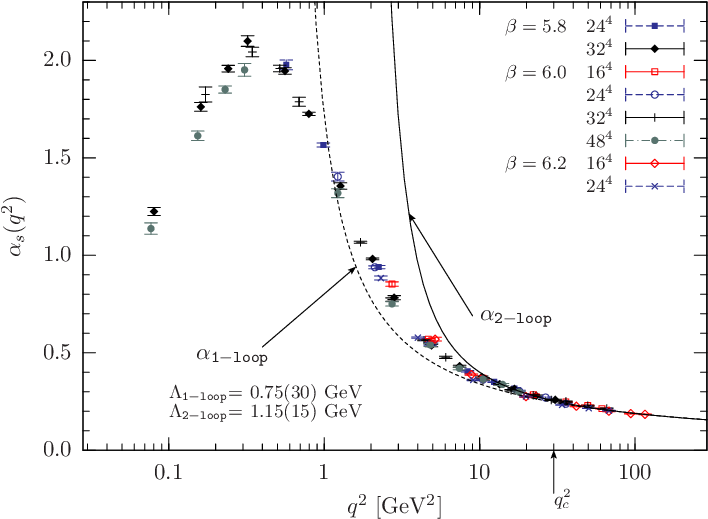}
  \caption{The running coupling $\alpha_s(q^2)$ as a function of the 
    momentum $q^2$. The data refer to first (\fc{}) gauge copies of
    quenched configurations.} 
  \label{fig:alpha_qq_fc}
\vspace{0.5cm}
  \includegraphics[width=0.8\textwidth]{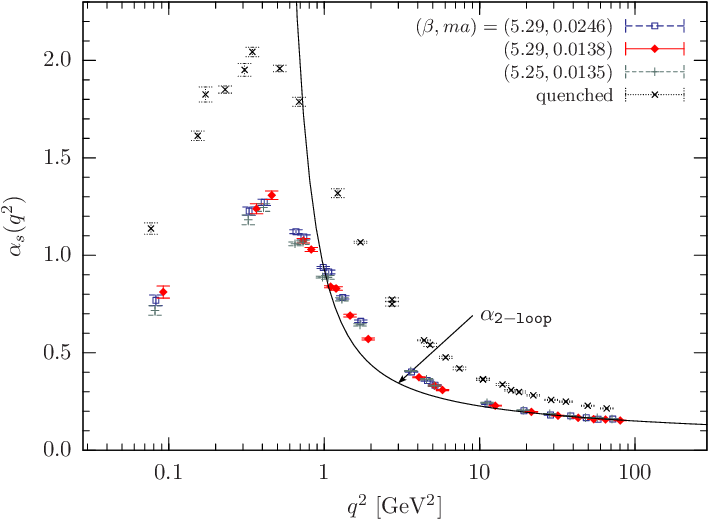}
  \caption{The running coupling $\alpha_s(q^2)$ as a function of
    momentum $q^2$ determined on gauge-fixed configurations in the
    unquenched case. For comparison, we also show quenched data
    (crosses) obtained at $\beta=6.0$ on a $32^4$ and $48^4$ lattice.}
  \label{fig:alpha_qq_dyn}
\end{figure*}

The results are shown in \Fig{fig:alpha_qq_fc}. There also the \mbox{1-loop}
contribution is shown where we used the lower bound
$q^2_c=50~\GeV^2$. The best fit of the \mbox{2-loop} expression to the
data gives $\Lambda_{\texttt{2-loop}}=1.15(15)~\GeV$ ($\chi^2/{\ndf}=7.5$), while
$\Lambda_{\texttt{1-loop}}=0.75(30)~\GeV$ is obtained ($\chi^2/{\ndf}=5.2$)
using just the \mbox{1-loop} part. For the \mbox{2-loop} expression we
used $q^2_c=30~\GeV^2$. The value for $\Lambda_{\texttt{2-loop}}$ is
similar within errors to the $SU(2)$ result given
in Ref.~\cite{Bloch:2003sk}. Note that we have imposed 
again a cone cut to data obtained on a $16^4$ and $24^4$ lattice. 

Approaching the infrared limit in \Fig{fig:alpha_qq_fc} one clearly sees
$\alpha_s(q^2)$ increasing for
\mbox{$q^2>0.4$~GeV$^2$}. However, after passing a maximum at
\mbox{$q^2\approx0.4$~GeV$^2$} $\alpha_s(q^2)$ decreases again. The
same behavior is found on our sets of dynamical gauge
configurations. These results are presented in \Fig{fig:alpha_qq_dyn}
where quenched data ($\beta=6.0$: $32^4$ and $48^4$) are also shown for
comparison. In fact, looking at \Fig{fig:alpha_qq_dyn} we clearly see
that the same turnover, as found in the quenched case, can be assumed for
the unquenched case, even though the data at the lowest (on-axis)
momenta have to be taken with special care. (See our discussion
concerning systematic effects due to asymmetric lattices.) 
Therefore, the data points at $q^2 \simeq 0.1 \mbox{GeV}^2$ have to be checked
carefully on larger symmetric lattices, in order to eliminate the last
doubts. 

We observe a quite clear unquenching effect for $\alpha_s(q^2)$ 
extending from the perturbative range
down to the infrared region. We think that this is
caused in the majority due to unquenching effects as found for the gluon 
propagator in \Sec{sec:gluon_results}.

\begin{figure}
\centering
  \includegraphics[width=0.8\textwidth]{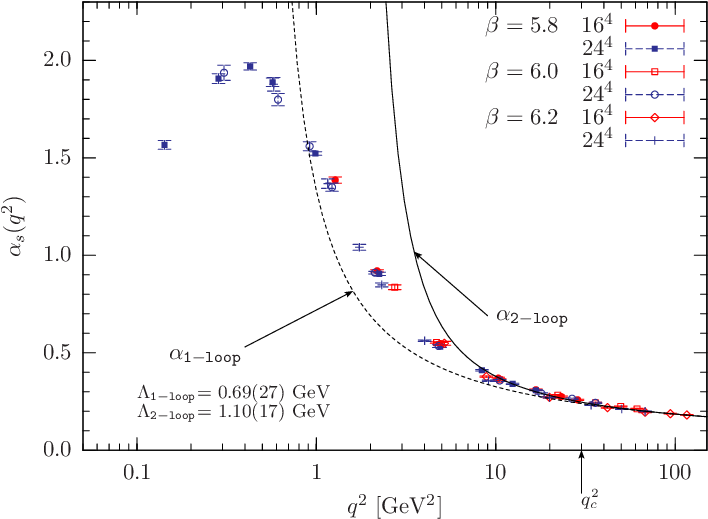}
  \caption{The running coupling $\alpha_s(q^2)$ as a function of
    momentum $q^2$ determined on best (\bc{}) gauge copies of quenched
    configurations.} 
  \label{fig:alpha_qq}
\end{figure}

Note that a similar infrared behavior of $\alpha_s(q^2)$ 
as presented here has been observed in other lattice studies
\cite{Furui:2003jr,Furui:2004cx}. But opposed to Ref.~\cite{Furui:2004cx}
we argue that the existence of a turnover is independent on the
choice of Gribov copies. In fact, qualitatively we have found the same
behavior if $\alpha_s(q^2)$ is calculated on \bc{} gauge copies.
To illustrate this, in \Fig{fig:alpha_qq} we show the corresponding
data extracted in \bc{} copies. There we have not impose the cone cut
on data associated with the $24^4$ lattice. Obviously, $\alpha_s(q^2)$
decreases also in this case and a decreasing running coupling constant
is not due to restricting to \fc{} gauge copies.
The values for $\Lambda_{\texttt{1-loop}}$ and $\Lambda_{\texttt{2-loop}}$ are
the same within error bars as obtained on $\fc{}$ gauge copies.

We should remark here that our observation of an infrared decreasing
running coupling is in agreement with recent studies of DSE results
obtained on a torus
\cite{Fischer:2002eq,Fischer:2002hn,Fischer:2005ui,Fischer:2005nf}. In
those studies $\alpha_s(q^2)$ was shown to tend to zero for $q^2 \to
0$ in one-to-one correspondence with what one finds on the
lattice. This would indicate very strong finite-size effects and a
slow convergence to the infinite-volume limit. However, on the lattice
we do not find any indication for such a strong finite-size effect,
except the convergence to the infinite-volume limit would be extremely
slow. An alternative resolution of this problem has been proposed by
Boucaud \etal \cite{Boucaud:2005ce}. These authors have argued\footnote{We thank
  A.~Lokhov for bringing us the arguments in \cite{Boucaud:2005ce} to
  our attention.} that the ghost-gluon vertex in the
infrared might contain \mbox{$q^2$-dependent} contributions which
could modify the DSE results for the mentioned propagators.
Note, however, that recent DSE studies of the ghost-gluon vertex did
not provide hints for such a modification
\cite{Schleifenbaum:2004id,Alkofer:2004it}. Thus, at present there
seems to be no solution of the puzzle.

For completeness we mention that 
running coupling constants decreasing in the infrared have also been found 
in lattice studies of the \mbox{3-gluon}
vertex~\cite{Boucaud:1998bq,Boucaud:2002fx} and the quark-gluon
vertex~\cite{Skullerud:2002ge}.

%-------------------------------------------------------------------------
\subsection{The vertex renormalization constant}
\label{sec:z1_results}

For discussing the infrared behavior of the running coupling it is
interesting to have an independent check of whether the
renormalization constant $\widetilde{Z}_1$ really stays constant
for all renormalization points.  Remember, this is always
assumed if renormalization-group invariance is demonstrated for the product
of gluon and ghost dressing functions in \Eq{eq:running_coup2} and hence
is important for this particular definition of a nonperturbative
running coupling constant. 

In principle the finiteness of $\widetilde{Z}_1$ has to be checked in
each renormalization scheme separately. A first attempt to calculate
$\widetilde{Z}_1$ on the lattice was made in \cite{Cucchieri:2004sq} for
quenched $SU(2)$ gauge theory using the particular renormalization 
scheme of zero gluon momentum (see also \cite{Bloch:2003sk} for an indirect
determination). To confirm these findings also for the gauge
group $SU(3)$ in the quenched and unquenched case, we have used 
the same renormalization scheme as in \cite{Cucchieri:2004sq} for a
calculation of $\widetilde{Z}_1$. The necessary observable to be estimated on
the lattice has been derived in \Sec{sec:ghost-gluon-vertex_latt} (see
\Eq{eq:Z^{-1}}). Since this is a combination of MC
averages for the ghost and gluon propagators and for the 3-point
function $G^{abc}_{\mu}$ of gluon, ghost and anti-ghost fields (see
\Eq{eq:G^{abc}_{mu}} and (\ref{eq:mathsf_G_mu})) we used again the
Bootstrap method to estimate errors. 

\begin{figure}
\begin{minipage}[t]{0.48\textwidth}
    \includegraphics[height=5.9cm]{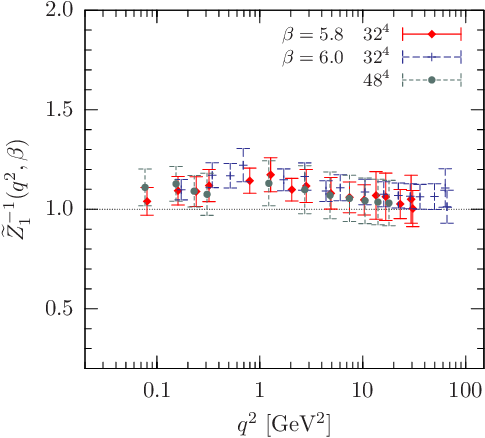}
  \caption{The (inverse of the) ghost-gluon-vertex renormalization constant
    $\widetilde{Z}_1$ is shown versus the momentum $q^2$. The data refer to
    quenched simulations at $\beta=5.8$, and 6.0 using a $32^4$ and
    $48^4$ lattice. Only data on \fc{} gauge copies are available.}
     \label{fig:z1_qq_fc}
\end{minipage}
\hfill
\begin{minipage}[t]{0.48\textwidth}
    \includegraphics[height=5.9cm]{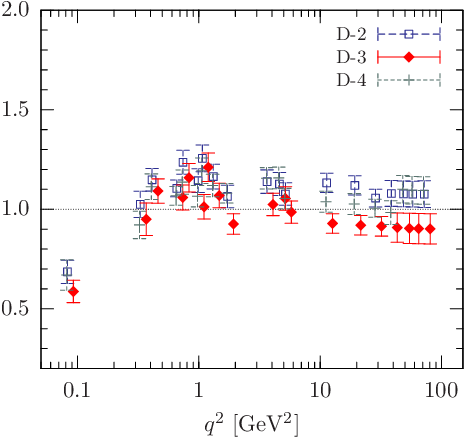}
  \caption{The same as in \Fig{fig:z1_qq_fc}, but the data are
    obtained on unquenched gauge configurations using a
    $24^3\times48$ lattice. The points at the lowest momentum refer to
    the on-axis momentum $k=(0,0,0,1)$ on that lattice. Labels
    refer to the parameters $\kappa$ and $\beta$ given in \Tab{tab:stat_dyn}.}  
     \label{fig:z1_qq_dyn}
\end{minipage}
\end{figure}

Our data for the inverse of $\widetilde{Z}_1$ from quenched
simulations are shown in \Fig{fig:z1_qq_fc}. To simplify matters, we
have restricted ourselves to present only data obtained on the largest lattice
available at $\beta=5.8$ and 6.0. For data on smaller lattices we
refer to our recent conference proceeding
\cite{Sternbeck:2005re}. The data there agree within errors with those
presented here. 

In \Fig{fig:z1_qq_fc} we clearly see that $\widetilde{Z}_1$ stays constant
in the region of momenta considered. This holds for the data at both
$\beta=5.8$ and $\beta=6.0$. Only a slight variation is
visible in the interval $0.3~\GeV^2 \le q^2 \le 2~\GeV^2$, 
but this remains within error bars. The same can be concluded from our
data obtained on unquenched configurations using a $24^3\times48$
lattice. These are shown in \Fig{fig:z1_qq_dyn} for three 
different settings of $\kappa$ and $\beta$ (see the data set entries \mbox{D-2},
\mbox{D-3}, \mbox{D-4} in \Tab{tab:stat_dyn}). For all three cases, the
renormalization constant $\widetilde{Z}_1$ does not differ beyond error
bars from being constant, but there is a certain trend of deviation
from unity which systematically depends on the parameter setting
$(\beta,\kappa)$. 

For each tuple there is also one data point at the lowest (on-axis)
momentum available on a $24^3\times48$ lattice (\ie $k=(0,0,0,1)$)
which is much lower than unity. This particular deviation we also find at the
lowest on-axis momentum in our data obtained on quenched
configurations at $\beta=6.0$ when using a $24^3\times48$ and
$32^3\times64$ lattice (not shown). Since the data taken in
simulations on a
$48^4$ lattice at the same $\beta$ do not show such a deviation, see
\Fig{fig:z1_qq_fc}, this effect is most probably caused by the
asymmetric lattice geometry as 
it was reported also for the propagators in
\Sec{sec:systematic}. Data at this momentum on an asymmetric lattice
would have to be ignored. To eliminate the last doubts we have also
inspected our data for $\widetilde{Z}_1$ obtained at the lowest 
on-axis momenta on a $24^3\times128$ lattice at $\beta=6.0$. We find
for these data that this particular deviation is even more dramatic. Hence,
asymmetric lattices are not appropriate for studying the
ghost-gluon-vertex renormalization constant at very low momentum.

In summary, our results for the quenched and unquenched case of
$SU(3)$ are in full agreement with those presented in
\cite{Cucchieri:2004sq} for the case of $SU(2)$. Even though there is
a weak deviation of $\widetilde{Z}_1$ from being constant this will
not have a dramatic influence on the running coupling.
Together with our data for the running coupling we can thus confirm 
that in the special $\MOM$ scheme considered here (gluon momentum equals
zero) the product in \Eq{eq:running_coup2} is indeed renormalization-group
invariant and thus defines a nonperturbative running coupling which
monotonously decreases with decreasing momentum in the range $q^2 <
0.3~\GeV^2$.

It is worthwhile to continue the lattice calculations of $\widetilde{Z}_1$
using other renormalization schemes, for example the scheme with
a symmetric subtraction point. This has been undertaken in a recent study
\cite{Schleifenbaum:2004id} where a semiperturbative calculation of
$\widetilde{Z}_1$ within the DSE approach has been presented. There are also data
shown for the same renormalization scheme as used
by us which qualitatively agree with our results and those in
\cite{Cucchieri:2004sq}.

%===============================================================================
%%% Local Variables: 
%%% mode: latex
%%% TeX-master: "Sternbeck"
%%% End:

\chapter{Confinement criteria under the lattice microscope}
\label{ch:confcriteria}

\begin{chapterintro}{H}
  ere we discuss data for the $SU(3)$ Landau gauge gluon and ghost
  propagators in the light of the Gribov-Zwanziger horizon condition and the
  Kugo-Ojima confinement scenario. For the latter we also
  present recent data for the function $u(q^2)$ whose zero-momentum
  limit $u(q^2=0)$, the Kugo-Ojima confinement
  parameter, is expected to be minus one. Note that our lattice estimate of the
  function $u(q^2)$ is (to our knowledge) the first presented in the
  literature. The data have been renormalized using
  a minimization process, developed here for the first
  time. We show that, under the assumption the ghost dressing
  function being divergent in the infrared, the function $u(q^2)$ will
  reach minus one at zero momentum. Finally, we 
  discuss numerical evidence for the gluon propagator
  violating reflection positivity explicitly.
\end{chapterintro}

%---------------------------------------------------------------------------
\section{Is the Gribov-Zwanziger horizon condition satisfied?} 
\label{sec:gribov_zwanziger_results}

The Gribov--Zwanziger horizon condition has been introduced
in \Sec{sec:horizon_condition}. It states that on the one hand
the ghost propagator in Landau gauge diverges in the limit of
vanishing momenta more rapidly than $1/q^2$, whereas on the other hand
the gluon propagator vanishes in the same limit. 

It should be clear from the discussion in the previous chapter that our data
support the picture of a diverging ghost propagator, 
even though we cannot confirm an infrared exponent
$\kappa_G>0.5$, a value expected from studies of the ghost DS
equation. Therefore, the Gribov--Zwanziger horizon condition appears to be
satisfied with respect to our data of the ghost propagator in Landau gauge.

Concerning the gluon propagator, however, we cannot give a
conclusive statement of whether it vanishes or stays finite in
the infrared. We have tried to fit the ansatz given in
\Eq{eq:infr_gluon_fit} to our data for the gluon dressing
function. As mentioned in \Sec{sec:gluon_results} this ansatz is
reasonable only if applied to data at the lowest momenta available to
us. We also could not confirm the corresponding value of the infrared 
exponent as expected from studies of the gluon DS
equation. In any case, with respects to the 
range of momenta available to us we cannot state without doubt whether
a power law as that given in \Eq{eq:infr_gluon_fit} really describes the
behavior of the gluon dressing function at much lower
momenta. Therefore, we cannot judge on the existence of an infrared
finite or vanishing gluon propagator, even though our fit with ansatz
\Eq{eq:infr_gluon_fit} supports the latter option.

\begin{figure}[t]
  \centering
  \includegraphics[width=0.8\textwidth]{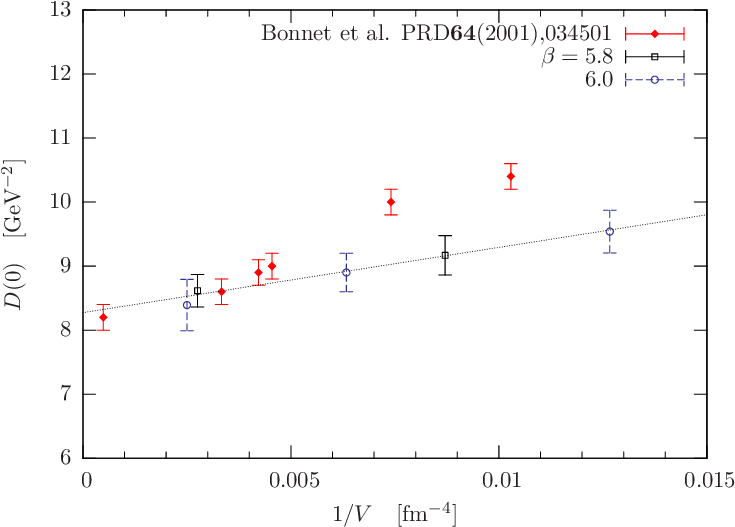}
  \caption{The gluon propagator at zero four-momentum, $D(0)$, plotted
    as a function of the inverse lattice volume. For comparison, data
    of the \name{Adelaide group} given in \cite{Bonnet:2001uh} have
    been included into this figure, too. The line refers to a fit of
    the linear ansatz given in \Eq{eq:gl_00_linear_ansatz} to the data
    at $\beta=5.8$ and 6.0.} 
  \label{fig:gl_00}
\end{figure}

There have been attempts in the recent literature
\cite{Bonnet:2001uh,Boucaud:2006pc} to argue for a finite gluon
propagator in the infrared  based on lattice data for the
gluon propagator at zero four-momentum, \ie 
\begin{displaymath}
  D(0) = \frac{1}{4V(N_c^2-1)}\sum_{\mu,a}\sum_{x,y}\left\langle
    A^a_{x,\mu}A^a_{x+y,\mu}\right\rangle_U.
\end{displaymath}
Here $A^a_{x,\mu}$ refers to the lattice gluon field as defined in
\Eq{eq:A_a-definition} and $V$ represent the volume in lattice units. For
notations we refer to \Ch{chap:latticeQCD}. In order to estimate the
corresponding value $D_{\infty}(0)$ in the infinite 
volume limit, in the Refs. \cite{Bonnet:2001uh,Boucaud:2006pc} the
ansatz \cite{Bonnet:2001uh} 
\begin{equation}
 \label{eq:gl_00_linear_ansatz}
  D(0) = \frac{c}{V} + \widehat{D}_{\infty}(0) 
\end{equation}
has been applied to the data of $D(0)$ for different
volumes\footnote{Note that we distinguish between $\widehat{D}_{\infty}(0)$
and $D_{\infty}(0)$. See below for a discussion.}. To set
the scale for $D(0)$, the gluon propagator $D(q^2,\mu^2)$
has been renormalized in $\MOM$ scheme choosing as renormalization
point $\mu=4~\GeV$. In doing so, both studies
\cite{Bonnet:2001uh,Boucaud:2006pc} independently confirm a
manifest finite value, namely
\begin{displaymath}
 \widehat{D}_{\infty}(0;\mu=4~\GeV)=\left\{
     \begin{array}{rl}
       7.95(13)~\GeV^{-2} & \text{\cite{Bonnet:2001uh}}\\
       9.10(30)~\GeV^{-2} & \text{\cite{Boucaud:2006pc}}\\
     \end{array}\right.\;.
\end{displaymath}
An extrapolation of our data yields the estimate
\begin{displaymath}
 \widehat{D}_{\infty}(0;\mu=4~\GeV)=8.27(10)~\GeV^{-2} 
\end{displaymath}
that approximately agrees with those of these studies. See also
\Fig{fig:gl_00}. Note that for $c$ we obtain
$c=102(18)~\fm^4\GeV^{-2}$ which agrees within errors with that in 
\cite{Boucaud:2006pc}, but not with \cite{Bonnet:2001uh}. In
\cite{Bonnet:2001uh} the tree-level, mean-field improved gauge action
of L\"uscher and Weisz \cite{Weisz:1982zw,Weisz:1983bn,Luscher:1984xn}
has been employed, while in \cite{Boucaud:2006pc} and in this thesis
the standard Wilson gauge action has been used. This might be the
reason for the deviation in $c$.

It is important to note that estimates of
$\widehat{D}_{\infty}$ have to be taken with caution, because these
actually do \emph{not} represent the true infinite volume limit
$D_{\infty}(0)$. It is also not clear whether a linear ansatz is
correct \cite{Bonnet:2001uh}. In this context, it is worthwhile to recall
a comment already made in \cite{Bonnet:2001uh} concerning the
infinite volume limit. This cannot be taken such simple as outlined
above. In fact, the infinite volume limit is given by
\begin{equation}
 \label{eq:D0_limit}
  D_{\infty}(0)  = \lim_{V\rightarrow\infty}
  \lim_{\myover{a\rightarrow0}{V=\const}} D(0).
\end{equation}
That is, the continuum limit (at fixed physical volume) has to be
taken before the infinite volume limit. This was not done for the
given estimates of $\widehat{D}_{\infty}$ and thus a complete systematic
extrapolation to the infinite volume limit remains to be carried out
yet \cite{Bonnet:2001uh}. For this one would first need to calculate $D(0)$
using different lattice spacings $a$, while keeping the physical
volume fixed. The results would then have to be extrapolated to the continuum
limit. Repeating this for a variety of lattice volumes the different
continuum limits obtained one could try to extrapolate to the infinite
volume limit $D_{\infty}(0)$. This, of course, is the essence of the limit
in \Eq{eq:D0_limit}. 

Therefore, we think a conclusive statement of whether
in Landau gauge the gluon propagator at zero momentum is finite or not 
cannot be made neither from our data at zero and non-zero momentum nor
from the data available in the literature at present.

%---------------------------------------------------------------------------
\section{The Kugo-Ojima confinement parameter}
\label{sec:ko_results}

In section \Sec{sec:kugo_ojima_def} we have introduced the Kugo-Ojima
confinement scenario. According to this scenario, colored asymptotic
states, if any, are confined from the physical state space of QCD in
covariant gauges by the quartet mechanism, if for the function 
$u(q^2)\delta^{ab}:=u^{ab}(q^2)$, with $u^{ab}(q^2)$ defined in
\Eq{eq:def_u}, the zero-momentum limit
\begin{equation}
 \label{eq:u_lim_0}
  u:=\lim_{q^2\rightarrow 0} u(q^2) = -1 
\end{equation}
is realized. As pointed out by \name{Kugo} \cite{Kugo:1995km}, in
Landau gauge this limit is connected to an infrared diverging 
ghost dressing function $J$. This can be easily seen from
\Eq{eq:ghost_ku_q} which yields
\begin{displaymath}
  \frac{1}{J(q^2)} = 1 + u(q^2) + q^2v(q^2)\quad
  \stackrel{q^2\rightarrow0}{\longrightarrow}\quad 1 + u(0)  \;.
\end{displaymath}

We have made an attempt to confirm the realization of the limit in
\Eq{eq:u_lim_0} --- here for lattice QCD in Landau gauge --- not
only by giving numerical evidence for a diverging ghost dressing function (see
\Sec{sec:ghost_results}), but also by estimating the function $u(q^2)$
for different momenta $q^2$ from our lattice simulations.

%--------------------------------------------------------------------------
\subsection{Expected infrared behavior}

\begin{figure}[t]
  \centering
  \mbox{\includegraphics[height=5cm]{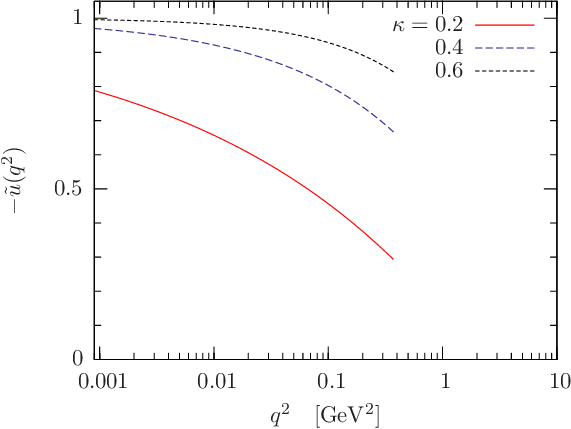}
  \includegraphics[height=5cm]{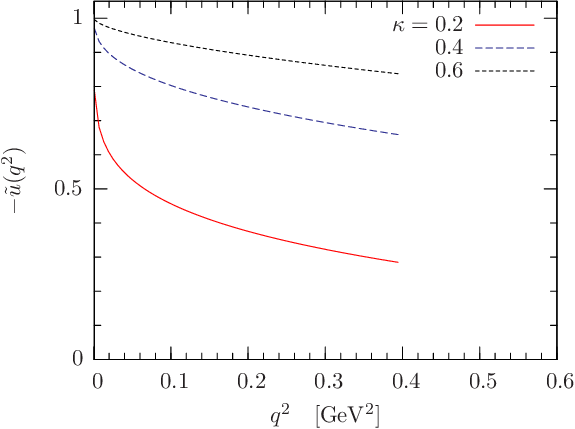}}
  \caption{The asymptote $-\tilde{u}(q^2,\mu^2)$ at low momentum (see
    \Eq{eq:u_asymptotic}) is illustrated in logarithmic and linear
    momentum scale. As ansatz for the ghost dressing function we used
    $J(q^2,\mu^2)=(q^2/\mu^2)^{-\kappa}$ with $\mu=4~\GeV$. The
    $\kappa$ values are 0.2, 0.4 and 0.6.} 
  \label{fig:ku_qq_illustr}
\end{figure}

Before discussing numerical results it is interesting to figure
out first what can be expected for the momentum dependence of $u(q^2)$
in the infrared. In fact, due to \Eq{eq:ghost_ku_q} we
know that in the limit of vanishing momenta the function $u(q^2)$
approaches the asymptote
\begin{equation}
  \label{eq:u_asymptotic}
  \tilde{u}(q^2,\mu^2) := \frac{1}{J(q^2,\mu^2)} - 1\;.
\end{equation}
If we assume that, for example, the power law $J(q^2,\mu^2)
\propto(q^2/\mu^2)^{-\kappa}$ describes the ghost dressing function at very low
momenta, then the asymptote $\tilde{u}(q^2,\mu^2)$, and so the
function $u(q^2,\mu^2)$, should expose an infrared behavior as
illustrated in \Fig{fig:ku_qq_illustr} for different infrared
exponents $\kappa$. Even if such a power law will turn out
not to be appropriate for $J(q^2,\mu^2)$ --- the results presented
in \Sec{sec:ghost_results} at least give rise to some doubt --- it is
commonly believed (and in 
agreement with our results) that the ghost dressing function diverges
in the infrared; and this of course independent of the renormalization point
$\mu$ chosen. Therefore, based on \Eq{eq:ghost_ku_q} the function
$u(q^2,\mu^2)$ is expected to reach minus one
and to join $\tilde{u}(q^2,\mu^2)$ at vanishing momentum,
irrespective of the chosen $\mu^2$. 

\begin{figure}[ht]
  \centering
  \includegraphics[width=0.8\textwidth]{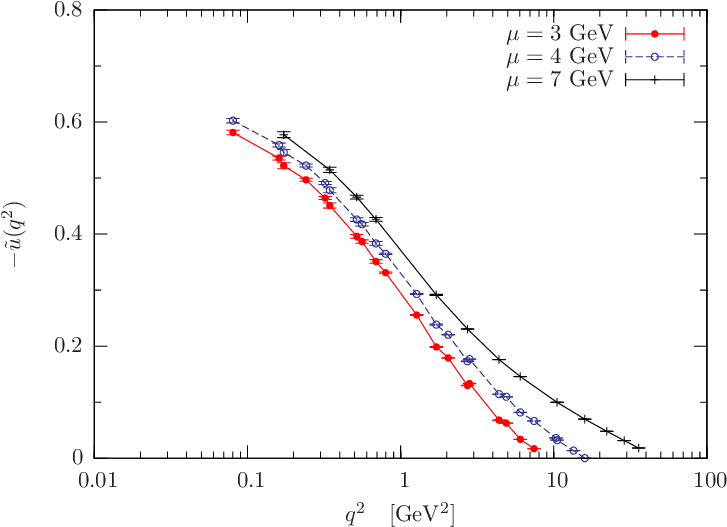}
  \caption{The asymptote $-\tilde{u}(q^2,\mu^2)$ as defined in
    \Eq{eq:u_asymptotic} is shown as a function of momentum $q^2$.
    For the ghost dressing function we used our data at $\beta=5.8$ and $6.0$
    renormalized either at $\mu=3$, $4$ or $7~\GeV$. The lattice size is
    $32^4$. Lines are drawn to guide the eye.} 
  \label{fig:ku_qq_mu}
\end{figure}

To confirm this point with numerical data, in \Fig{fig:ku_qq_mu} we
show $\tilde{u}(q^2,\mu^2)$ as obtained from our data for the ghost dressing
functions renormalized at different momenta $\mu$. In this figure we
clearly see that the different curves referring to different $\mu$
approach each other slowly with decreasing momentum. With respect to
\Fig{fig:ku_qq_illustr} we expect that all these curves run slowly 
towards minus one in the zero momentum limit by definition (see
\Eq{eq:u_asymptotic}) if the ghost dressing function diverges in the
infrared no matter how. 

%--------------------------------------------------------------------------
\subsection[Explicit lattice data for the function $u(q^2)$]
{Explicit lattice data for the function $\boldsymbol{u(q^2)}$}
\label{sec:lattice_data_for_u}

Apart from the asymptotes $\tilde{u}$ relying solely on data for
the ghost dressing functions we have also calculated on the lattice
the function
\begin{displaymath}
  u_L(q^2) = \frac{1}{N_c^2-1}\sum_{a=1}^{N_c^2-1}u^{aa}(q^2)
\end{displaymath}
explicitly in terms of MC averages for $u^{ab}(q^2)$
at different $q^2$ (see \Eq{eq:ku_trU}). These estimates were obtained
on $32^4$ lattice from our (quenched) gauge-fixed configurations thermalized at
$\beta=5.8$ and 6.0 (see runs labeled as S-3 and S-6 in \Tab{tab:stat_que}).

\subsubsection{Renormalization}

As for other observables the lattice data of $u_L(q^2)\equiv u_L(q^2,a^2)$
have to be renormalized yet, \ie assuming multiplicative
renormalizability we have to define a factor $Z_u$ that
relates the bare estimate $u_L$ to a renormalized one  
\begin{displaymath}
  u(q^2,\mu^2) = Z_u(\mu^2,a^2)\cdot u_L(q^2,a^2)\;.
\end{displaymath}
For this we have made use again of \Eq{eq:ghost_ku_q}. It relates the
renormalized ghost dressing function $J(q^2,\mu^2)$ to the
renormalized function $u(q^2,\mu^2)$ at finite~$q^2$. 
\begin{equation}
  \label{eq:ghost_ku_q_2}
  \frac{1}{J(q^2,\mu^2)} = 1 + u(q^2,\mu^2) + q^2v(q^2) 
\end{equation}
As in \Sec{sec:ghost_results}, we have renormalized the ghost dressing
function at the renormalization point \mbox{$\mu=4~\GeV$} and so with
\Eq{eq:ghost_ku_q_2} we can 
renormalize $u(q^2,\mu^2)$ at the same point. However, 
the function $v(q^2)$ (see \cite{Kugo:1995km}) is not
available to us, causing some difficulties for the determination
of~$Z_u(\mu^2,a^2)$.  

Nevertheless, we know that for small momenta $q^2$ the term
$q^2v(q^2)$ becomes less dominant in \Eq{eq:ghost_ku_q_2}. With the
ansatz of a Taylor expansion 
\begin{displaymath}
  v(q^2)  = A + Bq^2 + Cq^4 + \ldots
\end{displaymath}
we have made a $\chi^2$ fit using \footnote{For
minimization we used the new \Cpp{} implementation of the 
library \code{MINUIT} \cite{MINUIT}}
\begin{equation}
 \label{eq:chi2_ku}
  \chi^2 := \sum_{q^2_i}\left( 1 + Z_u u_L(q_i^2) -
  \frac{1}{J(q_i^2,\mu^2)} + Aq_i^2 + Bq_i^4 + Cq_i^6\right)^2
\end{equation}
and our lattice data for $u_L(q^2)$ and for $J(q^2,\mu^2)$
renormalized at $\mu=4~\GeV$. By applying the momentum cut $q_i^2<2~\GeV^2$
we have gained in this way $Z_u$ and the other parameters $A$, $B$ and
$C$. The parameters are given in \Tab{tab:chi2_ku} together with a value for
$\chi^2/\ndf$. 

Of course, the number of terms in the Taylor expansion depends on the
region of momenta considered. We have found that the given order is
necessary, but also sufficient in our case. Note that due to the
number of parameters and the amount of data available for
$q^2<2~\GeV^2$ in the minimization process we have not distinguished
between data at $\beta=5.8$ and 6.0. In general, the factor
$Z_u(\mu^2,a^2)$ differs for the two $\beta$ values, but due to
multiplicative renormalization there is an additional factor that
relates both. From data inspection we found that we can approximate
this to be one in our case.

\begin{table}
  \centering
  \begin{tabular}{c@{\qquad}ccc@{\qquad}c}
 \hline\hline\rule{0pt}{2.5ex}
    $Z_u$ & $A$ & $B$ & $C$ & $\chi^2/\ndf$\\
 \hline\rule{0pt}{2.5ex}
   1.146(7) &  0.22(3) & -0.21(5) & 0.06(3) & 0.59\\
 \hline\hline
  \end{tabular}
  \caption{The parameters obtained by \code{MINUIT} after minimization of the 
    function $\chi^2$ defined in \Eq{eq:chi2_ku}. For this we used
    our data of $u_L(q^2)$ and $J(q^2,\mu^2=(4~\GeV)^2)$ for $q^2<2~\GeV^2$
    at $\beta=5.8$ and 
    $6.0$. The value $Z_u$ has been used to renormalize $u$ at 
    $\mu=4~\GeV$ in \Fig{fig:ku_log_qq}.}  
  \label{tab:chi2_ku}
\end{table}

\begin{figure*}
  \centering
  \includegraphics[width=0.8\textwidth]{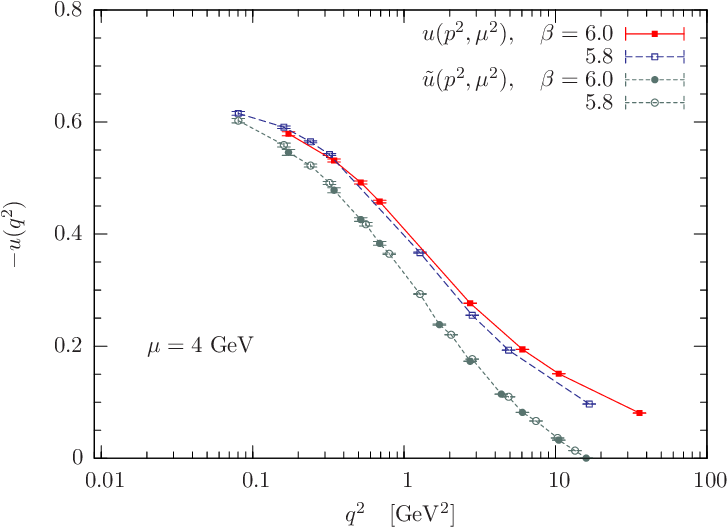}
  \caption{Data for the function $u(q^2,\mu^2)$ at
    $\beta=5.8$ and 6.0 are shown using full and open squares. Additionally,
    data of the asymptote $\tilde{u}(q^2,\mu^2)$ are shown at the same
    $\beta$ values (circles). All data refer to the same quenched
    configurations on a $32^4$ lattice and are renormalized at
    $\mu=4~\GeV$ as described in the text. Lines are drawn to
    guide the eye.} 
  \label{fig:ku_log_qq}
\end{figure*}

%-------------------------------------------------------------------------
\subsubsection{Discussion of numerical results}

In \Fig{fig:ku_log_qq} we show data for the
renormalized function $u(q^2,\mu^2)$ at $\beta=5.8$ and 6.0 as a
function of momentum $q^2$. Since the ghost dressing function was
renormalized at $\mu=4~\GeV$ so was the function $u(q^2,\mu^2)$
mediated by the minimization process mentioned above. In this figure
we also show data for the asymptote $\tilde{u}(q^2,\mu^2)$
renormalized at the same $\mu$. Note that only $Z_u$ has been of
relevance for providing this figure.

We clearly see that $\tilde{u}(q^2,\mu^2)$ and
$u(q^2,\mu^2)$ approach each other with decreasing momenta.
Therefore, as the ghost dressing function is expected to diverge in
the infrared, we expect $u(q^2,\mu^2)$ to approach minus one as does
the asymptote $\tilde{u}(q^2,\mu^2)$ by definition. In \Fig{fig:ku_log_qq} the
growth of $-u(q^2,\mu^2)$ becomes slower as the momentum decreases, but
this behavior we anticipate from our discussion above  (see
\Fig{fig:ku_qq_illustr}) concerning the infrared behavior expected
for~$\tilde{u}$. 

Although we cannot give a reasonable extrapolation of $u(q^2,\mu^2)$
towards the zero momentum limit, our results for both
$\tilde{u}(q^2,\mu^2)$ and $u(q^2,\mu^2)$ confirm that in the infrared
the realization of the limit given in \Eq{eq:u_lim_0} can be studied solely in
terms of the ghost propagator. As this propagator we have found to diverge
stronger than $1/q^2$, the function $u(q^2,\mu^2)$ will
reach minus one at zero momentum. This is because in the infrared the difference
$\abs{u(q^2,\mu^2)-\tilde{u}(q^2,\mu^2)}$ seems to vanish as we have
confirmed in this study for the first time (see \Fig{fig:ku_log_qq}).

Here a remark is in order. The studies of \name{Furui} and
\name{Nakajima}
\cite{Nakajima:1999dq,Nakajima:2000mp,Furui:2003jr,Furui:2004cx}) 
yield as a limit for $u(q^2,\mu^2)$ values ranging between
$-0.7$ and $-0.83$. We cannot confirm these results in our study,
because with respect to our data such limits would correspond to
linear or quadratic extrapolations towards the zero momentum limit in a
figure using a linear momentum scale. However, linear or quadratic
extrapolations would not be valid in this context. 

%------------------------------------------------------------------------------
\section{The gluon propagator explicitly violates reflection positivity}
\label{sec:pos_vio}

In this section we will show that our data for the gluon
propagator presented in \Sec{sec:infrared_ghost_gluon} show evidence
for a violation of reflection positivity. To substantiate this, a
subset of data obtained on our larger lattices has been selected for both
the quenched and the unquenched case and then the 
temporal correlator $C(t,\vec{p}^2=0)$ of the gluon propagator
has been calculated. The continuum expression of this correlator is
defined in \Eq{eq:C_t}; and the lattice equivalent is given by
\begin{equation}
 \label{eq:C_t_latt}
  C(t) \equiv C(t,\vec{p}^2=0) := \frac{1}{\sqrt{V}}\sum_{k_4=0}^{L_T-1}
  D(k_4,\vec{0}) \exp\left\{ \frac{2\pi ik_4t}{L_T}\right\} 
\end{equation}
where $L_T$ denotes the number of lattice points in $\mu=4$
(time) direction and $D$ refers to the gluon propagator in momentum space.

\begin{figure}[t]
\begin{minipage}[t]{0.48\textwidth}
  \includegraphics[width=\textwidth]{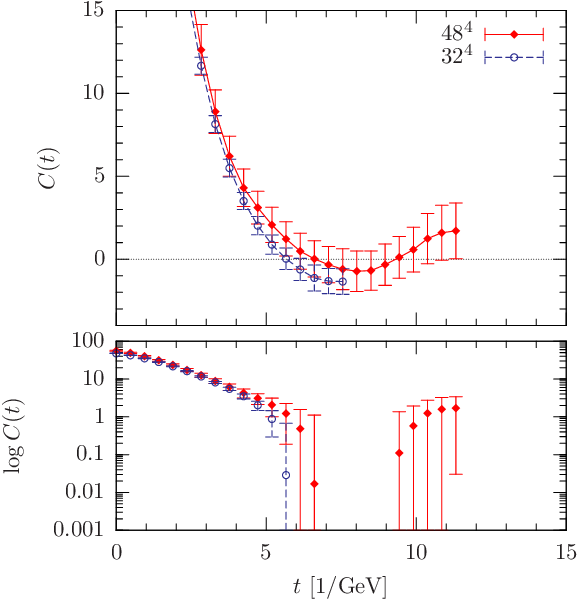}
  \caption{The upper panel shows the real space propagator $C(t)$
    of the gluon fields in the quenched approach for
    different lattice sizes at $\beta=6.0$. In the lower panel the
    same data are shown, however, as $\log C(t)$ for $C(t)>0$.}
\label{fig:c_t}
\end{minipage} 
\hfill
\begin{minipage}[t]{0.48\textwidth}
  \includegraphics[width=\textwidth]{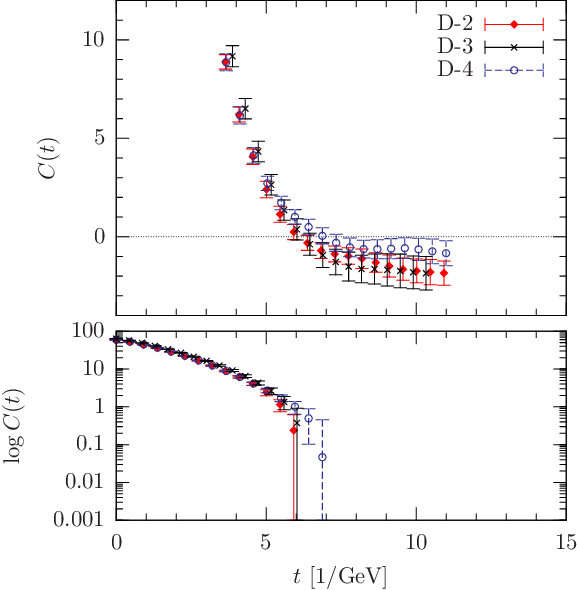}
  \caption{The same as in \Fig{fig:c_t}, however, for the unquenched
    gluon propagator on a $24^3\times48$ lattice. The labels refer to
    the corresponding set of parameters in \Tab{tab:stat_dyn}.} 
\label{fig:c_t_dyn}
\end{minipage}
\end{figure}

Starting with the quenched gluon propagator at $\beta=6.0$, the results
of the correlator are shown in \Fig{fig:c_t} for the lattice sizes $32^4$ and
$48^4$. It is obvious from this figure that the gluon propagator
violates reflection positivity in a finite range of~$t$. 
The same holds for the gluon propagator in the unquenched case as can
be seen in \Fig{fig:c_t_dyn}. There we made use of our measurements of the
gluon propagator on a $24^3\times48$ lattice at different values of~$\beta$
and~$\kappa$. These values are given in \Tab{tab:stat_dyn} and the
labels D-2, D-3 and D-4 refer to these.

\begin{figure}[t]
 \centering
  \mbox{\includegraphics[width=0.45\textwidth]{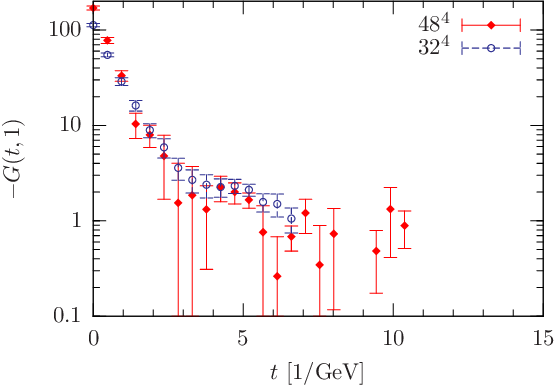}\quad
    \includegraphics[width=0.45\textwidth]{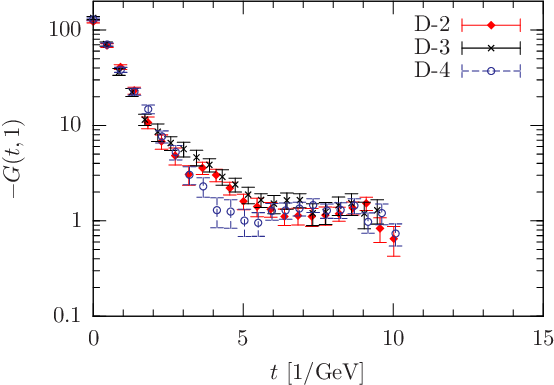}}
  \caption{The quantity $-G(t,a)$ as defined in \Eq{eq:G_t} is shown
    as a function of $t$ at $a=1$. Here we used the same quenched
    (left) and unquenched (right) data as in \Fig{fig:c_t} and
    \ref{fig:c_t_dyn}, respectively.}  
\label{fig:G_t}
\bigskip\bigskip
  \mbox{\includegraphics[width=0.45\textwidth]{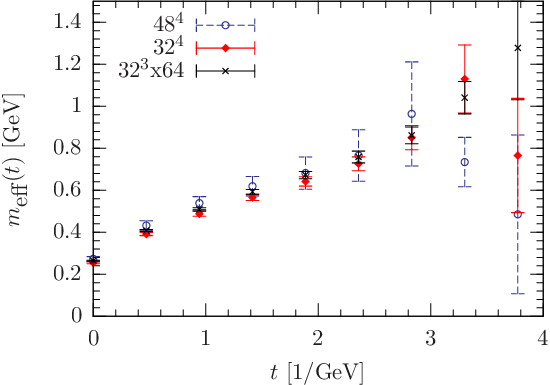}\quad
        \includegraphics[width=0.45\textwidth]{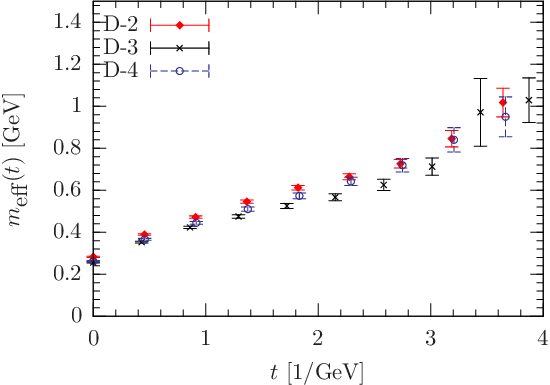}}
  \caption{The effective gluon mass $m_{\eff}$ of the quenched (left)
    and unquenched (right) gluon propagator.}
\label{fig:m_t}
\end{figure}

Following Refs.~\cite{Alkofer:2000wg,Aiso:1997au,Cucchieri:2004mf},
the statement of reflection-posi\-tivity violation can be made even more
clear by considering the quantity 
\begin{equation}
 \label{eq:G_t}
  G(t,a) := \frac{C(t)C(t+2a) - C^2(t+a)}{a^2}\; .
\end{equation}
This is a discretized expression of the continuum quantity
$G(t)[C(t)]^2$ where 
\begin{displaymath}
  G(t):=\frac{d^2}{dt^2}\ln C(t) =
  \frac{C(t)C''(t)-[C'(t)]^2}{[C(t)]^2}= \left\langle (\omega^2 -
  \langle\omega\rangle^2)\right\rangle\; .
\end{displaymath}
Here we adopted the notation \cite{Aiso:1997au,Cucchieri:2004mf}
\begin{displaymath}
  \langle\cdot\rangle := \int (\cdot) e^{-\omega t} \rho(\omega^2)
  d\omega\;.
\end{displaymath}
If the spectral density $\rho(\omega^2)$ is positive definite, so is
$G(t)$. But if instead $G(t)$ is found to be negative then there
cannot be a positive definite spectral density, which entails violation
of reflection positivity. This can easily be seen in \Fig{fig:G_t}
where values of $-G(t,a=1)$ are shown for the data 
considered above. There $G(t,1)$ is clearly negative for
all $t$ and therefore reflection positivity must be violated.

Beside of $G(t)$, one usually also discusses in this context the rise
of an \emph{effective gluon mass}. On the lattice this can be
defined as
\begin{equation}
 \label{eq:m_eff}
  m_{\eff}(t) := -\log\left\{\frac{C(t+a)}{C(t)}\right\}
\end{equation}
where $C(t)$ again refers to the real space propagator of the gluon
fields (\Eq{eq:C_t_latt}). Already in the first numerical
study of the gluon propagator \cite{Mandula:1987rh} this effective
mass was observed to rise with increasing distance $t$ (see also 
\cite{Bernard:1992hy,Marenzoni:1993td}). In fact, the
definition of $m_{\eff}$ in \Eq{eq:m_eff} yields that
\cite{Cucchieri:2004mf} 
\begin{displaymath}
  e^{m_{\eff}(t)} - e^{m_{\eff}(t+a)} = a^2 G(t,a) e^{m_{\eff}(t+a)}.
\end{displaymath}
If reflection positivity were satisfied, then $G(t,a)$ would be
positive semi-definite and therefore $m_{\eff}(t)\ge m_{\eff}(t+a)$. However, for
the gluon propagator in Landau gauge the opposite was found. This has
suggested from the beginning \cite{Mandula:1987rh} that the $SU(3)$ lattice
gluon propagator violates reflection positivity. With our data in
\Fig{fig:c_t} and \ref{fig:c_t_dyn} we show this \emph{explicitly} for
the quenched and unquenched $SU(3)$ gauge theory.

For the sake of completeness we have calculated the effective gluon
mass also for our data. The results are plotted in \Fig{fig:m_t} where
$m_{\eff}$ is shown as a function of $t$ for the 
quenched and unquenched case, respectively. The effective gluon mass
is clearly rising within error bars for moderate values of~$t$. 

Summarizing, our study has found a clear signal for the violation of
reflection positivity for both the quenched and the unquenched gluon
propagator. If we consider this statement in conjunction with the
results presented in the former section which support the Kugo-Ojima
confinement criterion to be satisfied for lattice QCD in Landau gauge, then
this supports the assumption that the transverse gluon states
are confined by the quartet mechanism.

%===============================================================================
%%% Local Variables: 
%%% mode: latex
%%% TeX-master: "../Sternbeck"
%%% End:

\chapter[Spectral properties of the FP operator]%
{Spectral properties of the Faddeev-Popov operator}
\label{ch:spec_FP_operator}

\begin{chapterintro}{S}
  pectral properties of the Landau gauge \FP operator are
  important for understanding many aspects of QCD in Landau
  gauge. In this chapter we report on a study of some of these properties,
  restricting ourselves to the quenched 
  approximations of lattice QCD. We shall start with a discussion of the
  low-lying eigenvalue distribution where the impact of
  the Gribov ambiguity is shown in particular.  Concerning the
  infrared behavior of the ghost propagator discussed above here we
  will analyze the contribution of the different eigenvalues and
  eigenmodes of the \FP operator to the ghost propagator at low momentum.
  Localization properties of the eigenmodes analyzed in addition.
\end{chapterintro}

%--------------------------------------------------------------------------
\section{Specification of lattice samples}

Our investigation of the infrared behavior of ghost and gluon
propagators (see \Sec{sec:infrared_ghost_gluon}) has revealed that
unquenching effects are negligible within errors for the ghost
propagator. We expect the same to hold for the \FP operator
itself. Therefore, we have analyzed its spectral properties solely in
the quenched approximation of QCD.

For this analysis we have used a subset of our pure $SU(3)$ gauge
configurations thermalized with the standard Wilson action at
$\beta=5.8$ and $6.2$ using the lattice sizes $12^4$, $16^4$ and
$24^4$. To study the influence of the Gribov ambiguity we
followed again our \fc{}-\bc{} strategy introduced in
\Sec{sec:fcbc_strategy} and have generated two ensembles of
\emph{first} (\fc{}) and \emph{best} (\bc{}) gauge-fixed configurations.
On those ensembles the low-lying eigenvalues $\lambda$ of the \mbox{FP}
operator and the corresponding eigenmodes have been separately
extracted. For this we used the parallelized version of the
\code{ARPACK} package \cite{arpack}, \code{PARPACK}. 

To be specific, the 200 lowest (non-trivial) eigenvalues and
their corresponding eigenfunctions have been calculated at $\beta=6.2$
using the lattice sizes $12^4$ and $16^4$ (see
\Tab{tab:statistics}). Due to restricted amount of 
computing time only 50 eigenvalues and eigenmodes have been extracted 
on the $24^4$ lattice at the same $\beta$. 
In addition, 90 eigenvalues have been calculated on a $24^4$ lattice
at $\beta=5.8$ providing us with an even larger physical volume. This
allows us to check whether low-lying eigenvalues are shifted towards
$\lambda \to 0$ as the physical volume is increased. The eight (trivial) zero 
eigenvalues with the corresponding constant zero modes have always
been discarded. 

\begin{table*}
  \centering
  \begin{tabular}{c@{\quad}cc@{\quad}c@{\quad}c@{\qquad}c}
\hline\hline\rule{0pt}{2.5ex}
    No. & $\beta$ & lattice & \# conf & \# copies & \# eigenvalues\\
\hline\rule{0pt}{2.5ex}
    F-1 & 5.8  & $24^4$  & 25  & 40 & 90\\
    F-2 & 6.2  & $12^4$  & 150 & 20 & 200\\
    F-3 & 6.2  & $16^4$  & 100 & 30 & 200\\
    F-4 & 6.2  & $24^4$  & 35  & 30 & 50\\
\hline\hline
  \end{tabular}
\caption{Statistics of data used in our analysis. The last
  column lists the number of eigenvalues extracted separately on \fc{}
  and \bc{} copies of $U$. At $\beta=6.2$ the corresponding eigenmodes
  were calculated, too. Labels given in the first row refer to the
  corresponding entries in \Tab{tab:stat_que}.}
  \label{tab:statistics}
\end{table*}

%----------------------------------------------------------------------------
\section{The low-lying eigenvalue spectrum}
\label{eq:lowlying_eigenvalues}

We start the discussion with the two lowest (nontrivial)
eigenvalues of the \FP operator and then give an estimate for the
density of low-lying eigenvalues. 

%---------------------------------------------------------------------
\subsection{The lowest and second lowest eigenvalues}

\begin{figure}[t]
 \centering
  \includegraphics[width=0.77\textwidth]{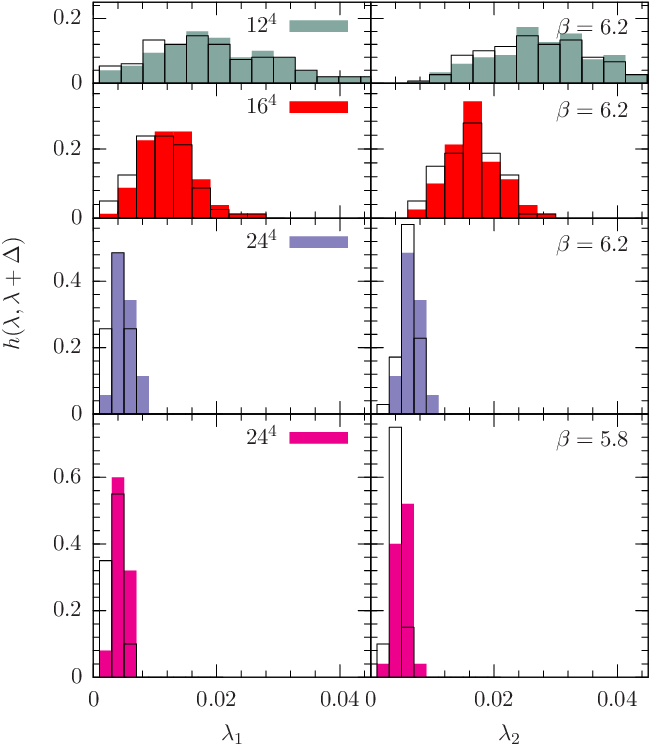}
  \caption{The frequency $h(\lambda)$ per configuration of the
    lowest (left panels) and second lowest (right panels)
    eigenvalue $\lambda$ of the Faddeev-Popov operator
    is shown. Filled boxes represent the distribution obtained
    on best (\bc{}) gauge copies, while open ones represent
    those on first (\fc{}) copies.}
  \label{fig:fps_lowest}
\end{figure}

The distributions of the two lowest-lying eigenvalues, $\lambda_1$ and
$\lambda_2$, of the \FP operator are shown for different volumes in
\Fig{fig:fps_lowest}. There $h(\lambda,\lambda+\Delta\lambda)$ represents
the average number (per configuration) of eigenvalues found in the
interval $[\lambda,\lambda+\Delta\lambda]$. To disentangle the
distributions for the two different sets of gauge copies, open (full)
bars refer to the distribution on \fc{} (\bc{}) gauge copies.

It is obvious from this figure that both eigenvalues, $\lambda_1$
and $\lambda_2$, are shifted to lower values as the physical volume is
increased. In conjunction the spread of $\lambda$ values is
shrinking. This would be even more obvious, if we had shown 
both distributions as functions of $\lambda$ in physical units.

\begin{figure}[htb]
  \centering 
  \includegraphics[width=0.9\textwidth]{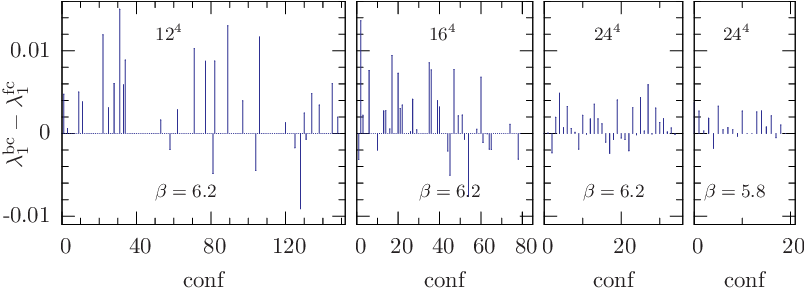}
  \caption{The differences $\lambda^{\bc}_1-\lambda^{\fc}_1$
    of the lowest FP eigenvalues calculated on \bc{} and \fc{}
    representatives for each gauge configuration are shown.
    From left to right the lattice sizes are $12^4$, $16^4$ and
    $24^4$ at $\beta=6.2$ and $24^4$ at $\beta=5.8$.}
  \label{fig:diff_lambda}
\end{figure}

It is also visible that the two low-lying eigenvalues
$\lambda^{\fc}_i$ ($i=1,2$) on \fc{} gauge copies tend to be lower
than those on \bc{} copies. However, this holds \emph{only on average}
as can be seen from \Fig{fig:diff_lambda}. There the differences
$\lambda^{\bc}_1-\lambda^{\fc}_1$ of the lowest eigenvalues on \fc{}
and \bc{} gauge copies are shown for different lattice sizes at
$\beta=6.2$ and $5.8$. It is evident that there are only few cases
where \mbox{$\lambda^{\bc}_1<\lambda^{\fc}_1$}, even though
$F^{\bc}\ge F^{\fc}$ always holds for the gauge functional.

In addition we have checked how the average values $\langle\lambda\rangle$
of the respective eigenvalue distributions tends towards zero as the linear
extension $aL$ of the physical volume is growing. For this the
lattice spacing~$a$ has been specified in physical units. As in
\Ch{ch:prop_results} we followed Ref.~\cite{Necco:2001xg} to
fix $a$. For $\beta=5.8$ and 6.2 we used $a^{-1}$=1.446~GeV and 2.914~GeV,
respectively, using the Sommer scale $r_0=0.5$~fm.

\begin{figure}
\begin{minipage}[t]{6.5cm}
 \includegraphics[height=6cm]{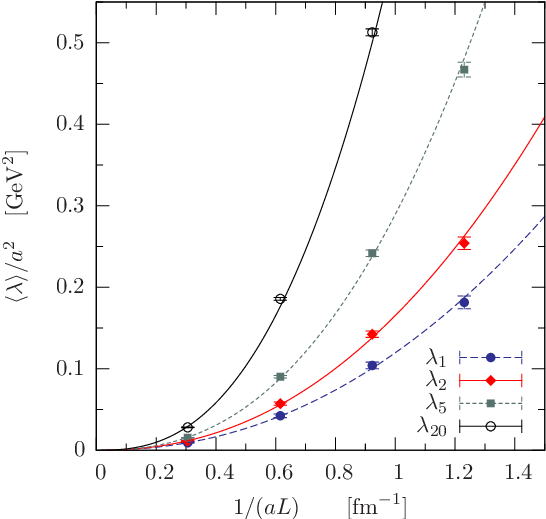}
  \caption{The average values $\langle\lambda_i\rangle/a^2$ (scaled to
    physical units) of the eigenvalues $\lambda_i$
    $(i=1,2,5,20)$ are shown vs.\ the inverse of the linear lattice
    extension $aL$. Only eigenvalues on
    \bc{} copies are shown. The lines represent fits
    to the data using the ansatz $a^{-2}\langle\lambda_i\rangle =
    C_i/(aL)^{2+\epsilon_i}$.}
\label{fig:fps_mean}
\end{minipage} 
\hfill
\begin{minipage}[t]{6.5cm}
\includegraphics[height=6cm]{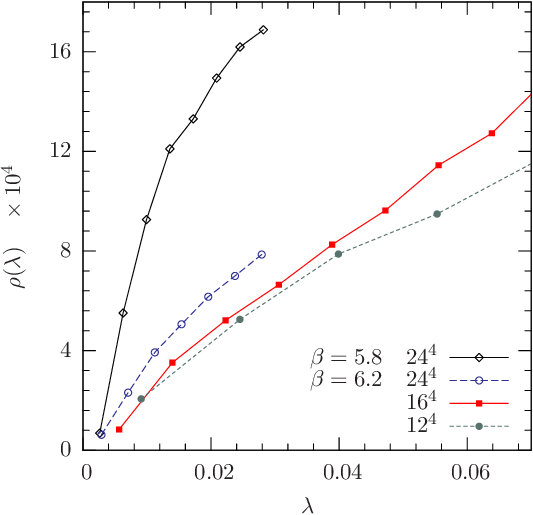}
  \caption{The eigenvalue density $\rho$ for  \bc{} copies
    as a function of $\lambda$ estimated on $12^4$ and $16^4$
    lattices at $\beta=6.2$ and on a $24^4$ lattice for $\beta=6.2$
    and $5.8$. Bin sizes have been chosen as small as possible for
    each lattice size. Symbols mark the middle of each bin and 
    lines are to guide the eye.}
\label{fig:fps_rho}
\end{minipage}
\end{figure}

If the low-lying eigenvalues are supplemented with physical units
it turns out that the average values of their distributions tend
towards zero stronger than by volume scaling proportional to
$1/(aL)^{2}$. In fact, using the ansatz 
\begin{equation}
 f(aL) = \frac{C}{(aL)^{2+\varepsilon}}
 \label{eq:lambda_aL_ansatz}
\end{equation}
to fit the data of $\langle\lambda_i\rangle/a^2$ for different $(aL)$,
always a positive $\varepsilon$ is found. The parameter of these fits
are given in \Tab{tab:fps_fit} and in \Fig{fig:fps_mean} we show the
data and the corresponding fitting functions.  There one clearly sees,
the low-lying eigenvalues not only approach zero, but also become
closer to each other with increasing $aL$. The latter issue has been
addressed by fitting the differences
$(\langle\lambda_{i+1}\rangle-\langle\lambda_i\rangle)/a^2$ of
adjacent average values using the same ansatz
\Eq{eq:lambda_aL_ansatz}. In \Tab{tab:fps_fit} we give the parameter
of such fits.

\begin{table}[t]
  \centering
  \begin{tabular}{c@{\qquad\qquad\quad}c@{\qquad\quad}c@{\qquad\quad}c}
\hline\hline\rule{0pt}{2.5ex}
     $a^2f(aL)$    & $C$ & $\epsilon$ & $\chi^2/\textsc{ndf}$\\
\hline\rule{0pt}{2.5ex}
    $\langle\lambda_1\rangle$ & 0.120(3) & 0.16(4) & 0.7\\
    $\langle\lambda_2\rangle$ & 0.165(4) & 0.24(5) & 1.8\\
    $\langle\lambda_5\rangle$ & 0.290(1) & 0.45(4) & 3.5\\*[0.1cm]
    $\langle\lambda_2\rangle-\langle\lambda_1\rangle$ & 0.045(2) &
    0.47(9) & 0.4\\
    $\langle\lambda_3\rangle-\langle\lambda_2\rangle$ & 0.051(1) &
    0.88(8) & 0.2\\
    $\langle\lambda_4\rangle-\langle\lambda_3\rangle$ & 0.033(1) &
    0.62(33) & 2.0\\
    $\langle\lambda_{5}\rangle-\langle\lambda_4\rangle$ & 0.037(1) &
    0.89(1) & 0.003\\
\hline\hline
  \end{tabular}
  \caption{The parameters $C$ and $\epsilon$ from fitting either the
    averages $\langle\lambda_i\rangle/a^{2}$ or the differences
    of adjacent average values
    $\langle\lambda_{i+1}\rangle/a^{2}-\langle\lambda_i\rangle/a^{2}$
    of the corresponding eigenvalue distributions to the ansatz
    $f(aL) = C_i/(aL)^{2+\epsilon_i}$.}
  \label{tab:fps_fit}
\end{table}

%----------------------------------------------------------------------------
\subsection{An estimate for the density of low-lying eigenvalues}

The eigenvalue density $\rho(\lambda)$ is of particular interest. At
small $\lambda$ this quantity has been estimated here by
\begin{equation}
  \label{eq:rho_lambda} \rho(\lambda) = \frac{h(\lambda,
\lambda+\Delta\lambda)}{N \Delta\lambda}\, ,
\end{equation} i.e. the average number $h$ of eigenvalues per
gauge-fixed configuration within the interval
$[\lambda,\lambda+\Delta\lambda]$ divided by the bin size
$\Delta\lambda$. For normalization the denominator $N=8V$ has been
chosen, since the FP matrix is a $N \times N$ sparse symmetric matrix
with $N$ linearly independent eigenstates. The trivial zero
modes would be described by a term $8\delta(\lambda)$ in $\rho(\lambda)$
(not shown). 

The estimates for the density $\rho$ are shown in \Fig{fig:fps_rho}
for the different volumes used. The bin sizes have been reasonably
adjusted for each volume separately. In \Fig{fig:fps_rho} one clearly
sees the eigenvalue density close to $\lambda=0$ becomes steeper as a
function of $\lambda$ as the physical volume becomes larger. It is
remarkable that the increase going from $\beta=6.2$ to $\beta=5.8$ on
a $24^4$ lattice is larger than going from $12^4$ to $24^4$ at
$\beta=6.2$ fixed, although in both cases the physical volume is
increased by a factor of about 16.

%-----------------------------------------------------------------
\section{Eigenmode expansion of the ghost propagator}

Along with the calculation of the low-lying eigenvalues, the
corresponding eigenvectors $\vec{\phi}(x)$ have been determined as
well. These are of particular importance for the infrared behavior of
the ghost propagator as it becomes clear from \Eq{eq:contribution}. In
fact, if all 
eigenvalues $\lambda_i$ of the FP operator and the corresponding
eigenvectors $\vec{\Phi}_i(k)$ in momentum space were available the
ghost propagator could be constructed out of them according to
\Eq{eq:contribution} and
(\ref{eq:Def-ghost-q-by-spectrum}). Unfortunately, their determination 
for each configuration is numerically too demanding.

However, restricting the sum in \Eq{eq:contribution} to the $n$ lowest
eigenvalues and eigenvectors ($n \ll N=8V-8$), we can figure out
to what extent these modes, \ie the corresponding estimator
\Eq{eq:Def-ghost-q-by-spectrum} and (\ref{eq:contribution})
\begin{displaymath}
   G_n(q^2(k)) = \langle G(k|n) \rangle_{\textrm{MC}}
\end{displaymath}
where
\begin{displaymath}
   G(k|n) = \frac{1}{8}~\sum_{i=1}^n
   \frac{1}{\lambda_i}\,\vec{\Phi}_i(k)\cdot\vec{\Phi}_i(-k)
\end{displaymath}
saturates the \emph{full} ghost propagator $G(q^2)$. The latter is obtained,
of course, independently for a set of momenta by inverting the FP matrix on
a set of plane waves. See \Sec{sec:ghost_results} for the
corresponding data of $G(q^2)$. 

\begin{figure}[t]
  \centering
  \includegraphics[width=8cm]{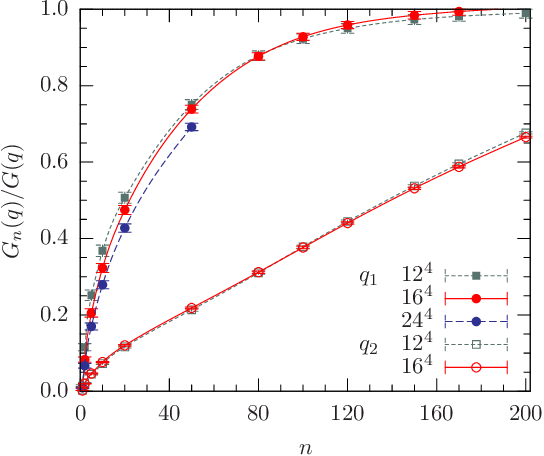}
  \caption{The ratio of the truncated ghost propagator $G_n(q^2)$
   (in terms of the $n$ lowest FP eigenmodes and eigenvalues) to
   the full estimate $G(q^2)$ (taken from \cite{Sternbeck:2005tk})
   shown as a function of $n$ for the lowest ($q^2_1$) and second
   lowest ($q^2_2$) momentum. The inverse coupling is $\beta=6.2$
   and the lattice size ranges from $12^4$ to $24^4$.
   All data refer to \bc{} copies.}
  \label{fig:gh_vs_eigenmodes}
\end{figure}

The degree of saturation is shown in \Fig{fig:gh_vs_eigenmodes} for the lowest
$q^2_1$ and the second lowest momentum $q^2_2$ available on different
lattice sizes for $\beta=6.2$. There the values of $G_n(q^2)$ have been
presented relative to the values for the full propagator $G(q^2)$ in order to
assess the saturation for different volumes. Since $\vec{\Phi}_i(k)$
has been obtained by a fast Fourier transformation of the eigenvector
$\vec{\phi}_i(x)$, all lattice momenta $k$ are available. Thus
$G_n(q^2)$ refers to the average over all $k$ giving raise to the same
momentum $q^2$. The full propagator values $G(q^2)$ at $q^2_1(k)$ and
$q^2_2(k)$, however, refer to the averages over lattice momenta
$k=([1,0],0,0)$ and to $k=(1,1,0,0)$, respectively.

Let us consider first the lowest momentum $q^2_1$. We observe from
\Fig{fig:gh_vs_eigenmodes} that the approach to convergence differs,
albeit slightly, for the three different lattice sizes. The relative
deficit for $n<50$ rises with the lattice volume. For $n>100$ the rate
on a $16^4$ lattice is even a bit larger than that on a $12^4$
lattice. We could not afford to get data for $n>50$ on the
$24^4$ lattice. However, for the $12^4$ and $16^4$ lattices the rates
of convergence are about the same.  For example, taking
only 20 eigenmodes one is definitely far from saturation (by about
50\%) whereas 150 to 200 eigenmodes are sufficient to reproduce
the ghost propagator within a few percent. In other words, the ghost
propagator at lowest momentum on a $12^4$ ($16^4$)
lattice is formed by the lowest 0.12\% (0.03\%) of the
eigenvalues and eigenfunctions of the \FP operator.

For the second lowest momentum $q^2_2$ the contribution of even 200
eigenmodes is far from being sufficient to approximate the propagator.

%----------------------------------------------------------------------------
\section{The problem of exceptional configurations}
\label{sec:exceptional}

We turn now to a peculiarity of the ghost propagator at larger $\beta$
of which has been reported already in
\cite{Sternbeck:2005tk,Sternbeck:2005vs}. It was also seen 
in an earlier $SU(2)$ study \cite{Bakeev:2003rr}. While inspecting our
data we found, though rarely, that there are exceptionally
large values in the Monte Carlo (MC) time histories of the ghost
propagator at lowest momentum. 

\begin{figure*}
  \centering
  \mbox{\includegraphics[width=0.48\textwidth]{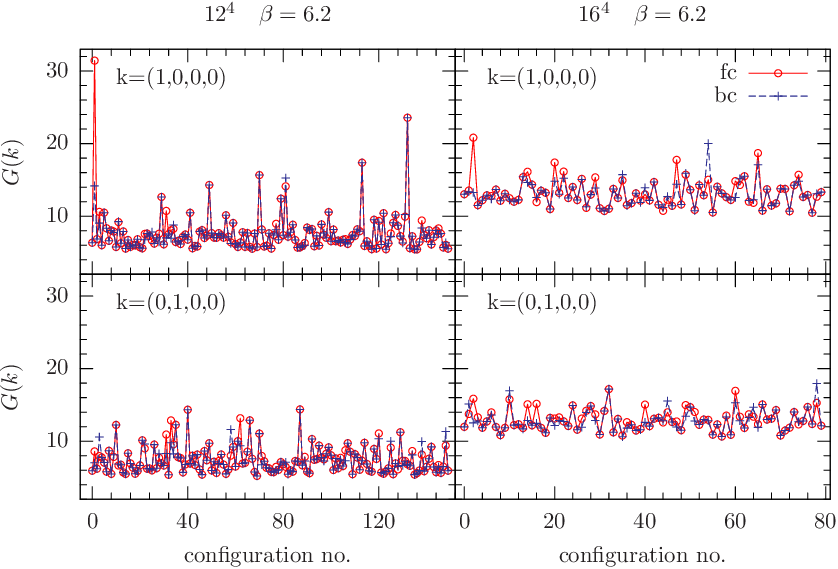}\quad
    \includegraphics[width=0.48\textwidth]{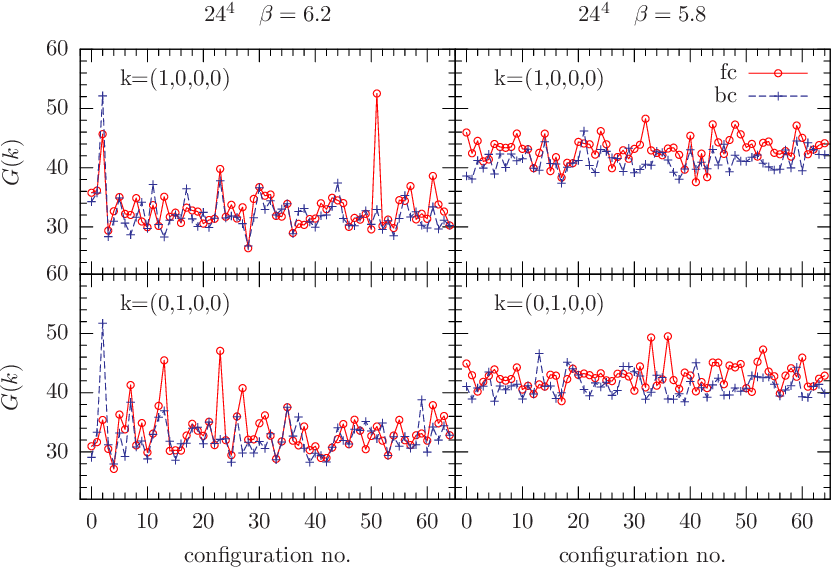}}
  \caption{The MC time histories of the ghost propagator calculated on
    first (\fc) and best (\bc) gauge copies on a $12^4$,
           $16^4$ and $24^4$ lattice at \mbox{$\beta=6.2$} and $24^4$
           lattice at \mbox{$\beta=5.8$}. From left to right the
           corresponding runs listed in \Tab{tab:statistics} and
           \Tab{tab:stat_que} are F-2, F-3, F-4/S-9 and F-1/S-2. The
           upper and lower panels 
           show data for the lowest momentum realization
           \mbox{$k=(1,0,0,0)$} and \mbox{$k=(0,1,0,0)$}, respectively.}
  \label{fig:ghost_history}
\end{figure*}

In \Fig{fig:ghost_history} such time histories are shown of the ghost
propagator on \fc{} and \bc{} gauge copies for the two smallest
momentum realizations \mbox{$k=(1,0,0,0)$} and
\mbox{$k=(0,1,0,0)$}. From left to right the panels are ordered in
ascending order with the (physical) lattice sizes $12^4$, $16^4$ and
$24^4$ at $\beta=6.2$ and $24^4$ at $\beta=5.8$.

As can be seen from this figure in the majority extreme spikes are
reduced (or even not seen) when the ghost propagator could be afforded
to be measured on a better gauge copy (\bc{}) for a particular
configuration. Furthermore, it is obvious that the
\emph{exceptionality} of a given gauge copy is exhibited not
simultaneously for all different realizations of the lowest
momentum. Consequently, to reduce the impact of such large values on the
evaluation of a statistical average for the ghost propagator one
should average, if possible, over all momentum realizations giving rise
to the same momentum~$q^2$. 

We have tried to find a correlation of such outliers in the
history of the ghost propagator with other quantities measured
in our simulations. For example we have checked whether there is a
correlation between the values of the ghost propagator $G(k)$ as they
appear in \Fig{fig:ghost_history} and the lowest eigenvalue
$\lambda_1$ of the FP operator. 

\begin{figure*}
  \centering
  \includegraphics[width=0.8\textwidth]{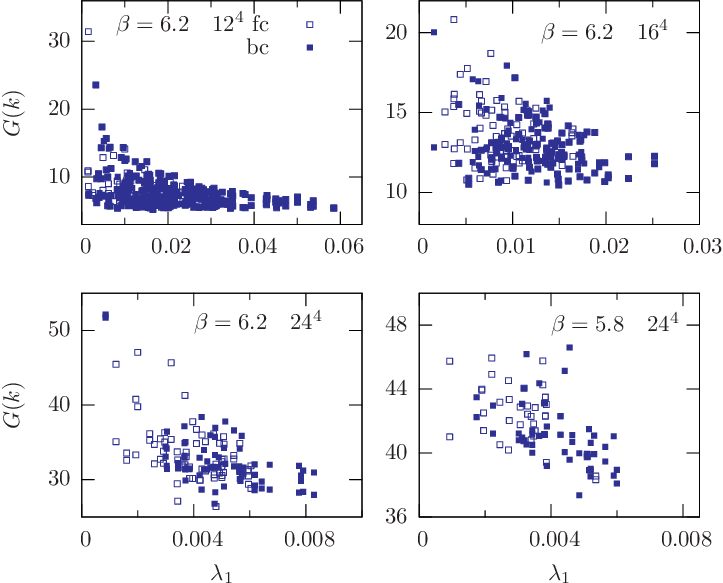}
  \caption{Scatter plots of MC time history values of the ghost
    propagator $G(k)$ at \mbox{$k=([1,0],0,0)$} vs.\ the lowest FP
    eigenvalue $\lambda_1$ are shown. The upper panels show data at
    $\beta=6.2$ on a $12^4$ (left) and $16^4$ (right) lattice,  the
    lower ones on a $24^4$ lattice at $\beta=6.2$ (left) and
    $\beta=5.8$ (right). Open symbols refer to \fc{} gauge copies and
    filled symbols to \bc{} copies.} 
\label{fig:fps_gh}
\medskip
  \centering
  \includegraphics[width=0.8\textwidth]{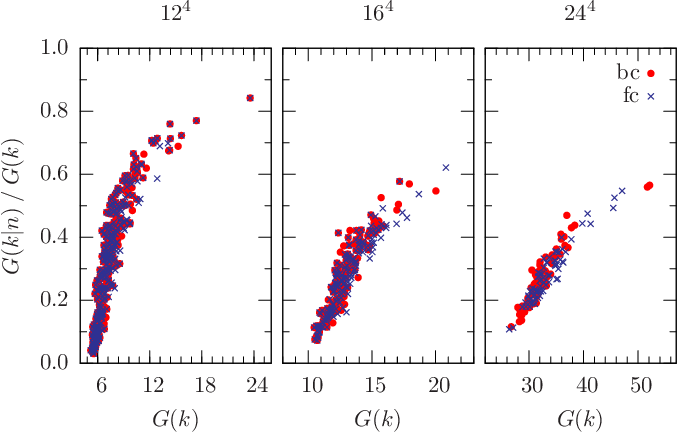}
  \caption{Scatter plot of the relative contribution of the truncated
  sums
  $G(k|n)$ over eigenmodes (see \Eq{eq:contribution}) to the full
  ghost propagator values $G(k)$ versus $G(k)$ for lattice momenta
  \mbox{$k=(1,0,0,0)$} and \mbox{$k=(0,1,0,0)$}. From left to
  right the lattice sizes are $12^4$, $16^4$ and
  $24^4$ all for $\beta=6.2$. Data for \fc{} and \bc{} gauge copies
  have been plotted separately.}
\label{fig:gh_vs_lowest_ev}
\end{figure*}

In \Fig{fig:fps_gh} we show such correlation in a scatter plot for
different lattice sizes at $\beta=5.8$ and $6.2$. There each entry
corresponds to a pair $[\lambda_1,G(k)]$ measured on a given gauge
copy of our sets of \fc{} and \bc{} copies. It is visible in this
figure, gauge copies giving rise to an extremely large MC value for
the ghost propagator are those with very low values for
$\lambda_1$. This holds for the $12^4$ and $16^4$ lattice. However, a
very low eigenvalue is \emph{not sufficient} to obtain large MC values
for $G(k)$ as can be seen in the same figure. It is possible, of
course, that such gauge copies with extremely small eigenvalues would
turn out to be \emph{exceptional} for another realization of lowest
momentum $q^2(k)$ than those two we have used. This might explain why
some configurations with extremely small lowest eigenvalues were not
found to be exceptional with respect to the ghost propagator at
$k=(1,0,0,0)$ and $k=(0,1,0,0)$.

In the light of \Eq{eq:contribution} it is not adequate to
concentrate just on the lowest eigenvalues. Instead, one can monitor
the contribution of a certain number of eigenvalues $\lambda_i$ and
eigenmodes $\vec{\Phi}_i(k)$ to the ghost propagator at some momentum
in question. Therefore, we have compared the truncated sums $G(k|n)$
according to \Eq{eq:contribution}
with the MC history values of the full ghost
propagator $G$. In fact, we show in the scatter plots in
\Fig{fig:gh_vs_lowest_ev} the ratios $G(k|n)/G(k)$ versus $G(k)$ for
$n=10$ and for various lattice sizes.

Obviously there is a strong correlation between the chosen group of
low-lying modes and the MC time history values of the full ghost
propagator.  Indeed, if we consider values $G(k)>15$ to be
\emph{exceptional} in the left-most panel ($12^4$ lattice) we find
that the contribution of the 10 lowest modes amounts to more than 75\%
of the actual value of the ghost propagator. On the opposite, for low
$G(k)$ values the main contributions come necessarily from the higher
eigenmodes, while the 10 lowest modes contribute a minor part only. A
similar but less dominant contribution of the 10 lowest modes is found
for the time histories produced on larger lattices ($16^4$, $24^4$).

%-------------------------------------------------------------------------
\section{Localization properties of low-lying eigenvectors}
\label{sec:ipr}

In recent years a good deal of attention has been directed to the
localization properties of various operators (Dirac operator,
covariant Laplacian) in the hope to understand more about confinement
beyond heavy quark probes. The first quantity of interest is the inverse
participation ratio (\IPR{}). Given an eigenvector $\vec{\phi}(x)$ the
\IPR{} is defined as 
\begin{displaymath}
   \IPR = V \sum_{x} |\vec{\phi}(x)|^4 \qquad\textrm{with}\; V=L^4.
\end{displaymath}
Although a direct physical meaning for this quantity is lacking yet,
it is a measure for the localization of an eigenvector. It enables
us to distinguish between eigenmodes with approximately uniformly
distributed modulus squared $|\vec{\phi}(x)|^2$ \mbox{($\IPR\approx1\ldots
2$)} and more specific ones with a small number of sites $x$ having large
intensity $|\vec{\phi}(x)|^2$ \mbox{($\IPR \sim
\textrm{O}(100)$)} where they might be pinned down by special local gauge field 
excitations. Note, the eight (trivial) zero modes ($\lambda=0$)
of the FP operator are constant and have $\IPR{}=1$.

\begin{figure}
\centering
  \includegraphics[width=0.95\textwidth]{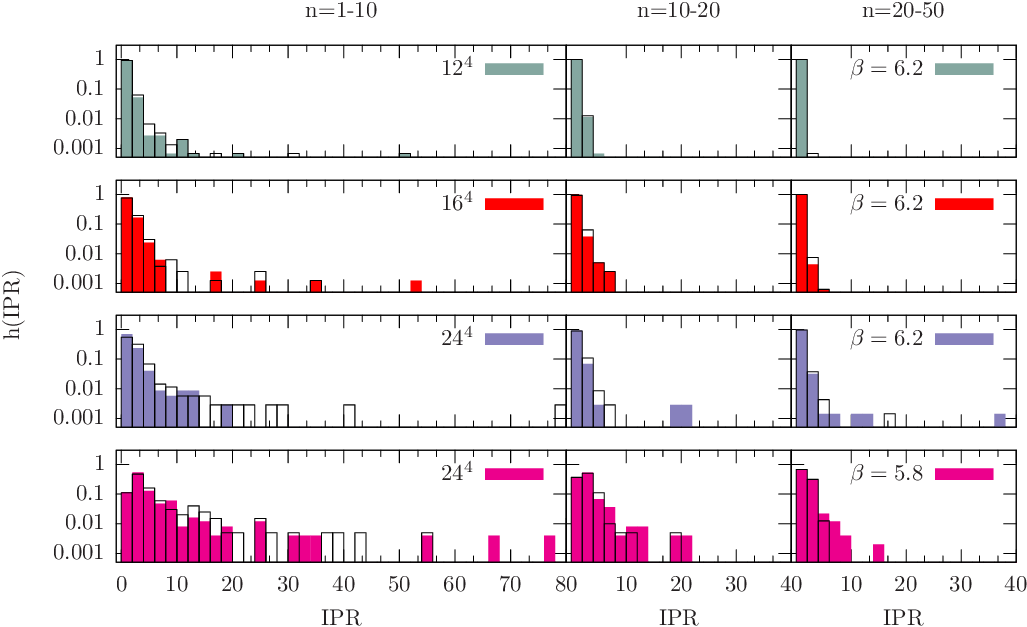}
  \caption{The relative distribution $h$ of \IPR{} values of the 10
    (left), the 10 to 20 (middle) and the 20 to 50 (right) lowest
    eigenmodes
    are shown. Note there is a logarithmic scale for $h(\IPR{})$. Each
    row
    corresponds to one pair of $\beta$ and lattice size. Filled boxes
    refer to distributions on \bc{} gauge copies, while open
    ones correspond to \fc{} copies.}
  \label{fig:ipr}
\end{figure}

In \Fig{fig:ipr} the relative distribution $h$ of \IPR{} values per
gauge-fixed configuration are shown, separately for certain groups of
eigenstates. Again open (full) histogram bars refer to the
distribution on \fc{} (\bc{}) gauge copies. From this figure we learn
that the majority of eigenvectors of the Faddeev-Popov operator is
\emph{not localized} independent of the choice of gauge
copies. In any case, the rare cases of large $\IPR$ values have been
found among the 10 lowest non-zero eigenmodes. This becomes more
likely as the physical volume is increased. So far we have no physical
interpretation what causes the stronger localization in these rare
cases.

%===============================================================================
%%% Local Variables: 
%%% mode: latex
%%% TeX-master: "Sternbeck"
%%% End:

%-- Conclusions ----------------------------------------------------------------

\titleformat{\chapter}
            [display]
	    {\Huge\scshape\flushright}
	    {\filleft\MakeUppercase{\chaptertitlename}\Huge\thechapter}
	    {4ex}
            {\vspace{3ex}\titlerule\vspace{0.5ex}\filleft}[\vspace{0.5ex}%
            \titlerule]
\chapter*{Conclusions and outlook}
\addcontentsline{toc}{chapter}{Conclusions and outlook}
\markboth{Conclusions and outlook}{}
\markright{Conclusions and outlook}{}

{\font\yn=cmr17 scaled \magstep4
 \setlength{\parindent}{0pt}
 \setlength{\parskip}{.66\baselineskip}
\begin{window}[0,l,{\yn I},{}]
  n this study we have focused on the infrared properties of $SU(3)$
  gluodynamics in Landau gauge using the framework of lattice QCD.
  We have tried to cover as much as possible the different aspects relevant
within this context and to verify several assumptions made in recent
years. In doing so, we hope to have provided a consistent analysis of
several issues that have an effect on the infrared behavior of gluon
and ghost propagators, and that, in turn, the behavior we have found 
satisfies necessary criteria for confinement which apply to QCD
in Landau gauge. 
\end{window}
}
\bigskip

\subsection*{The method}

For this study we used the Wilson formulation of lattice QCD with and
without dynamical clover-improved Wilson fermions. The gauge group was fixed to
$SU(3)$. The quenched gauge configurations were generated
at the three values $\beta=5.8$, 6.0 and 6.2 using a variety of
different symmetric and asymmetric lattice geometries. In the
symmetric case we used the lattice sizes $12^4$, $16^4$, $24^4$,
$32^4$ and $48^4$, whereas the asymmetric lattices were chosen to be of 
size $24^3\times48$, $32^3\times64$, $16^3\times128$ and $24^3\times128$.
To study unquenching effects we have analyzed gauge
configurations provided to us by the $\QCDSF$ collaboration. These
configurations were thermalized on a $24^3\times48$ lattice in
the presence of two flavors of clover-improved Wilson fermions using three
different pairs of $\beta$ and $\kappa$. The corresponding lattice spacings are
comparable to that at $\beta=6.0$ in the quenched case. 

Both the quenched and unquenched configurations were transformed such
that they satisfy the lattice Landau gauge condition. For gauge-fixing
we used either over--relaxation or Fourier--accelerated
gauge-fixing. A subset of our quenched configurations was gauge-fixed
even more than once, always starting from a different random gauge copy of the
initial (unfixed) configuration. This has allowed us to examine how 
the Gribov ambiguity affects the gluon and ghost propagators
and the eigenvalue spectrum of the FP operator.

\subsection*{The results}

In the following we give a summary of our results presented in
previous chapters. Then we draw our conclusions and give
recommendations for future studies.

\subsubsection{The influence of the Gribov ambiguity}

We have demonstrated that the presence of Gribov copies systematically  
affects the ghost propagator at low momentum, whereas
for the gluon propagator such an effect stays within error bars (see also
\cite{Sternbeck:2005tk}). To be specific: 
Measuring the ghost propagator and ignoring the Gribov ambiguity, the
ghost propagator near the momentum $q^2=0.2~\!\GeV^2$ ($1~\!\GeV^2$)
turns out to be overestimated by about 5\% (2\%) compared 
to an estimate obtained on a ensemble of \emph{best} gauge
copies. As \emph{best} we have considered that gauge-fixed copy 
which gave rise to the largest gauge functional value for a particular gauge
configuration. Our results corroborate previous findings for the
$SU(2)$ gluon and ghost propagators
\cite{Cucchieri:1997dx,Bakeev:2003rr,Nakajima:2003my}, but cast doubt on 
those for the $SU(3)$ gluon propagator  ($\beta=5.8$, $12^4$ lattice)
in \cite{Silva:2004bv}. 

Additionally, our data are in favor of the picture promoted in
\cite{Zwanziger:2003cf}. According to this, continuum vacuum
expectation values of correlation functions obtained from a functional 
integration over the fundamental modular region $\Lambda$ are equal to
those over the Gribov region $\Omega$. Gribov copies inside $\Omega$ 
should not affect expectation values in the continuum, because
functional integrals are dominated by the common boundary of
$\Lambda$ and $\Omega$. We have found some numerical evidence that 
the influence of Gribov copies on the ghost propagator decreases 
at the same (physical) momentum if the physical volume is enlarged.
Note that very recently \cite{Bogolubsky:2005wf} similar indications
have been found for the $SU(2)$ gauge group taking non-periodic $Z(2)$
transformations into account. 
In order to eliminate the last doubts, a future study should continue
and follow our \fc{}-\bc{} strategy (explained in the text) on lattice
sizes larger than $24^4$. 

To make the study of the Gribov-copy dependence more
complete, we have also shown that the Gribov ambiguity
is reflected in the low-lying eigenvalue spectrum of the FP operator.
We have found that, on average, the low-lying eigenvalues extracted on
\bc{} gauge copies are larger than those on \fc{} copies. Thus better
gauge-fixing (in terms of the gauge functional) has the
tendency to keep gauge-fixed configurations
slightly away from the Gribov horizon. 

\medskip

\subsubsection{Other systematic effects on the propagators}

Using different lattices sizes at three different $\beta$ values, we have tried
to analyze the systematic effects on the gluon and ghost propagators of
changing either the lattice spacing $a$ or the physical
volume~$V$. We have found that for both, the gluon and ghost dressing
functions, finite volume effects are clearly visible at volumes smaller
than $(2.2~\fm)^4$, which corresponds to a $16^4$ lattice at $\beta=5.8$.
The effect grows with decreasing momentum or decreasing lattice size.
At larger volumes, however, the data for $q>1~\GeV$ coincide within
errors for the different lattice sizes. For $q<1~\GeV$ we have found only
small finite volume effects for both dressing functions at the lowest
momentum. This is based on data obtained on the lattice sizes $24^4$,
$32^4$ and $48^4$ using $\beta=5.8$ and 6.0. Concerning discretization effects
our study is only partial and limited to a region of intermediate momenta.
In any case, fixing the physical volume to
\mbox{$V\approx(2.2~\fm)^4$}
we have found that the gluon dressing function at constant physical
momentum increases with decreasing the lattice spacing. A similar
effect (beyond error bars) is not observable for the ghost dressing function.

We have combined this investigation of lattice artifacts with an analysis of
effects caused by asymmetric lattice geometries. We could demonstrate that the
more asymmetric the lattice size has been chosen, the larger are the systematic
errors induced by that. In fact, data obtained
for the ghost (gluon) propagator on asymmetric lattices are less
enhanced (suppressed) than those obtained on symmetric lattices geometries.
The same effects have been reported recently for the gluon propagator
in three-dimensional pure $SU(2)$ gauge theory
\cite{Cucchieri:2006za}. Therefore,
extractions of an infrared exponent for the gluon propagator
using only on-axis momenta on asymmetric lattices 
\cite{Silva:2005hb,Silva:2005hd} 
(without adapting the lattice spacing in the different directions to
compensate this) should be taken with caution.

\subsubsection{Infrared behavior of gluon and ghost propagators}

Studying the momentum dependence of the gluon and ghost
dressing functions, we have applied cuts on our data, namely the
cylinder and the cone cut \cite{Leinweber:1998uu}. They have much reduced
the lattice artifacts mentioned above. In fact, our data
for the renormalized dressing functions surviving these cuts lie on
smooth curves if considered as functions of the momentum. 

We have compared these
renormalized data with recent solutions of truncated system of DSEs 
for the gluon and ghost propagators and have tried to extract the infrared
exponents, $\kappa_D$ and $\kappa_G$, for the gluon and ghost
dressing functions, respectively. From fits with the corresponding
power laws to our data we \emph{cannot} confirm the relation
$\kappa_G=2\kappa_D$ as expected from DSE studies in the
continuum \cite{vonSmekal:1997is,vonSmekal:1998yu}. Also, both
exponents are found to be significantly lower than expected
in \cite{Lerche:2002ep,Zwanziger:2001kw}. Such a conclusion was also
drawn in other lattice studies
\cite{Boucaud:2005ce,Boucaud:2006if,Furui:2004cx}.
 
Moreover, our fits suggest that power-behaved gluon and ghost
propagators are not the best description of the momentum dependence of
our data, at least in the region of lower momenta
considered here. The data for the ghost propagator seem to depend
logarithmically on the momenta in this region. This observation has been
confirmed very recently in 
\cite{Boucaud:2006if}, even though in this reference also arguments for
an infrared finite ghost dressing function have been put forward.

Studying unquenching effects on the gluon and ghost propagators
we have found that these effects are small for the ghost propagator, but
are clearly visible for the gluon propagator at intermediate
momenta. Concerning the infrared limit 
of both propagators, the influence of two fermion flavors seems to
become less towards lower momenta. This all agrees with the findings in DSE
studies (\eg \cite{Fischer:2003rp,Fischer:2005wx}).

From our data, we cannot judge without doubt on the existence of an infrared
vanishing gluon propagator.

\subsubsection{The running coupling constant and the ghost-gluon vertex}

In connection with the infrared behavior of the gluon and ghost
dressing functions, we have determined the running coupling constant
based on the ghost-gluon vertex. This coupling constant is found to
match the RG-invariant two-loop expression considering data
at large momenta. However, for $q^2<0.4~\GeV^2$ the running coupling
constant decreases with decreasing momentum. The same is observed
considering data for the unquenched case. Also an influence of Gribov
copies cannot be made responsible for this. Therefore, we  
cannot confirm an infrared fixed point for this coupling constant
as it has been proposed in DSE studies
\cite{vonSmekal:1997is,vonSmekal:1998yu} (see also
\cite{Alkofer:2004it}). This reflects once more the  
different infrared exponents we have found for the gluon and
ghost propagators. 

In any case, our data are in qualitative agreement with recent studies
of DSEs on a torus
\cite{Fischer:2002eq,Fischer:2002hn,Fischer:2005ui,Fischer:2005nf}. There
a similar behavior for the gluon and ghost propagators and for the
running coupling constant at low momentum has been presented. Also
other lattice studies agree with our results for this coupling constant
\cite{Furui:2003jr,Furui:2004cx,Boucaud:2005ce}.

In order to verify that the assumption of a bare ghost-gluon vertex is
valid beyond perturbation theory, we have calculated the corresponding 
renormalization constant $\widetilde{Z}_1$ using a $\MOM$ scheme with
zero gluon momentum. Our data show that in this
scheme $\widetilde{Z}_1$ is approximately equal to one within error bars for
all momenta considered here. Only a slight deviation is visible in the
interval $0.3~\GeV^2 \le q^2 \le 2~\GeV^2$. Unquenching effects are
not resolvable. We thus agree with a recent DSE study
\cite{Schleifenbaum:2004id} where a semiperturbative calculation of
$\widetilde{Z}_1$ has been presented using a $\MOM$ scheme with asymmetric
(the same as used by us) and symmetric subtraction points. Furthermore,
our results are in full agreement with those presented in
\cite{Cucchieri:2004sq} for the case of $SU(2)$.
It is worthwhile to continue the lattice calculation of $\widetilde{Z}_1$
using other renormalization schemes, for example with
a symmetric subtraction point.

Together with our data for the running coupling constant we can thus confirm 
that the product in \Eq{eq:running_coup} is renormalization-group
invariant in the particular $\MOM$ scheme considered here and
defines a nonperturbative running coupling constant which,
as shown here, decreases monotonously with decreasing momentum
$q^2<0.4~\GeV^2$. 

\subsubsection{Results on the confinement criteria}

In addition we have presented results which support the Kugo--Ojima
confinement scenario \cite{Kugo:1979gm} to be realized for lattice QCD
in Landau gauge. To confirm this, one has to show that the function
$u^{ab}(q^2)=u(q^2)\delta^{ab}$ defined via the Green's function in
\Eq{eq:def_u} \cite{Kugo:1995km,Alkofer:2000wg} has the zero-momentum
limit: $u(0)=-1$. Alternatively, if this limit is realized then the
(renormalized) ghost dressing function $J(q^2)$ must diverge in the same
limit. This is because at zero momentum it holds that $1+u(0)=1/J(0)$
\cite{Kugo:1995km}. 

In this thesis we have shown that the ghost dressing function seems to
diverge at zero momentum which is in favor of the Kugo--Ojima
confinement scenario and also satisfies the Gribov-Zwanziger horizon
condition
\cite{Gribov:1977wm,Zwanziger:1992qr,Zwanziger:2001kw}. Moreover, we
have measured the function $u(q^2)$ at  
different momenta $q^2$. To our knowledge a direct calculation
of $u(q^2)$ has never been done before. Conclusions in other studies
\cite{Furui:2003jr,Furui:2004cx,Watson:2001yv,Alkofer:2001iw} 
were drawn on results obtained for the ghost dressing functions only.
For the renormalization of $u(q^2)$ we have   
developed a new ansatz using a minimization process (see
\Sec{sec:lattice_data_for_u} for details). Our renormalized data for
both the ghost dressing function and the function $u(q^2)$ 
are consistent with $u(q^2) - J^{-1}(q^2)$ approaching minus
one in the limit of vanishing momenta, even though this convergence is
very slow. Therefore, as the ghost dressing
function diverges we expect $u(q^2)$ to reach minus one. Hence the
Kugo-Ojima confinement criterion is realized.

Based on our data obtained on larger lattice sizes we are also in the
fortunate position to present 
numerical evidence that not only the quenched but also the unquenched $SU(3)$
gluon propagator in Landau gauge violates reflection positivity
\emph{explicitly}. Thus, on one hand, we agree with 
other lattice studies for quenched $SU(3)$ 
\cite{Mandula:1987rh,Bernard:1992hy,Marenzoni:1993td,Furui:2004cx} and for
three-dimensional $SU(2)$ 
\cite{Cucchieri:2004mf} gauge theory. On the other hand,
we confirm presently available solutions to the corresponding DSE
of the gluon propagator \cite{Alkofer:2003jj}, even though our data
suggest a different infrared exponent for the gluon propagator. 

Since reflection positivity is violated by the gluon
propagator and the Kugo--Ojima confinement criterion seems to be
satisfied, we agree with the conjecture that transverse gluon states
are confined by the quartet mechanism \cite{Kugo:1979gm}.

\subsubsection{Spectral properties of the FP operator}

We have also investigated the spectral properties of the FP
operator and their relation to the ghost propagator in $SU(3)$ Landau
gauge \cite{Sternbeck:2005vs}. As expected from
Ref.~\cite{Zwanziger:2003cf} we have found 
that the low-lying eigenvalues are shifted towards $\lambda=0$
and the eigenvalue density $\rho(\lambda)$ becomes a steeper rising
function as the volume is increased. 

On average, the corresponding FP eigenmodes are not localized,
however, a few large $\IPR$ values have been seen among the lowest eigenmodes.

We could demonstrate that the ghost propagator at low momentum is
dominated by the low-lying eigenvalues and eigenmodes of the FP
operator. For example, the value of the ghost propagator at lowest
momentum (on a $12^4$ lattice at $\beta=6.2$) can be estimated using
a number of 200 low-lying eigenvalues and eigenmodes of the FP
operator. In other words, 
a fraction of about 0.12\% of the whole spectrum is sufficient to
reconstruct the asymptotic result. For larger volumes the
number of necessary eigenmodes seems to be somewhat larger. Also for the
next higher momentum, saturation needs a much bigger part of the
low-lying spectrum.

\subsection*{Concluding remarks}

We hope to have presented a careful numerical study of different
aspects of $SU(3)$ Landau gauge gluodynamics, a subject which has been
the focus of much attention in recent years. We have concentrated on
the low momentum region and have clarified the influence of different
systematic effects on the infrared behavior of gluon and ghost
propagators in Landau gauge. We have shown that the momentum
dependence of both propagators is consistent with different criteria for
confinement. Considering also our results for the running coupling
constant and for the ghost-gluon vertex we agree with the findings of
other lattice studies and of DSE studies on a torus. However, there is some
disagreement with DSE studies in the continuum. At present there seems
to be no solution of this puzzle. 

We think, it is interesting to look in a future study also at the
momentum dependence of other vertex functions and of the quark
propagator in Landau gauge. In particular, for clover-improved
Wilson fermions this has not been done before. It is also
worthwhile to perform a similar 
study using another gauge condition, for example, the Coulomb
gauge. Such investigations might provide further valuable information
towards a full understanding of nonperturbative QCD (see \eg
\cite{Zwanziger:1998ez,Cucchieri:2000gu,Zwanziger:2002sh,Greensite:2003xf}).

In the same way, the larger momentum region should be investigated
further. Note that our study has focused only on the infrared
momentum region. Since computing time was restricted, the amount of
data for the ghost propagator and for the running coupling constant in
the ultraviolet momentum region is not as large as in the low
momentum region. It is worthwhile, however, to continue and to perform
additional measurements at larger momenta considering different
loop-expansions of the corresponding asymptotic form. In particular, an
investigation of nonperturbative power corrections to the ghost and
gluon propagators in Landau gauge (see \eg
\cite{Boucaud:2000nd,Boucaud:2000ey,Boucaud:2001st,Boucaud:2005xn})
due to non-zero values of QCD condensates could be an interesting
topic in future lattice studies.

%===============================================================================
%%% Local Variables: 
%%% mode: latex
%%% TeX-master: "../Sternbeck"
%%% End:

%-- Appendix ----------------------------------------------------------------

\appendix
\titleformat{\chapter}
            [display]
	    {\Huge\scshape\flushright}
	    {\vspace{4ex}\titlerule\vspace{0.5ex}%
              \filleft\LARGE\scshape{\chaptertitlename}%
             \quad\thechapter\filleft}
	    {0.5ex}
            {\titlerule\vspace{3ex}\filleft}[\vspace{-2ex}]
%============================================================================
\chapter{Some details on algorithms and 
performance}

\begin{chapterintro}{W}
  e list  algorithms and numerical libraries used in this
  study for the different purposes and we give references for further
  details. Then we report on some experiences we gained with
  gauge-fixing. In particular, we compare the performance of
  over-relaxation and Fourier-accelerated 
  gauge-fixing. After this, we analyze gauge functional values obtained by
  following our \fc{}-\bc{} strategy and demonstrate that the final
  ranking of functional values is already visible at an intermediate
  iteration step. We show that the inversion of the FP matrix
  can be accelerated by using a pre-conditioned conjugate
  gradient algorithm.
\end{chapterintro}

%----------------------------------------------------------------------------
\section{A note on the algorithms used}
\label{app:algorithms_used}

For the generation of our quenched configurations we employed
the \emph{hybrid over-relaxation} algorithm. It has become standard for the
simulation of pure gauge theory and is a combination of several, say
$N_{ov}$, \emph{micro-canonical over-relaxation} steps
\cite{Brown:1987rr,Creutz:1987xi} and one
\emph{heatbath} \cite{Fabricius:1984wp,Kennedy:1985nu} step. In both
steps a decomposition of $SU(3)$ link variables into $SU(2)$ matrices, 
as proposed by \name{Cabbibo} and \name{Marinari} \cite{Cabibbo:1982zn},
was applied. We have always used $N_{ov}=4$. Further details on these
algorithms can be found, for examples, in the PhD theses
\cite{Knechtli:1999tw,Gehrmann:2002pn}.

All sets of our unquenched $SU(3)$ gauge configurations were provided
to us through the $\QCDSF$ collaboration which generated them using the
\emph{hybrid Monte Carlo} algorithm \cite{Duane,Gottlieb} with
even-odd preconditioning \cite{Gupta}. 

Both the quenched and unquenched configurations were transformed such
that they satisfy the lattice Landau gauge condition (see
\Eq{eq:stop_crit}). For gauge-fixing we have employed two popular algorithms:
\emph{over-relaxation} (RLX) \cite{Mandula:1990vs} and
\emph{Fourier--accelerated gauge-fixing} (FAG) \cite{Davies:1987vs}. For
Fourier--accelerated gauge-fixing we used the implementation as
introduced in \cite{Davies:1987vs}. For future studies, however, we
recommend to use the multigrid implementation of FAG as proposed
in \cite{Cucchieri:1998ew}. This implementation is better suited for a
parallel computing environment with distributed memory (see
\App{sec:fag_vs_rlx}).

Fourier transforms, for example of the gluon fields, were calculated
using an algorithm that performs a Fast-Fourier transformation (FFT). For all
FFTs, we have employed the \code{FFTW}-library \cite{FFTW98}. We can
recommend this library (see also online: \link{http://www.fftw.org}).

For observables which involve the inverse of the FP operator we applied
the pre-conditioned conjugate gradient algorithm to solve the
corresponding linear systems. As pre-conditioning matrix
we used the inverse Laplacian operator $\Delta^{-1}$ with diagonal
color substructure. This significantly has reduced the amount of computing
time as it is discussed in more detail in \App{sect:pcg}.

The eigenvalues and eigenmodes of the FP operator were calculated
using the \code{ARPACK} package, \code{PARPACK} \cite{arpack}.
For minimization of $\chi^2$ functions we used the new \Cpp{} implementation
of the library \code{MINUIT} \cite{MINUIT}. We can recommend both
libraries too.

%----------------------------------------------------------------------------
\section{Experience report on lattice Landau gauge fixing}

In this section we report on some experiences we made with
gauge-fixing. Since efficiency of gauge-fixing algorithms is an 
important issue in studying gauge-dependent quantities, this report
might help in future studies.
\begin{figure}[t]
  \centering
  \includegraphics[width=0.8\textwidth]{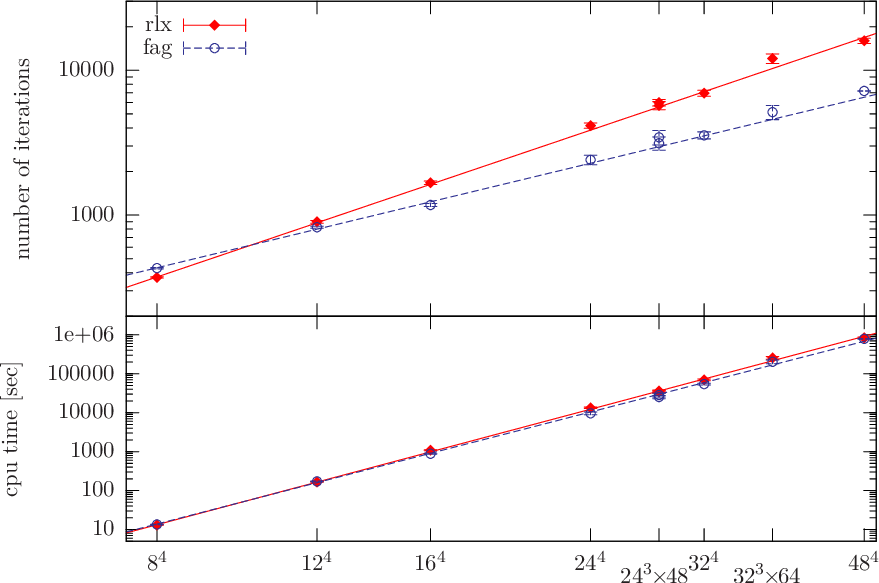}
  \caption{The average number of iterations for both gauge-fixing
    using over-relaxation (rlx) and gauge-fixing using the
    Fourier-accelerated method (fag) is shown in the upper
    panel. The lower panel shows the corresponding CPU time in seconds
    (sum over all processors).}
  \label{fig:rlx_fag_L}
\end{figure}

%----------------------------------------------------------------------------
\subsection{Over-relaxation versus Fourier--accelerated gauge-fixing}
\label{sec:fag_vs_rlx}

For the purpose of this study we have employed two different
algorithms for fixing 
thermalized gauge configurations to Landau gauge. These are
\emph{over-relaxation} (RLX) \cite{Mandula:1990vs} and
\emph{Fourier--accelerated gauge-fixing} (FAG) \cite{Davies:1987vs}.
Both algorithms are commonly used in the lattice community, but
usually either of them is chosen. The reason why we used two is
simple: Initially, we had started with over-relaxation, but later we
checked whether computing time can be saved using FAG instead. Now
this puts us in the fortunate position to provide an explicit
cross-check on the performance of both algorithms. This might be
interesting for future studies, in particular when writing
applications for computing time.

Even though both algorithms are quite different, they are both
iterative in nature. After each iteration-cycle the functional
$F_U[g]$ (\Eq{eq:functional}) is increased a bit until a (local)
maximum is reached. As stopping criterion not the functional itself,
but the violation of transversality (see \Eq{eq:transcondition}) is
used. In our implementation the iteration process stopped as soon as
$\Theta< 10^{-14}$ with
\begin{equation}
 \label{eq:theta_x}
  \Theta := \max_{x}\Re\Tr\left[(\nabla_{\mu} {}^g\!\! A_{x,\mu})(\nabla_{\mu}
  {}^g\!\! A_{x,\mu})^{\dagger}\right]
\end{equation}
was fulfilled at each lattice site $x$. The number of necessary
iterations depends on the lattice sizes, but usually it also varies  
quite strongly for different configurations using the same lattice size. To
get an impression about these variations have a look at
\Fig{fig:func_maxdAdA_iter}. There $\Theta$ is shown versus the
numbers of iterations for two different random gauge 
copies of the same gauge configuration.

\newcommand{\Nit}{N_\textrm{iter}}
\newcommand{\Tit}{T_\textrm{iter}}
In any case, the mean iteration number $\Nit$ and the mean iteration time $\Tit$
(until the final precision is reached) scale with the lattice size
$V=L_S^3\times L_T$ according to 
\begin{subequations}
 \label{eq:Nit_Tit}
 \begin{eqnarray}
   \label{eq:Nit}
   \Nit(V;a,b) &=& a \cdot V^b\;, \\ 
   \label{eq:Tit}
   \Tit(V;c,d) &=& c \cdot V^d\;.
 \end{eqnarray}
\end{subequations}
$L_S$ and $L_T$ denote the number of lattice points in spatial
and temporal direction, respectively. 
That this scaling holds for both algorithms can be seen in
\Fig{fig:rlx_fag_L}. There in the upper (lower) panel we show $\Nit$
($\Tit$) as a function of the lattice size. The parameters from fits
of the Ans\"atze in \Eq{eq:Nit_Tit} to the data are given in
\Tab{tab:rlx_fag_L}. Obviously, FAG performs better on larger lattice
sizes than RLX. 
\begin{table}
  \centering
  \begin{tabular}{c@{\quad}ccc@{\qquad}ccc}
\hline\hline\rule{0pt}{2.5ex}
        & \mc{2}{c}{$\Nit$}& & \mc{2}{c}{$\Tit$} &\\
        &  a       &  b      & $\chi^2/ndf$ & c & d & $\chi^2/ndf$\\
\hline
\rule{0pt}{3ex}
\texttt{rlx} & 4.5(2) & 0.53(1) & 1.6 & $3.2(4)\cdot10^{-5}$ & 1.55(1) &
    5.6\\
   \texttt{fag} &     19(2)    & 0.38(1) & 3.0 & $5.0(9)\cdot10^{-5}$ & 1.51(2) &
    8.1\\
\hline\hline
  \end{tabular}
  \caption{The parameters of the functions $ \texttt{iter}(V;a,b)$
    and $\texttt{cpu}(V;c,d)$ fitted to the average number of iterations and
    average CPU time, respectively.}
  \label{tab:rlx_fag_L}
\end{table}

The reason why in our implementation the actual amount of computing
time for FAG is comparable to that of RLX is just because we have performed 
two Fast-Fourier transformations (FFTs) in each iteration cycle of FAG. Of
course, in a parallel computing environment with distributed memory
FFT routines usually do not scale well if the number of processors
in increased. Therefore, for future studies we rather recommend to use
the multigrid implementation of FAG as proposed by \name{Cucchieri} and
\name{Mendes} \cite{Cucchieri:1998ew}.

%---------------------------------------------------------------------------
\subsection{A way to preselect \textit{best} gauge copies}

\sloppy
\begin{floatingfigure}[r]{7.3cm}
 \centering
  \includegraphics[width=6.5cm]{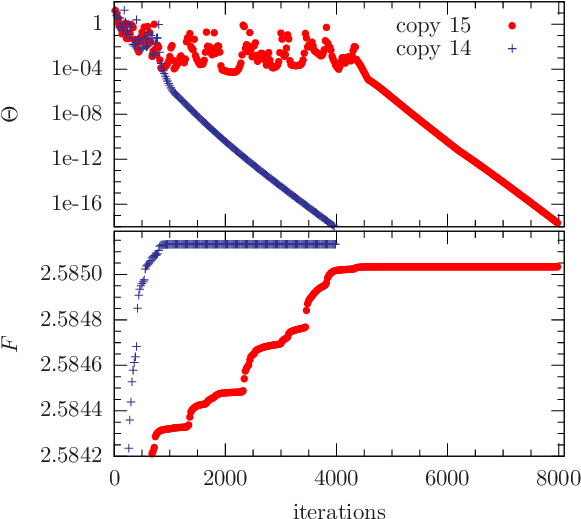}
  \caption{The values of $F$ and $\Theta$
    (see \Eq{eq:functional} and (\ref{eq:theta_x})) and are shown as
    functions of the iteration number for two 
    sample gauge copies of the same configurations
    ($\beta=6.0,24^4$).\medskip}
  \label{fig:func_maxdAdA_iter}
\end{floatingfigure}
We have seen above that fixing gauge configurations to Landau
gauge might become a CPU time intensive task on 
large lattices. This was even more intensive for us when we followed our
\fc{}-\bc{}-strategy (see \Sec{sec:fcbc_strategy}). For this,
several random gauge copies for each gauge
configuration $U$ were gauge-fixed. Then we have selected the 
first gauge copy and that with the largest (final)
functional value (\Eq{eq:functional}) to study the influence of Gribov
copies on different observables. Obviously, it would be quite helpful
if the final ranking of functional values were known, without actually
doing all the necessary iterations.
\begin{floatingfigure}[r]{8.2cm}
  \centering
  \includegraphics[width=8cm]{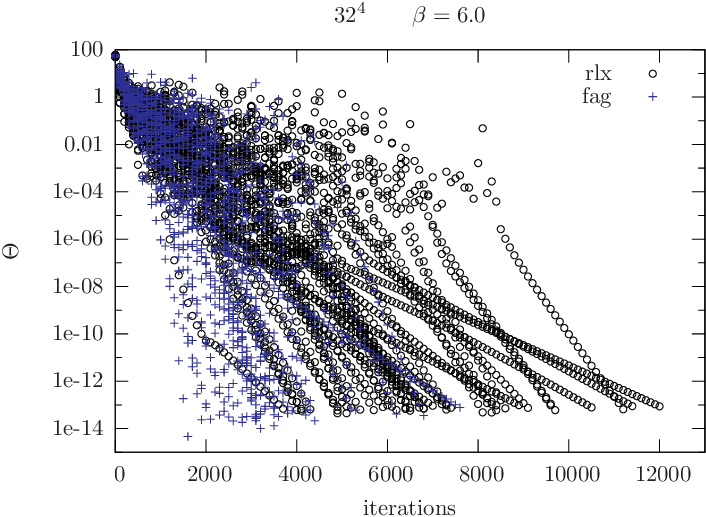}
  \caption{The local maximum of violation of transversality, $\Theta$,
    is shown as a function of the number of iterations for
    over-relaxation (rlx) and  Fourier accelerated gauge-fixing
    (fag). The lattice size is $32^4$ and $\beta=6.0$.\medskip}
  \label{fig:maxdAdA_irlx_ifag}
\end{floatingfigure}

Starting just from a random gauge copy of $U$ we
cannot forecast the final functional value, because those iterative 
gauge-fixing algorithms behave like a chaotic system under change of initial
conditions. For example, if the order of going through 
the lattice is changed, a different functional value might been reached.
Also checking the ranking of functional values after a fixed
number of iterations is not very reasonable, since iteration processes
are quite different as is illustrated in
\Fig{fig:func_maxdAdA_iter}. There the functional values $F$ and the
maximal violation of transversality $\Theta$ (\Eq{eq:theta_x}) are
shown at intermediate 
iteration steps for two sample gauge copies. Although both iteration
processes started from a (different) random gauge copy of the same
configuration, the ways to convergence differ substantial. For one
process, labeled as \texttt{copy 15}, the functional $F$
reveals more often irregular jumps to larger values during the iteration
loop than for the other process. This happens always in
conjunction with sudden increases of $\Theta$. 
However, inspecting \Fig{fig:maxdAdA_irlx_ifag} such
non-monotonous behavior is seen in the majority only for larger values of
$\Theta$. Thus it seems that for $\Theta$ below some threshold the final
ranking of functional values is already fixed. 

From our study we know the final and a list of intermediate functional
values for each gauge-fixed copy of $U$, because we actually have
gauge-fixed each copy until the gauge condition was reached. 
Therefore, we can check now if the
\emph{final} ranking of functional values is already visible at
an intermediate iteration state, \ie at fixed, but larger values of 
$\Theta$. 

\begin{figure}[h]
  \centering
  \includegraphics[width=0.8\textwidth]{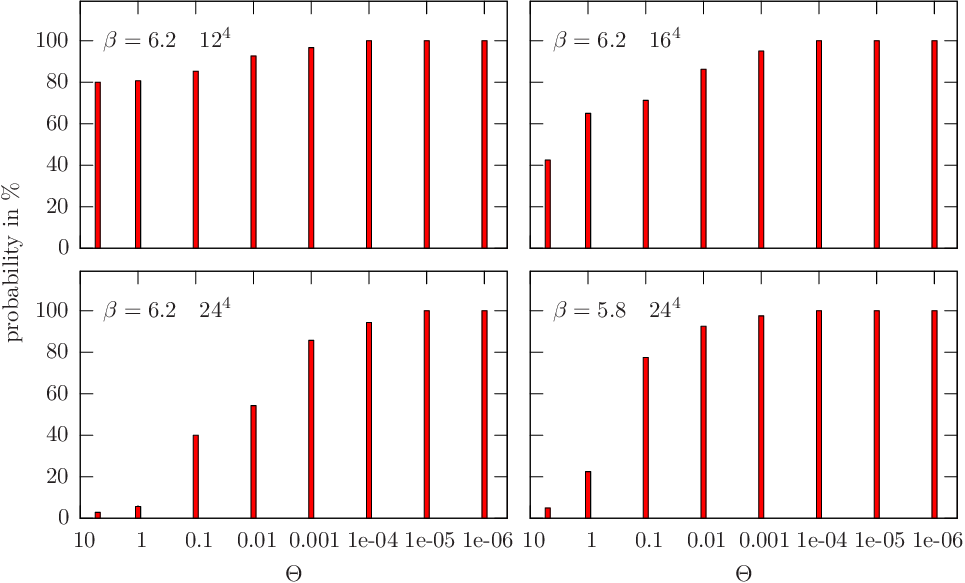}
  \caption{In this figure we show the probability of having found
    (at intermediate values of $\Theta$) those copies which will
    result in the largest functional values after the gauge-fixing
    process has converged.} 
  \label{fig:hist_preselect}
\end{figure}

To be specific, in \Fig{fig:hist_preselect} we show the probability 
of whether that copy with the largest functional value at
an intermediate value of $\Theta$ is that with the largest
functional value (best copy) after gauge-fixing has finished.
Of course, for $\Theta$ below the convergence criteria this holds and 
for $\Theta>1$ (not fixed) it does not.

Looking at \Fig{fig:hist_preselect} we easily see that
the probability of having found those copies which will result in the
largest functional values increases with 
lowering $\Theta$. For $\Theta\approx 10^{-4}$ the
probability is almost 100\%, independent of $\beta$ and the
lattice size. In other words, to select the best gauge-fixed copy for each
configuration $U$, it is enough to gauge-fix several random copies of
$U$ until the value $\Theta=10^{-4}$ has been reached and then continue the
process only for that with the largest intermediate
functional value among all. In this way copies are singled out which
will definitely not reach a larger functional value compared to others.
If we had used this strategy we would have saved about 40\% (57\%)
of the total number of iterations for the $24^4$ lattice at
$\beta=5.8$ (6.2), which is quite a lot.

\section{Speeding up the inversion of the FP operator}
\label{sect:pcg}

For the solution of the linear system $M\vec{\phi}=\vec{\psi}_c$
with symmetric matrix $M$, the conjugate gradient (CG) algorithm is
the method of choice. Its
convergence rate depends on the condition number, the ratio of largest
to lowest eigenvalue of $M$. When all $U_{x,\mu}=\identity$ obviously the
FP operator is minus the Laplacian $\Delta$ with a diagonal color
substructure. Thus instead of solving
$M\vec{\phi}=\vec{\psi}_c$ one rather solves the
transformed system
\begin{displaymath}
  [M\Delta^{-1}]\,(\Delta\vec{\phi}\,)=\vec{\psi}_c
\end{displaymath}
In this way the condition number is reduced, however,
the price to pay is one extra matrix multiplication by
$\Delta^{-1}$ per iteration cycle. In terms of CPU time this should be
more than compensated by the reduction of iterations.

The pre-conditioned CG algorithm (PCG) can be described as follows:
\begin{eqnarray*}
  \textit{initialize:}\hfill \\
  \vec{r}^{\,(0)} &=& \vec{\psi} - M\vec{\phi}^{\,(0)},\quad
  \vec{p}^{\,(0)} = \Delta^{-1}\,\vec{r}^{\,(0)},\\
  \gamma^{(0)} &=& ( \vec{p}^{\,(0)},\vec{r}^{\,(0)})\\*[0.1cm]
  \textit{start do loop:} && k=0,1,\ldots\\
  \vec{z}^{\,(k)} &=&  M\vec{p}^{\,(k)},\quad
  \alpha^{(k)}=\gamma^{\,(k)}/(\vec{z}^{\,(k)},\vec{p}^{\,(k)})\\
  \vec{\phi}^{\,(k+1)} &=& \vec{\phi}^{\,(k)} +
  \alpha^{(k)}\vec{p}^{\,(k)}\\
  \vec{r}^{\,(k+1)} &=& \vec{r}^{\,(k)} -
  \alpha^{(k)}\vec{z}^{\,(k)}\\
  \vec{z}^{\,(k+1)} &=& \Delta^{-1}\,\vec{r}^{\,(k+1)}\\
  \gamma^{(k+1)} &=& ( \vec{z}^{\,(k+1)}, \vec{r}^{\,(k+1)})\\
  &\textit{if}&(\gamma^{(k+1)}<\varepsilon)\quad\textit{exit do loop}\\
  \vec{p}^{\,(k+1)} &=& \vec{z}^{\,(k+1)} +
  \frac{\gamma^{(k+1)}}{\gamma^{(k)}}\vec{p}^{\,(k)}\\
  \textit{end do loop}
\end{eqnarray*}
Here $(\cdot,\cdot)$ denotes the scalar product.

\begin{table}[t]
  \centering
  \begin{tabular}{c@{\quad}cc@{\quad}cc@{\quad}cc}
    \hline\hline\rule{0pt}{3ex}
            & \mc{2}{l}{CG}  & \mc{2}{l}{PCG} & \mc{2}{l}{speed up}\\
    lattice & iter & CPU[sec] & iter & CPU[sec] & iter & CPU[sec]\\
    \hline
\rule{0pt}{3ex}
    $8^4$   &   1400 &  3.7   & 570  & 2.4  & 60\%  & 35\% \\
    $16^4$  &   3900 & 240    & 1050 & 130  & 73\%  & 46\% \\
    $32^4$  &   9900 & 13400  & 2250 & 3900 & 77\%  & 71\% \\
    \hline\hline
  \end{tabular}
  \caption{The average number of iterations and CPU time per
            processor (PE) using the CG and PCG algorithm to invert
            the FP operator are given for different lattice sizes. All
            inversions have been 
            performed at $\beta=5.8$ with source
            $\delta^{bc} \exp (i\,k\!\cdot\! y)$ where
            $k=(1,0,0,0)$. To compare the different
            lattice sizes 4 PEs have always been used.}
  \label{tab:speedup}
\end{table}

To perform the additional matrix multiplication with $\Delta^{-1}$ we
used two fast Fourier transformations $\mathcal{F}$, due to
\mbox{$(-\Delta)^{-1}=\mathcal{F}^{-1}\,q^{-2}(k)\,\mathcal{F}$}.
The performance we achieved is presented in \Tab{tab:speedup}. We conclude
that on larger lattice sizes the reduction of
iterations is about 70-75\%, while the resulting reduction of CPU time depends
on the lattice size. This is because we are using the fast Fourier
transformations in a parallel CPU environment. If the ratio
of used processors to the lattice size is small (see \eg
the data for $32^4$ lattice at this table),
almost the same reductions of CPU time as for the number of iterations
is achieved.

Further improvement may be achieved by using the multigrid Poisson solver to
solve \mbox{$\Delta\vec{z}^{\,(k)}=\vec{r}^{\,(k)}$}. This method
is supposed to perform better on parallel machines. Perhaps a further
improvement is possible by using as pre-conditioning matrix
\mbox{$\widetilde{M}^{-1}=-\Delta^{-1}-\Delta^{-1}M_1\Delta^{-1} + \ldots$}
which is an approximation of the FP operator \mbox{$M=-\Delta+M_1$} to a
given order \cite{Zwanziger:1993dh} (see also
\cite{Furui:2003jr}). However, the larger the
order, the more matrix multiplications per iteration cycle are
required. This may reduce the overall performance.
We have not checked so far which is the optimal order.

%===============================================================================
%%% Local Variables: 
%%% mode: latex
%%% TeX-master: "Sternbeck"
%%% End:

%-Literaturverzeichnis--------------------------------------------

\cleardoublepage
\titleformat{\chapter}
            [display]
	    {\Huge\scshape\flushright}
	    {\vspace{4ex}\filleft\MakeUppercase{\chaptertitlename}%
             \Huge\thechapter}
	    {4ex}
            {\vspace{3ex}\titlerule\vspace{0.5ex}\filleft}[\vspace{0.5ex}%
             \titlerule]

\addcontentsline{toc}{chapter}{Bibliography}
\nocite{*}

\bibliography{bibliography}
\bibliographystyle{alphadidiEN}
\cleardoublepage

%-Danksagung*-----------------------------------------------------
\pagestyle{plain}
% acknowledgement
%-----------------------------------------------------------------------
\chapter*{Acknowledgments}
\addcontentsline{toc}{chapter}{Acknowledgments}

First and foremost, I would like to thank my Ph.D.~advisor Michael
M\"uller-Preu\ss{}ker for the opportunity to work on this subject and
the good support over the last years. The support from him stems from
the time I~began studying for my degree at the Humboldt-University Berlin.
Also, I would like to thank him for the possibility to extend my stay
within the PHA group half a year longer than planned before and for
reading the manuscript. 

Furthermore, I owe deep gratitude to Ernst--Michael Ilgenfritz for
many helpful discussions and for careful reading several drafts
of this thesis. His questioning mind contributed greatly to this project.

I also thank Giuseppe Burgio, Hilmar Forkel and Jan Volkholz for
enlightening discussions and for proof reading parts of the manuscript.
They delivered many useful comments and improving
remarks. Furthermore, I would like to thank my former and present
office mates Dirk Peschka and Peter Schemel for discussions and
inspiring coffee breaks. It was also always a pleasure for me to
have discussions with Dietmar Ebert all the years at
Humboldt-University. 

Also I am indebted to Hinnerk Stüben for contributing parts of the program
code. Due to his expert knowledge I have learned a lot about efficient
programming for high performance computing. This has enabled me to develop the
simulation program for this study.

I also acknowledge valuable discussions and correspondence with Reinhard
Alkofer, Christian Fischer, Valentin K.~Mitrjushkin, Arwed Schiller and
Lorenz von Smekal.  

\medskip

I wish to send a special thank to Anett for all her love she gave me
during the years and her support while writing this thesis. I thank
Nikola and wish the best for her degree.

I would also like to thank the Magisters for providing me
Internet access at home all the time, in particular on weekends. This
helped me to keep the simulations running all the time.

\bigskip

This work was supported by the DFG-funded graduate school GK 271 and
by the SFB/Tr9. All simulations have been done on the IBM pSeries 690 at HLRN. 
I used configurations generated by the $\QCDSF$ collaboration which I
could access in the framework of the I3 Hadron-Physics initiative (EU
contract RII3-CT-2004-506078). I thank Gerrit Schierholz, Dirk Pleiter
and Stefan Wollny for their help.

\cleardoublepage

%-----------------------------------------------------------------
\end{document}